\documentclass[12pt]{book}

\usepackage{pdfpages}
\usepackage[french]{babel}
\usepackage{graphicx}
\usepackage{epsf,amsfonts,hyperref}
\renewcommand{\appendix}[1]{
    \addtocounter{section}{1}
    \setcounter{equation}{0}
    \renewcommand{\thesection}{\Alph{section}}
    \section*{Appendix \thesection\protect\indent #1}
    \addcontentsline{toc}{section}{Appendix \thesection\ \ \ #1}
}
\newcommand\encadremath[1]{\vbox{\hrule\hbox{\vrule\kern8pt
\vbox{\kern8pt \hbox{$\displaystyle #1$}\kern8pt}
\kern8pt\vrule}\hrule}}
\def\enca#1{\vbox{\hrule\hbox{
\vrule\kern8pt\vbox{\kern8pt \hbox{$\displaystyle #1$}
\kern8pt} \kern8pt\vrule}\hrule}}

\newcommand\figureframex[3]{
\begin{figure}[bth]
\hrule\hbox{\vrule\kern8pt
\vbox{\kern8pt \vbox{
\begin{center}
{\mbox{\epsfxsize=#1.truecm\epsfbox{#2}}}
\end{center}
\caption{#3}
}\kern8pt}
\kern8pt\vrule}\hrule
\end{figure}
}
\newcommand\figureframey[3]{
\begin{figure}[bth]
\hrule\hbox{\vrule\kern8pt
\vbox{\kern8pt \vbox{
\begin{center}
{\mbox{\epsfysize=#1.truecm\epsfbox{#2}}}
\end{center}
\caption{#3}
}\kern8pt}
\kern8pt\vrule}\hrule
\end{figure}
}

\renewcommand{\thesection}{\arabic{section}}

\makeatletter
\@addtoreset{equation}{section}
\makeatother
\newtheorem{theorem}{Th\'eor\`{e}me}[section]

\newtheorem{remark}{Remarque}[section]
\newtheorem{proposition}{Proposition}[section]
\newtheorem{exemple}{Exemple}[section]
\newtheorem{lemma}{Lemme}[section]
\newtheorem{corollary}{Corollaire}[section]
\newtheorem{definition}{Definition}[section]
\def\br{\begin{remark}\rm\small}
\def\er{\end{remark}}
\def\bex{\begin{exemple} \small}
\def\eex{\end{exemple}}
\def\bt{\begin{theorem}}
\def\et{\end{theorem}}
\def\bd{\begin{definition}}
\def\ed{\end{definition}}
\def\bp{\begin{proposition}}
\def\ep{\end{proposition}}
\def\bl{\begin{lemma}}
\def\el{\end{lemma}}
\def\bc{\begin{corollary}}
\def\ec{\end{corollary}}
\def\beaq{\begin{eqnarray}}
\def\eeaq{\end{eqnarray}}
\newcommand{\proof}[1]{{\noindent \bf Preuve:}\par
{#1} $\square$}

\newcommand{\CA}{{\cal A}}
\newcommand{\CB}{{\cal B}}
\newcommand{\CC}{{\cal C}}
\newcommand{\CE}{{\cal E}}
\newcommand{\CF}{{\cal F}}
\newcommand{\CG}{{\cal G}}

\newcommand{\CM}{{\cal M}}
\newcommand{\CN}{{\cal N}}
\newcommand{\CO}{{\cal O}}
\newcommand{\CP}{{\cal P}}

\newcommand{\CS}{{\cal S}}

\newcommand{\CW}{{\cal W}}

\newcommand{\CZ}{{\cal Z}}

\newcommand{\eq}[1]{Eq.~(\ref{#1})}

\newcommand{\beq}{\begin{equation}}
\newcommand{\eeq}{\end{equation}}

\newcommand{\bea}{\begin{eqnarray}}
\newcommand{\eea}{\end{eqnarray}}

\newcommand\eol{\hspace*{\fill}\linebreak}
\newcommand\eop{\vspace*{\fill}\pagebreak}

\newcommand{\vs}{\vspace{0.7cm}}
\renewcommand{\and}{{\qquad {\rm and} \qquad}}

\newcommand{\virg}{{\qquad , \qquad}}

 \newcommand{\Tr}{{\,\rm Tr}\:}
\newcommand{\tr}{{\,\rm tr}\:}

\newcommand{\Res}{\mathop{\,\rm Res\,}}

\newcommand{\ri}{{\rm i}}

\newcommand{\td}[1]{{\tilde{#1}}}

\renewcommand{\l}{\lambda}
\newcommand{\om}{\omega}

\newcommand{\ee}[1]{{{\rm e}^{#1}}}

\renewcommand{\d}{{{\partial}}}

\newcommand{\Pint}{{\int\kern -1.em -\kern-.25em}}

\newcommand{\moy}[1]{\left<{#1}\right>}

\renewcommand{\Im}{{\mathrm{Im}}}

\renewcommand{\l}{\lambda}
\renewcommand{\L}{\Lambda}

\newcommand{\ovl}{\overline}
\newcommand{\bfx}{{\mathbf x}}

\newcommand{\acycle}{{\cal A}}
\newcommand{\bcycle}{{\cal B}}

\newcommand{\curve}{{\cal E}}

\newcommand{\pbar}{\ovl{p}}
\newcommand{\qbar}{\ovl{q}}

\newcommand{\bfa}{{\mathbf{a}}}

\newcommand{\primef}{{E}}

\newcommand{\Bergmann}{{\underline{B}}}

\newcommand{\EOBethe}{{\bf [I]}}
\newcommand{\eyno}{{\bf [II]}}
\newcommand{\CEO}{{\bf [III]}}
\newcommand{\EMO}{{\bf [V]}}
\newcommand{\EOSym}{{\bf [VI]}}
\newcommand{\EOinvariants}{{\bf [IV]}}
\newcommand{\EOvolume}{{\bf [VII]}}

\begin{document}

\pagestyle{empty}
\thispagestyle{empty}

\begin{center}
Universit\'e Paris 6 -- Pierre et Marie Curie

et

Service de Physique Th\'eorique, Commissariat \`a l'\'Energie Atomique

\vspace{1.5cm}

{\bf \large TH\`ESE DE DOCTORAT DE L'UNIVERSIT\'E PARIS 6}

\vspace{1cm}

Sp\'ecialit\'e: {\bf Physique th\'eorique}

\vspace{1cm}

pr\'esent\'ee par {\bf Nicolas ORANTIN}

\vspace{1cm}

pour obtenir le grade de docteur de l'Universit\'e Paris 6

\vspace{2cm}

{\Large \bf Du d\'eveloppement topologique des mod\`eles de matrices \`a la th\'eorie des cordes topologiques:
combinatoire de surfaces par la g\'eom\'etrie alg\'ebrique.}

\vspace{6cm}

\end{center}

Soutenue le 13 Septembre 2007 devant le jury compos\'e de

\vs

M. AKEMANN Gernot, rapporteur

M. BR\'EZIN \'Edouard, pr\'esident

M. EYNARD Bertrand, directeur

M. KONTSEVICH Maxim

M. KOSTOV Ivan

M. NEKRASOV Nikita, rapporteur




\vspace{26pt}
\pagestyle{headings}
\setcounter{page}{1}


\newpage

\thispagestyle{empty}

\null

\newpage

{\bf \huge R\'esum\'e.}
\vs

Le mod\`ele \`a deux matrices a \'et\'e introduit pour \'etudier le mod\`ele d'Ising sur surface al\'eatoire. Depuis,
le lien entre les mod\`eles de matrices et la combinatoire de surfaces discr\'etis\'ees s'est beaucoup d\'evelopp\'e.
Cette th\`ese a pour propos d'approfondir ces liens et de les \'etendre au del\`a des mod\`eles de matrices
en suivant l'\'evolution de mes travaux de recherche.
Tout d'abord, je m'attache \`a d\'efinir rigoureusement le mod\`ele \`a deux matrices hermitiennes formel
donnant acc\`es aux fonctions g\'en\'eratrices de surfaces discr\'etis\'ees portant une structure de spin. Je montre
alors comment calculer, par des m\'ethodes de g\'eom\'etrie alg\'ebrique, tous les termes du d\'eveloppement topologique des observables comme
formes diff\'erentielles d\'efinies sur une courbe alg\'ebrique associ\'ee au mod\`ele: la courbe spectrale.
Dans un second temps, je montre comment, imitant la construction du mod\`ele \`a deux matrices,
on peut d\'efinir de telles formes diff\'erentielles sur n'importe quelle courbe alg\'ebrique poss\'edant de nombreuses
propri\'et\'es d'invariance sous les d\'eformations de la courbe alg\'ebrique consid\'er\'ee.
En particulier, on peut montrer que si cette courbe est la courbe spectrale d'un mod\`ele de matrices, ces
invariants reconstituent les termes des d\'eveloppements topologiques des observables du mod\`ele. Finalement,
je montre que pour un choix particulier des param\`etres, ces objets peuvent \^etre rendus invariants modulaires
et sont solutions des \'equations d'anomalie holomorphe de la th\'eorie de Kodaira-Spencer donnant un nouvel
\'el\'ement vers la preuve de la conjecture de Dijkgraaf-Vafa.

{\em Mots cl\'es:} matrices al\'eatoires, combinatoire, th\'eorie des cordes, espaces des modules, int\'egrabilit\'e, g\'eom\'etrie alg\'ebrique

\vs

{\bf \huge Summary}
\vs

The 2-matrix model has been introduced to study Ising model on random surfaces. Since then, the link between matrix models
and combinatorics of discrete surfaces has strongly tightened. This manuscript aims to investigate these deep links
and extend them beyond the matrix models, following my work's evolution. First, I take care to define properly
the hermitian 2 matrix model which gives rise to generating functions of discrete surfaces equipped with a spin structure.
Then, I show how to compute all the terms in the topological expansion of any observable by using algebraic geometry tools.
They are obtained as differential forms on an algebraic curve associated to the model: the spectral curve. In a second part,
I show how to define such differentials on any algebraic curve even if it does not come from a matrix model. I then study
their numerous symmetry properties under deformations of the algebraic curve. In particular, I show that these objects
coincide with the topological expansion of the observable of a matrix model if the algebraic curve is the spectral curve of this
model. Finally, I show that fine tuning the parameters ensure that these objects can be promoted to modular invariants and
satisfy the holomorphic anomaly equation of the Kodaira-Spencer theory. This gives a new hint that the
Dijkgraaf-Vafa conjecture is correct.


\newpage

\thispagestyle{empty}

\null

\newpage

{\huge \bf Remerciements}

\vs

Je tiens avant toutes choses \`a remercier mon directeur de th\`ese Bertrand Eynard.
Il a pass\'e \'enorm\'ement de temps \`a me transmettre une partie de ses connaissances sur ce tr\`es vaste domaine que repr\'esentent
les matrices al\'eatoires tout au long de l'\'evolution de mes travaux. J'ai \'egalement pu profiter de sa large connaissance du sujet, de son impressionnante intuition et de sa vision globale des
applications possibles \`a travers les sujets qu'il m'a propos\'es: tous tr\`es riches, passionnants et pleins de perspectives. Je tiens aussi \`a le
remercier pour sa disponibilit\'e, sa gentillesse et sa patience.
Je suis \'egalement tr\`es reconnaissant envers
Kirone Mallick pour ses nombreux conseils depuis que j'ai eu le plaisir de travailler avec lui il y a maintenant 4 ans. Il
a toujours \'et\'e l\`a pour me guider lorsque j'ai eu des choix \`a faire.

\vs

Je remercie \'Edouard Br\'ezin et Maxim Kontsevich d'avoir accept\'e de faire partie de mon jury et tout particuli\`erement
Nikita Nekrasov et Gernot Akemann qui ont accept\'e la lourde t\^ache de rapporter mon manuscript ainsi
que Ivan Kostov avec qui j'ai eu la chance de pouvoir interagir tr\`es fr\'equemment.

\vs

Merci \'egalement \`a Michel Berg\`ere, Philippe Di Francesco et
Aleix Prats-Ferrer d'avoir accept\'e de relire des versions pr\'eliminaires de ma th\`ese. J'ai aussi pris beaucoup de plaisir \`a discuter
r\'eguli\`erement avec eux et les nombreuses questions de Michel m'ont souvent permis de clarifier mes id\'ees.
Je voudrais \'egalement remercier mes collaborateurs Leonid Chekhov et Marcos Mari\~no qui m'ont aussi beaucoup
appris dans leurs sujets respectifs et se sont toujours montr\'es disponibles pour r\'epondre \`a mes questions.

\vs

Merci enfin au SPhT, qui m'a permis d'effectuer cette th\`ese dans un cadre toujours agr\'eable et stimulant sans
avoir \`a me pr\'eoccuper d'autre chose que mes travaux de recherche, ainsi qu'\`a tous ses membres et plus particuli\`erement \`a
J\'er\'emie Bouttier, Fran\c{c}ois David, Vincent Pasquier, Pierre Vanhove et
Jean-Bernard Zuber que j'ai r\'eguli\`erement d\'erang\'es par mes questions
ainsi qu'au Niels Bohr Institute qui m'a h\'eberg\'e pendant deux mois durant lesquels j'ai eu le plaisir
de pouvoir apprendre aupr\`es de Charlotte Kristjansen.

\newpage

\thispagestyle{empty}

\null

\newpage

\section*{Avant propos et plan.}

Les mod\`eles de matrices al\'eatoires, depuis leur apparition dans le monde de la physique en 1951, n'ont eu de cesse
de trouver des applications dans des domaines tr\`es vari\'es des sciences parmi lesquels on peut citer quelques \'exemples en vrac:
la physique nucl\'eaire, la physique statistique, l'\'etude des syst\`emes chaotiques, la th\'eorie des cordes,
les th\'eories conformes,
la combinatoire, la th\'eorie des nombres, la th\'eorie des noeuds, l'int\'egrabilit\'e, l'\'etude des
probl\`emes de Riemann Hilbert ...

La richesse et la vari\'et\'e des domaines li\'es au matrices al\'eatoires en font un objet d'\'etude passionnant et enrichissant,
particuli\`erement pour un \'etudiant, tant les perspectives de recherche sont nombreuses. Au cours de ma th\`ese j'ai ainsi
p\^u toucher \`a des probl\`emes tr\`es diff\'erents les uns de autres et cotoyer des gens d'horizons tr\`es divers.
Ce m\'emoire n'a pas pour but de cataloguer l'\'etat de l'art dans l'ensemble des domaines o\`u les mod\`eles de matrices
ont apport\'e une contribution, le volume n\'ecessaire \`a un tel expos\'e \'etant bien trop grand
et mes connaissances trop limit\'ees.

Je me restreindrai donc \`a un domaine que j'ai plus particuli\`erement \'etudi\'e et sur lequel ont port\'e mes travaux au cours
de cette th\`ese: le lien entre les mod\`eles de matrices hermitiennes et la combinatoire de surfaces. J'essaierai de
montrer que ces deux probl\`emes sont profond\'ement reli\'es et combien la structure sous-jacente est riche, m\'elangeant
g\'eom\'etrie alg\'ebrique et in\'egrabilit\'e. J'ai essay\'e de rendre ce m\'emoire aussi accessible que possible,
quel que soit le domaine d'expertise du lecteur, qu'il soit math\'ematicien ou physicien, et j'esp\`ere avoir r\'eussi
\`a faire un expos\'e ne n\'ecessitant que des connaissances \'el\'ementaires de physique et math\'ematiques. Je suis en
effet tr\`es heureux que ce domaine puisse donner naissance \`a des discussions entre math\'ematiciens et physiciens et
j'esp\`ere donc que ce m\'emoire puisse \^etre lisible par les diff\'erentes communaut\'es li\'ees, de pr\`es ou de loin, \`a
ce vaste sujet.

Les r\'esultats que j'ai obtenu au cours de ma th\`ese ont tous \'et\'e le fruit de travaux avec mon directeur de th\`ese,
B. Eynard. Ils d\'ecoulent pour la plupart d'un r\'esultat qu'il a obtenu dans le cadre du mod\`ele \`a une matrice hermitienne
au moment o\`u j'ai entam\'e ma th\`ese sous sa direction: il a montr\'e dans \cite{eynloop1mat}, comment calculer tout le
d\'eveloppement topologique des fonctions de corr\'elations de ce mod\`ele gr\^ace \`a des outils de g\'eom\'etrie alg\'ebrique
et \`a une repr\'esentation diagrammatique \'el\'egante du r\'esultat. Mes travaux ont essentiellement consist\'e en l'exploration
de certaines des nombreuses perspectives ouvertes par cette construction.

Cette th\'ese suit l'\'evolution de mes travaux au cours de ces trois derni\`eres  ann\'ees:
\begin{itemize}

\item Le chapitre 1 consiste en une courte introduction sur les mod\`eles de matrices pr\'esentant un historique
et quelques applications motivant les travaux pr\'esent\'es dans la suite;

\item Le chapitre 2 est enti\`erement consacr\'e au sujet principal de ma th\`ese: le mod\`ele \`a deux matrices hermitiennes.
Apr\`es avoir d\'efini sans ambiguit\'e le mod\`ele et les observables \'etudi\'ees, je montre comment, en g\'en\'eralisant
la m\'ethode de \cite{eynloop1mat}, on peut calculer le d\'eveloppement topologique complet de toutes
les observables en termes de la courbe spectrale classique. Ce chapitre est bas\'e sur mes travaux avec Chekhov et Eynard  \eyno\CEO\EOBethe\cite{EOmix} .

\item Le chapitre 3 montre comment cette m\^eme m\'ethode permet de r\'esoudre le mod\`ele \`a une matrice ainsi que le mod\`ele
\`a une matrice en champ ext\'erieur. La d\'erivation des r\'esultats du mod\`ele de matrice en champ ext\'erieur est bas\'e
sur mon travail avec Eynard \EOinvariants.

\item Dans le chapitre 4, je pr\'esente une g\'en\'eralisation de cette construction au dela des mod\`eles de matrices.
Je montre comment, partant d'une courbe alg\'ebrique quelconque, on peut construire une famille infinie de formes diff\'erentielles
imitant les r\'esultats obtenus dans les deux chapitres pr\'ec\'edents. Lorsque la courbe consid\'er\'ee est la courbe
spectrale d'un mod\`ele de matrices, ces formes diff\'erentielles coincident avec les termes du d\'eveloppement topologique
des observables de ce mod\`ele. J'\'etudie \'egalement le comportement de ces formes diff\'erentielles sous les d\'efor- mations
de la courbe ainsi que les propri\'et\'es de sym\'etrie induites et comment ces propri\'et\'es permettent de comparer diff\'erents
mod\`eles et de retrouver facilement des r\'esultats classiques dans l'\'etude d'int\'egrales matricielles. Ces r\'esultats
ont \'et\'e obtenus avec Eynard dans \EOinvariants \EOSym.

\item Apr\`es une tr\`es br\`eve introduction entre le lien entre mod\`eles de matrices et th\'eorie des cordes, je montre
que les objets introduits dans le chapitre pr\'ec\'edent satisfont les \'equations d'anomalie holomorphe pour les cordes ferm\'ees de BCOV \cite{BCOV}
et je propose des \'equations pour les cordes ouvertes satisfaites par les fonctions de corr\'elation d\'efinies pr\'ec\'edemment.
Ces r\'esultats viennent d'une collaboration avec Eynard et Mari\~no \EMO.

\item Le chapitre 6 est une conclusion o\`u je r\'esume les principaux r\'esultats pr\'esent\'es dans cette th\`ese et certaines
perspectives ouvertes par ces travaux.

\end{itemize}

Ces chapitres correspondant au coeur de la th\`ese sont suivis d'un chapitre comportant quelques appendices techniques et
d'une partie o\`u j'ai regroup\'e mes articles. Ces deux derniers chapitres permettent, entre autres choses, de regrouper toutes
les d\'emonstrations n\'ecessaires aux r\'esultats pr\'esent\'es ici. Ainsi, j'ai pr\'esent\'e les r\'eultats sans d\'emonstration
dans le corps du texte. Lorsque cel\`a peut aider \`a une meilleure compr\'ehension du r\'esultat, j'ai cependant parfois
pr\'esent\'e une id\'ee des d\'emonstrations.

\eop

{\bf \large Liste d'articles utilis\'es pour ce m\'emoire.}

\vs

\EOBethe  B.Eynard, N.Orantin, ``Mixed correlation functions in the 2-matrix model, and the Bethe Ansatz'',
{\bf JHEP} 0508 028 (2005), math-ph/0504058.

\vs

\eyno  B.Eynard, N.Orantin,
``Topological expansion of the 2-matrix model correlation functions: diagrammatic rules for a residue formula'',
{\em J. High Energy Phys.} {\bf JHEP12}(2005)034, math-ph/0504058.

\vs

\CEO  L.Chekhov, B.Eynard and N.Orantin,
``Free energy topological expansion for the 2-matrix model'',
{\em J. High Energy Phys.} {\bf JHEP12} (2006) 053, math-ph/0603003.

\vs

\EOinvariants  B.Eynard, N.Orantin, ``Invariants of algebraic curves and topological expansion'',
\`a paraitre dans {\em Communication in Number Theory and Physics}, volume 1, n° 2, math-ph/0702045.

\vs

\EMO  B.Eynard, M.Mari\~{n}o, N.Orantin,
``Holomorphic anomaly and matrix models'', {\em J. High Energy Phys.} {\bf JHEP06} (2007) 058, hep-th/0702110.

\vs

\EOSym  B.Eynard, N.Orantin,
``Topological expansion of mixed correlations in the hermitian 2 Matrix Model and x-y symmetry of the $F_g$ invariants'',
soumis \`a {\em JHEP}, arXiv:0705.0958 .

\vs

\EOvolume  B.Eynard, N.Orantin, ``Weil-Petersson volume of moduli space, Mirzakhani's recursion and matrix models'',
arXiv:0705.3600.

\newpage

\tableofcontents


\chapter{Introduction.}

Avant toute intrusion dans le monde des matrices al\'eatoires, il semble indispensable de rappeler leur surprenante
apparition dans le monde de la physique ainsi que leurs si nombreuses et non moins surprenantes applications par la suite.
La premi\`ere partie de ce chapitre a donc pour but de rappeler l'origine des mod\`eles de matrices al\'eatoires en g\'en\'eral
et des objets \'etudi\'es dans la suite plus particuli\`erement. Dans la seconde partie de ce chapitre, nous pr\'esenterons
divers probl\`emes math\'ematiques et th\'eories physiques li\'es aux mod\`eles de matrices, que ce soit des applications directes
ou bien simplement des probl\`emes comportant une structure similaire. Nous nous attacherons particuli\`erement \`a montrer
les points communs \`a ces diff\'erentes th\'eories.

\section{Petit historique des mod\`eles de matrices.}

\subsection{Matrices al\'eatoires et physique nucl\'eaire.}

Les matrices al\'eatoires sont arriv\'ees en physique presque par hasard. En effet, au cours des ann\'ees 50, les physiciens
nucl\'eaires tentent d'\'etudier des noyaux de plus en plus gros, c'est-\`a-dire de d\'eterminer le spectre de l'Hamiltonien
caract\'erisant chaque noyau. La complexit\'e du probl\`eme et la taille de l'op\'erateur \`a diagonaliser croissant avec la taille du noyau,
ils se sont finalement trouv\'es face \`a un probl\`eme qu'ils ne pouvaient pas r\'esoudre ni m\^eme d\'ecrire th\'eoriquement.
Ils \'etaient cependant capables d'avoir acc\`es \`a ce spectre exp\'erimentalement en \'etudiant la diffusion de neutrons lanc\'es sur ce noyau.
Ainsi l'exp\'erience donnait acc\`es \`a la r\'epartition des niveaux d'\'energie, la mani\`ere dont ils sont corr\'el\'es
entre eux, leur densit\'e $\rho(E)$ (i.e. le nombre moyen de niveaux dans un intervalle d'\'energie)...

L'\'etude de ces r\'esultats exp\'erimentaux a fait appara\^itre une structure tr\`es robuste et universelle de la
r\'epartition statistique des niveaux d'\'energie ne d\'ependant pas directement du noyau \'etudi\'e ni de la r\'egion du spectre observ\'ee (voir \cite{Mehta}
et les nombreuses r\'ef\'erences \`a l'int\'erieur).
En effet, si les niveaux avaient \'et\'e d\'ecorr\'el\'es, on se serait par exemple attendu \`a trouver une distribution de
Poisson qui ne fut pas du tout observ\'ee. Au contraire, la distribution observ\'ee a \'et\'e retrouv\'ee par Wigner
en partant d'une statistique bien diff\'erente. Puisque l'Hamiltonien d'un noyau lourd est trop complexe pour \^etre
d\'ecrit explicitement, Wigner a propos\'e de le mod\'eliser par une matrice de taille $N\times N$ dont les coefficients
sont des variables al\'eatoires d\'ecor\'el\'ees suivant chacune une loi gaussienne adoptant ainsi une approche statistique.
Lorsque la taille de la matrice $N$
tend vers l'infini, il a observ\'e que le spectre  devient (quasi-)continu sur un intervalle $[a,b]$ et que la densit\'e
d'\'energie suit la c\'el\`ebre loi du "demi-cercle" de Wigner \cite{Wigner}:
\beq
\rho(E) = {4 \over \pi (b-a)} \sin \phi \;\;\; \hbox{avec} \;\;\; E = {a+b\over 2} + {b-a\over 2} \cos \phi.
\eeq
Or, les corr\'elations issues de cette densit\'e correspondent exactement aux observations exp\'erimentales obtenues par diffusion de neutrons lents.
L'existence d'une approche statistique de ce probl\`eme n'est pas due au hasard mais bien aux propri\'et\'es
particuli\`eres de ce type de diffusion. En effet, les neutrons bombardant le noyau \`a \'etudier apportent ici chacun une
faible \'energie. Il faut alors en accumuler beaucoup pour pouvoir arracher un neutron au noyau. L'\'energie individuelle
des neutrons incidents est donc totalement absorb\'ee par le noyau au cours de cette r\'eaction avant d'\^etre restitu\'ee \`a travers un neutron
diffus\'e. L'\'etude des neutrons diffus\'es ne doit donc plus
caract\'eriser les comportements individuels des neutrons incidents mais leur comportement collectif impliquant un caract\`ere
statistique.

Cette approche statistique peut s'appliquer \`a d'autres probl\`emes physiques li\'es \`a un Hamiltonien trop compliqu\'e pour \^etre d\'ecrit exactement. On a alors pu observer que les
caract\'eristiques g\'en\'erales du spectre ne d\'ependent pas des d\'etails du probl\`eme \'etudi\'e mais simplement de
ses sym\'etries, chaque cas correspondant \`a l'int\'egration sur un ensemble de matrices particulier (\cite{Mehta} est une nouvelle
fois une source formidable d'informations sur ce point). En effet, l'\'etude des
matrices al\'eatoires elles-m\^emes a permis d'observer des propri\'et\'es d'universalit\'es: lorsque la taille des matrices tend
vers l'infini, on peut trouver certaines caract\'eristques du spectre (par exemple l'espacement entre deux valeurs propres
successives) qui ne d\'ependent absolument pas des d\'etails de la mesure d'int\'egration mais seulement de l'ensemble des matrices
sur lequel on int\'egre.

Il peut donc \^etre int\'eressant d'\'etudier les matrices al\'eatoires en posant une mesure sur l'ensemble de matrices pr\'esentant des
caract\'eristiques communes plut\^ot que de les voir comme un ensemble de coefficients al\'eatoires r\'eels corr\'el\'es. C'est cette approche
que nous consid\'ererons dans cette th\`ese en consid\'erant une mesure d'int\'egration sur l'ensemble des matrices hermitiennes.

\subsection{Mod\`eles de matrices et combinatoire.}

Pour bien comprendre comment le mod\`ele qui nous int\'eresse a vu le jour, il est n\'ecessaire de rappeler comment
un r\'esultat de Chromodynamique Quantique (QCD) s'est av\'er\'e \^etre fondamental dans le cadre des matrices al\'eatoires.
Cette th\'eorie est une th\'eorie des champs o\`u les particules sont des quarks portant une charge de couleur qui est un objet
tridimensionnel. Ils interagissent donc \`a travers des matrices de jauge de taille $3 \times 3$, les gluons, pr\'esentant
une sym\'etrie $SU(3)$. La r\'esolution de cette th\'eorie est un probl\`eme extr\`emement complexe et si l'on peut obtenir
des r\'esultats perturbatifs aux petites \'echelles o\`u la constante de couplage est faible, l'\'etude aux grandes
\'echelles ne peut \^etre d\'ecrite de la m\^eme fa\c{c}on. En 74, 't Hooft \cite{thooft} a propos\'e d'aborder le probl\`eme diff\'eremment
en le g\'en\'eralisant: au lieu de consid\'erer seulement trois couleurs, on peut \'etudier le m\^eme probl\`eme avec un nombre
de couleurs arbitraire $N$.  On peut alors faire tendre ce nombre $N \to \infty$ et exprimer les observables du mod\`ele \`a
travers leur d\'eveloppement en ${1 \over N}$ pour obtenir finalement des informations sur la valeur $N=3$. 't Hooft a alors
d\'ecouvert que les graphes de Feynman contribuant \`a un ordre donn\'e dans ce d\'eveloppement ont une topologie fix\'ee.

Le type de mod\`eles que nous allons \'etudier dans ce m\'emoire a \'et\'e introduit bien plus tard. Partant des techniques classiques
de th\'eorie des champs, Br\'ezin, Itzykson, Parisi et Zuber \cite{BIPZ} ont interpr\'et\'e les int\'egrales matricielles issues de
la th\'eorie des matrices al\'eatoires comme des fonctions g\'en\'eratrices de graphes \'epais lorsque la taille des matrices tend
vers l'infini. Pour cela, ils ont d\'evelopp\'e les int\'egrales consid\'er\'ees autour d'un col pour se ramener au calcul de
valeurs moyennes d'op\'erateurs par rapport \`a une mesure gaussienne. L'application du th\'eor\`eme de Wick (voir l'appendice \ref{appwick}) permet
alors de repr\'esenter le r\'esulat comme une somme sur un ensemble de graphes o\`u les vertex sont orient\'es.
Nous reviendrons plus pr\'ecis\'ement sur cette construction dans le d\'ebut du chapitre 2 puisqu'elle est la base des objets
\'etudi\'es ici.

L'\'etude des mod\`eles de matrices a connu un regain d'inter\^et quelques ann\'ees plus tard gr\^ace \`a une simple observation.
Plusieurs domaines tr\`es diff\'erents de la physique, tels que la th\'eorie des cordes, la gravitation quantique ou bien
l'\'etude de membranes, n\'ecessitent de pouvoir caract\'eriser des surfaces et, plus particuli\`erement, de pouvoir mettre une
mesure sur cet ensemble. En 1985, plusieurs chercheurs \'etudiant de telles surfaces ont observ\'e que l'interpr\'etation introduite par
\cite{BIPZ} permet d'approcher ce type de probl\`eme \`a partir des int\'egrales de matrices \cite{ambjornrmt,David,Kazakov}. En effet,
consid\'erant un graphe \'epais, il existe une proc\'edure tr\`es simple permettant de remplacer tout vertex \`a $k$ pattes
par un polyg\^one \`a $k$ c\^ot\'es. A un graphe consistant en un recollement de vertex, on fait donc correspondre une surface
form\'ee de polyg\^ones coll\'es entre eux par leurs ar\^etes. La th\'eorie des matrices al\'eatoires permet ainsi
de calculer des objets de la forme:
\beq
\CZ = \sum_{\CS} \CP(\CS)
\eeq
o\`u $\CS$ est une surface discr\'etis\'ee construite comme un recollement de polyg\^ones et $\CP$ est un poids associ\'e \`a
toute surface de ce type. Pour passer \`a de "vraies" surfaces, c'est-\`a-dire continues, il faut alors faire tendre le nombre de
polyg\^ones vers l'infini et r\'eduisant leur taille de mani\`ere \`a garder l'aire de la surface consid\'er\'e finie.

Dans ce cadre, la propri\'et\'e d\'ecouverte par 't Hooft en QCD signifie que, lorsque la taille des matrices tend vers l'infini, le d\'eveloppement en ${1 \over N}$
permet de s\'electionner le genre des surfaces g\'en\'er\'ees. Ainsi, on peut \'ecrire
\bea\label{introdevtopo}
\CZ &=& \sum_{g=0}^\infty N^{2 -2g} \sum_{\CS_g} \CP(\CS_g) = \sum_{g=0}^\infty N^{2 -2g} \CZ^{(g)} \cr
&=&\begin{array}{r}
{\includegraphics[width=10cm]{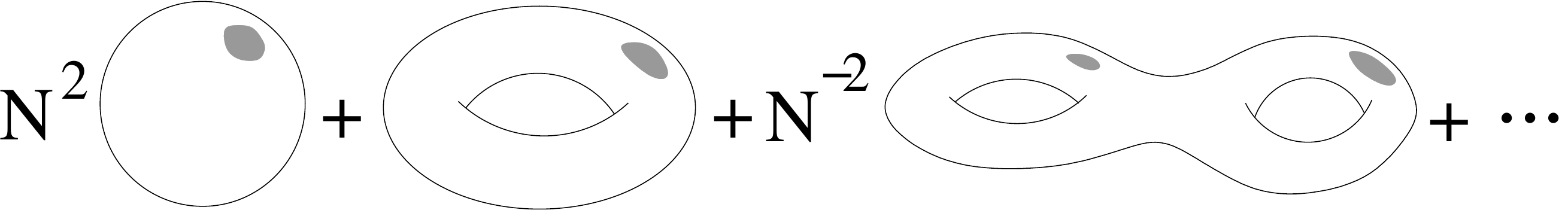}}
\end{array}\cr
\eea
o\`u l'on somme maintenant sur l'ensemble des surfaces $\CS_g$ de genre $g$ donn\'e pour obtenir chaque terme de ce d\'eveloppement
topologique. Cette propri\'et\'e s'est av\'er\'ee fondamentale dans l'\'etude des matrices al\'eatoires et a permis de faire
de grands progr\`es.

\subsection{Matrices al\'eatoires et surfaces continues.}

Revenons maintenant \`a une \'eventuelle proc\'edure permettant d'atteindre des surfaces continues \`a partir des mod\`eles de matrices.
Une telle limite est atteinte en faisant tendre le nombre moyen de polyg\^ones constitutifs des surfaces vers l'infini. Ceci
est en g\'en\'eral obtenu en approchant un point critique dans l'espace des param\`etres du mod\`ele. En effet, on peut montrer \cite{courseynard} que
le nombre moyen de polyg\^ones $<n>$ est donn\'e par une relation de la forme
\beq
<n> = g {\partial \ln \CZ \over \partial g}
\eeq
o\`u $g$ est un param\`etre du mod\`ele de matrices et $\CZ$ la fonction g\'en\'eratrice de surfaces discr\'etis\'ees issue
de ce mod\`ele. Il faut donc atteindre une singularit\'e de $\ln \CZ$ comme fonction de $g$ pour faire diverger cette quantit\'e.
On d\'etermine ainsi un point critique $g_c$ tel que pour tout terme du d\'eveloppement topologique $\CZ^{(h)}$ d\'efini dans \eq{introdevtopo}, il existe un
exposant critique $\alpha_h$ tel que
\beq
\CZ^{(h)}  = (g - g_c)^{- \alpha_h} \CZ_{sing}^{(h)} + \hbox{termes sous dominants} .
\eeq
On peut alors d\'efinir une constante cosmologique liant l'aire moyenne des polyg\^ones $\epsilon^2$ et la distance au point
critique:
\beq
\L \epsilon^2  := (g - g_c).
\eeq
La fonction g\'en\'eratrice totale est alors donn\'ee par le comportement
\beq
\CZ = \sum_{h=0}^\infty N^{2-2h} (g-g_c)^{- \alpha_h} \CZ_{sing}^{(h)}
\eeq
au voisinage de la singularit\'e $g_c$. Il a \'et\'e montr\'e \cite{Brezindsl,Douglas,Migdal} que l'exposant critique $\alpha_h$ est lin\'eaire
en le genre $h$: il existe un exposant $\gamma$ tel que
\beq\label{ah}
\alpha_h = (\gamma - 2) (1-h).
\eeq
Les exposants de ${1 \over N^2}$ et $(g-g_c)$ sont donc comparables et ceci permet de d\'efinir une variable combinant les deux
$\kappa := N (g- g_c)^{1 - {\gamma  \over 2}}$ permettant d'exprimer:
\beq
\CZ = \sum_{h=0}^\infty \kappa^{2 -2h} \CZ_{sing}^{(h)}
\eeq
et ainsi de m\'elanger les contributions des surfaces de genres diff\'erents \`a travers une double limite d'\'echelle
obtenue par $N \to \infty$ et $g \to g_c$ en gardant $\kappa$ fini. Cette double limite d'\'echelle est int\'eressante
puisqu'elle permet d'avoir acc\`es aux surfaces continues depuis les mod\`eles de matrices et de m\'elanger les surfaces de genres
diff\'erents ce qui est un \'el\'ement fondamental de la th\'eorie des cordes o\`u ces derni\`eres int\'eragissent en se
s\'eparant et se recollant cr\'eent des surfaces de genre plus \'elev\'e. Nous reviendrons sur ce point dans la partie suivante
d\'edi\'ee aux applications des matrices al\'eatoires.

\section{Applications des mod\`eles de matrices al\'eatoires.}

Les matrices al\'eatoires sont un outil formidable tant ses applications sont nombreuses aussi bien en physique qu'en math\'ematiques.
Dans cette partie nous pr\'esentons seulement une petite partie de celles-ci plus particuli\`erement reli\'ees aux r\'esultats
pr\'esent\'es dans ce m\'emoire.

\subsection{Volume symplectique de l'espace des modules de surfaces de Riemann.}

En math\'ematiques, on peut \^etre int\'eress\'e par d\'ecrire et compter des surfaces partageant une m\^eme topologie. Lorsque
ces surfaces sont discr\'etis\'ees, ceci se ram\`ene \`a de la pure combinatoire mais lorsque les surfaces sont continues, on ne peut
plus aborder le probl\`eme de la m\^eme mani\`ere. Il faut d\'efinir un certain nombre de param\`etres, appel\'es modules, dont la
valeur permet de caract\'eriser chaque surface ou type de surface. L'ensemble des surfaces est donc mis en bijection avec l'espace
de toutes les valeurs possibles de ces param\`etres: l'espace des modules \cite{Mumford}. Un premier ensemble de modules peut \^etre identifi\'e
comme le genre $g$ et le nombre de bords $k$ d'une surface. On peut alors d\'ecouper l'ensemble des surfaces en des sous-ensembles
$\CS_k^{(g)}$ contenant toutes les surfaces de genre $g$ donn\'e et ayant $k$ bords. A chacun de ces ensembles on peut associer un
ensemble de modules tels que la longueur de chaque bord par exemple dont les diff\'erentes valeurs forment un espace $\CM_k^{(g)}$.
Alors que dans le cas des surfaces discr\'etis\'ees, on les compte en associant \`a chaque assemblage de polyg\^ones un poids
et en effectuant la somme de ces poids, l'\'equivalent continu consiste \`a munir l'espace des modules d'une mesure et \`a calculer
le volume de ce dernier par rapport \`a cette mesure. En effet, chaque point de l'espace des modules correspond \`a une surface
et imposer une mesure sur cet espace consiste \`a associer un poids \`a chaque surface, l'int\'egration rempla\c{c}ant alors la somme
discr\`ete.

Dans l'\'etude de ces espaces, Riemann lui-m\^eme a \'et\'e le premier \`a \'etudier l'espace $\CM_g$ de toutes les structures complexes sur une surface orient\'ee de genre $g$ modulo
l'ensemble des diff\'eormorphismes pr\'eservant l'orientation. Il a ainsi pu montrer que l'espace $\CM_g$ a pour dimension r\'eelle
$6g-6$. Beaucoup plus r\'ecemment, d\'emontrant ainsi une conjecture pr\'ec\'edement \'etablie par Witten \cite{Witten}, Kontsevich \cite{kontsevich}a
montr\'e que l'on pouvait calculer le volume de tels espaces par le calcul d'int\'egrales de matrices du type
\beq
\CZ(\L) = {\int dM exp -\Tr \left( {\L M^2 \over 2}- i {M^3 \over 6} \right) \over \int dM -\Tr {\L M^2 \over 2}}
\eeq
o\`u $\L$ est une matrice de champ ext\'erieur en utilisant la propri\'et\'e disant que les volumes $v_n^{(g)}:= \hbox{Vol}\left(\CM_n^{(g)}\right)$
satisfont les \'equations de Korteweg-de Vries (KdV).

Le fait que cette hierachie int\'egrable d'\'equations apparaisse dans l'\'etude des matrices al\'eatoires est connu depuis longtemps
et beaucoup de travaux ont montr\'e que de telles hierarchies int\'egrables sont fondamentalement li\'ees \`a diff\'erents mod\`eles
de matrices al\'eatoires \cite{Brezindsl,Migdal,Douglas,Banks}. Au contraire, l'\'emergence de telles contraintes dans le cadre de l'\'etude de l'espace des modules
de surfaces continues etait plus surpenante et n'est toujours pas comprise et fait l'objet de nombreux travaux. Tr\`es r\'ecemment,
Mirzakhani \cite{Mirz1,Mirz2} a propos\'e une relation de r\'ecurrence permettant de retrouver la valeur de ces volumes avec une approche
m\'elangeant g\'eom\'etrie alg\'ebrique et g\'eom\'etrie hyperbolique. Ces relations semblent \^etre un premier pas vers une meilleure
compr\'ehension de ce lien entre hierarchies int\'egrables et espaces de modules.

\subsection{Th\'eorie de la gravitation quantique bidimensionnelle.}

La gravitation est s\^urement la force fondamentale la plus facilement observable \`a l'\'echelle macroscopique puisqu'elle
a toujours \'et\'e ressentie par tout le monde et a \'et\'e formalis\'ee tr\`es t\^ot \`a travers les lois de Newton. Cependant
son comportement \`a petites \'echelles est plus probl\'ematique. Il para\^itrait en effet naturel de vouloir quantifier cette
int\'eraction fondamentale comme l'\'electromagn\'etisme et les autres forces, mais les techniques de th\'eorie des champs utilis\'ees
ne peuvent s'appliquer dans ce cas l\`a puisque la gravitation est une th\'eorie non renormalisable.

Comment quantifier cette th\'eorie en pratique? On peut tenter d'imiter ce qui se fait d\'ej\`a dans les autres th\'eories
en utilisant les int\'egrales de chemins, c'est-\`a-dire sommer sur tous les chemins d'\'etats possibles entre un \'etat initial
et un \'etat final donn\'e. On sait depuis Einstein que le champ gravitationnel est un champ courbant l'espace temps. Pour le caract\'eriser
il faut donc se donner une vari\'et\'e $\CE$ de dimension 4, des coordonn\'ees $\{x_i\}_{i=1}^4$ sur celle-ci et une m\'etrique
$g_{ij}$ donn\'ee par une matrice $4 \times 4$. Son action est alors donn\'ee par la courbure de $\CE$:
\beq
S:= {-c^3 \over 16 \pi G} \int_{\CE} dx_1 dx_2 dx_3 dx_4 \sqrt{- det \, g} R
\eeq
o\`u $R$ est la courbure scalaire de la m\'etrique $g$ et $G$ un nombre appel\'e constante gravitationnelle. Une int\'egrale de chemin sera alors de la forme:
\beq
\CZ = \sum_{\CE} exp \left(-i {S(\CE) \over \hbar} \right)
\eeq
o\`u la somme porte sur toutes les vari\'et\'es $\CE$ satisfaisant les conditions initiale et finale. Le calcul d'une telle somme
est extr\`emement compliqu\'e et un tel probl\`eme n'a pas encore pu \^etre r\'esolu, en particulier parce qu'on ne sait pas bien
d\'ecrire les diff\'erentes topologies de $\CE$ intervenant dans cette somme.

On peut cependant essayer de simplifier le probl\`eme pour tenter de le r\'esoudre en diminuant la dimension de l'espace temps:
on peut consid\'erer l'espace temps comme une vari\'et\'e de dimension 2: une dimension d'espace et une dimension de temps. On va \'egalement faire une rotation de Wick
$t \to i t$, faisant passer dans l'espace des temps complexes et rendant la m\'etrique Euclidienne. Dans ce cas l\`a, le probl\`eme se rapproche d'un probl\`eme
de calcul de volume d'espace de modules de surfaces continues d\'ecrit plus haut. On doit en effet calculer des int\'egrales de
chemin de la forme:
\beq
\CZ = \sum_{\CE} exp \left(- {\int dx^2 \sqrt{- det \, g} (\L + G R) \over \hbar} \right)
\eeq
o\`u la somme porte sur toutes les surfaces satisfaisant les conditions aux limites et $\L$ est la constante cosmologique.
En fait, le symbole de somme est ici un peu abusif puisqu'il sous entend une somme discr\`ete alors que l'espace des surface
n'est pas \`a priori discret. En fait, il faudrait plut\^ot r\'e\'ecrire $\CZ$ comme une int\'egrale sur l'espace des
modules des surfaces de Riemann $\CE$.

C'est justement sur ce point que les int\'egrales de matrices s'av\`erent efficaces
\`a condition de donner un sens pr\'ecis \`a cette somme discr\`ete. Une id\'ee naturelle pour d\'ecrire une surface, fr\'equemment
utilis\'ee par exemple en m\'ecanique dans les m\'ethodes de mod\'elisation en \'el\'ements finis, consiste \`a discr\'etiser cette
derni\`ere en la rempla\c{c}ant par une surface "proche" compos\'ee uniquement de polyg\^ones coll\'es entre eux par leurs ar\^etes.
La somme $\CZ$ consiste alors \`a sommer sur de telles surfaces discr\'etis\'ees en donnant un poids particulier \`a
chacun des polyg\^ones les composant ainsi qu'\`a l'op\'eration de recollement d'ar\^etes: on peut donc voir $\CZ$ comme une fonction g\'en\'eratrice de telles surfaces. Or
comme nous le verrons, c'est exactement le type de fonctions g\'en\'eratrices que l'on peut obtenir \`a partir des int\'egrales de
matrices. Notons qu'il faut alors faire tr\`es attention \`a cette discr\'etisation des surfaces et pouvoir caract\'eriser
son impact sur les observables du mod\`ele.

Pour la suite, remarquons que l'on peut \'egalement introduire des champs de "mati\`ere" scalaires $X_i: \CE \to \mathbb{C}$ coupl\'es \`a la gravit\'e, par exemple
\`a travers l'action modifi\'ee:
\beq\label{gravitematiere}
S = \int dx^2 \sqrt{- det \, g} (\L + G R + \sum_i g^{\alpha \beta} \partial_\alpha X_i \partial_\beta X_i ).
\eeq

\subsection{Th\'eories des cordes.}

Une autre th\'eorie peut se mettre sous une forme similaire et se ramener au m\^eme probl\`eme consistant en
une int\'egrale sur l'espace des modules des surfaces de Riemann: la th\'eorie des cordes \cite{green}. Celle-ci est une g\'en\'eralisation
des th\'eories des champs habituelles o\`u les particules ne sont plus des points de l'espace temps
mais des objets \`a une dimension, i.e. des cordes param\'etr\'ees par des coordonn\'ees ${X_i(s,t)}_{i=1,\dots,3}$ o\`u
$s$ est une abscisse curviligne sur la corde. Pour d\'efinir une int\'egrale de chemin il faut alors int\'egrer sur tous les \'etats $X_i(t)$
entre une corde initiale $X_i(s,t_0)$ et une corde finale $X_i(s,t_f)$. Or, lorsqu'une corde \'evolue dans le temps, elle
dessine une surface (bidimensionnelle) dans l'espace temps. L'int\'egrale de chemin correspond donc comme dans le paragraphe
pr\'ec\'edent \`a sommer sur toutes les surfaces entre les \'etats initiaux et finaux:
\beq
\CZ = \sum_{\CE} exp \left(-i {S(\CE) \over \hbar} \right).
\eeq
Plusieurs actions $S(\CE)$ peuvent alors \^etre choisies et plusieurs mesures pour la sommation.
Nous n'alllons pas nous attarder sur toutes les actions possibles. Notons que parmi toutes celles-ci,
Polyakov \cite{Polyakov} a propos\'e une action invariante sous les changements de coordonn\'ees de la surface:
\beq\label{Polyakov}
S:=- {\mu c \over 2} \int ds dt \sqrt{- det \, g} ( \sum_i G_{\mu \nu} g^{\alpha \beta} \partial_\alpha X_\mu \partial_\beta X_\nu )
\eeq
o\`u $G$ est la m\'etrique de Minkowski sur l'espace-temps et $g$ la m\'etrique induite sur la feuille d'univers (i.e. la surface
dessin\'ee par l'\'evolution de la corde dans le temps). Notons que l'on reconnait ici exactement la m\^eme somme sur
les surfaces que celle introduite dans le paragraphe pr\'ec\'edent dans le cadre de la gravitation quantique bidimensionnelle
dans \eq{gravitematiere}.

Cependant, cette th\'eorie des cordes pr\'esente de nombreux d\'efauts. Non seulement, elle ne peut \^etre v\'erifi\'ee exp\'erimentalement
mais au niveau th\'eorique, les sym\'etries n\'ecessaires impliquent de fortes contraintes. En effet, on ne peut pas construire une telle
th\'eorie dans un espace temps de dimension 4 mais 26! Il faut donc \'eliminer 22 dimensions. Une partie du chemin a d\'ej\`a
\'et\'e fait par Ramond \cite{Ramond} et Neveu et Schwarz \cite{Neveu} en introduisant la supersym\'etrie: ils ont ramen\'e le probl\`eme
\`a une th\'eorie \`a dix dimensions. Il reste donc 6 dimensions \`a \'eliminer pour revenir aux 4 dimensions de l'espace temps r\'eel.
Ceci peut \^etre fait en r\'eduisant la taille des 6 directions suppl\'ementaires, en les compactifiant. Les nombreux travaux dans
ce domaine font appara\^itre une tr\`es riche structure int\'eressante au moins au niveau math\'ematique si ce n'est pour retrouver
les propri\'et\'es physiques du monde r\'eel.

\br
Nous n'avons \'evoqu\'e ici qu'un exemple parmi bien d'autres de mod\`ele de th\'eorie des cordes li\'e aux mod\`eles
de matrices al\'eatoires. Nous verrons un autre exemple dans le chapitre 5 qui lui est totalement d\'edi\'e:
les th\'eories des cordes topologiques. Ce cas correspond \`a une th\'eorie des cordes o\`u seule la topologie de
la feuille d'univers (i.e. la surface g\'en\'er\'ee par l'\'evolution temporelle de la corde) entre en compte dans le calcul de la fonction de partition. Si deux types de th\'eories des cordes
apparaissent dans ce cadre, tous deux sont fortement reli\'es aux mod\`eles de matrices \cite{MMtopo}: que ce soit par les propri\'et\'es
de variation de la fonction de partition du type A sous les d\'eformation de la feuille d'univers ou par la conjecture de Dijkgraaf
et Vafa \cite{DV} identifiant directement un mod\`ele de matrice \'equivalent au type B, les points communs entre mod\`eles de matrices et
th\'eories des cordes topologiques sont frappants.
\er

\subsection{Th\'eories conformes et mod\`eles minimaux.}

Explicitons un peu plus comment ces dimensions critiques pour l'espace temps sont obtenues. Au niveau local, l'action de Polyakov
a une sym\'etrie sous le changement de la m\'etrique:
\beq
g_{\alpha \beta} \to e^\varphi g_{\alpha \beta},
\eeq
la sym\'etrie conforme. Cependant cette sym\'etrie est bris\'ee par la quantification: la sommation sur les surfaces
introduit une anomalie. On peut caract\'eriser cette anomalie par un unique scalaire: la charge centrale $c$. Celle-ci
s'annule lorsque la th\'eorie quantique est invariante conforme. Pour la d\'efinir, il est n\'ecessaire d'\'etudier plus pr\'ecis\'ement
les transformations conformes de la m\'etrique. Ce sont les transformations qui changent la m\'etrique par un facteur scalaire
et donc laissent invariante l'action de la th\'eorie \cite{Kac}: elles sont donn\'ees par
\beq
z \to z (1 + \epsilon f(z))
\eeq
o\`u $f$ est une fonction analytique de la variable complexe $z = x+i y$
avec $x$ et $y$ des coordonn\'ees sur la surface consid\'er\'ee. Elles sont donc g\'en\'er\'ees par les op\'erateurs
$l_n := z^{n+1} {\partial \over \partial z}$. En fait, comme souvent, pour obtenir les bons g\'en\'erateurs des transformations
conformes, on doit introduire un ordre normal pour pr\'eciser l'ordre dans lequel les diff\'erents op\'erateurs agissent. On obtient ainsi les g\'en\'erateurs $L_n:= :l_n:$ qui satisfont
les relations de commutation:
\beq
[L_n,L_m]=(m-n)L_{m+n}+ {c \over 12} (n^3-n) \delta_{m+n} Id,
\eeq
c'est-\`a-dire qu'ils forment une extension de l'alg\`ebre de Virasoro avec une charge centrale $c$.

On peut construire une repr\'esentation de cette alg\`ebre en partant d'un op\'erateur primaire $|0>$:
\beq
L_0 |0> = h |0> \;\;\; \hbox{et} \;\;\; \forall n >0 \, , \; L_n |0> = 0
\eeq
consid\'er\'e comme le vide de la repr\'esentation. L'ensemble des \'etats engendr\'es par les autres op\'erateurs $L_{-n}$:
\beq
|n_1 n_2 \dots n_k>:= L_{- n_1} L_{- n_2} \dots L_{- n_k} |0>
\eeq
engendrent alors, en g\'en\'eral, une repr\'esentation de dimension infinie irr\'eductible de l'alg\`ebre de Virasoro.
Cependant ceci n'est plus vrai dans certains cas particuliers o\`u la repr\'esentation engendr\'ee est r\'eductible puisque
les \'etats engendr\'es ne sont pas lin\'eairement ind\'ependants. En effet, pour les valeurs de $h$ particuli\`eres:
\beq
h_{r,s} = {1 \over 48} \left( 12(r-s)^2+ (1-c)(r^2-s^2-2) +(r^2-s^2)\sqrt{(25-c)(1-c)}\right)
\eeq
o\`u $r$ et $s$ sont entiers, on obtient de telles repr\'esentations r\'eductibles.

\br
Certains mod\`eles ne font intervenir qu'un nombre fini de repr\'esentations r\'eductibles de ce type et sont ainsi enti\`erement solubles:
les mod\`eles minimaux index\'es par deux entiers $p$ et $q$ premiers entre eux. Dans ce cas, la charge centrale est de la
forme:
\beq
c= 1 - 6 {(p-q)^2 \over pq}
\eeq
et les dimension des champs primaires sont donn\'ees par la table de Kac \cite{Kac}:
\beq
h_{r,s} = {(rq-sp)^2 -(p-q)^2 \over 4pq}
\eeq
o\`u $s$ et $r$ sont deux entiers strictement positifs et strictement inf\'erieurs respectivement \`a $q$ et \`a $p$.

Nous verrons comment r\'esoudre ces mod\`eles gr\^ace aux mod\`eles de matrice dans le chapitre 4 de cette th\`ese.
\er

On peut montrer que l'on peut associer une telle charge centrale \`a chaque champ suivant sa
statistique: \`a un boson, on associe une charge 1 tandis qu'\`a un fermion on associe une charge ${1 \over 2}$. On peut \'egalement
montrer que le champ de jauge $g_{\alpha \beta}$ a une dimension n\'egative \'egale \`a $-26$ due, entre autres choses, \`a l'introduction
de fant\^omes de Fadeev-Popov.
La charge totale d'une th\'eorie des cordes $D$ dimensionnelle avec l'action de Polyakov est donc \'egale \`a:
\beq
c = D-26
\eeq
puisqu'il y a $D$ champs bosoniques dans cette th\'eorie. On voit ainsi qu'il est n\'ecessaire d'avoir une th\'eorie avec un
espace temps de dimension $D=26$ pour que la sym\'etrie conforme soit pr\'eserv\'ee au niveau quantique.
Pour r\'eduire la dimension critique, on peut introduire la supersym\'etrie et donc $D$ fermions. Ceci induit de plus de nouveaux
fant\^omes de Fadeev-Popov avec une charge $+11$. On a alors la charge centrale
\beq
c = D-26 + {D \over 2} +11
\eeq
ce qui donne une dimension critique $D=10$ comme annonc\'e plus haut.

En fait, les mod\`eles de matrices permettent non seulement d'avoir acc\`es aux th\'eories des cordes critiques mais \'egalements
\`a celles qui ont une dimension non critique et donc une charge centrale non nulle. En particulier, les mod\`eles minimaux
correspondent \`a une charge centrale $c<1$.

Dans tous les cas, pour comparer les r\'esultats des th\'eories conformes introduites ici et des
mod\`eles de matrices vus comme fonctions g\'en\'eratrices de surfaces discr\'etis\'ees, il faut pouvoir trouver un language
commun. Dans les deux cas on peut se ramnener \`a une unit\'e de comparaison possible: l'aire des surfaces g\'en\'er\'ees.
On peut se demander comment se comportent les observables de ces diff\'erents mod\`eles en termes de l'aire $\CA$ des surfaces
consid\'er\'ees. Plus particuli\`erement comment divergent ces observables quand $\CA \to \infty$. En g\'en\'eral,
pour les th\'eories conformes, on peut montrer que pour un genre $h$ et une aire $\CA$ fix\'ee pour les surfaces $\CE$,
on a un comportement du type \cite{KPZ}:
\beq
\CZ^{(h)} = \sum_{\CE} exp \left(-i {S(\CE) \over \hbar} \right)\sim_{\CA \to \infty} \CA^{\gamma_h-3}
\eeq
o\`u l'exposant $\gamma_h$ est lin\'eaire en le genre et donn\'e par\footnote{On peut noter la similitude entre cette \'equation et \eq{ah}.}:
\beq
\gamma_h = 2h + \gamma_{string} (1-h).
\eeq
Le coefficient $\gamma_{string}$, appel\'e susceptibilit\'e de corde d\'epend de la th\'eorie consid\'er\'ee. Par exemple,
pour un mod\`ele minimal de type $(p,q)$, on peut montrer q'il vaut:
\beq
\gamma_{string} = -2 {|p-q| \over p+q-|p-q|}.
\eeq
On peut calculer un tel exposant critique pour n'importe quelle autre observable $<\CO>$ du mod\`ele et ainsi le comparer
avec les r\'esultats de mod\`eles de matrices.

Nous verrons ainsi que la comparaison de ces exposants, \`a condition de bien normaliser les observables et variables, permet
de montrer que les mod\`eles de matrices donnent acc\`es aux th\'eories conformes (en particulier en prenant certaines limites
o\`u les surfaces discr\'etis\'ees deviennent continues).

\subsection{De l'\'emergence de la g\'eom\'etrie alg\'ebrique et des syst\`emes int\'egrables.}

Si les applications pr\'esent\'ees jusqu'ici ont toutes un lien avec la combinatoire de surfaces discr\'etis\'ees ou continues,
les mod\`eles de matrices sont li\'es \`a de nombreux autres probl\`emes qui ne seront pas ou tr\`es peu abord\'es dans cette th\`ese.
Si elle ne semble pas aussi directement reli\'ee \`a notre probl\`eme, l'une des propri\'et\'es des mod\`eles de matrices
qui semble fondamentale dans toutes les construction introduites ici est son int\'egrabilit\'e. En effet, cette propri\'et\'e,
que l'on retrouve sous diff\'erentes formes suivant l'aspect des mod\`eles de matrices \'etudi\'es, assure que le probl\`eme est soluble:
cela signifie que l'on est en principe capable de calculer toutes les observables du mod\`ele consid\'er\'e (ce qui ne signifie
pas que l'on puisse le faire facilement en pratique).

Il a \'et\'e observ\'e que les observables des mod\`eles de matrices sont, en g\'en\'eral, solutions d'\'equations diff\'erentielles appartenant
\`a des hierarchies int\'egrables telles que les hierarchies KP, la hierarchie de Toda 2 dimensionnelle, celle de Whitham ...
En fait, lorsque l'on int\`egre sur une matrice de taille finie $N$, on trouve un syst\`eme int\'egrable quantique caract\'eris\'e
par $\hbar \sim {1 \over N}$. On s'attend donc \`a retrouver un syst\`eme int\'egrable classique lorsque la taille $N  \to \infty$
et ainsi tous les ingr\'edients le caract\'erisant comme une courbe alg\'ebrique appel\'ee courbe spectrale
ou la fonction $\tau$ satisfaisant des relations bilin\'eaires.

Nous verrons que nous retrouvons effectivement tous ces ingr\'edients dans le cadre des int\'egrales formelles qui sont li\'ees
\`a la limite des int\'egrales de matrices convergentes lorsque la taille des matrices tend vers l'infini. On ira m\^eme
plus loin puisque nous montrerons que ces \'el\'ements classiques permettent de calculer toutes les corrections en ${1 \over N}$,
i.e. toutes les corrections semi-classiques\footnote{Nous verrons que ceci n'est pas exact puisque le mod\`ele
\'etudi\'e n'est pas exactement l'int\'egrale de matrice dont on s'attend \`a ce qu'elle soit la fonction tau d'un syst\`eme
int\'egrable quantique. Il reste dons une derni\`ere \'etape \`a franchir pour pouvoir reconstruire le syst\`eme quantique lui-m\^eme
par resommation des termes semi-classiques.}.

Si le lien entre int\'egrabilit\'e et mod\`eles de matrices semble bien compris \`a pr\'esent, le lien direct entre
combinatoire de surfaces et int\'egrabilit\'e restait jusqu'\`a r\'ecemment assez obscur. Pourtant cet outil semble fondamental dans l'\'etude
de l'espace des modules de surfaces de Riemann. En effet, si l'approche de Kontsevich a \'et\'e de r\'einterpr\'eter ce probl\`eme
en termes d'int\'egrales matricielles, ceci est principalement d\^u au fait que les volumes d'espaces de modules sont solutions
d'\'equations diff\'erentielles de la hierarchie KdV et donc \`a l'int\'egrabilit\'e.
Si des travaux r\'ecents \cite{OP,LNM,NO} ont permis de comprendre un peu mieux ce lien \`a la lumi\`ere de l'\'etude de partitions
al\'eatoires et leur lien avec les fonctions tau de syst\`emes int\'egrables,
j'esp\`ere que cette th\`ese permettra
de donner de premiers \'el\'ements de compr\'ehension utilisant le formalisme des matrices al\'eatoires.



\chapter{Mod\`ele \`a deux matrices hermitiennes.}

Dans ce chapitre, nous pr\'esentons une m\'ethode de r\'esolution du mod\`ele \`a deux matrices hermitiennes formel. Ce mod\`ele
est souvent d\'ecrit comme la limite d'une int\'egrale matricielle convergente o\`u la taille des matrices tend vers $\infty$ puisqu'il
correspond \`a un d\'eveloppement perturbatif de cette int\'egrale par rapport \`a un param\`etre qui est de l'ordre de l'inverse de la taille des matrices
consid\'er\'ees. Nous allons donc montrer comment calculer tous les termes du d\'eveloppement de toutes les observables de
ce mod\`ele par rapport \`a ce param\`etre.
Cette m\'ethode, bas\'ee sur les travaux de Eynard pour le mod\`{e}le \`{a} une matrice \cite{eynloop1mat} et ses
m\'ethodes g\'en\'erales de r\'esolution des \'equations de boucles, montrera toute sa puissance dans les chapitres suivants
o\`{u} nous la g\'en\'eraliserons au del\`{a} des mod\`{e}le de matrices al\'eatoires.

Le mod\`{e}le \`{a} deux matrices hermitiennes a fait l'objet de diff\'erentes approches dans la litt\'erature. En effet,
les nombreux domaines o\`{u} ce probl\`{e}me, qui consiste essentiellement \`{a} \'evaluer des int\'egrales matricielles,
semble appara\^{i}tre ont permis \`{a} diff\'erentes sp\'ecialit\'es de se retrouver sur un m\^{e}me terrain. Ainsi chacun
a pu utiliser ses m\'ethodes favorites et l'on peut facilement imaginer que les contributions de physiciens, probabilistes ou
combinatoriciens vont apporter des points de vue bien diff\'erents. Le mod\`ele \'etudi\'e ici a \'et\'e introduit en premier
par Kazakov comme un mod\`ele de physique statistique \cite{KazakovIsing}: il a \'et\'e utilis\'e pour mod\'eliser le mod\`ele d'Ising
sur surface al\'eatoire, i.e. d\'ecrire un syst\`eme de spins en int\'eractions vivant sur un r\'eseau triangulaire al\'eatoire.

La m\'ethode d\'evelopp\'ee ici consiste
\`{a} r\'esoudre un ensemble d'\'equations satisfaites par les observables du syst\`{e}me. Ces \'equations sont
connues sous le nom d'\'equations de boucles \cite{Migdalloop,staudacher,ZJDFG,ACKM,eynm2m} dans le milieu des matrices al\'eatoires mais elles se retrouvent en fait dans diverses
sp\'ecialit\'es de physique ou de math\'ematiques sous des noms aussi divers que contraintes de Virasoro, identit\'es de Ward,
\'equations de Schwinger Dyson,
\'equations de Tutte \cite{tutte,tutte2} ou m\^{e}me \'equations de Baxter \cite{Babelon}. Si des m\'ethodes ont \'et\'e propos\'ees
pour les r\'esoudre dans des cas particuliers \cite{AkAm,Ak96,akeman} ou m\^{e}me totalement \cite{ACKM} dans le cas du mod\`{e}le \`{a}
une matrice, l'approche de Eynard \cite{eynloop1mat,ec1loopF} a pour avantage la simplicit\'e et la g\'en\'eralit\'e. En effet,
en utilisant un langage appropri\'e, celui de la g\'eom\'etrie alg\'ebrique, on peut r\'esoudre ces \'equations de
mani\`{e}re intrins\`{e}que de fa\c{c}on \`{a} ne plus faire appara\^itre les specificit\'es de tel ou tel mod\`{e}le
en se ramenant \`a l'\'etude d'une courbe alg\'ebrique associ\'ee au mod\`ele \cite{KazMar}.

Cependant, ces \'equations sont tr\`{e}s g\'en\'erales et admettent des solutions pr\'esentant des structures bien
diff\'erentes les unes des autres. Il existe ainsi plusieurs d\'efinitions non \'equivalentes de ce qui est usuellement
d\'enomm\'e "int\'egrale matricielle" dans la litt\'erature. Toutes satisfont les \'equations de boucles mais elles
ne coinc\"ident pas de mani\`{e}re g\'en\'erale. Cette impr\'ecision dans la d\'efinition, r\'ecurrente dans la litt\'erature,
a men\'e \`a des contradictions flagrantes particuli\`{e}rement mises en avant par \cite{AkAm} et \cite{Kanzieper,Deo,BrezinDeo} et expliqu\'ees plus tard par \cite{BDE}.

Nous d\'ebuterons donc ce chapitre en pr\'ecisant ce que nous entendons par "int\'egrale de matrice". Ceci nous
permettra ensuite d'interpr\'eter ces int\'egrales comme fonctions g\'en\'eratrices de surfaces
constitu\'ees de polygones de couleurs diff\'erentes. Enfin, je pr\'esenterai le principal r\'esultat de cette partie:
comment, en utilisant des concepts de g\'eom\'etrie alg\'ebrique, on peut calculer de mani\`ere explicite
toutes ces "int\'egrales de matrices formelles" en termes d'objets fondamentaux d\'efinis sur une courbe alg\'ebrique appel\'ee courbe
spectrale.

\section{D\'efinition du mod\`ele.}

Consid\'erons deux potentiels polyn\^omiaux\footnote{Toute la m\'ethode pr\'esent\'ee ici peut s'\'etendre au cas de potentiels
dont la d\'eriv\'ee est une fonction rationnelle. Cependant, les quelques subtilit\'es techniques induites par cette g\'en\'eralisation
n'apportent rien de nouveau \`{a} la compr\'ehension du probl\`{e}me et entrent dans le cadre
du chapite 4. Le lecteur int\'eress\'e pourra ais\'ement adapter la description faite du mod\`{e}le \`{a}
une matrice avec potentiel rationnel ou bien se reporter \`{a} la d\'emonstration de \EOinvariants.}:
\beq\label{defV}
V_1(x) := \sum_{k=0}^{d_1+1} {t_k \over k} x^{k}
\;\;\;
\hbox{et}
\;\;\;
V_2(x) := \sum_{k=0}^{d_2+1}{\tilde{t}_k \over k} x^{k}
\eeq
et d\'efinissons l'int\'egrale
\beq
\label{defZ}
\CZ_{2MM}:=\int_{H_N\times H_N} dM_1 dM_2\, e^{-{1\over
\hbar} Tr(V_1(M_1) + V_2(M_2) - M_1 M_2 )},
\eeq
o\`u $M_1$ et $M_2$ sont deux matrices hermitiennes de taille $N \times N$,
$dM_1$ et  $dM_2$ sont les produits des mesures de Lebesgues des composantes r\'eelles de $M_1$ et $M_2$
\beq
dM:= \prod_i dM_{ii} \prod_{i<j} dRe(M_{ij}) dIm(M_{ij})
\eeq
et $\hbar={T\over N}$ est un param\`etre de d\'evelopement perturbatif.
Une telle int\'egrale ne semble a priori pas souffrir de d\'efaut de d\'efinition. Cependant, suivant la sp\'ecialit\'e dans laquelle
elle est \'etudi\'ee ou bien m\^{e}me le r\'egime d'\'etude, le signe int\'egrale a plusieurs significations
diff\'erentes dans la litt\'erature.
Il est donc important de
pr\'eciser quel objet se cache vraiment sous cette d\'enomination.

\subsection{Int\'egrale sur l'ensemble des matrices hermitiennes.}
La d\'efinition la plus naturelle consiste \`{a} diagonaliser les matrices $M_1= U_1^\dagger \L_1 U_1$ et $M_2= U_2^\dagger \L_2 U_2$ pour ensuite int\'egrer
sur les valeurs propres r\'eelles,
\beq
\L_i = diag(\lambda_{i,1}, \dots, \lambda_{i,N}),
\eeq
d'une part et le groupe unitaire,
$U_i^\dagger U_i = 1$, d'autre part. D\'es lors l'int\'egrale \ref{defZ} s'\'ecrit
\beq\begin{array}{l}
\CZ_{herm}:=\cr
:= \int_{U(N) \times U(N)}dU_1 dU_2 \int_{\mathbb{R}^n \times \mathbb{R}^n}d\L_1 d\L_2 \Delta(\lambda_1)^2 \Delta(\lambda_2)^2 e^{-{1\over
\hbar} Tr(V_1(\L_1) + V_2(\L_2) - U_1^\dagger \L_1 U_1 U_2^\dagger  \L_2 U_2)}\cr
\end{array}
\eeq
o\`{u} le d\'eterminant de Vandermonde
\beq
\Delta(\lambda_{1}):= \prod_{i>j} (\lambda_{1,i}-\lambda_{1,j})
\eeq
vient du Jacobien du changement de variable:
\beq
dM_1 = \Delta(\lambda_1)^2 dU_1 d\L_1
\eeq
avec $d\L_1 = \prod_{i=1}^N d\lambda_i$ et $dU$ est la mesure de Haar sur le groupe $U(N)$.

Sous cette forme, un seul terme contient encore un couplage entre les valeurs propres et les matrices unitaires de changement de base par
l'interm\'ediaire de la combinaison $U_1 U_2^\dagger$.
Cette contribution peut \^{e}tre totalement explicit\'ee en utilisant la
\bt
Formule d'Harish Chandra-Itzykson-Zuber (HCIZ) \cite{HC,IZ}:
\beq
\int_{U(N)}dU e^{\Tr \L_1 U \L_2 U^\dagger} = {\det E \over 2^{N(N-1) \over 2} \Delta(\lambda_1) \Delta(\lambda_2)} \prod_{k=1}^{N-1} k! ,
\eeq
o\`{u} la matrice $E$, de taille $N \times N$, est d\'efinie par:
\beq
E_{ij} = \exp (\lambda_{1,i} \lambda_{2,j}).
\eeq
\et
On peut ainsi r\'eexprimer la fonction de partition comme:
\bea
\CZ_{herm} &=& \left( \pi \hbar \right)^{N(N-1) \over 2} {1 \over N!} \int_{\mathbb{R}^N \times \mathbb{R}^N} \det (e^{{1 \over \hbar} \lambda_{1,i} \lambda_{2,j}})
\Delta(\lambda_1) \Delta(\lambda_2) \times \cr
&& \phantom{ {1 \over N!} \int_{\mathbb{R}^N \times \mathbb{R}^N} \det (e^{{1 \over \hbar}})} \times \prod_{i=1}^N d\lambda_1 d\lambda_2 e^{-{1 \over \hbar}\left[ V_1(\lambda_{1,i}) + V_2(\lambda_{2,i})\right]} \cr
&=& \left( \pi \hbar \right)^{N(N-1) \over 2} \int_{\mathbb{R}^N \times \mathbb{R}^N}
\Delta(\lambda_1) \Delta(\lambda_2) \prod_{i=1}^N d\lambda_{1,i} d\lambda_{2,i} e^{-{1 \over \hbar}\left[ V_1(\lambda_{1,i}) + V_2(\lambda_{2,i})- \lambda_{1,i} \lambda_{2,i} \right]}. \cr
\eea

Cette expression permet, entre autres choses, d'identifier des crit\`eres de convergence de cette int\'egrale: il faut que les potentiels
$V_1$ et $V_2$ soient pairs et que leur termes de plus haut degr\'es $t_{d_1+1}$ et $\tilde{t}_{d_2+1}$
aient une partie r\'eelle strictement positive si $d_1>1$ ou $d_2>1$. Plus g\'en\'eralement, il faut que la partie r\'eelle de la fonction
$V_1(x) + V_2(y) -xy$ soit born\'ee inf\'erieurement sur $\mathbb{R}^2$.

Le calcul d'une telle int\'egrale peut \^{e}tre ramen\'e \`{a} l'\'etude de deux familles de polyn\^{o}mes
moniques
\beq
\pi_n(x) = x^n + \dots \;\;\; \hbox{et} \;\;\; \sigma_n(y) = y^n +\dots
\eeq
biorthogonaux par rapport \`{a} la mesure:
\beq
\int_{\mathbb{R}^2} dx dy \pi_n(x) \sigma_n(y) e^{-{1 \over \hbar} \left[  V_1(x) + V_2(y) -xy \right]} = h_n \delta_{nm}.
\eeq

Cette m\'ethode de r\'esolution a \'et\'e beaucoup \'etudi\'ee dans la litt\'erature \cite{Mehta2,Boulatov} et a permis de faire le lien
avec certains probl\`{e}mes de Riemann-Hilbert (voir \cite{Eynhab} pour une revue sur le sujet). Cependant, je voudrais encore rappeler  que cette d\'efinition de
l'int\'egrale de matrices ne coincide pas avec l'objet d'\'etude de cette th\`{e}se et les r\'esultats obtenus dans ce contexte
ne peuvent s'appliquer aux probl\`{e}mes de combinatoire de cartes.

\br
Cette int\'egrale n'a pas forc\'ement de d\'eveloppement en $\hbar^2$!
\er

\br
Un r\'esultat r\'ecent permet de calculer une grande classe d'observables de ce mod\`ele  gr\^ace \`a une formule de type
HCIZ. Cette formule, dont une premi\`ere extension a \'et\'e mise en avant par Morozov \cite{Morozov} suivant
Shatashvili \cite{Shat},
a \'et\'e fortement g\'en\'eralis\'ee par \cite{Eynpratts}:
\bt
Formule de Eynard-Prats Ferrer:
pour toute fonction $F$ compos\'ee d'un produit de traces contenant chacune un produit de deux matrices:
\beq
\begin{array}{l}
\int_{U(N)} F(\L_1,U\L_2U^\dagger) e^{-\Tr \L_1 U \L_2 U^\dagger} dU = \cr
= {\prod_{i=1}^N k! \over N! (-2 \pi)^{N(N-1) \over 2} \Delta(\lambda_1) \Delta(\lambda_2)} \sum_{\sigma,\tau \in S_N}
(-1)^{\sigma \tau} e^{- \Tr \L_{1,\sigma} \L_{2,\tau}} \times \cr
 \qquad \qquad \qquad \qquad \qquad \qquad \times \int_{T_N} dT F(\L_{1,\sigma}+T,\L_{2,\tau}+T^\dagger) e^{-\Tr T T^\dagger} \cr
\end{array}
\eeq
o\`{u} $T_N$ est l'ensemble des matrices complexes triangulaires sup\'erieures strictes muni de la mesure $dT$
du produit des mesures de Lebesgues des parties r\'eelles et imaginaires de tous les \'el\'ements de la matrice,
$S_N$ est l'ensemble des permutations de taille $N$ et
\beq
\L_{i,\sigma} = \hbox{diag} (\lambda_{i,\sigma(1)}, \lambda_{i,\sigma(2)}, \dots, \lambda_{i,\sigma(N)})
\eeq
pour toute permutation $\sigma \in S_N$.
\et
Cette g\'en\'eralisation donne acc\`{e}s \`{a} toute les int\'egrales matricielles du type:
\beq
\int_{H_N\times H_N} dM_1 dM_2\, F(M_1,M_2) e^{-{1\over
\hbar} Tr(V_1(M_1) + V_2(M_2) - M_1 M_2 )}
\eeq
o\`{u} $F$ est contraint par les propri\'et\'es de la formule de Eynard-Prats Ferrer.
\er

\subsection{Int\'egrale sur un chemin de valeurs propres.}\label{secchem}

Dans le paragraphe pr\'ec\'edent, nous avons impos\'e aux valeurs propres un crit\`{e}re de r\'ealit\'e contraignant
ainsi fortement les potentiels acceptables. On peut retourner le probl\`{e}me en fixant des potentiels arbitraires et
contraindre les chemins d'int\'egration possibles pour les valeurs propres par ces potentiels. L'int\'egrale
\beq\label{chemin}
\int_{ \CC}
\Delta(\lambda_1) \Delta(\lambda_2) \prod_{i=1}^N d\lambda_1 d\lambda_2 e^{-{1 \over \hbar}\left[ V_1(\lambda_{1,i}) + V_2(\lambda_{2,i})- \lambda_{1,i} \lambda_{2,i} \right]}
\eeq
est convergente si les chemins d'int\'egration $\CC$ pour les couples de valeurs propres $(\lambda_{1,i},\lambda_{2,i})$ sont choisis de mani\`{e}re appropri\'ee, c'est-\`{a}-dire
tels que $Re( V_1(x) + V_2(y) - xy)$ est born\'ee inf\'erieurement sur tout chemin $\CC_{1,i} \times \CC_{2,i}$ parcouru par
le couple $(\lambda_{1,i},\lambda_{2,i})$. Si les potentiels
ne sont pas triviaux ($d_1>1$ ou $d_2>1$), cette condition revient \`{a} rechercher les chemins $\CC_1$ et $\CC_2$ tels que
\beq
\int_{\CC_1} e^{-{1 \over \hbar} V_1(x)} dx \;\;\; \hbox{et} \;\;\; \int_{\CC_2} e^{-{1 \over \hbar} V_2(y)} dy
\eeq
soient absolument convergentes. G\'en\'eriquement, il existe $d_1$ (resp. $d_2$) chemins homologiquement ind\'ependants $\CC_{1,i},i=1 \dots d_1$
(resp. $\CC_{2,j},j=1 \dots d_2$) allant de l'infini \`{a} l'infini satisfaisant cette propri\'et\'e.
On peut donc d\'ecomposer n'importe quel chemin $\CC$ acceptable en:
\beq
\exists! \, \kappa , \; \; \CC = \sum_{i,j} \kappa_{i,j} \, \CC_{1,i} \times \CC_{2,j}
\eeq
o\`{u} $\kappa$ est une matrice \`{a} coefficients complexes. Ce qui nous permet de r\'eexprimer la fonction de partition sur le
chemin correspondant comme
\beq
\begin{array}{rcl}
\CZ_{conv}(\kappa) &=& {1 \over N!} {\displaystyle \sum_{\sigma \in S_N} (-1)^\sigma \sum_{k_1,\dots,k_N} \sum_{l_1, \dots,l_N} \prod_{i=1}^N
\int_{\l_{1,i}\in \CC_{1,i},\l_{2,\sigma(i)} \in \CC_{2,\sigma(i)}}} \cr
&& {\displaystyle \qquad \Delta(\l_1) \Delta(\l_2) \prod_{i} \kappa_{k_i,l_{\sigma(i)}} e^{-{1 \over \hbar} \left[V_1(\l_{1,i}) + V_2(\l_{2,i})-\l_{1,i} \l_{2,\sigma(i)}\right]} d\l_{1,i} d\l_{2,i} }\cr
&=& {\displaystyle \sum_{k_1,\dots,k_N} \sum_{l_1, \dots,l_N} \prod_i \kappa_{k_i,l_i} \prod_{i=1}^N \int_{\l_{1,i}\in \CC_{1,k_i},\l_{2,i} \in \CC_{2,l_i}}} \cr
&& {\displaystyle \qquad \Delta(\l_{1,i}) \Delta(\l_{2,i}) \prod_i e^{-{1 \over \hbar} \left[V_1(\l_{1,i})+V_2(\l_{2,i})- \l_{1,i} \l_{2,i}\right]} d\l_{1,i} d\l_{2,i}} .\cr
\end{array}
\eeq

La m\'ethode de polyn\^{o}mes orthogonaux peut se g\'en\'eraliser directement \`{a} ce cadre en changeant simplement
le chemin d'int\'egration dans la condition d'orthogonalit\'e des polyn\^{o}mes:
\beq
\int_{\CC^2} dx dy \pi_n(x) \sigma_n(y) e^{-{1 \over \hbar} \left[  V_1(x) + V_2(y) -xy \right]} = h_n \delta_{nm}.
\eeq

\subsection{Int\'egrale formelle.}

Si les deux d\'efinitions donn\'ees pr\'ec\'edemment sont intimement li\'ees, l'objet de notre \'etude est d'une toute autre nature,
inspir\'ee par les techniques de calcul de th\'eorie des champs utilis\'ees par les physiciens. On la d\'efinit comme
une {\bf int\'egrale de matrice formelle}, i.e. une s\'erie formelle dont chacun des termes est bien d\'efini et
correspond \`a la fonction g\'en\'eratrice de cartes bicolori\'ees.

Dans un premier temps, je t\^acherai d'expliquer comment des approximations (non rigoureusement justifi\'ees) inspir\'ees
de la physique ont naturellement men\'e \`{a} voir les int\'egrales de matrices comme des objets combinatoires. Je
m'appliquerai ensuite \`{a} pr\'esenter la d\'efinition tout \`{a} fait rigoureuse de ce que l'on entend par {\bf int\'egrale formelle}.

\subsubsection{Lien intuitif entre int\'egrale de matrice et combinatoire.}

Na\"{i}vement, pour calculer une int\'egrale de la forme
$I := \int_\mathbb{R} dx \, e^{-{1 \over \hbar} V(x)}$ o\`{u} $V$ est un polyn\^{o}me, un physicien est tent\'e d'utiliser la m\'ethode dite du col.
Celle-ci consiste, en gros, \`{a} dire que l'int\'egrale est domin\'ee par la valeur de l'int\'egrand en son maximum,
le reste \'etant consid\'er\'e comme une pertubation sous dominante du fait de la d\'ecroissance exponentielle de
$e^{-{1 \over \hbar} V(x)}$. Plus pr\'ecis\'ement, on choisit un "point col" $s_i$ comme solution de:
\beq
V'(s_i) = 0.
\eeq
Le d\'eveloppement de Taylor du potentiel au voisinage  de ce col peut alors s'\'ecrire:
\beq
V(x+ s_i) = V(s_i) + g x^2 + \delta V(x)
\eeq
o\`{u} $\delta V(x)$ ne contient que des termes de degr\'e $\geq 3$ en $x$ et $g ={ V''(s_i) \over 2}$. Le changement de variable
$x \to x+s_i$ donne alors:
\beq
I = e^{-{1 \over \hbar} V(s_i)} \int_\mathbb{R} dx e^{-{g \over \hbar} x^2} e^{-{1 \over \hbar} \delta V(x)}.
\eeq
Le d\'eveloppement de Taylor de la partie non quadratique implique alors:
\beq
I = e^{-{1 \over \hbar} V(s_i)} \int_\mathbb{R} \sum_k {(\delta V(x))^k \over k!} e^{-{g \over \hbar} x^2} dx
\eeq
qui peut \^{e}tre ramen\'ee \`{a} une somme d'int\'egrales gaussiennes si l'on inverse la sommation et l'int\'egration:
\beq
I = e^{-{1 \over \hbar} V(s_i)} \sum_k  \int_\mathbb{R} {(\delta V(x))^k \over k!} e^{-{g \over \hbar} x^2} dx.
\eeq
Si l'on repr\'esente chaque int\'egrale $\int_\mathbb{R} x^k e^{-{g \over \hbar} x^2} dx$ par un vertex \`{a} $k$
pattes, le th\'eor\`{e}me de Wick\footnote{Voir l'appendice \ref{appwick} d\'edi\'e \`a ce th\'eor\`eme dans le cadre des int\'egrales
de matrices gaussiennes.} nous dit que $I$ est alors la fonction g\'en\'eratrice de
tous les graphes compos\'es de vertex de valence au moins 2 et au plus $deg(V)$ dont les poids respectifs d\'ependent des coefficients
du potentiel $V$ et du point col autour duquel on d\'eveloppe.

Comment s'\'etend une telle m\'ethode dans le cas qui nous int\'eresse?

En premier lieu, on doit d\'eterminer les extrema $(\CM_1,\CM_2)$ de l'action
\beq
\CS(M_1,M_2) = V_1(M_1) + V_2(M_2) -  M_1 M_2.
\eeq
Ceux-ci sont donn\'es par les solutions du syst\`eme
\beq\label{col1}
\left\{ \begin{array}{l}
V_1'(\CM_1) = \CM_2 \cr
V_2'(\CM_2) = \CM_1 \cr
\end{array}
\right. .
\eeq
Sans perte de g\'en\'eralit\'e, on peut consid\'erer que ces matrices sont des matrices diagonales dont les valeurs propres
sont solutions du syst\`eme (\ref{col1}) vu comme un syt\`eme num\'erique.
On peut r\'esoudre  celui-ci en \'eliminant $\CM_2$ gr\^ace \`a la premi\`ere \'equation.
On obtient ainsi une \'equation polynomiale de degr\'e $d_1 d_2$ en $\CM_1$
\beq
V_2'\left(V_1'(\CM_1)\right) = \CM_1.
\eeq
Il existe donc de mani\`{e}re g\'en\'erique $d_1 d_2$ "points cols" $(\xi_i,\eta_i)$\footnote{Chaque point col est un couple
de nombres complexes.} solution de cette \'equation o\`{u} $\xi_i$ est une valeur propre de $\CM_1$ et $\eta_i$ la valeur
popre de $\CM_2$ correspondante. Pour construire un point col dans l'espace des couples de matrices hermitiennes $H_N\times H_N$, il faut alors distribuer
leurs $N$ valeurs propres entre ces diff\'erents points cols:
\beq
\CM_1 = {\rm diag}\,(\mathop{\overbrace{\xi_1,\dots,\xi_1}}^{n_1\,{\rm fois}},\mathop{\overbrace{\xi_2,\dots,\xi_2}}^{n_2\,{\rm fois}},\dots,\mathop{\overbrace{\xi_{d_1 d_2},\dots,\xi_{d_1 d_2}}}^{n_{d_1 d_2}\,{\rm fois}})
\eeq
et
\beq
\CM_2 = {\rm diag}\,(\mathop{\overbrace{\eta_1,\dots,\eta_1}}^{n_1\,{\rm fois}},\mathop{\overbrace{\eta_2,\dots,\eta_2}}^{n_2\,{\rm fois}},\dots,\mathop{\overbrace{\eta_{d_1 d_2},\dots,\eta_{d_1 d_2}}}^{n_{d_1 d_2}\,{\rm fois}})
\eeq
avec $\sum_{i=1}^{d_1 d_2} n_i = N$.  Ainsi, un point col est caract\'eris\'e par des {\bf fractions de remplissage} ${n_i \over N}$ qui repr\'esentent la proportion de valeurs propres $(\CM_1,\CM_2)$ situ\'ees
au point col $(\xi_i,\eta_i)$: on a choisi la configuration o\`{u} le comportement de $(\xi_i,\eta_i)$ est dominant pour
$n_i$ valeurs propres. En fixant une configuration, on a bris\'e la sym\'etrie $U(N)$ de l'int\'egrale en la
sym\'etrie $\prod_{i=1}^{d_1 d_2} U(n_i)$ chaque \'el\'ement  du produit se rapportant \`{a} la sym\'etrie entre les
valeurs propres au voisinage du m\^eme point col.

\br
On consid\`{e}re tous les extrema de l'action et non pas seulement ses minima comme on devrait intuitivement le faire.
\er

\br
On peut voir le syst\`{e}me d'\'equations \ref{col1} comme la condition d'\'equilibre des valeurs propres de $\CM_1$
et $\CM_2$ soumis au potentiels $V_1(x) + V_2(y) -xy$.
Le choix de fractions de remplissage consiste \`{a} distribuer les valeurs propres dans les diff\'erents puits form\'es par
ce potentiel (voir l'exemple \ref{ex1}).
\er

\br
Notons que ces fractions de remplissage ont une influence sur le r\'esultat de l'approximation du col.
Nous verrons que ce sont effectivement des param\`{e}tres qu'il faut se donner pour d\'efinir l'int\'egrale formelle  de mani\`{e}re unique.
On retrouvera ces param\`{e}tres tout au long de ce chapitre et il est important de garder \`{a} l'esprit que ce sont des
donn\'ees du probl\`{e}me au m\^{e}me titre que les potentiels $V_1$ et $V_2$.
\er

\bex \label{ex1}

Consid\'erons les potentiels
\beq
\left\{
\begin{array}{l}
V_1(x) = x^3-x \cr
V_2(y) = {y^3 \over 3} - 1.01 y \cr
\end{array} \right.
.
\eeq

La surface $z = V_1(x) +V_2(y) - xy$ est alors:
\beq
\begin{array}{r}
{\rm \includegraphics[width=14cm]{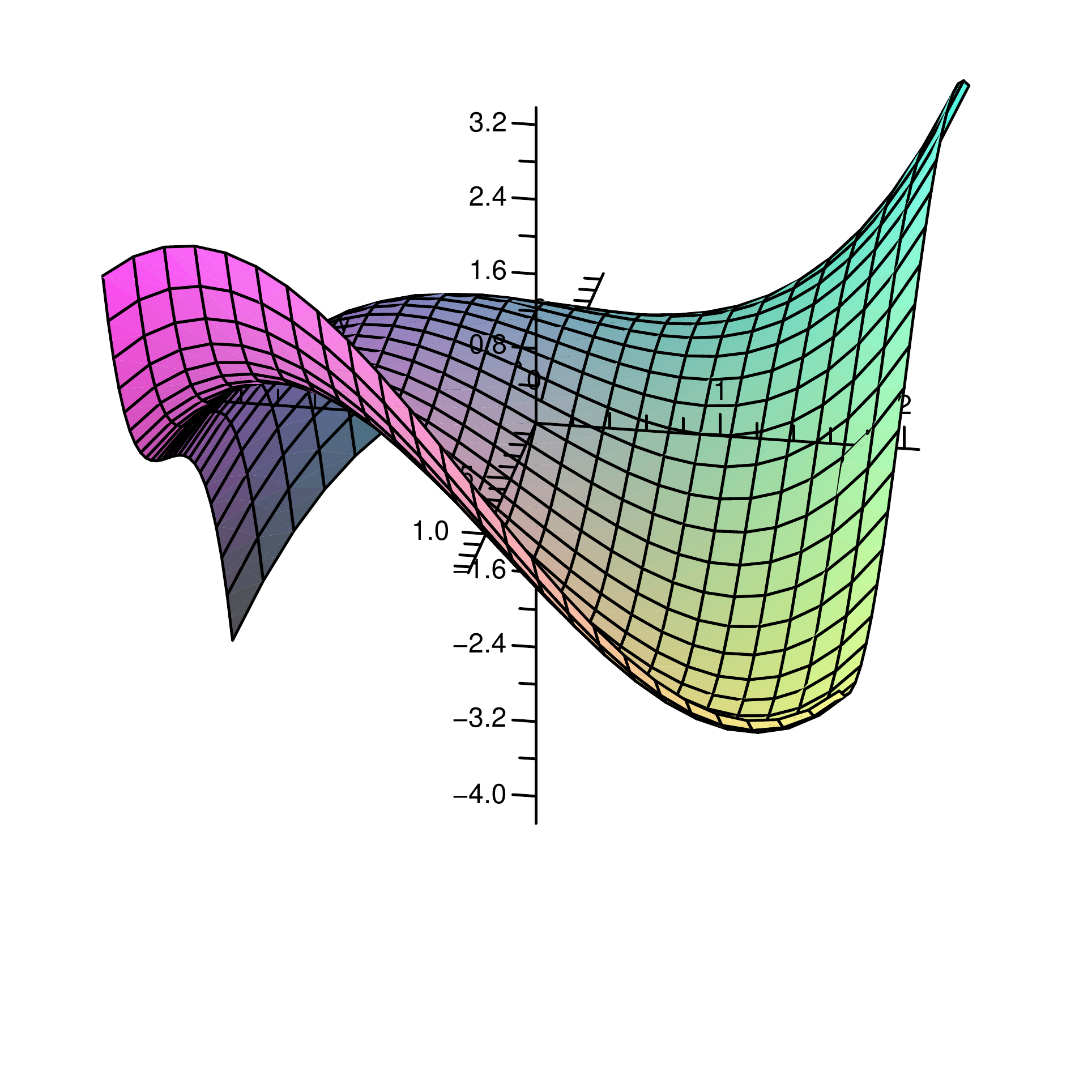}}
\end{array}
\eeq
et l'on  a quatre points cols:
\bea
p_1 = (0,8904;1,3786) \virg p_2 = (-0,0107;-0,9997) \cr
 p_3 = (-0,7168;0,5415) \virg p_4=(-0,1629;-0,9204) . \cr
\eea
On peut voir que parmi ceux-ci, seul $p_1$ est un vrai minimum alors que $p_2$ et $p_3$ sont des points selle
et $p_4$ est un maximum.

\eex

Une fois les fractions de remplissage fix\'ees, l'approximation du col consiste \`{a} effectuer un d\'eveloppement autour du col correspondant en r\'e\'ecrivant
les matrices \`{a} int\'egrer comme une perturbation autour de celui-ci:
\beq
M_1 = \CM_1 + m_1 \virg M_2 = \CM_2 + m_2
\eeq
puis \`{a} consid\'erer que l'int\'egrale est domin\'ee par la valeur de l'int\'egrand en ce point. Le d\'eveloppement de
Taylor de l'action au voisinage de ce point col permet de mettre \`{a} part sa partie quadratique.
Si l'on commute les signes d'int\'egration et de sommation du d\'eveloppement de Taylor, on peut r\'e\'ecrire cette int\'egrale comme une s\'erie
dont chacun des terme est une fonction de corr\'elation par rapport \`{a} une mesure gaussienne:
\beq
\CZ = e^{-{1 \over \hbar}\Tr \left[ V_1(\CM_1) + V_2(\CM_2) - \CM_1\CM_2 \right]}
\sum_{k,l} {(- \hbar)^{-k+l} \over k! l!}  \left<\left[\Tr \delta_{\CM_1} V_1(m_1)\right]^k \left[\Tr \delta_{\CM_2} V_2(m_2)\right]^l\right>_{g}
\eeq
o\`u $\delta_{\CM_1} V_1(m_1)$ et $\delta_{\CM_2} V_2(m_2)$ repr\'esentent les parties non quadratiques de $V_1$ et $V_2$ dans
le d\'eveloppement de Taylor autour du point col et $<.>_g$ la valeur moyenne par rapport \`a une mesure gaussienne.
Le th\'eor\`{e}me de Wick (voir l'appendice \ref{appwick}) et la repr\'esentation du r\'esultat
sous forme de diagrammes de Feynman impliquent alors que $\CZ$ est la fonction g\'en\'eratrice de graphes ferm\'es colori\'es
portant une structure de spin.

C'est cette approche que nous allons rendre rigoureuse dans le paragraphe suivant pour d\'efinir les int\'egrales de matrices
formelles.

%
%


\subsubsection{D\'efinition de l'int\'egrale de matrices formelle.}

En se basant sur l'intuition fournie par le paragraphe pr\'ec\'edent, on d\'efinit l'int\'egrale de matrice formelle par la
fonction de partition r\'esultant de la d\'ecomposition de l'int\'egrale convergente en int\'egrales gaussiennes par d\'eveloppement
en s\'erie de Taylor au voisinage d'un point col. Pour cela, il est n\'ec\'essaire de donner un sens \`{a} ce d\'eveloppement de Taylor
en pr\'ecisant l'interaction entre les diff\'erentes solutions du syst\`{e}me d'\'equations num\'eriques:
\beq
\left\{
\begin{array}{l}
V_1'(\xi_i) = \eta_i \cr
V_2'(\eta_i) = \xi_i \cr
\end{array}
\right.
\eeq

\bd
Soit un entier $N$ et deux potentiels polyn\^omiaux\footnote{On pourrait en pratique seulement demander que ces potentiels
aient une d\'eriv\'ee rationnelle et tous les r\'esultats resteront valables (voir par exemple \EOinvariants). Cependant,
par soucis de simplicit\'e je me restreind ici au cas poynomial.} $V_1$ et $V_2$ d\'efinis comme pr\'ec\'edemment par \eq{defV}.

Soit $\vec{n}$ une $d_1 d_2$-partition de $N$:
\beq
\vec{n}:= \{n_1, n_2, \dots, n_{d_1 d_2}\} \qquad \qquad \hbox{tels que} \qquad \qquad \sum_{i=1}^{d_1 d_2} n_i = N.
\eeq

Soient $\{(\xi_i,\eta_i)\}_{i=1}^{d_1 d_2}$, les $d_1 d_2$ solutions du syst\`{e}me de deux \'equations num\'eriques:
\beq
\left\{
\begin{array}{l}
V'_1(\xi_i)=\eta_i\cr
V'_2(\eta_i)=\xi_i
\end{array}\right. .
\eeq

On d\'efinit les parties non quadratiques des d\'eveloppements autour de ces points cols par
\beq
\delta V_{1,i}(x) = V_1(x) - V_1(\xi_i) - {V_1''(\xi_i)\over 2}(x-\xi_i)^2
\eeq
et
\beq
\delta V_{2,i}(y) = V_2(y) - V_2(\eta_i) - {V_2''(\eta_i)\over 2}(y-\eta_i)^2.
\eeq

Pour tout $l$, on d\'efinit le polyn\^ome en $T$:
\beq\label{Akl}
\begin{array}{l}
\sum_{k=l/2}^{dl/2} A_{k,l} T^k = \cr
= {(-1)^l N^l\over l!\, T^l}\,\int dM_1 \dots dM_d d\td{M}_1 \dots d\td{M}_d (\sum_i \Tr \delta V_{1,i}(M_i)+ \delta V_{2,i}(\td{M}_i))^l \cr
 \,\, \prod_{i=1}^{d} \ee{-{N \over T}\left(\Tr {V_1''(\xi_i)\over 2}(M_i-\xi_i\,{\bf 1}_{n_i})^2+{V_2''(\eta_i)\over 2}(\td{M}_i-\eta_i\,{\bf 1}_{n_i})^2 - (M_i-\xi_i\,{\bf 1}_{n_i})(\td{M}_i-\eta_i\,{\bf 1}_{n_i}) \right)} \cr
 \,\,\,\prod_{i>j} \det(M_i\otimes {\bf 1}_{n_j} - {\bf 1}_{n_i} \otimes M_j)
\,\,\prod_{i>j} \det(\td{M}_i\otimes {\bf 1}_{n_j} - {\bf 1}_{n_i} \otimes \td{M}_j) \cr
\end{array}
\eeq
comme une int\'egrale gaussienne sur les matrices {\bf hermitiennes} $M_i$ et $\td{M}_i$ de taille $n_i \times n_i$.

La fonction de partition du mod\`ele \`a deux matrices formel est alors d\'efinie comme la s\'erie en $T$ (\cite{eynform,Eynhab}):
\beq\label{defZform}\encadremath{
\CZ_{form} = \sum_{k=0}^\infty T^k \big( \sum_{j=0}^{2k} A_{k,j} \big).
}\eeq

\ed

\subsubsection{Lien avec l'int\'egrale normale \`{a} sym\'etrie bris\'ee.}

Ecrivons explicitement la fonction de partition formelle:
\bea
\CZ_{form}& =& \sum_{l=0}^\infty {(-1)^l N^l\over l!\, T^l}\,\int {\displaystyle \prod_{\alpha=1}^d} dM_\alpha d\td{M}_\alpha  \left(\sum_i \Tr \delta V_{1,i}(M_i)+ \delta V_{2,i}(\td{M}_i)\right)^l \cr
&& \,\, \prod_{i=1}^{d} \ee{-{N \over T}\left(\Tr {V_1''(\xi_i)\over 2}(M_i-\xi_i\,{\bf 1}_{n_i})^2+{V_2''(\eta_i)\over 2}(\td{M}_i-\eta_i\,{\bf 1}_{n_i})^2 - (M_i-\xi_i\,{\bf 1}_{n_i})(\td{M}_i-\eta_i\,{\bf 1}_{n_i})\right)} \cr
&& \,\,\,\prod_{i>j} \det(M_i\otimes {\bf 1}_{n_j} - {\bf 1}_{n_i} \otimes M_j)
\,\,\prod_{i>j} \det(\td{M}_i\otimes {\bf 1}_{n_j} - {\bf 1}_{n_i} \otimes \td{M}_j). \cr
\eea

Si l'on \'echange l'ordre de la sommation sur $l$ et de l'int\'egration sur les matrices, on reconnait le d\'eveloppement de Taylor
de
\beq
\begin{array}{l}
\int {\displaystyle \prod_{\alpha=1}^d} dM_\alpha d\td{M}_\alpha \prod_{i=1}^{d} \ee{-{N \over T}\left( \Tr V_1(M_i-\xi_i\,{\bf 1}_{n_i})+ \Tr V_2(\td{M}_i-\eta_i\,{\bf 1}_{n_i}) - (M_i-\xi_i\,{\bf 1}_{n_i})(\td{M}_i-\eta_i\,{\bf 1}_{n_i})\right)} \cr
\prod_{i>j} \det(M_i\otimes {\bf 1}_{n_j} - {\bf 1}_{n_i} \otimes M_j)
\,\,\prod_{i>j} \det(\td{M}_i\otimes {\bf 1}_{n_j} - {\bf 1}_{n_i} \otimes \td{M}_j) \cr
\end{array}
\eeq
qui par changement des variable d'int\'egration $X_i:= M_i-\xi_i\,{\bf 1}_{n_i}$ et $Y_i:=\td{M}_i-\eta_i\,{\bf 1}_{n_i}$
prend la forme de la fonction de partition du {\bf mod\`{e}le \`{a} deux matrices \`{a} sym\'etrie bris\'ee} \cite{Eynhab}:
\beq
\begin{array}{rcl}
\CZ_{normb} &:=&
\int dX_1 \dots dX_d dY_1 \dots  dY_d \prod_{i=1}^{d} \ee{-{N \over T}\left( \Tr V_1(X_i)+ \Tr V_2(Y_i) - X_i Y_i)\right)} \cr
&& \prod_{i>j} \det(X_i\otimes {\bf 1}_{n_j} - {\bf 1}_{n_i} \otimes X_j)
 \,\,\prod_{i>j} \det(Y_i\otimes {\bf 1}_{n_j} - {\bf 1}_{n_i} \otimes Y_j). \cr
\end{array}
\eeq
Cette derni\`{e}re consiste \`{a} briser la sym\'etrie $U(N)$ en une sym\'etrie $\prod_i U(n_i)$ en consid\'e- rant l'int\'egrale
matricielle comme l'int\'egrale sur $d_1 d_2$ ensembles de couples de matrices $(X_i,Y_i)$ dont les valeurs propres
appartiennent respectivement \`{a} des couples de contours $(\CC_{1,i},\CC_{2,i})$ acceptables au sens de \eq{chemin}.
Ainsi, par opposition au mod\`{e}le de la partie \ref{secchem} o\`{u} toutes les matrices avaient des valeurs propres conditionn\'ees
\`{a} vivre sur un seul chemin acceptable $\CC:=\kappa_{ij} \, \CC_{1,i} \times \CC_{2,j} $ quelconque, les diff\'erentes
matrices ont des valeurs propres sur un chemin diff\'erent.

{\bf Attention:}

Malgr\'e cette proximit\'e entre les deux mod\`{e}les, le mod\`{e}le normal \`{a} sym\'etrie bris\'ee et le mod\`ele
formel ne coincident pas
en g\'en\'eral, ne serait-ce que par le probl\`{e}me pos\'e par l'\'echange de la sommation des termes de la s\'erie de
Taylor et de l'int\'egration gaussienne.

\br
Le lien exact entre ces diff\'erents mod\`{e}les ainsi que leurs diff\'erences sont peu connus au moment de la r\'edaction de ce
m\'emoire et vont bien au del\`{a} de la port\'ee de ce dernier. Je ne m'attarderai donc pas plus sur ceux-ci me concentrant
\`{a} pr\'esent uniquement sur le mod\`{e}le formel. Le lecteur int\'eress\'e par ces aspects peut trouver une plus longue discussion
dans \cite{Eynhab} par exemple.
\er

\br
On abusera d\'es \`{a} pr\'esent et tout au long de cette th\`{e}se de la notation intuitive:
\beq\label{noteint}
\CZ_{form} = \int_{form} e^{-{1 \over \hbar} \Tr \left( V_1(M_1) + V_2(M_2) -M_1 M_2\right)} dM_1 dM_2
\eeq
tout en gardant \`{a} l'esprit que l'int\'egrale n'est ici qu'une notation rappelant l'origine de cette fonction de partition.
On ne devra jamais oublier que ce n'est pas une int\'egrale sur l'ensemble des matrices hermitiennes \`{a} proprement parler.
\er

\subsubsection{Interpr\'etation combinatoire.}

Comme nous l'avons vu plus haut, le mod\`{e}le formel a \'et\'e d\'efini de mani\`{e}re \`{a} correspondre au d\'eveloppement
de Feynman de l'int\'egrale du mod\`{e}le normal \`{a} sym\'etrie bris\'ee. A ce titre, il est naturel d'en chercher une
interpr\'etation combinatoire comme fonction g\'en\'eratrice de graphes et, par extension, de surfaces discr\'etis\'ees.
C'est d'ailleurs cette interpr\'etation qui a motiv\'e l'introduction de ce mod\`{e}le dans le cas de l'\'etude du mod\`{e}le
d'Ising sur surface al\'eatoire \cite{KazakovIsing}, ainsi que l'engouement des physiciens pour les mod\`{e}les de matrices al\'eatoires
en g\'en\'eral dans le cadre des th\'eories de gravitation quantique.

Etant donn\'es les deux potentiels $V_1$ et $V_2$ et un point col dans l'espace $\mathbb{C}^N\times \mathbb{C}^N$, c'est-\`{a}-dire l'ensemble des $d_1 d_2$ solutions
$(\xi_i,\eta_i)$ du syst\`{e}me
\beq
\left\{
\begin{array}{l}
V_1'(\xi_i) = \eta_i \cr
V_2'(\eta_i) = \xi_i \cr
\end{array}
\right.
\eeq
et un ensemble de fractions de remplissage
$\{\epsilon_i\}_{i=1, \dots, d_1 d_2}$ telles que
\beq
\sum_{i=1}^{d_1 d_2} \epsilon_i = 1,
\eeq
on d\'efinit les poids:
\bd
Pour tout $i = 1, \dots , d_1 d_2$, on d\'efinit:
\beq
\forall k=1, \dots, d_1+1 \, , \;\; t_{k,i}:= {V_1^{(k)}(\xi_i) \over (k-1)!}
\eeq
et
\beq
\forall k=1, \dots, d_2+1 \, , \;\; \tilde{t}_{k,i}:= {V_2^{(k)}(\eta_i) \over (k-1)!}.
\eeq
Pour tout couple $i,j = 1, \dots , d_1 d_2$ avec $i\neq j$ et tout couple $k,l = 1, \dots, \infty$, on d\'efinit:
\beq
h_{k,i}:= \sum_{j\neq i} {\epsilon_j \over (\xi_j-\xi_i)^k}
\virg
\tilde{h}_{k,i}:= \sum_{j\neq i} {\epsilon_j \over (\eta_j-\eta_i)^k},
\eeq
\beq
h_{k,i;l,j}:= {(k+l-1)! \over (k-1)! (l-1)!} {1 \over (\xi_j-\xi_i)^k(\xi_i-\xi_j)^l}
\eeq
et
\beq
\tilde{h}_{k,i;l,j}:= {(k+l-1)! \over (k-1)! (l-1)!} {1 \over (\eta_j-\eta_i)^k(\eta_i-\eta_j)^l}.
\eeq
\ed

\br
Les poids $t_{k,i}$ (resp. $\td{t}_{k,i}$) ne sont rien d'autre que les coefficients de la partie non-quadratique du d\'eveloppement
de Taylor de $V_1$ (resp. $V_2$) autour du point $\xi_i$ (resp. $\eta_i$) introduite dans le paragraphe
pr\'ec\'edent:
\beq
\delta V_{1,i}(x) = \sum_{k = 3}^{d_1 +1} {t_{k,i}\over k} (x- \xi_i)^k
\virg
\delta V_{2,i}(y) = \sum_{k = 3}^{d_2 +1} {\td{t}_{k,i}\over k} (y- \eta_i)^k,
\eeq
alors que les $h_{k,i;l,j}$ (resp. $\td{h}_{k,i;l,j}$) viennent du d\'eveloppement de
$\prod_{i>j} \det(M_i\otimes {\bf 1}_{n_j} - {\bf 1}_{n_i} \otimes M_j)$ (resp. $\prod_{i>j} \det(\td{M}_i\otimes {\bf 1}_{n_j} - {\bf 1}_{n_i} \otimes \td{M}_j)$)
dans \eq{Akl}.
\er

Ces poids nous permettent d'associer une valeur \`{a} chaque graphe de l'ensemble:
\bd\label{definterpretsurf}
Soit $\CG$ l'ensemble des graphes ferm\'es form\'es de vertex \'epais\footnote{Un vertex \'epais est un vertex dont les pattes
sont des rubans comme pr\'esent\'es dans l'appendice \ref{appwick}. Ce sont donc des vertex orient\'es.} de valence $k \geq 1$, portant une "couleur"
$i = 1, \dots, d_1 d_2$ et un "spin" (i.e. un signe + ou -) et coll\'es selon les prescriptions suivantes:
\begin{itemize}
\item Deux vertex peuvent \^{e}tre coll\'es par leurs pattes si et seulement si ils ont la m\^{e}me couleur;

\item Deux vertex peuvent \^{e}tre coll\'es par leurs centres si et seulement si ils sont de m\^eme spin et de couleur diff\'erente.

\end{itemize}

\ed

On peut r\'esumer les \'el\'ements apparaissant dans la construction de ces graphes par le tableau suivant:
\vs

\begin{tabular}{|c|c|}
\hline
{\bf El\'ement} & {\bf Figure} \\
\hline
vertex $k$-valent de spin + et couleur $i$ &
$\begin{array}{r}
{\rm \includegraphics[width=3.2cm]{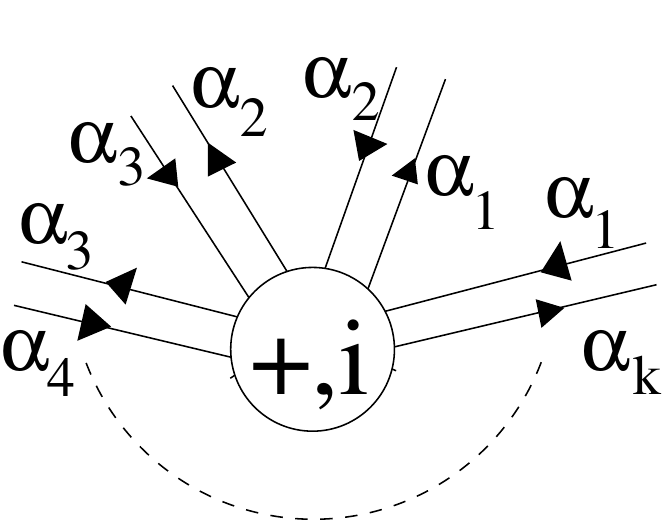}}
\end{array}$ \\
\hline
vertex $k$-valent de spin - et couleur $i$ &
$  \begin{array}{r}
{\rm \includegraphics[width=3.2cm]{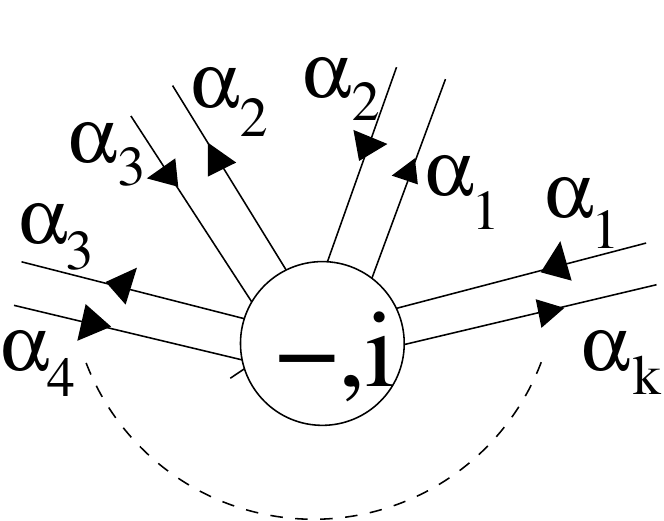}}
\end{array}$\\
\hline
\begin{tabular}{c}
lien entre deux pattes de vertex \\ de spin + et couleur i\\ \end{tabular} & $  \begin{array}{r}
{\rm \includegraphics[width=3.2cm]{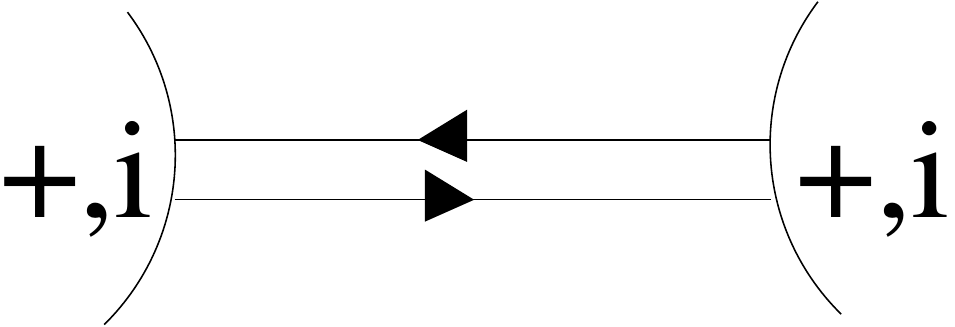}}
\end{array}$ \\
\hline
\begin{tabular}{c}
lien entre deux pattes de vertex \\ de spin - et couleur i\\ \end{tabular} & $  \begin{array}{r}
{\rm \includegraphics[width=3.2cm]{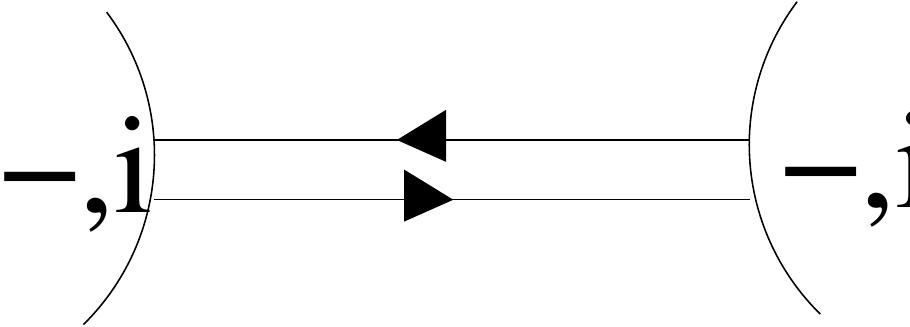}}
\end{array}$ \\
\hline
\begin{tabular}{c}
lien entre deux pattes de vertex \\ de couleur i et spins diff\'erents \\ \end{tabular} & $  \begin{array}{r}
{\rm \includegraphics[width=3.2cm]{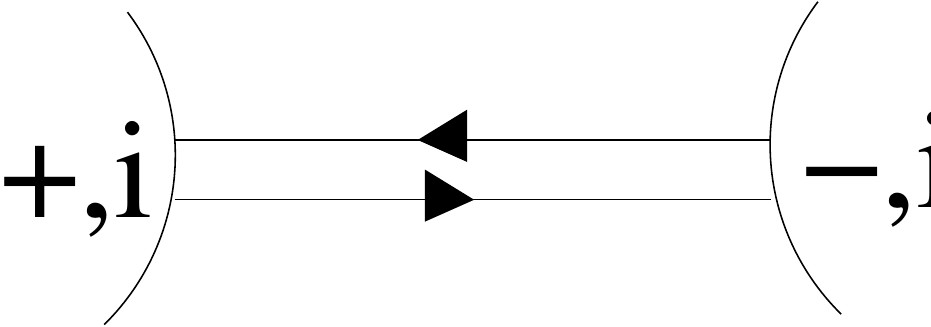}}
\end{array}$ \\
\hline
\begin{tabular}{c}
lien entre les centres d'un vertex $k$-valent\\ de couleur $i$ et de spin +
et d'un vertex\\ $l$-valent de couleur $j$ et de spin + \\
\end{tabular} & $  \begin{array}{r}
{\rm \includegraphics[width=3.2cm]{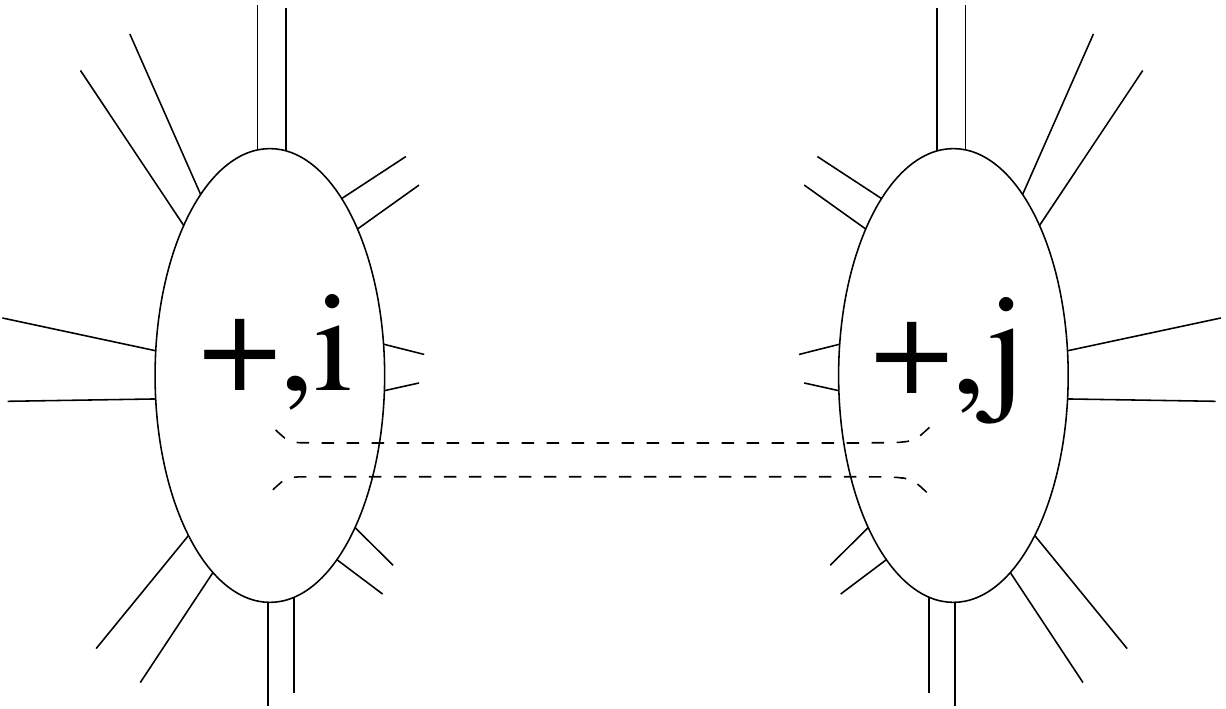}}
\end{array}$ \\
\hline
\begin{tabular}{c}
lien entre les centres d'un vertex $k$-valent\\ de couleur $i$ et de spin -
et d'un vertex\\ $l$-valent de couleur $j$ et de spin - \\
\end{tabular} & $  \begin{array}{r}
{\rm \includegraphics[width=3.2cm]{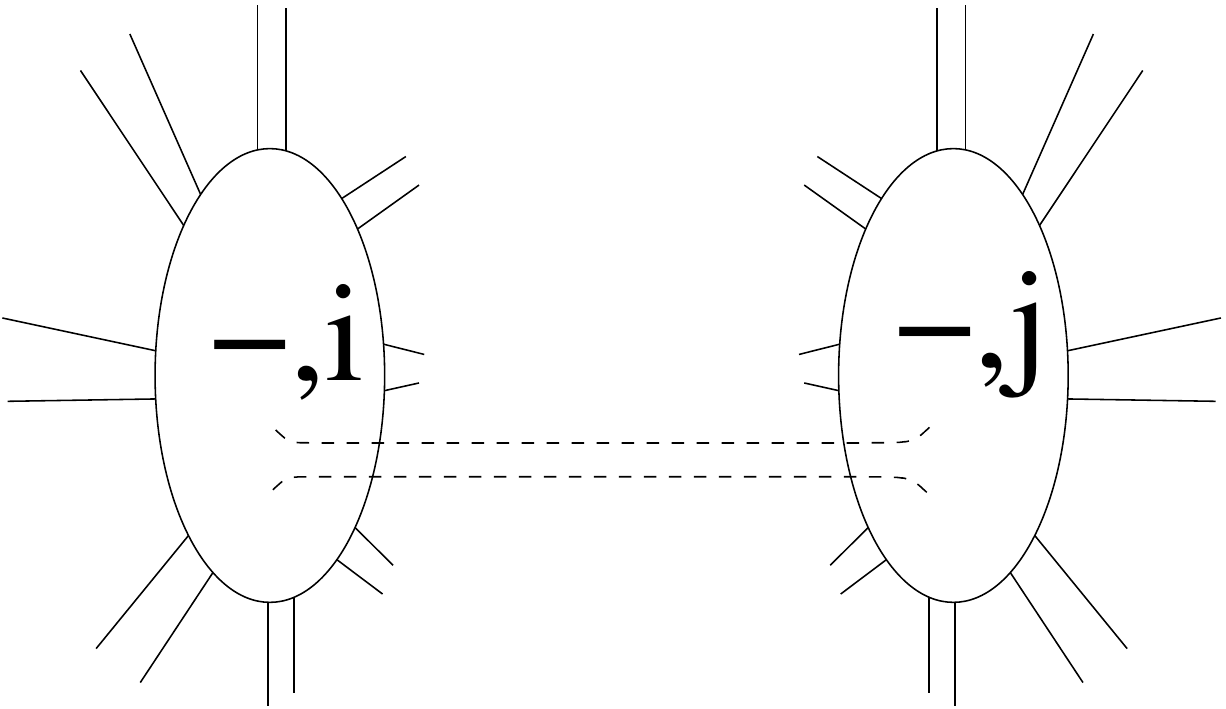}}
\end{array}$ \\
\hline
Boucle ind\'ependante de couleur $i$ & $  \begin{array}{r}
{\rm \includegraphics[width=3.2cm]{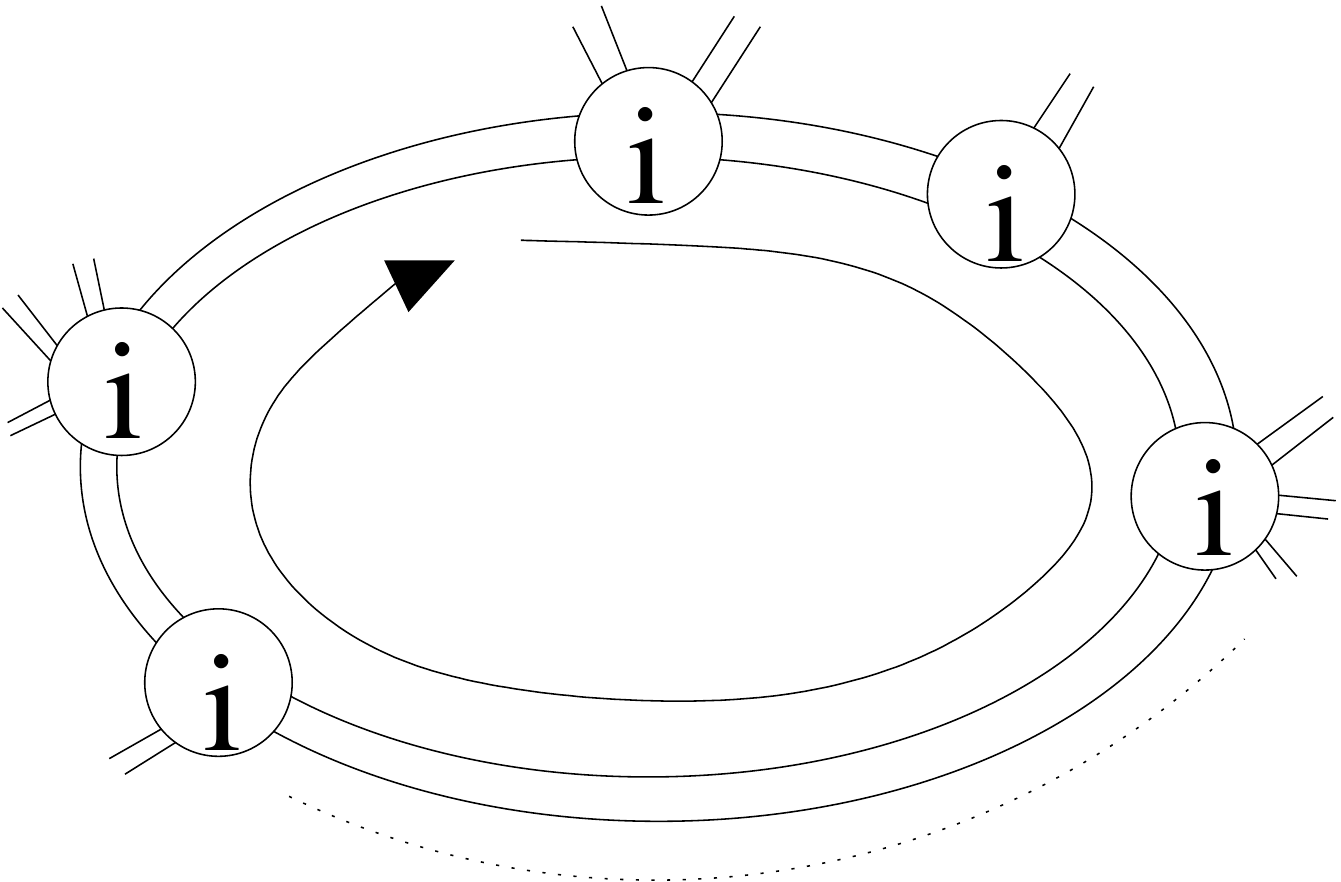}}
\end{array}$ \\
\hline
\end{tabular}
\vs

Un tel graphe peut \^etre vu comme un ensemble de $d_1 d_2$ graphes non colori\'es construits comme dans l'appendice \ref{appwick} et li\'es
entre eux par le centre de certains vertex de m\^{e}me spin. Ainsi, pour obtenir tous les graphes de $\CG$, on
commence par dessiner tous les graphes ferm\'es form\'es de vertex \'epais de valence $k \geq 1$ et portant chacun un spin et
coll\'es entre eux par leurs pattes. On recolle alors de toutes les mani\`{e}res possibles $d_1 d_2$ de ces graphes
(en autorisant la pr\'esence de plusieurs exemplaires du m\^eme graphe et le graphe vide) par le centre de vertex de m\^eme spin.

Pour caract\'eriser un graphe de $\CG$, on utilise  les notations suivantes:

\bd
Soit un graphe $G \in \CG$. On note:
\begin{itemize}
\item $n_{k,i}(G)$ := nombre de vertex de valence k, de spin + et de couleur $i$ dont le centre est libre\footnote{Un centre est
dit libre si il n'est pas reli\'e \`a un autre centre par un lien.};

\item $\tilde{n}_{k,i}(G)$ := nombre de vertex de valence k, de spin - et de couleur $i$ dont le centre est libre;

\item $n_{++,i}(G)$ := nombre d'ar\^etes liant deux vertex de couleur $i$ et de spin $+$ par leurs pattes;

\item $n_{--,i}(G)$ := nombre d'ar\^etes liant deux vertex de couleur $i$ et de spin $-$ par leurs pattes;

\item $n_{+-,i}(G)$ := nombre d'ar\^etes liant deux vertex de couleur $i$ et de spins diff\'erents par leurs pattes;

\item $n_{a,i}(G) = n_{+-,i}(G) + n_{++,i}(G) + n_{--,i}(G)$ := nombre total d'ar\^etes liant deux vertex de couleur $i$ par leurs pattes quelque soit leurs spins;

\item $n_{k,i;l,j}(G)$ := nombre de paires de vertex de spin + coll\'es par leurs centres de valences et de couleurs
 respectives $(k,i)$ et $(l,j)$\footnote{Le nombre total de vertex de spin + est donn\'e par $\sum_{k,l,i,j} 2 n_{k,i;l,j}(G)+n_{k,i}(G)$.};

\item $\tilde{n}_{k,i;l,j}(G)$ := nombre de paires de vertex de spin - coll\'es par leurs centres de valences et de couleurs
 respectives $(k,i)$ et $(l,j)$\footnote{Le nombre total de vertex de spin - est donn\'e par $\sum_{k,l,i,j} 2 \tilde{n}_{k,i;l,j}(G)+\tilde{n}_{k,i}(G)$.};

\item $G_i$ := composante de couleur $i$ du graphe obtenue en brisant tous les liens entre centres de polygones;

\item $l_i(G)$ := nombre de boucles ind\'ependantes dans le graphe $G_i$;

\item $\hbox{\#Aut}(G)$ := cardinal du groupes d'automorphismes de $G$;

\item $\chi(G)$ := caract\'eristique d'Euler-Poincar\'e de $G$:
\beq
\begin{array}{rcl}
\chi(G) &=& \sum_i \left[ l_i(G) - n_{a,i}(G) + \sum_k \left(n_{k,i}(G) + \tilde{n}_{k,i}(G)\right) \right]\cr
&=& \sum_{i=1}^{d_1 d_2} \chi(G_i) - 2 \sum_{i,j,k,l} \left(n_{k,i;l,j}(G)+ \tilde{n}_{k,i;l,j}(G)\right).\cr
\end{array}
\eeq

\end{itemize}
\ed

\bd
A chaque graphe $G \in \CG$, on associe un poids:
\beq
\begin{array}{rcl}
\CW(G) &:=& {\displaystyle {N^{\chi(G)} \over \hbox{\#Aut(G)}} T^{n_T(G)} \prod_{i=1}^{d_1 d_2} \prod_{k=3}^{d_1+1} (t_{k,i}+Th_{k,i})^{n_{k,i}(G)}
\prod_{k=3}^{d_2+1} (\tilde{t}_{k,i}+T\tilde{h}_{k,i})^{\tilde{n}_{k,i}(G)} }\cr
&& {\displaystyle \prod_{i= 1}^{d_1d_2} h_{1,i}^{n_{1,i}(G)} \tilde{h}_{1,i}^{\tilde{n}_{1,i}(G)} h_{2,i}^{n_{2,i}(G)} \tilde{h}_{2,i}^{\tilde{n}_{2,i}(G)}
\prod_{k>d_1+1} h_{k,i}^{n_{k,i}(G)} \prod_{l>d_2+1} \tilde{h}_{l,i}^{\tilde{n}_{l,i}(G)} }\cr
&& {\displaystyle \prod_{i=1}^{d_1 d_2} \epsilon_i^{l_i(G)} t_{2,i}^{n_{++,i}(G)} \tilde{t}_{2,i}^{n_{--,i}(G)} (t_{2,i} \tilde{t}_{2,i}-1)^{-n_{v,i}(G)}
\prod_{j>i} \prod_k \prod_l h_{k,i;l,j}^{n_{k,i;l,j}(G)} \tilde{h}_{k,i;l,j}^{\tilde{n}_{k,i;l,j}(G)} }\cr
\end{array}
\eeq
o\`{u}
\beq
\begin{array}{rcl}
n_T(G)&:=& \sum_i\left( \sum_k {k \over 2} (n_{k,i}(G)+\tilde{n}_{k,i}(G)) - \sum_{k=3}^{d_1+1} n_{k,i}(G) - \sum_{l=3}^{d_12+1} \tilde{n}_{l,i}(G)\right) \cr
&& + \sum_i \sum_{j>i} \sum_k \sum_l {k+l \over 2} (n_{k,i;l,j}(G) + \tilde{n}_{k,i;l,j}(G)) .\cr
\end{array}
\eeq
\ed

Associer un tel poids \`{a} un graphe $G$ correspond \`{a} associer un poids \`{a} chacun des \'el\'ements composant le graphe
comme suit:
\vs

\begin{tabular}{|c|c|}
\hline
{\bf El\'ement} & {\bf Poids} \\
\hline
\begin{tabular}{c}
vertex $k$-valent de spin + et couleur $i$ \\
dont le centre est libre\\
\end{tabular}&
  $\left\{ \begin{array}{l} {N \over T} t_{k,i} + N h_{k,i} \; \hbox{si} \; k \in [3,d_1 +1]  \cr N h_{k,i} \; \hbox{sinon} \end{array} \right.$\\
\hline
\begin{tabular}{c}
vertex $k$-valent de spin - et couleur $i$ \\
dont le centre est libre\\ \end{tabular}&
  $\left\{ \begin{array}{l} {N \over T} \tilde{t}_{k,i} + N h_{k,i} \; \hbox{si} \; k \in [3,d_2 +1]  \cr N h_{k,i} \; \hbox{sinon} \end{array} \right.$\\
\hline
\begin{tabular}{c}
lien entre deux pattes de vertex \\ de spin + et couleur i\\ \end{tabular} & ${1 \over N}{t_{2,i} \over t_{2,i} \tilde{t}_{2,i} -1}$ \\
\hline
\begin{tabular}{c}
lien entre deux pattes de vertex \\ de spin - et couleur i\\ \end{tabular} & ${1 \over N}{\tilde{t}_{2,i} \over t_{2,i} \tilde{t}_{2,i} -1}$ \\
\hline
\begin{tabular}{c}
lien entre deux pattes de vertex \\ de couleur i et spins diff\'erents \\ \end{tabular} & ${1 \over N}{1 \over t_{2,i} \tilde{t}_{2,i} -1}$ \\
\hline
\begin{tabular}{c}
Paire compos\'ee d'un vertex $k$-valent\\ de couleur $i$ et de spin + \\
et d'un vertex $l$-valent\\ de couleur $j$ et de spin + \\
\end{tabular} & $h_{k,i;l,j}$ \\
\hline
\begin{tabular}{c}
Paire compos\'ee d'un vertex $k$-valent\\ de couleur $i$ et de spin - \\
et d'un vertex $l$-valent\\ de couleur $j$ et de spin - \\
\end{tabular} & $\tilde{h}_{k,i;l,j}$ \\
\hline
Boucle ind\'ependante de couleur $i$ & $N \epsilon_i$ \\
\hline
\end{tabular}
\vs

\bex
Consid\'erons un exemple de graphe apparaissant lorsque l'on a deux couleurs: bleu, repr\'esent\'e par l'indice 1 et rouge repr\'esent\'e
par l'indice 2:
\beq \label{exdiag}
\begin{array}{r}
{\rm \includegraphics[width=7cm]{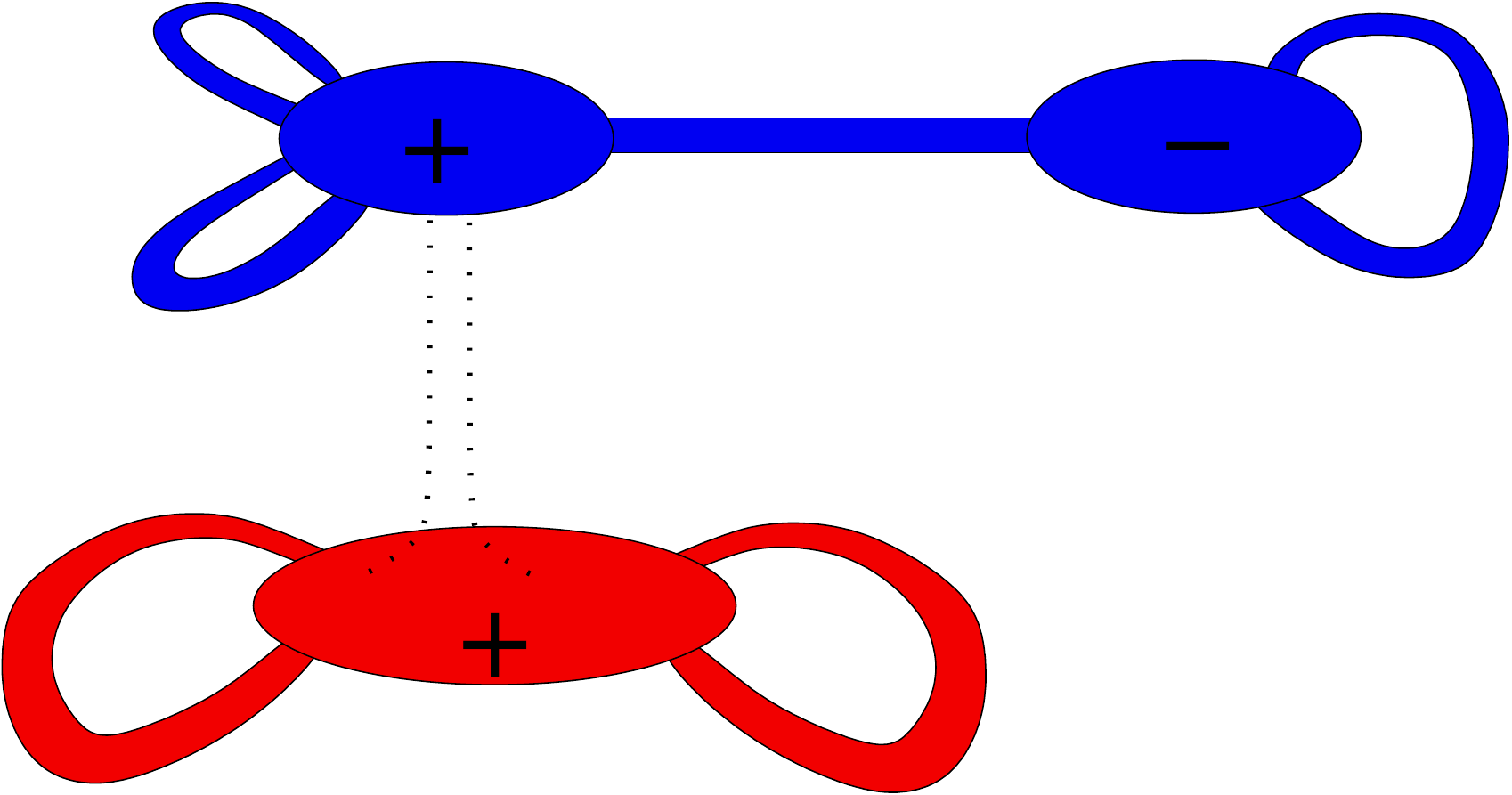}}
\end{array}
\eeq

Il est constitu\'e d'un vertex bleu de valence 5 et de spin +, d'un vertex bleu de valence 3 et de spin - et d'un
vertex rouge de valence 4 et de spin + reli\'es entre eux par 6 ar\^etes formant $3+4$ boucles ind\'ependantes, i.e.:
\beq
\begin{array}{l}
\tilde{n}_{3,1} = n_{--,1} = n_{+-,1}=n_{5,1;4,2} = 1
\virg \cr
n_{++,1} = \tilde{n}_{++,2} = 2
\virg
l_1 = 4
\;\;\;\;\;
\hbox{et}
\;\;\;\;\;
l_2 = 3.
\end{array}
\eeq
et son poids est donc donn\'e par:
\beq
N^2 T^{15 \over 2} (\tilde{t}_{3,1} + T \tilde{h}_{3,1})\epsilon_1^4 \epsilon_2^3t_{2,1}^2t_{2,2}^2
\tilde{t}_{2,1} (t_{2,1}\tilde{t}_{2,1} -1)^4 (t_{2,2}\tilde{t}_{2,2} -1)^2 h_{5,1;4,2}.
\eeq

\eex

Ces poids nous permettent de controler la quantit\'e de chaque \'el\'ement dans un graphe, et l'on d\'efinit
la fonction g\'en\'eratrice correspondante comme:
\bt\label{thcombi}
La {\bf fonction de partition du mod\`{e}le \`{a} deux matrices hermitiennes formel} est \'egale \`{a}
\beq\label{eqZ}\encadremath{
\CZ_{form} = \prod_i e^{-{N^2 \over T} \epsilon_i ( V_1(\xi_i) + V_2(\eta_i) - \xi_i \eta_i)} \prod_{j>i} \left[(\xi_i-\xi_j)(\eta_i-\eta_j)\right]^{N^2 \epsilon_i \epsilon_j}
\sum_{G\in\CG} \CW(G).}
\eeq
\et

\proof{
Ce th\'eor\`{e}me r\'esulte directement du th\'eor\`{e}me de Wick d\'ecrit en appendice.

On peut cependant remarquer que cette fonction est bien une s\'erie formelle en puissances de $T$. En effet, l'in\'egalit\'e:
\beq
n_T(G) \geq {1 \over 2} \sum_i \left(\sum_k (n_{k,i}+\tilde{n}_{k,i}) + \sum_{j>i} \sum_{k,l} (n_{k,i;l,j}+\tilde{n}_{k,i;l,j}) \right)
\eeq
assure que seul un nombre fini de graphes contribue \`{a} une puissance donn\'ee de $T$. On peut donc bien \'ecrire:
\beq
\CZ_{form} = \sum_{m=0}^\infty A_m T^m
\eeq
o\`{u} chaque $A_m$ est une somme finie de termes donn\'ee dans le langage du paragraphe pr\'ec\'edent par:
\beq
A_m = \sum_{j=0}^{2m} A_{m,j}
\eeq
o\`{u} les $A_{m,j}$ sont donn\'es par \eq{Akl}.}

On d\'efinit \'egalement la fonction g\'en\'eratrice des graphes connexes par la proc\'edure classique:
\bd
L'{\bf \'energie libre}:
\beq\label{F2MM}\encadremath{
\CF = - {1 \over N^2} \ln \CZ_{form}}
\eeq
est la fonction g\'en\'eratrice des graphes {\bf connexes} $G$ de $\CG$ compt\'es avec un poids $\CW(G)$.
\ed
 C'est \'egalement une s\'erie formelle en $T$ qui prend la forme:
\beq
\CF = \sum_{n=0}^\infty B_n T^n.
\eeq

\section{Combinatoire des cartes, surfaces discr\'etis\'ees.}

Nous avons d\'efini la fonction g\'en\'eratrice de graphes \'epais d\'eriv\'es des diagrammes de Feynman. On peut
ais\'ement voir que ces derniers sont en bijection avec un ensemble de cartes (ou surfaces discr\'etis\'ees dans le langage
des physiciens) colori\'ees portant une structure de spins semblable \`{a} celle d'un mod\`{e}le d'Ising \cite{David,Kazakov,KazakovIsing}.

\subsection{G\'en\'eration de surfaces discr\'etis\'ees bicolores ferm\'ees.}

En effet, on peut utiliser une description duale en rempla\c{c}ant simplement tout vertex $k$-valent par le $k$-gone
dont les c\^ot\'es sont les perpendiculaires aux pattes du vertex. Ainsi, les graphes sont remplac\'es par des surfaces
compos\'ees de polygones suivant:
\bd
Soit $\CS$ l'ensemble des surfaces ferm\'ees form\'ees de polygones orient\'es \`{a} $k \geq 1$ c\^{o}t\'es, portant une "couleur"
$i = 1, \dots, d_1 d_2$ et un "spin" (i.e. un signe + ou -) et coll\'es selon les prescriptions:
\begin{itemize}
\item Deux vertex peuvent \^{e}tre coll\'es le long de leurs ar\^etes si et seulement si ils ont la m\^{e}me couleur;

\item Deux vertex peuvent \^{e}tre coll\'es par leurs centres si et seulement si ils sont de m\^eme spin et de couleur diff\'erente.

\end{itemize}
\ed

La bijection \'el\'ement par \'el\'ement est r\'esum\'ee dans le tableau:
\vs

\begin{tabular}{|c|c|}
\hline
{\bf Diagramme de Feynmann} & {\bf Surface discr\'etis\'ee} \\
\hline
vertex $k$-valent de spin + et couleur $i$ &
  $k$-gone de spin + et couleur $i$\\
\hline
vertex $k$-valent de spin - et couleur $i$ &
 $k$-gone de spin - et couleur $i$ \\
\hline
\begin{tabular}{c}
lien entre deux pattes de vertex \\ de spin + et couleur i\\ \end{tabular} &
\begin{tabular}{c} ar\^ete commune \`{a} deux polygones \\
de spin + et couleur $i$ \end{tabular} \\
\hline
\begin{tabular}{c}
lien entre deux pattes de vertex \\ de spin - et couleur i\\ \end{tabular} &
\begin{tabular}{c} ar\^ete commune \`{a} deux polygones \\
de spin - et couleur $i$ \end{tabular} \\
\hline
\begin{tabular}{c}
lien entre deux pattes de vertex \\ de couleur i et spins diff\'erents \\ \end{tabular} &
\begin{tabular}{c} ar\^ete commune \`{a} deux polygones \\
de spin diff\'erents et couleur $i$ \end{tabular} \\
\hline
\begin{tabular}{c}
lien entre les centres d'un \\vertex $k$-valent\\ de couleur $i$ et de spin + \\
et d'un vertex $l$-valent\\ de couleur $j$ et de spin + \\
\end{tabular} &
\begin{tabular}{c}
centre commun \`{a} un $k$-gone\\ de couleur $i$ et de spin + \\
et un $l$-gone\\ de couleur $j$ et de spin + \\
\end{tabular}\\
\hline
\begin{tabular}{c}
lien entre les centres d'un \\vertex $k$-valent\\ de couleur $i$ et de spin - \\
et d'un vertex $l$-valent\\ de couleur $j$ et de spin - \\
\end{tabular} &
\begin{tabular}{c}
centre commun \`{a} un $k$-gone\\ de couleur $i$ et de spin - \\
et un $l$-gone\\ de couleur $j$ et de spin - \\
\end{tabular}\\
\hline
Boucle ind\'ependante de couleur $i$ & Sommet de couleur $i$ \\
\hline
\end{tabular}

\bex
Le diagramme \ref{exdiag} est ainsi envoy\'e sur la surface:
\beq
\begin{array}{r}
{\rm \includegraphics[width=4cm]{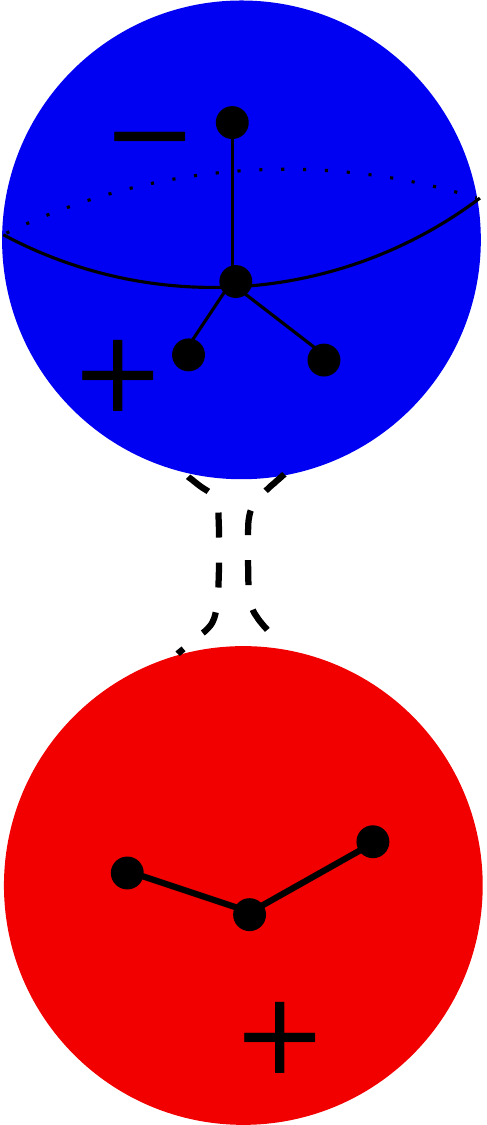}}
\end{array}
\eeq
et l'on peut v\'erifier que l'on obtient bien le m\^eme poids que pr\'ec\'edement. Notons que l'on a pu dessiner chacun des deux
graphes sur une sph\`{e}re, leurs caract\'eristiques d'Euler \'etant toutes deux \'egales \`{a} 2. On peut \'egalement voir que la
sph\`{e}re bleue est divis\'ee en deux faces de spins oppos\'es s\'epar\'ees par un \'equateur correspondant au seul lien
$+-$ du graph dual.

\eex

La fonction de partition du mod\`{e}le formel peut donc \^etre vue comme la fonction g\'en\'eratrice des surfaces de l'ensemble $\CS$,
c'est-\`{a}-dire des recollements de surfaces discr\'etis\'ees monocolores sur lesquelles vit une structure de spin,
en associant \`{a} chaque surface un poids qui peut \^{e}tre facilement retrouv\'e par l'utilisation de la bijection entre graphes
et surfaces discr\'etis\'ees. La description compl\`{e}te de cette proc\'edure est d\'ecrite dans l'appendice \ref{appsurf}.

\subsection{Surfaces ouvertes et conditions de bord.}

Il est \'egalement int\'eressant de pouvoir g\'en\'erer des surfaces ouvertes (par exemple pour des applications \`{a}
la th\'eorie des cordes ou bien \`{a} l'\'etude des th\'eories conformes ou simplement avec un objectif combinatoire d'\'enum\'eration de cartes).
Ces fonctions peuvent \^{e}tre obtenues simplement par variation des poids associ\'es aux diff\'erents \'el\'ements des graphes.
En effet, la fonction g\'en\'eratrice des surfaces dont on a enlev\'e (ou marqu\'e) un $k$-gone de spin + (resp. de spin -) et de couleur $i$
est donn\'ee par $k {\partial\over \partial t_{k,i}} \ln \CZ_{form}$ (resp. $k {\partial \over \partial \tilde{t}_{k,i}} \ln \CZ_{form}$). Or, compter toutes les
surfaces auxquelles on a enlev\'e un $k$-gone revient \`{a} compter toutes les surfaces ouvertes avec un bord de longueur
$k$ et une condition de bord impos\'ee par le spin et la couleur du $k$-gone \`a retirer.

Cependant, ces fonctions g\'en\'eratrices ne sont pas pratiques \`{a} manipuler dans les faits et par soucis de simplicit\'e ainsi
que pour des raisons historiques issues de la repr\'esentation sous forme d'int\'egrale matricielle de $\CZ_{form}$, il est pr\'ef\'erable
de travailler avec les fonctions de corr\'elations de la forme:
\beq
k_1 \dots k_m l_1 \dots l_n {\partial \over \partial t_{k_1}} \dots {\partial \over \partial t_{k_m}} {\partial\over \partial \tilde{t}_{l_1}} \dots {\partial\over \partial \tilde{t}_{l_n}} \ln \CZ_{form}
\eeq
o\`u les $t_i$ et $\tilde{t}_j$ sont les coefficients des potentiels $V_1$ et $V_2$ respectivement (voir \eq{defV}).

En effet, le poids d'un
$k$-gone de spin + (resp. de spin -) et de couleur $i$ est encod\'e dans les coefficient $t$ du potentiel $V_1$ et $\tilde{t}$
du potentiel $V_2$ par l'interm\'ediaire de $t_{k,i}$ (resp. $\tilde{t}_{k,i}$).

Il est \'egalement utile de d\'efinir les r\'esolvantes du type
\beq
\sum_{k} {k \over x^{k+1}} \partial_{t_k} \ln \CZ_{form} \virg \sum_{k} {k \over y^{k+1}} \partial_{\tilde{t}_k} \ln \CZ_{form}
\eeq
pour consid\'erer toutes les longueurs de bords possibles. Les param\`{e}tres $x$ et $y$ sont alors des fugacit\'es associ\'ees aux bords.
On a ici introduit les {\bf op\'erateurs d'insertion de boucle}\footnote{Ce nom provient du fait qu'un tel op\'erateur agit sur
les surfaces g\'en\'er\'ees en ajoutant un bord, c'est-\`a-dire en inserrant une boucle dans le graphe dual. Nous reviendrons plus longuement sur
la description de son action dans la partie \ref{partdevtopo} de ce chapitre.} consistant en une d\'eriv\'ee formelle par rapport \`{a} tous les
coefficients des potentiels $V_1$ et $V_2$ respectivement:
\beq\label{loopinsertion}\encadremath{
{\partial \over \partial V_1(x)}:= \sum_{k} {k \over x^{k+1}} \partial_{t_k}
\virg
{\partial \over \partial V_2(y)}:= \sum_{k} {k \over y^{k+1}} \partial_{\tilde{t}_k}.}
\eeq

Les fonctions de corr\'elation ainsi d\'efinies:
\beq\label{succinsert}
\overline{W}_{k,l}({\bf x_K},{\bf y_L}):= - {\partial \over \partial V_1(x_1)} \dots {\partial \over \partial V_1(x_k)} {\partial \over \partial V_2(y_1)} \dots {\partial \over \partial V_2(y_l)} \CF
\eeq
g\'en\`{e}rent donc des surfaces ouvertes \`{a} $k+l$ bords dont chacun a une condition de bord homog\`{e}ne: $k$ d'entre eux
viennent de l'extraction d'un polygone de spin + et les $l$ autres de l'extraction d'un polygone de spin -.

\br
On utilisera ici encore abondament la notation d'int\'egrale matricielle pour d\'esigner les fonctions de corr\'elation. En effet,
l'action de la d\'erivation par rapport \`{a} l'un des coefficients des potentiels est facile \`{a} repr\'esenter:
\bea
k {\partial \over \partial t_k} \ln \CZ_{form}&=&
{k\over \CZ_{form}} {\partial \over \partial t_k} \int dM_1 dM_2e^{-{1\over
\hbar} \Tr (V_1(M_1) + V_2(M_2) - M_1 M_2 )} \cr
&=& -{1 \over\hbar \CZ_{form}} \int dM_1 dM_2\, \Tr M_1^k  e^{-{1\over
\hbar} \Tr (V_1(M_1) + V_2(M_2) - M_1 M_2 )} \cr
&=&{1 \over \hbar} \left< \Tr M_1^k \right> \cr
\eea
L'action de l'op\'erateur d'insertion de boucles peut alors \^etre repr\'esent\'ee par:
\beq
{\partial \over \partial V_1(x)} \CF = - \hbar \left< \Tr {1 \over x-M_1} \right>,
\eeq
et plus g\'en\'eralement les fonctions de corr\'elations $\overline{W}_{k,l}$ sont donn\'ees par:
\beq\encadremath{
\overline{W}_{k,l}({\bf x_K}, {\bf y_L})  = \hbar^{2-k-l} \left< \prod_{i=1}^k \Tr {1 \over x_i -M_1} \prod_{j=1}^l \Tr {1 \over y_j -M_2} \right>_c,}
\eeq
o\`u l'indice $c$ signifie que l'on ne tient compte que de la partie connexe. Ces fonctions de corr\'elation sont dites "simples"
car elles ne font appara\^itre qu'un seul type de matrice ($M_1$ ou $M_2$) \`{a} l'int\'erieur de chaque trace.

Il est naturel de vouloir \'etendre cette d\'efinition \`{a} la valeur moyenne d'une classe plus grande de fonctions de $M_1$ et
$M_2$ invariantes par action du groupe $U(N)$ en m\'elangeant les deux types de matrices:
\beq\encadremath{
\begin{array}{l}
\overline{H}_{k_1,\dots,k_l;m;n}(S_1,S_2, \dots, S_l;x_1 ,\dots, x_m;y_1,\dots,y_n) := \hbar^{2-l-m-n} \cr
\;  \left<
{\displaystyle \prod_{i=1}^l } \Tr \left({1 \over x_{i,1}-M_1}{1 \over y_{i,1}-M_2} {1 \over x_{i,2}-M_1}{1 \over y_{i,2}-M_2}
\dots {1 \over x_{i,k_i}-M_1}{1 \over y_{i,k_i}-M_2}\right) \times \right.\cr
 \left. \qquad \; \; \times {\displaystyle \prod_{j=1}^m \Tr { 1 \over x_j-M_1} \prod_{s=1}^n \Tr {1 \over y_s-M_2}}
\right>_c \cr
\end{array}}
\eeq
o\`{u} $S_i$ repr\'esente la suite de longueur $2k_i$
\beq
S_i:=[x_{i,1},y_{i,1},x_{i,2},y_{i,2},x_{i,3},y_{i,3},\dots,x_{i,k},y_{i,k}]
\eeq
dans laquelle les variables de type $x$ et $y$ alternent. A cause de l'invariance cyclique de la trace, ce n'est pas $S_i$ \`a proprement
parler qui intervient dans la fonction de corr\'elation mais sa classe d'\'equivalence sous les permutations cycliques et il
sera utile de repr\'esenter cette derni\`{e}re graphiquement par
\beq\label{cycle}
S_i =
\begin{array}{r}
{\rm \includegraphics[width=3.5cm]{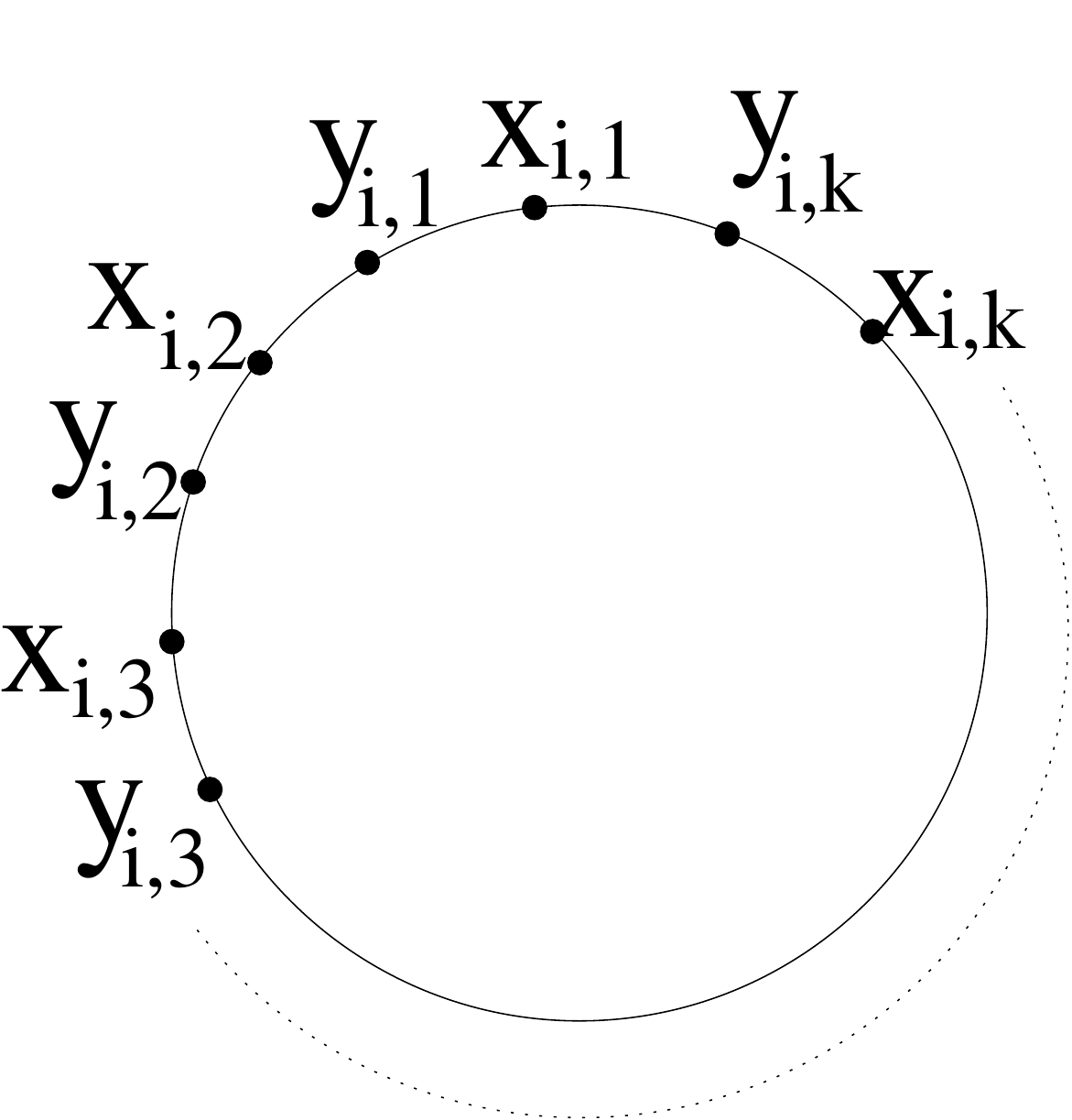}}
\end{array}
.
\eeq
Ces fonctions de corr\'elation plus compliqu\'ees sont dites "mixtes" puisqu'elles m\'elangent des matrices $M_1$ et $M_2$
\`{a} l'int\'erieur d'une m\^eme trace.

Que signifient ces nouvelles fonctions en termes combinatoires? Consid\'erons le cas le plus simple:
$\left< \Tr M_1^k M_2^l \right>$. Si l'on d\'eveloppe en diagrammes de Feynman l'int\'egrale correspondante, on g\'en\`{e}re
des surfaces qui comportent toutes un polygone \`{a} $k+l$ c\^ot\'es, $k$ d'entre eux portant un spin $+$ suivis de $l$ spin $-$:
\beq
\begin{array}{r}
{\rm \includegraphics[width=3.5cm]{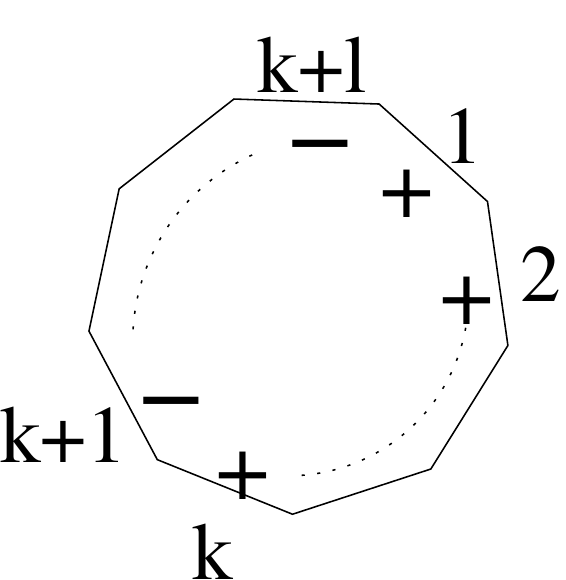}}
\end{array}
\eeq
Les surfaces g\'en\'er\'ees par cette fonction de corr\'elation sont donc des surfaces \`{a} un bord de longueur $k+l$
mais avec une condition de bord changeant deux fois. Cela signifie que, pour calculer les poids des surfaces g\'en\'er\'ees,
on fait comme si elles \'etaient entour\'ees d'une couronne de longueur $k+l$ compos\'ee de $k$ polygones de spin + successifs suivis
de $l$ polygones de spin $-$\footnote{On ne doit pas tenir compte du poids de la couronne elle m\^{e}me mais seulement des
contributions venant de l'interaction entre la courrone et la surface.}.

\er

D'un point de vue purement combinatoire, de telles fonctions de corr\'elation ne peuvent \^{e}tre simplement d\'efinies \`{a}
l'aide des poids $\CW$ consid\'er\'es jusqu'\`a pr\'esent: on doit consid\'erer une l\'eg\`{e}re g\'en\'eralisation des cartes. On doit introduire des polygones \`{a} $\sum_{\alpha=1}^K k_\alpha + \sum_{\beta=1}^L l_\beta$ c\^ot\'es de spins
diff\'erents donn\'es par la s\'equence:
\beq
\overbrace{+,+,\dots,+}^{k_1},\overbrace{-,-,\dots,-}^{l_1},\overbrace{+,\dots,+}^{k_2},\overbrace{-,\dots,-}^{l_2} \dots ,\overbrace{+,+,\dots,+}^{k_K},\overbrace{-,-,\dots,-}^{l_L},
\eeq
et de couleur $i$ avec un poids $t_{(k_1,l_1,k_2,\dots,k_K,l_L),i}$.
Les polygones consid\'er\'es jusqu\`{a} pr\'esent correspondent \`{a} $(K,L)=(1,0)$ pour ceux de spin + et $(K,L)=(0,1)$
pour ceux de spin -. D\'es lors, la g\'en\'eralisation de la fonction de partition est \'evidente: on consid\`{e}re l'ensemble
des graphes construits avec ces nouveaux polygones\footnote{Ces polygones portant plusieurs spins ne peuvent \^etre li\'es par leurs
centres \`{a} aucun autre \'el\'ement de la carte. Ils ne peuvent \^etre li\'es que par leurs c\^ot\'es \`{a} des polygones de m\^eme
couleur.} muni des poids $\CW_{general}$ induits par l'introduction des $t_{(k_1,l_1,k_2,\dots,k_K,l_L),i}$. La fonction de partition correspondante est alors
donn\'ee par:
\beq
\CZ_{general} := \prod_i e^{-{N^2 \over T} \epsilon_i ( V_1(\xi_i) + V_2(\eta_i) - \xi_i \eta_i)} \prod_{j>i} \left[(\xi_i-\xi_j)(\eta_i-\eta_j)\right]^{N^2 \epsilon_i \epsilon_j}
\sum_{G\in\CG_{general}} \CW_{general}(G).
\eeq
Les fonctions de corr\'elation mixtes sont alors obtenues en d\'erivant la fonction de partition par rapport aux nouveaux poids
avant de prendre ces derniers \'egaux \`{a} 0. Par exemple:
\beq
\left. {\partial \ln \CZ_{general} \over \partial t_{(k,l),i}}\right|_{t_{(k_1,l_1,k_2,\dots,k_K,l_L),i}=0}
\rightarrow \left< \Tr M_1^k M_2^l \right> .
\eeq

\br
Les r\'esolvantes introduites ici comme fonctions de corr\'elations, qu'elles soient mixtes ou non, sont des s\'eries formelles
en leurs param\`{e}tres $x$ et $y$, d\'efinies quand ces derniers tendent vers l'infini. Ainsi, en tant que fonctions de ces param\`{e}tres
complexes, elles sont en g\'en\'eral multivalu\'ees. Dans tous les cas, la bonne valeur est s\'el\'ectionn\'ee par la r\`{e}gle
suivante:
\beq
\left<\Tr {1 \over x-M_1} B \right> {\displaystyle \sim_{x \to \infty}} {1 \over x} \left< \Tr B \right>.
\eeq
Ceci permet \'egalement de voir les fonctions de corr\'elation simples comme des limites des fonctions de corr\'elations mixtes par:
\beq
\left<\Tr {1 \over x-M_1} {1 \over y-M_2} \right> \sim_{y \to \infty}  {1 \over y} \left< \Tr {1 \over x-M_1} \right>.
\eeq

\er

\section{D\'eveloppement topologique.}

Chaque terme $B_k$ de la s\'erie formelle en $T$, $\CF = {\displaystyle \sum_{k=0}^\infty} B_k T^k$, est un polyn\^ome en ${1 \over N^2}$ car
l'exposant $k$ majore le nombre de vertex formant les graphes $G$ contribuant \`{a} $B_k$ (voir la discussion dans la preuve du th\'eor\`{e}me
\ref{thcombi}).
Ainsi, on peut noter:
\beq
B_k = \sum_h B_{k,h} N^{-2h},
\eeq
et d\'efinir les termes :
\beq
\CF^{(h)} := \sum_{k=0}^\infty B_{k,h}T^k
\eeq
pour tout $h$ positif ou nul.

Or, en s'attardant sur la d\'efinition combinatoire de la fonction de partition \eq{eqZ}, on peut observer que ces termes correspondent au coefficient de
$N^{-2h}$ dans l'\'energie libre. Par ailleurs, l'exposant de $N$ dans le poids d'un graphe donn\'e est sa caract\'eristique d'Euler-Poincar\'e.
Ainsi, en s\'electionnant un exposant particulier de $N$ par $\CF^{(h)}$, on impose que les surfaces aient un genre $h$ fix\'e:
$\CF^{(h)}$ est la fonction g\'en\'eratrice des surfaces connexes de $\CS$ de genre $h$ en utilisant le m\^{e}me
poids que pour $\CZ_{form}$. On peut donc \'ecrire le {\bf d\'eveloppement topologique}:
\beq\encadremath{
\CF = \sum_{h=0}^{\infty} \CF^{(h)} N^{-2h}}
\eeq
comme une s\'erie formelle en ${ 1 \over N^2}$.

De mani\`{e}re analogue, on peut \'ecrire un d\'eveloppement topologique
\beq
\overline{H}_{k_1,k_2,\dots, k_l;m;n} = \sum_{h=0}^{\infty} N^{-2h} \overline{H}_{k_1,k_2,\dots, k_l;m;n}^{(h)}
\eeq
pour toutes les fonctions de corr\'elation s\'electionnant le genre $h$ des surfaces ouvertes g\'en\'er\'ees.

\br
Le d\'evellopement topologique est une s\'erie formelle en ${1 \over N^2}$ et n'a donc aucune raison de converger. Cependant, chacun
des coefficients $\CF^{(h)}$ est une fonction analytique de ses param\`{e}tres avec un rayon de convergence non nul comme nous
allons le montrer dans ce chapitre.
\er

L'essentiel du travail pr\'esent\'e dans ce chapitre consiste \`{a} calculer explicitement les diff\'erents termes $\CF^{(h)}$ et $\overline{H}^{(h)}$ du d\'eveloppement
topologique de n'importe quelle fonction de corr\'elation ainsi que de l'\'energie libre.

\section{Cas particulier: mod\`ele d'Ising.}\label{partcombi}

Nous avons consid\'er\'e dans la partie pr\'ec\'edente un mod\`{e}le de matrices formel pour deux potentiels $V_1$ et $V_2$
g\'en\'eriques et des fractions de remplissage quelconques. Dans cette partie nous
\'etudions le cas particulier o\`{u} les valeurs propres de la matrice col autour de laquelle on d\'eveloppe
sont toutes \'egales: la matrice est totalement d\'eg\'en\'er\'ee et une seule fraction de remplissage est non nulle et donc
\'egale \`a l'unit\'e.
Il n'y a alors qu'une seule couleur sur les surfaces g\'en\'er\'ees.
Sans perte de g\'en\'eralit\'e, on peut supposer que cet unique point col (i.e. l'unique couple de valeurs propres
des deux matrices col) se situe en $(\xi,\eta) = (0,0)$. Alors, la fonction
de partition $\CZ_{form}$ est la fonction g\'en\'eratrice des surfaces ferm\'ees (connexes ou non) form\'ees de $k_1$-gones
de spin +, avec $k_1\leq d_1$, et de $k_2$-gones de spin -, avec $k_2 \leq d_2$, recoll\'es par leurs ar\^etes, en utilisant
les poids suivants:
\begin{itemize}
\item les $k$-gones de spin + sont compt\'es avec un poids ${N \over T} t_k$;
\item les $k$-gones de spin - sont compt\'es avec un poids ${N \over T} \tilde{t}_k$;
\item les ar\^etes communes \`{a} deux polygones de spin + sont compt\'ees avec un poids ${1 \over N} {t_2\over t_2 \tilde{t_2}-1}$;
\item les ar\^etes communes \`{a} deux polygones de spin - sont compt\'ees avec un poids ${1 \over N} {\tilde{t}_2\over t_2 \tilde{t_2}-1}$;
\item les ar\^etes communes \`{a} deux polygones de spins diff\'erents sont compt\'ees avec un poids ${1 \over N} {1 \over t_2 \tilde{t_2}-1}$;
\item les sommets sont compt\'es avec un poids N.
\end{itemize}

Les fonctions de corr\'elation $\overline{H}_{k_1,\dots,k_l;m;n}(S_1,S_2, \dots, S_l;x_1 ,\dots, x_m;y_1,\dots,y_n)$ sont alors les
fonctions g\'en\'eratrices de telles surfaces connexes avec $l+m+n$ bords et des conditions de bord diff\'erentes\footnote{Il y a deux conditions de bord
possibles d\'ecrites dans la partie 2. On appelera condition de type + (resp. de type -) la condtion obtenue
en fixant un polygone de spin + (resp. -) \`{a} l'ext\'erieur de la surface}:
\begin{itemize}
\item Les $m$ bords associ\'es aux variables $x_i$ ont une condition homog\`{e}ne de type +;
\item Les $n$ bords associ\'es aux variables $y_i$ ont une condition homog\`{e}ne de type -;
\item Les $l$ bords associ\'es aux cycles $S_i$ ont une condition de bord non homog\`{e}ne donn\'ee par l'alternance
de condition + et de condition - suivant l'alternance des variables $x_{i,j}$ et $y_{i,j}$ dans le cycle $S_i$.
\end{itemize}

Leur d\'eveloppement topologique permet alors de s\'electionner le genre des surfaces g\'en\'er\'ees: $\overline{H}_{k_1,\dots,k_l;m;n}^{(g)}(S_1,S_2, \dots, S_l;x_1 ,\dots, x_m;y_1,\dots,y_n)$
est la fonction g\'en\'eratrice des surfaces d\'ecrites plus haut avec la condition suppl\'ementaire qu'elles soient de genre $g$.

Dans la suite, il sera parfois utile de repr\'esenter graphiquement ces fonctions de corr\'elation en r\'ef\'erence aux
surfaces g\'en\'er\'ees. On repr\'esente la fonction de corr\'elation:
\beq\label{diagmix}
\overline{H}_{k_1,\dots,k_l;m;n}^{(g)}(S_1,S_2, \dots, S_l;x_1 ,\dots, x_m;y_1,\dots,y_n):=
\begin{array}{r}
{\rm \includegraphics[width=5.9cm]{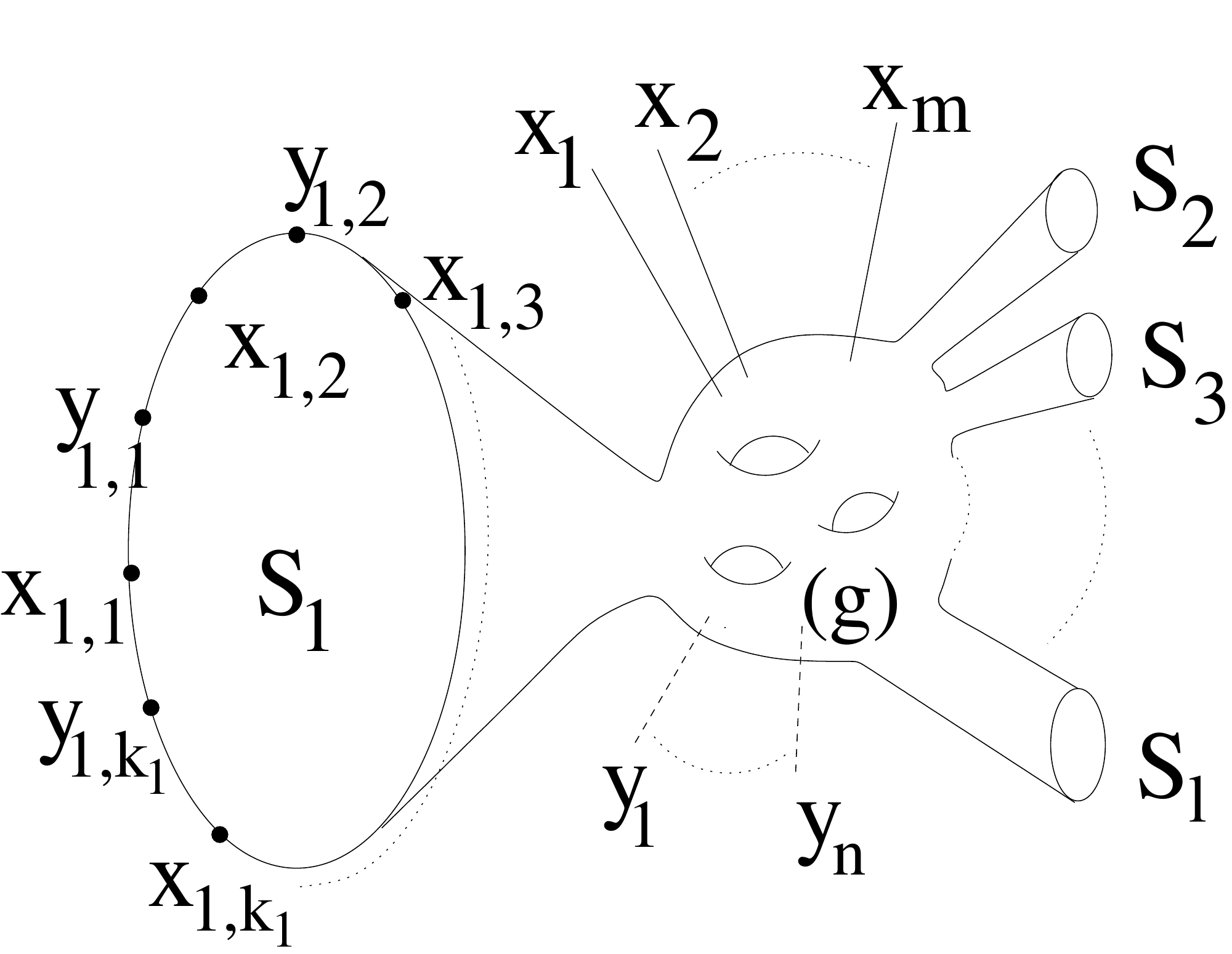}}
\end{array}
\eeq
par une "surface" de genre $g$ avec $l+m+n$ bords sur lesquels les conditions de bords + et - sont repr\'esent\'ees par les variables $x$ et $y$ respectivement.
Les bords o\`{u} la condition est homog\`{e}ne ont \'et\'e r\'eduits en un point par souci de simplicit\'e.
De m\^{e}me, les fonctions de corr\'elation simples sont repr\'esent\'ees par:
\beq
\overline{W}_{m,n}^{(g)}(x_1, \dots,x_m; y_1, \dots ,y_n):=\begin{array}{r}
{\rm \includegraphics[width=6cm]{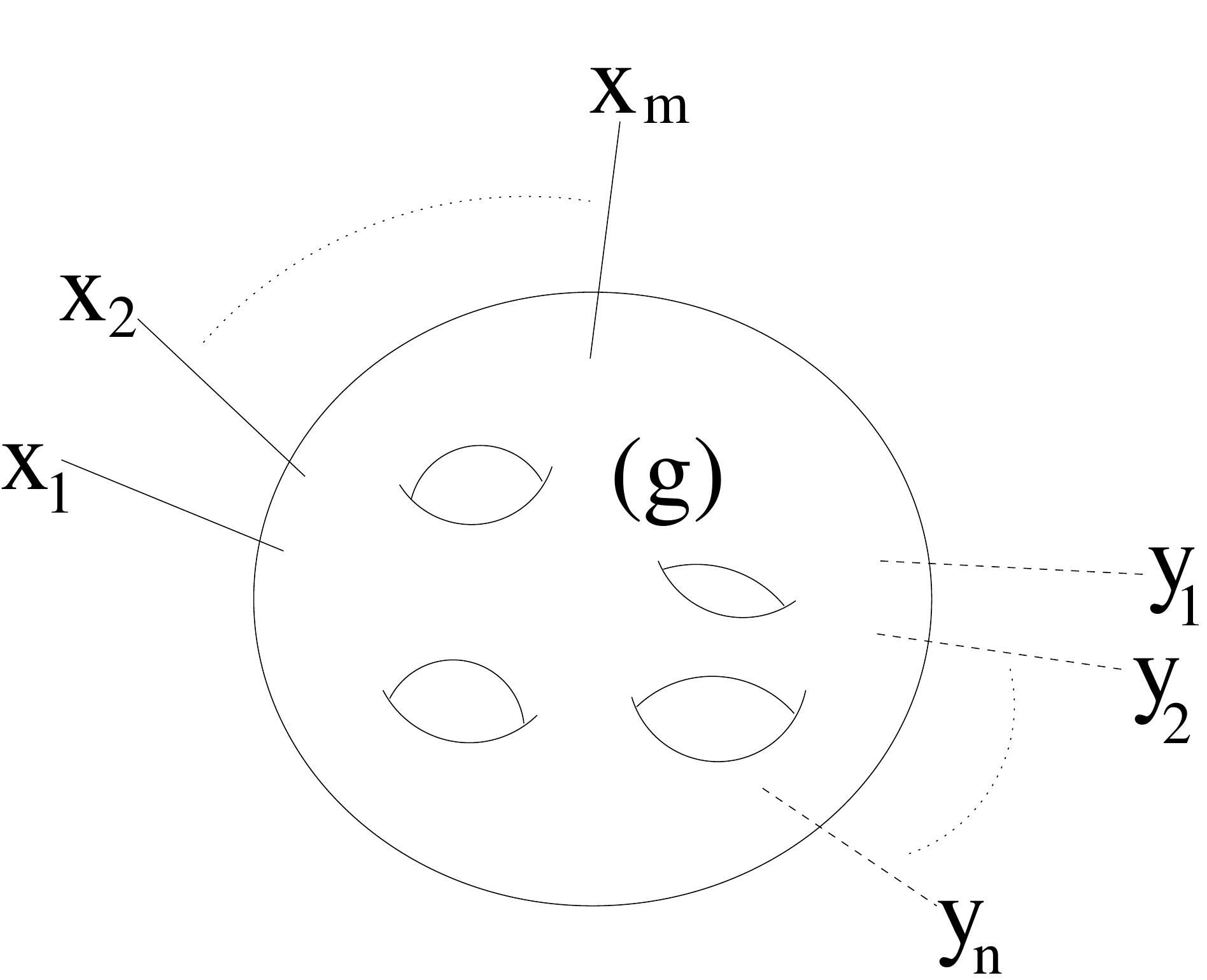}}
\end{array}.
\eeq

\br
Cette repr\'esentation prend tout son sens quand les fonctions de corr\'elation sont effectivement fonctions g\'en\'eratrices
de surfaces discr\'etis\'ees, c'est-\`{a}-dire lorsque l'on a un seul point col. Cependant, elle sera utile tout au long de cette
th\`{e}se et sera reprise m\^eme lorsque l'interpr\'etation combinatoire des fonctions de corr\'elation est l\'eg\`{e}rement
modifi\'ee par l'existence de fractions de remplissage.
\er

\section{Double limite d'\'echelle et limite continue.}\label{DSL}

Si le mod\`ele \`a deux matrices formel permet de compter les surfaces discr\'etis\'ees, leur succ\`es vient en partie de la conviction
que ces surfaces compos\'ees de polygones donnent acc\`es aux surfaces continues par une proc\'edure de passage \`a la limite.
Une telle proc\'edure peut \^etre obtenue intuitivement en faisant croitre le nombre de polygones composant les surfaces
tout en r\'eduisant leur taille de mani\`ere \`a garder l'aire de la surface constante.

Plus pr\'ecis\'ement, en gravitation quantique (ou th\'eorie des cordes), on veut pouvoir calculer une int\'egrale de la forme
\beq
Z_{cont} = \sum_g \int_{\Sigma_g} {\cal D} \Sigma_g e^{-E(\Sigma_g)}
\eeq
o\`u l'int\'egrale porte sur toutes les surfaces ferm\'ees de genre $g$ et $E(\Sigma_g)$ est l'action d'Einstein-Polyakov d\'efinie par
\eq{Polyakov} du chapitre 1:
\beq
E(\Sigma_g):= 4 \pi G (2-2g) + \L \, \hbox{Aire}(\Sigma_g) + \hbox{Champ de mati\`ere coupl\'e}.
\eeq
Pour pouvoir approcher cette quantit\'e par les int\'egrales de matrices formelles, on discr\'etise les surfaces $\Sigma_g$
en les d\'ecoupant en polygones d'aire moyenne $\epsilon^2$. En effet, on peut alors approcher chacun des termes constituant
$Z_{cont}$ par le terme de genre $g$, $\CF_{\rm 2MM}^{(g)}$ du d\'eveloppement topologique de l'\'energie libre d'un mod\`ele \`a deux matrices\footnote{ Lorsqu'il
pourra y avoir confusion sur le mod\`ele auquel se rapport l'\'energie libre, nous utiliserons l'indice $2MM$ pour
identifier le mod\`ele \`a deux matrices.} (si l'on a besoin d'imposer une structure de spin de type Ising pour la mati\`ere
coupl\'ee \`a la gravit\'e) \`a condition d'identifier les param\`etre des deux th\'eories:
\begin{itemize}

\item La taille des matrices \`a consid\'erer $N$ est li\'ee au param\`etre $G$ par:
\beq
\ln N \propto G;
\eeq

\item On associe une constante cosmologique coupl\'ee \`a chaque type de polygone\footnote{On consid\`ere ici le cas pr\'esent\'e
dans la partie pr\'ec\'edente o\`u il n'y a pas de fraction de remplissage \`a fixer.}:
\beq
\L_k \epsilon^2 = \ln(t_k) - {k \over 2}\ln(T)
\virg
\widetilde{\L}_k \epsilon^2 = \ln(\tilde{t}_k) - {k \over 2}\ln(T).
\eeq
\end{itemize}

Il faut maintenant faire tendre la taille des polygones $\epsilon^2$ vers 0 tout en gardant l'aire totale finie, i.e.
il faut faire tendre le nombre moyen de polygones vers l'infini de mani\`ere \`a compenser leur petite taille.
Or, on peut voir ais\'ement que le nombre moyen de polyg\^{o}nes dans une surface de genre $g$ est donn\'e par:
\beq
<n_k> = t_k {\partial \CF_{\rm 2MM}^{(g)} \over \partial t_k}
\virg
<\tilde{n}_k> = \tilde{t}_k {\partial \CF_{\rm 2MM}^{(g)} \over \partial \tilde{t}_k}.
\eeq

Il est donc n\'ecessaire que le nombre total de polyg\^{o}nes $\sum_k t_k {\partial \CF_{\rm 2MM} \over \partial t_k} + \tilde{t}_k {\partial \CF_{\rm 2MM} \over \partial \tilde{t}_k}$
diverge comme $\epsilon^{-2}$, ce qui peut \^etre obtenu en r\'eglant de mani\`ere appropri\'ee les coefficients des potentiels du mod\`ele
\`a deux matrices consid\'er\'e.

On voit en particulier qu'il existe plusieurs fa\c{c}ons d'approcher cette limite continue: on peut favoriser certains types
de polygones au d\'etriment des autres ou bien tous leur donner la m\^eme importance. Ces diff\'erentes limites correspondent
\`a approcher des points critiques ou multi-critiques correspondant \`a des singularit\'es de types diff\'erents dans l'espace
des modules du mod\`ele \`a deux matrices et permettent d'atteindre diff\'erentes th\'eories conformes coupl\'ees \`a
la gravit\'e.

Nous reviendrons en d\'etail sur cette proc\'edure de passage \`a la limite dans la partie \ref{partapplic} du chapitre 4 lorsque nous
aurons tous les outils n\'ecessaires pour la d\'ecrire rigoureusement.

\section{Equations de boucles.}

Parmi les nombreuses approches utilis\'ees pour tenter de r\'esoudre les mod\`{e}les de matrices\footnote{Ce terme g\'en\'erique
inclus et confond de mani\`{e}re abusive les diff\'erentes d\'efinitons de l'int\'egrale matricielle $\CZ$.}, l'une a apport\'e
des r\'esultats tr\`es concluants dans l'\'etude de diff\'erentes int\'egrales de matrices formelles: la m\'ethode dite
des \'equations de boucles. Celle-ci consiste \`{a} obtenir une hi\'erarchie d'\'equations satisfaites par les fonctions de corr\'elations
et l'\'energie libre par "int\'egration par partie" de $\CZ$. Ces \'equations portent en fait plusieurs noms suivant la mani\`{e}re
dont elles ont \'et\'e obtenues et surtout le contexte dans lequel elles ont \'et\'e d\'eriv\'ees. Du point de vue des physiciens,
elles sont appel\'ees \'equations de Schwinger-Dyson en th\'eorie des champs et furent obtenues en combinatoire des cartes
triangul\'ees sous le nom d'\'equations de Tutte (\cite{tutte,tutte2}).

Formellement, en utilisant les notations en termes d'int\'egrales matricielles, ces \'equations signifient que l'int\'egrale de matrice reste inchang\'ee par un changement de variable
$M_i \to M_i + \epsilon \, \delta M_i$. En consid\'erant un choix de $\delta M$ assez large, on peut refermer l'ensemble d'\'equations
obtenu et ainsi r\'esoudre le mod\`{e}le\footnote{En fait, les \'equations obtenues ne semblent pas \^etre ferm\'ees \`a premi\`ere
vue. Pour r\'eellement les fermer, il faut faire appelle \`a des notions de g\'eom\'etrie alg\'ebrique et \`a la courbe spectrale
d\'efinie dans la partie suivante.}.

\subsection{D\'erivation des \'equations de boucles.}

Dans ce paragrpahe, je pr\'esente la recette \`{a} utiliser pour d\'eriver les \'equations de boucles satisfaites par les fonctions
de corr\'elation et l'\'energie libre. La d\'erivation rigoureuse, plus technique, consiste \`a effectuer un changement de
variables dans les int\'egrales gaussiennes composant la fonction de partition.
Elle peut \^etre trouv\'ee dans \cite{eynform}.
Cependant, le lecteur uniquement int\'eress\'e par la r\'esolution de ce mod\`{e}le peut se contenter de cette partie.

Lorsque l'on effectue un changement de variable
\beq
M_1 \to \td{M}_1 := M_1 + \epsilon f(M_1,M_2),
\eeq
on change l'action dans $\CZ$, d'une part, ainsi que la mesure par l'interm\'ediaire du jacobien du changement de variable d'autre part:
\bea
\CZ &= & \int dM_1\, dM_2\, e^{- {1 \over \hbar} Tr\left(V_1(M_1) + V_2(M_2) -
M_1 M_2 \right)}\cr
& = & \int d\td{M}_1\, dM_2\, e^{- {1 \over \hbar} Tr\left(V_1(\td{M}_1) + V_2(M_2) -
\td{M}_1 M_2 \right)}\cr
& = & \int dM_1\, dM_2\,  e^{- {1 \over \hbar} Tr\left(V_1({M}_1) + V_2(M_2) -
{M}_1 M_2 \right)} \times \cr
&& \qquad \qquad \qquad \qquad \times (1 + \epsilon (J(M_1,M_2)- K(M_1,M_2))+ 0(\epsilon^2)) \cr
\eea
o\`{u} $J(M_1,M_2)$ et $K(M_1,M_2)$ correspondent \`{a} l'ordre 1 en $\epsilon$ respectivement du jacobien du changement de variable
et de la variation de l'action et sont d\'efinis par
\bea
\det \frac{\partial \td{M}_1}{ \partial M_1} & = & 1 + \epsilon J(M_1,M_2) + O(\epsilon^2) \cr
{1 \over \hbar} Tr\left(V_1(\td{M}_1) + V_2(M_2) -
\td{M}_1 M_2 \right) & = &{1 \over \hbar} Tr\left(V_1(M_1) + V_2(M_2) -
M_1 M_2\right) \cr
& & + \epsilon K(M_1,M_2) + O(\epsilon^2).\cr
\eea

A l'ordre 1 en $\epsilon$, cette invariance se traduit par
\beq
\moy{J(M_1,M_2)}  = \moy{K(M_1,M_2)}.
\eeq

\subsubsection{D\'{e}termination de $J(M_1,M_2)$ et de $K(M_1,M_2)$.}
\begin{itemize}
\item
Le terme correspondant \`{a} la variation de l'action est obtenu facilement et s'\'ecrit
\beq
K(M_1,M_2) = {1 \over \hbar}  Tr \left[ V_1'(M_1) f(M_1,M_2) - M_2 f(M_1,M_2) \right] .
\eeq

\item
La d\'{e}termination de $J(M_1,M_2)$ est quant \`{a} elle moins \'{e}vidente, mais peut \^{e}tre
r\'{e}sum\'{e}e par deux r\`{e}gles simples, que nous d\'esignerons par le noms de r\`{e}gles "split" et "merge" (\cite{courseynard}).
Celles-ci correspondent en fait aux deux types de changement de variables effectu\'{e}s en pratique. Il est int\'eressant de
red\'eriver ces r\`{e}gles comme un exercice par le calcul explicite du Jacobien du changement de variables. Pour ce faire, il
est important de noter que les variables ind\'ependantes de la matrices $M$ sont les \'el\'ements $M_{i,i}$ ainsi que
$Re(M_{i,j})$ et $Im(M_{i,j})$ pour $i<j$ (voir \cite{eynform}). A ces deux r\`egles vient \'egalement s'ajouter la r\`egle de cha\^{i}ne valable pour
tout type de changement de variable.
\vspace{0.5cm}

{\bf R\`egle de cha\^{i}ne.}

Pour un changement de variable faisant appara\^itre plusieurs facteurs, on applique les r\`egles Split et Merge successivement
\`a chacun des termes en suivant la r\`egle de Leibnitz: le r\'esultat est la somme des termes obtenus en utilisant les r\`egles
Split et Merge \`a chacun des facteurs en gardant le reste constant.

\vspace{0.5cm}

{\bf{R\`{e}gle Split.}}

Le premier type de changement de variable que nous devons consid\'erer correspond \`{a}
$f(M_1,M_2) = A \frac{1}{x-M_1} B$ o\`{u} A et B sont des
fonctions de $M_1$ et $M_2$. Alors la correction issue du Jacobien
s'\'ecrit :
\bea\label{split}
J(M_1,M_2) & = & Tr\left(A \frac{1}{x-M_1}\right) Tr\left(
\frac{1}{x-M_1} B\right) \cr
& & + \,\, \hbox{contributions venant de
$A(M_1)$ et $B(M_1)$.}
\eea
Ainsi, chaque fois que l'on rencontre un terme du type $\frac{1}{x-M_1}$
en dehors d'une trace, on "coupe" l'expression en deux traces en
introduisant un facteur $\frac{1}{x-M_1}$ dans chaque trace.
\vspace{0.5cm}

{\bf{R\`{e}gle merge.}}

On rencontre \'{e}galement des changements de variable du type
$f(M_1,M_2) = A Tr \left(\frac{1}{x-M_1} B \right)$. On a alors :
\bea\label{merge}
J(M_1,M_2) & = & Tr\left(A \frac{1}{x-M_1} B \frac{1}{x-M_1} \right) \cr
& & + \,\, \hbox{contributions venant de $A(M_1)$ et $B(M_1)$.}
\eea
C'est-\`{a}-dire que, chaque fois que l'on rencontre un terme
$\frac{1}{x-M_1}$ dans une trace, on regroupe toute l'expression
\`{a} l'int\'{e}rieur d'une m\^{e}me trace en rempla\c{c}ant "Tr" par
un duplicata du facteur $\frac{1}{x-M_1}$.
\end{itemize}

\section{Limite planaire.}

Chacune des fonctions de corr\'elation entrant dans les \'equations de boucles a un \eol
d\'eveloppement topologique en ${1 \over N^2}$ .
Ainsi chaque \'equation de boucle peut \^{e}tre \eol
d\'ecompos\'ee en une hierarchie infinie d'\'equations portant sur les termes
d\'eveloppement de 't Hooft de ces observables\footnote{C'est la mani\`ere intuitive de voir les choses. Ici, les \'equations
de boucles ressomm\'ees, o\`u tous les termes du d\'eveloppement topologique sont regroup\'es dans une s\'erie infinie, ne sont
qu'une notation pratique pour d\'ecrire de mani\`ere concise l'ensemble de la hierarchie d'\'equations correspondant \`a chaque coefficient
des puissance de ${1 \over N^2}$.}. Nous nous int\'eresserons dans cette partie aux fonctions g\'en\'eratrices
des surfaces de genre nul, c'est-\`{a}-dire aux termes dominants dans les d\'eveloppements en ${1 \over N^2}$. Nous n'\'etudierons
donc que la premi\`{e}re \'equation de chacune de ces hierarchies: celle qui correspondrait \`{a} la limite obtenue lorque la taille de la
matrice tend vers l'infini.

\vs

{\bf Attention.}

Consid\'erer la limite planaire des \'equations de boucles ne veut pas dire que les coefficients d'ordre inf\'erieur
dans le d\'eveloppement topologique sont n\'egligeables par rapport \`a ceux d'ordre sup\'erieur. Cela signifie
juste que l'on travaille avec les termes de l'ordre $0$ des s\'eries formelles d\'efinissant les observables du mod\`ele
et rien de plus. Il faut donc se m\'efier de ce terme "limite planaire" qui ne d\'esigne en fait aucune limite.

\subsection{Equation de boucle ma\^itresse et courbe spectrale.}\label{masterloop}

Consid\'erons dans un premier temps le changement de variables le plus simple faisant intervenir la matrice $M_1$:
\begin{equation}
M_2 \rightarrow M_2 + \epsilon \frac{1}{x-M_1}.
\end{equation}

L'\'{e}quation de boucle associ\'{e}e s'\'ecrit alors simplement en utilisant la r\`{e}gle split \eq{split}:
\begin{equation}
x \overline{W}_{1,0}(x) -1 = \frac{1}{N} \left\langle Tr \frac{1}{x-M_1}
V_2'(M_2)\right\rangle . \label{eqboucle1}
\end{equation}

Le changement de variable, polyn\^{o}mial en $y$:
\begin{equation}
M_1 \rightarrow M_1 + \epsilon \left( \frac{1}{x-M_1}
\frac{V_2'(y)-V_2'(M_2)}{y-M_2}\right)
\end{equation}
donne, en utilisant (\ref{eqboucle1}), l'\'{e}quation de boucle :
\begin{equation}
(y-Y(x))U(x,y) = V_2'(y) \overline{W}_{1,0}(x) - P(x,y) - x \overline{W}_{1,0}(x) +1 -
\frac{1}{N^2}U(x,y;x) \label{eqboucle2}
\end{equation}
o\`u l'on a not\'e:
\beq
U(x,y) := {1 \over N} \left< \Tr {1 \over x- M_1} {V_2'(y)-V_2'(M_2) \over y-M_2} \right>,
\eeq
\beq
P(x,y) := {1 \over N} \left< \Tr {V_1'(x)-V_1'(M_1) \over x- M_1} {V_2'(y)-V_2'(M_2) \over y-M_2} \right>,
\eeq
\beq
U(x,y) := {1 \over N} \left< \Tr {1 \over x- M_1} {V_2'(y)-V_2'(M_2) \over y-M_2} \Tr{1 \over x-M_1}\right>_c
\eeq
et
\beq
Y(x) := V_1'(x) - {1 \over N} \left<\Tr {1 \over x-M_1} \right>.
\eeq

$U(x,y)$ \'{e}tant un polyn\^{o}me en $y$, il n'a pas de singularit\'{e}
pour $y$ fini, et l'\'{e}quation (\ref{eqboucle2}) prise en $y =
Y(x)$ s'\'{e}crit :
\begin{equation}
\left\{
\begin{array}{l}
y = Y(x) \\
( V_2'(y) -x ) ( V_1'(x)-y) - P(x,y) + 1 = \frac{1}{N^2}
U(x,y;x) \\
\end{array}
\right. .
\end{equation}

En notant le polyn\^{o}me de degr\'{e}s ($d_1 + 1 $) en $x$ et ($d_2 + 1
$) en $y$ :
\begin{equation}
E(x,y) = ( V_2'(y) -x ) ( V_1'(x)-y) - P(x,y) + 1 ,
\end{equation}
on peut r\'{e}\'{e}crire l'\'{e}quation pr\'{e}c\'{e}dente en :
\begin{equation}\encadremath{
E(x, Y(x)) = \frac{1}{N^2} U(x,Y(x);x) \label{eqmaitresse1}
}\end{equation}
appel\'{e}e {\bf \'{e}quation de boucle ma\^itresse}.

Lorsque l'on consid\`{e}re la limite planaire, i.e. l'ordre 0 de cette \'equation, on obtient une \'equation
alg\'ebrique:
\beq\label{mastereq}
\CE(x,Y^{(0)}(x)) = 0
\eeq
o\`{u}
\beq\label{E2MM}\encadremath{
\CE(x,y):=( V_2'(y) -x ) ( V_1'(x)-y) - P^{(0)}(x,y) + 1}
\eeq
 est l'ordre dominant du d\'eveloppement topologique de $E$. On a ainsi associ\'e une courbe alg\'ebrique au
mod\`{e}le \`{a} deux matrices:
\beq
\CE(x,y)=0
\eeq
que nous nommerons {\bf courbe spectrale classique} puisqu'elle correspond \`{a} la limite planaire $N \to \infty$,
i.e. la limite classique $\hbar \to 0$.

Cette courbe alg\'ebrique est un objet fondamental de la r\'esolution des \'equations de boucles. En effet, tous
les termes du d\'eveloppement topologique des fonctions de corr\'elations (et pas seulement leur limite planaire)
seront obtenus en termes de fonctions d\'efinies sur la surface de Riemann compacte associ\'ee \`{a} $\CE$. En d'autres termes,
les modules de $\CE$ contiennent toute l'information n\'ecessaire \`{a} la d\'etermination de l'int\'egrale matricielle $\CZ$.

\vspace{1cm}

\textbf{Remarques:}
\begin{itemize}
\item L'\'equation $\CE(x,Y^{(0)}(x))=0$ permet de fixer de mani\`{e}re implicite la fonction $Y^{(0)}(x)$ et donc $\overline{W}_{1,0}^{(0)}(x)$.
Cependant, cette \'equation a $d_2+1$ solutions \`{a} $x$ donn\'e. Ceci montre que $\overline{W}_{1,0}(x)$ est \`{a} priori une fonction
multivalu\'ee dans $\mathbb{C}$. Cepedant celle-ci est monovalu\'ee en tant que fonction sur la surface de Riemann associ\'ee \`{a} $\CE$:
les diff\'erentes valeurs pour un $x$ donn\'e correpondent alors \`{a} la structure en feuillets de $\CE$ ( voir la partie \ref{geoalg}).
Il ne faut cependant pas oublier que $\overline{W}_{1,0}(x)$ est d\'efini comme une s\'erie infinie en $x$ qui n'a de sens que pour $x \to \infty$.
Cette contrainte suppl\'ementaire permet de retrouver quelle est la bonne valeur de $\overline{W}_{1,0}(x)$ pour un $x$ donn\'e.

\item Pour d\'eriver la courbe spectrale classique, nous avons pris le parti de privil\'egier $M_1$. Mais nous aurions pu effectuer
exactement la m\^{e}me d\'erivation en privil\'egiant $M_2$ dans les changements de variables, auquel cas la limite planaire de l'\'equation de boucle maitresse
aurait donn\'e:
\begin{equation}
\CE(X^{(0)}(y), y) =0,
\end{equation}
o\`u l'on note
\beq
X(y) := V_2'(y) - \left<\Tr {1 \over y-M_2} \right>.
\eeq
Ainsi, au lieu de travailler avec $\CE(x,y)$, il faudrait travailler avec $\CE(y,x)$. Cependant, par d\'efinition, les
r\'esultats ne doivent pas d\'ependre de ce choix et l'on voit apparaitre un premi\`{e}re propri\'et\'e de sym\'etrie
des fonctions de corr\'elations.

\end{itemize}

\subsection{Limite planaire des fonctions de corr\'elation mixtes.}

De mani\`{e}re g\'en\'erale, le d\'eveloppement topologique des fonctions de corr\'elation ne peuvent pas s'exprimer sans faire
appel \`{a} des notions de g\'eom\'etrie alg\'ebrique et aux sp\'ecificit\'es de la courbe spectrale $\CE$. Cependant, ceci n'est plus vrai si l'on se limite \`{a} l'\'etude
de la limite planaire, i.e. l'ordre dominant quand $N \to \infty$. Gr\^{a}ce \`{a} une jolie formule similaire \`{a} une Ansatz
de Bethe, il est possible d'exprimer n'importe quelle fonction de corr\'elation mixte en termes de produits de traces mixtes \`a
deux points $\left< \Tr {1 \over x-M_1} {1 \over y-M_2} \right>$ avec des coefficients rationnels.

Dans cette partie, par souci de brievet\'e dans les notations, on les simplifiera en notant l'ordre dominant des traces mixtes par
\beq
H_{k}^{(0)}:= {1 \over N} \left< \Tr {1\over x_1-M_1} {1 \over y_1 - M_2} {1\over x_2-M_1} {1 \over y_2 - M_2} \dots {1\over x_k-M_1} {1 \over y_k - M_2} \right> + \delta_{k1},
\eeq
i.e. $H_{k}^{(0)}$ est la fonction g\'en\'eratrice des surfaces de genre z\'ero  avec un bord et $2k$ op\'erateur de bords, i.e. des
disques avec $2k$ op\'erateur de bords que l'on repr\'esente comme dans \eq{cycle} par:
\beq
H_{k}^{(0)}:=\begin{array}{r}
{\rm \includegraphics[width=5cm]{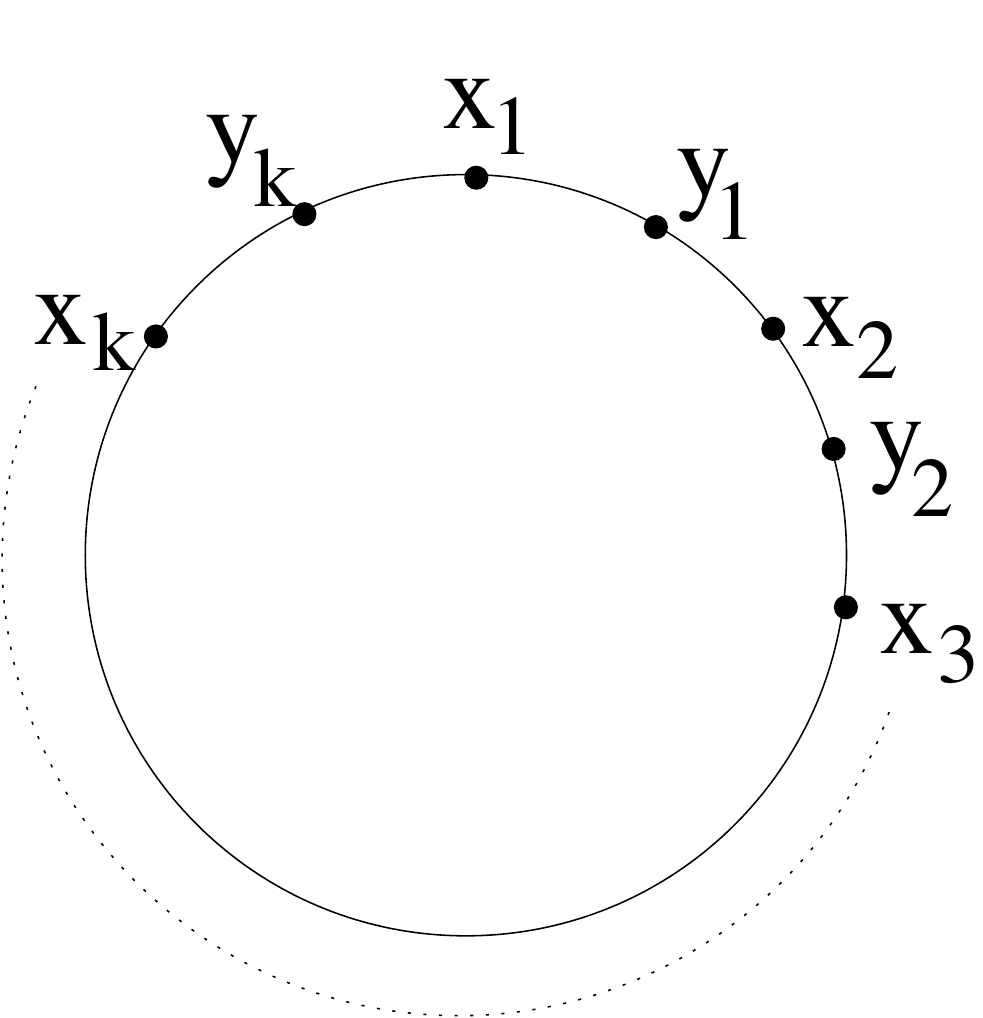}}
\end{array}
\eeq

Un r\'esultat classique donne la fonction de corr\'elation mixte la plus simple en termes de la courbe spectrale par:
\bl\label{mixsimple}
La fonction g\'en\'eratrice des disques avce deux op\'erateurs de bords est donn\'ee par:
\beq\encadremath{
H_1^{(0)}(x,y):={\CE(x,y) \over (X^{(0)}(y)-x)(y-Y^{(0)}(x))}.}
\eeq
\el

On peut alors montrer que pour tout k, $H_k^{(0)}$ est d\'ecomposable sur la base des $H_1^{(0)}$ suivant le th\'eor\`{e}me:

\bt
Pour tout $k\geq 1$, la fonction \`{a} $k$ points $H_k^{(0)}(x_1,y_1,x_2,y_2, \dots ,x_k,y_k)$
\beq\label{Ansatz}\encadremath{
H_k^{(0)}(x_1,y_1,\dots,x_k,y_k) = \sum_{\sigma\in \overline{S}_k}\, C^{(k)}_\sigma(x_1,y_1,\dots,x_k,y_k)\,\, \prod_{i=1}^k H_1^{(0)}(x_i,y_{\sigma(i)})}
\eeq
o\`{u} les coefficients $C_\sigma^{(k)}$ sont des fractions rationnelles des $x_i$ et des $y_i$ ind\'ependantes des potentiels
et avec des p\^{o}les, au plus simples, situ\'es \`{a} points coincidents et o\`{u} $\overline{S}_k$ est l'ensemble des permutations
planaires de $(1,2, \dots ,n)$.
\et

Pr\'ecisons tout d'abord ce que l'on entend par permutation planaire:

\bd
Une permutation $\sigma$ de $(1,2,\dots ,k)$ est dite planaire si
\beq
n_{cycles}(\sigma) + n_{cycles}(S\circ\sigma) = k+1
\eeq
o\`{u} $S$ est la permutation cyclique:
\beq
S(i) = i+1 \; [k]
\eeq
et $n_{cycles}(\sigma)$ est le nombre de cycles dans la d\'ecomposition de $\sigma$ en cycles irr\'eductibles.

On  note $\overline{S}_k$ l'ensemble des permutations planaires de $(1,2,\dots,k)$.
\ed

Cette notion de planarit\'e se r\'ef\`{e}re \`{a} une repr\'esentation des permutations par des syst\`{e}mes d'arches qui nous
sera utile tout au long de cette partie.

Soit une permutation $\sigma$ de $(1,2, \dots ,k)$. Pour la repr\'esenter, pla\c{c}ons les variables $(x_1,y_1,x_2,y_2, \dots ,x_k,y_k)$
sur un disque en pr\'eservant leur ordre:
\beq
\begin{array}{r}
{\rm \includegraphics[width=3.5cm]{disc.pdf}}
\end{array}
\eeq

La permutation $\sigma$ est alors simplement repr\'esent\'ee en reliant les points $x_i$ aux points $y_{\sigma(i)}$\footnote{Dans cette
repr\'esentation, les variables $x$ et $y$ marquent des d\'emarcations sur le bord alors qu'ils ont \'et\'e introduits comme
fugacit\'es associ\'ees \`{a} la longueur des diff\'erentes conditions de bords. Ainsi, en termes de surafces discr\'etis\'ees,
elles ne devraient pas se situer \`{a} des interfaces mais entre deux interfaces.}:
\beq
\sigma := \begin{array}{r}
{\rm \includegraphics[width=13cm]{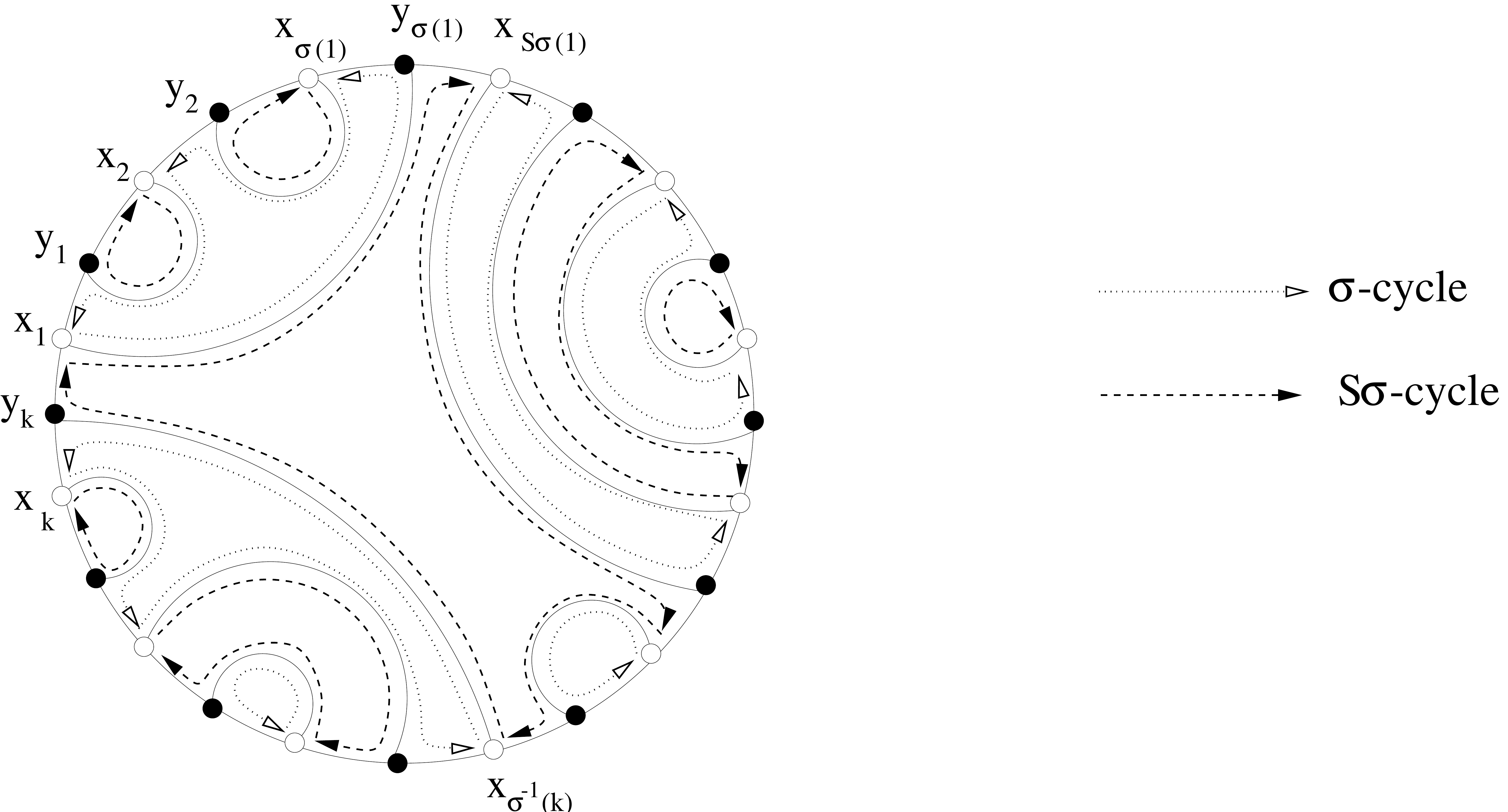}}
\end{array}
\eeq

La condition de planarit\'e signifie alors que le syst\`{e}me d'arches repr\'esentant $\sigma$ est planaire, i.e. ne pr\'esente
aucun croisement. En effet, chaque face ainsi dessin\'ee dans le disque correspond \`{a} un cycle de la d\'ecomposition
de $\sigma$ et de $S \circ \sigma$ et l'on a traduit la condition de de planarit\'e par la relation d'Euler-Poincar\'e. Le nombre d'\'el\'ements de $\overline{S}_k$ est donn\'e par les nombres de Catalan:
\beq
Card\left( \overline{S_k} \right) = Cat(k) = {(2k)! \over k! (k+1)!}.
\eeq

Cette repr\'esentation en termes de syst\`{e}mes d'arches est \'egalement utile pour calculer la valeur des termes $C^{(k)}_\sigma$.

\proof{
Je vais montrer ici d'o\`{u} vient cette d\'ecomposition sans montrer comment calculer les coefficients $C^{(k)}_\sigma$. La d\'erivation
montr\'ee ici est bien plus simple et \'el\'egante que celle pr\'esent\'ee dans \EOBethe. Elle est due \`{a}
L. Cantini \cite{Luigi}.

Les $H_{k}^{(0)}$ satisfont l'\'equation de boucles obtenue par le changement de variable $\delta M_2={1\over x_1-M_1}{1\over y_1-M_2}\,\dots \,{1\over x_{k}-M_1}{1\over y_{k}-M_2}$ en ne gardant que les
coefficients de $N$ dans le d\'eveloppement topologique:
\bea\label{loopeqH1}
&& (X(y_k)-x_1)\, H_k^{(0)}(x_1,y_1,x_2,\dots,x_k,y_k) \cr
&=& \sum_{j=1}^{k-1} {H_{k-j}^{(0)}(x_{j+1},\dots,y_k)-{H}_{k-j}^{(0)}(x_{j+1},\dots,x_k,y_j)\over y_k-y_j}\, {H}_j^{(0)}(x_1,\dots,y_j) \cr
&& - Pol_{y_k} V_2'(y(k)) H_k^{(0)}(x_1,\dots,y_k)  \cr
\eea
o\`{u} $Pol_{x} f(x)$ indique la partie polyn\^{o}miale de f(x).

Si l'on effectue le changement de variable $\delta M_2={1\over x_{k}-M_1}{1\over y_{k}-M_2}\,\dots \,{1\over x_1-M_1}{1\over y_1-M_2}$ qui consiste
\`{a} lire la suite $x_1,y_1,x_2,y_2, \dots,x_k, y_k$ dans l'autre sens, on obtient l'\'equation de boucles:
\bea\label{loopeqH2}
&& (X^{(0)}(y_k)-x_k)\, H_k^{(0)}(x_k,y_{k-1},x_{k-1},\dots,y_2,x_2,y_1,x_1,y_k) \cr
&=& \sum_{j=1}^{k-1} {H_{j}^{(0)}(x_{j},y_{j-1}\dots,y_1,x_1,y_j)-H_{j}^{(0)}(x_{j},y_{j-1}\dots,y_1,x_1,y_k)\over y_j-y_k} \times \cr
&& \qquad \qquad \times  {H}_{k-j}^{(0)}(x_{k},y_{k-1},x_{k-1},\dots,x_{j+1},y_j) \cr
&& - Pol_{y_k} V_2'(y(k)) H_k^{(0)}(x_k,y_{k-1},x_{k-1},\dots,y_2,x_2,y_1,x_1,y_k) . \cr
\eea

Or, $H_k^{(0)}(x_k,y_{k-1},x_{k-1},\dots,y_2,x_2,y_1,x_1,y_k)$ d\'ecrit les m\^{e}mes objets combinatoires que
$H_k^{(0)}(x_1,y_1,x_2,\dots,x_k,y_k)$ mais avec l'orientation du bord invers\'ee. En effet, ces deux fonctions sont les fonctions
g\'en\'eratrices des disques avec $2k$ op\'erateurs de bords mais si l'un les d\'ecrit vus du dessus, l'autre les d\'ecrit vus du
dessous:
\beq
H_k^{(0)}(x_k,y_{k-1},x_{k-1},\dots,y_2,x_2,y_1,x_1,y_k) = H_k^{(0)}(x_1,y_1,x_2,\dots,x_k,y_k)
\eeq
car
\beq
\begin{array}{r}
{\rm \includegraphics[width=3.5cm]{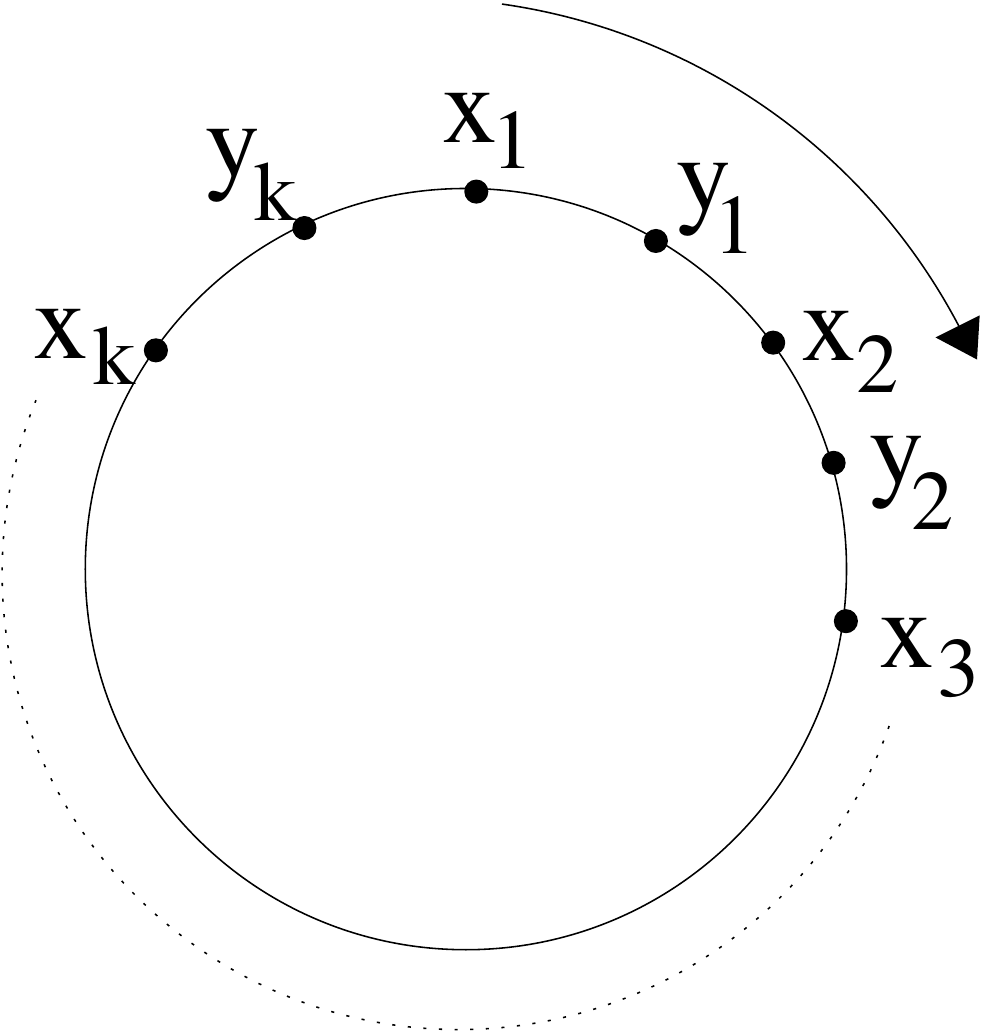}}
\end{array}
=
\begin{array}{r}
{\rm \includegraphics[width=3.5cm]{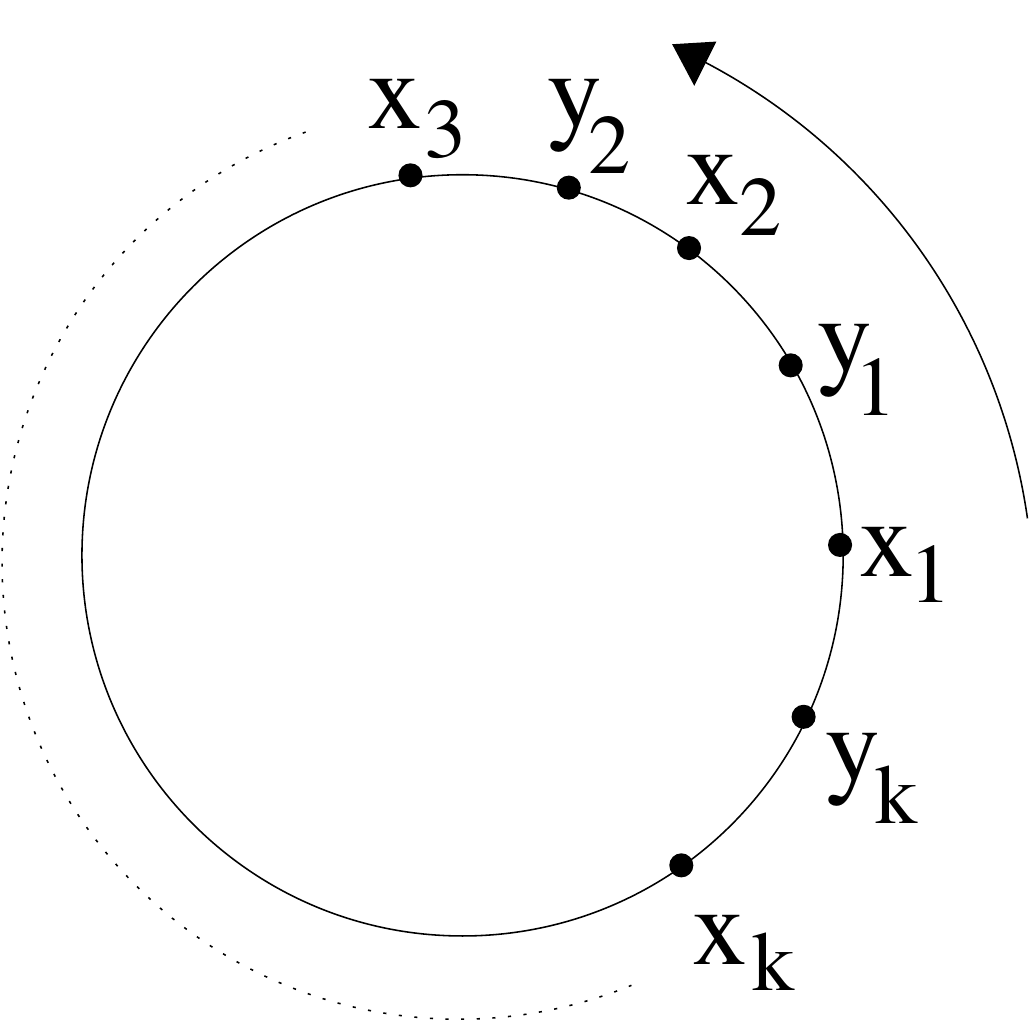}}
\end{array}.
\eeq
Cette propri\'et\'e permet d'\'ecrire la seconde \'equation de boucle \eq{loopeqH2} sous la forme:
\bea
&& (X^{(0)}(y_k)-x_k)\, H_k^{(0)}(x_1,y_1,x_2,\dots,x_k,y_k) \cr
&=& \sum_{j=1}^{k-1} {H_{j}^{(0)}(x_1,y_1,\dots,x_j,y_j)-H_{j}^{(0)}(x_1,y_1,\dots,x_j,y_k)\over y_j-y_k} \times \cr
&& \qquad \qquad \times  {H}_{k-j}^{(0)}(x_{j+1},y_{j+1},\dots , y_{k-1},x_k,y_j) \cr
&& - Pol_{y_k} V_2'(y(k)) H_k^{(0)}(x_1,y_1,x_2,\dots,x_k,y_k) . \cr
\eea
La diff\'erence avec \eq{loopeqH1} prouve le th\'eor\`{e}me:
\beq\label{Ansatz2}
\begin{array}{l}
H_k^{(0)}(x_1,y_1,x_2,\dots,x_k,y_k) =\cr
=  \sum_{j=1}^{k-1} {H_{j}^{(0)}(x_1,y_1,\dots,x_j,y_j) H_{k-j}^{(0)}(x_{j+1},y_{j+1},\dots,x_k,y_k)- \hbox{idem}(y_k \leftrightarrow y_j) \over (x_k-x_1)(y_k-y_j)} \cr
\end{array}
\eeq

}

Les coefficients rationnels $C^{(k)}_\sigma$ sont facilement d\'ecrits en termes des $C^{(k)}_{Id}$ uniquement par le th\'eor\`{e}me:
\bt
\label{defC}
Pout tout $k\geq 1$, et toute permutation $\sigma\in S_k$,
$C^{(k)}_\sigma$ est une fonction rationnelle de ses arguments $x_1,\dots,y_k$,
par:

$\bullet$
$\quad C^{(k)}_\sigma(x_1,y_1,x_2,\dots,x_k,y_k) := 0 $ si $\sigma$  n'est pas planaire;

$\bullet$
Si $\sigma$ est planaire, on d\'ecompose $\sigma$ et $S \circ \sigma$ en leurs produits de cycles:
\beq
\sigma=\sigma_1 \sigma_2 \dots \sigma_l
\virg
S\circ \sigma=\td\sigma_1 \td\sigma_2 \dots \td\sigma_{\td{l}}
\eeq
tels que:
\beq
\sigma_j = (i_{j,1},i_{j,2},\dots,i_{j,l_j}) \virg \sigma(i_{j,m})=i_{j,m+1}
\eeq
\beq
\td\sigma_j = (\td{i}_{j,1},\td{i}_{j,2},\dots,\td{i}_{j,\td{l}_j}) \virg \sigma(\td{i}_{j,m})=\td{i}_{j,m+1}-1
\eeq

\bea\label{defCprodF}
C^{(k)}_\sigma(x_1,y_1,x_2,\dots,x_k,y_k)
&:=&\prod_{j=1}^l C_{Id}^{(l_j)}(x_{i_{j,1}},y_{i_{j,2}},x_{i_{j,2}},y_{i_{j,3}},\dots,x_{i_{j,l_j}},y_{i_{j,1}}) \cr
&& \prod_{j=1}^{\td{l}} C_{Id}^{(\td{l}_j)}(x_{\td{i}_{j,1}},y_{\td{i}_{j,2}-1},x_{\td{i}_{j,2}},\dots,y_{\td{i}_{j,\td{l}_j}-1},x_{\td{i}_{j,\td{l}_j}},y_{\td{i}_{j,1}-1}) \cr
\eea
o\`{u} $C_{Id}^{(1)}:=1$.
\et

Ce th\'eor\`{e}me signifie que pour calculer un coefficient $C^{(k)}_\sigma$, ils suffit de dessiner les syt\`{e}me d'arches
correspondant \`{a} la permutation $\sigma$. Si deux arches se croisent, le coefficient a pour valeur 0,
sinon il est \'egal au produit des $C_{Id}$ pris sur chacune des faces\footnote{Attention! Les faces correspondant \`a $\sigma$
et $S \circ \sigma$ doivent \^etre orient\'ees de mani\`eres oppos\'ees. Par exemple, l'identit\'e et a permutaion cyclique ne
donnent pas le m\^eme r\'esultats \`a cause de cette orientation.}.

\bex
Consid\'erons la permutation $\sigma \in \overline{S}_{12}$ donn\'ee par:
\beq
\sigma = \left( \begin{array}{cccccccccccc}
1&2&3&4&5&6&7&8&9&10&11&12 \cr 3&1&2&7&6&5&4&8&12&10&9&11 \cr
\end{array}\right).
\eeq

Sa d\'ecomposition en cycles s'\'ecrit:
\beq
\sigma = (1,3,2)(4,7)(5,6)(8)(9,12,11)(10)
\eeq
et celle de $S \circ \sigma$ est donn\'ee par
\beq
 S \circ \sigma = (1,4,8,9)(2)(3)(5,7)(6)(10,11)(12).
\eeq

Il en r\'esulte le d\'ecoupage du disque en faces par le syst\`{e}me d'arches associ\'e \`{a} $\sigma$:
\beq
\begin{array}{r}
{\rm \includegraphics[width=8cm]{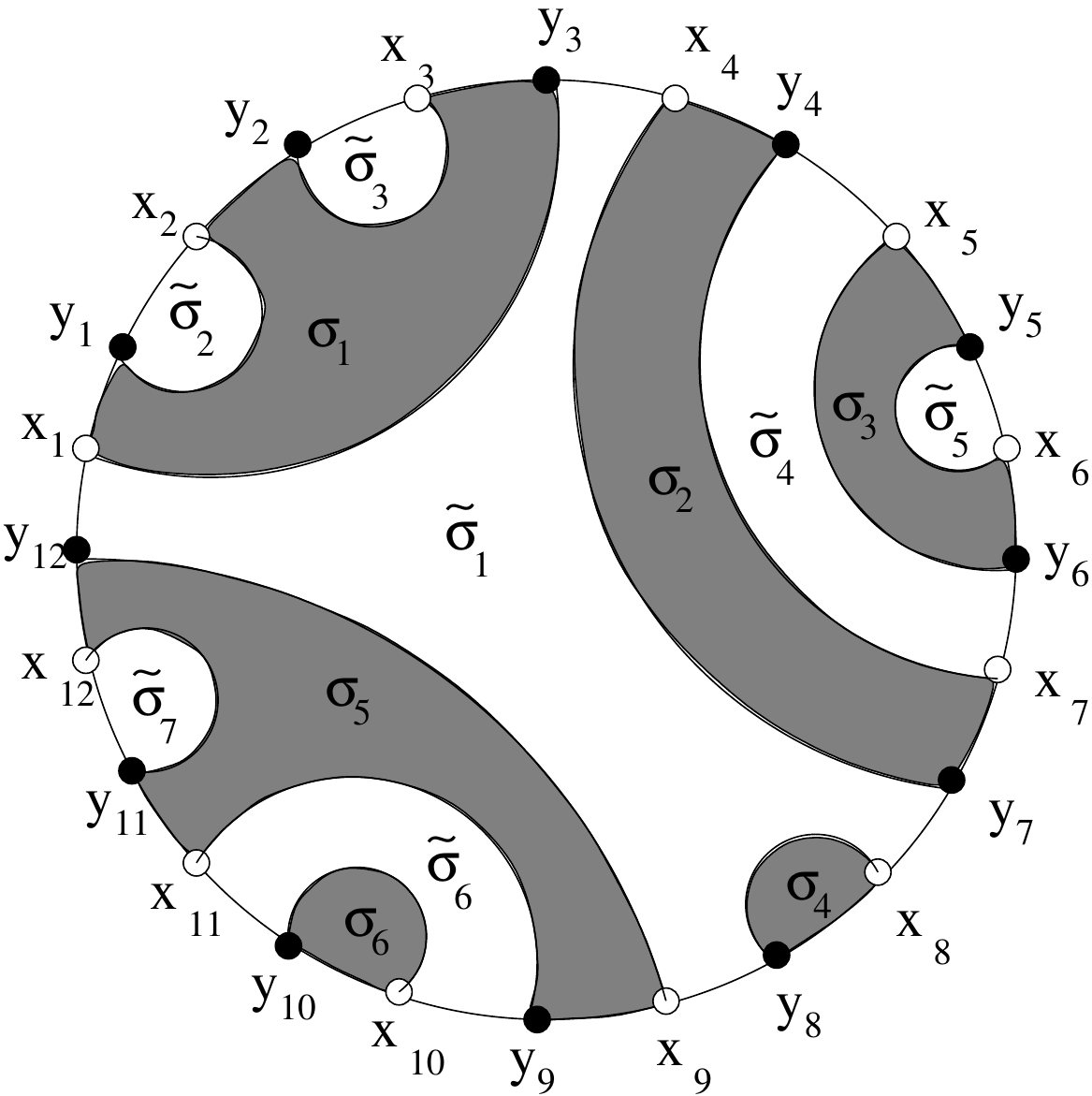}}
\end{array}.
\eeq

Celui-ci nous permet de d\'eterminer le coefficient:
\bea
C_{\sigma}^{(12)} &=&
 C_{Id}^{(3)}(x_1,y_3,x_3,y_2,x_2,y_1) C_{Id}^{(2)}(x_5,y_6,x_6,y_5)\cr
 && C_{Id}^{(3)}(x_9,y_{12},x_{12},y_{11},x_{11},y_9) C_{Id}^{(2)}(x_4,y_7,x_7,y_4)\cr
&&
C_{Id}^{(4)}(x_1,y_3,x_4,y_7,x_8,y_8,x_9,y_{12})
C_{Id}^{(2)}(x_5,y_6,x_7,y_4)\cr
&& C_{Id}^{(2)}(x_{10},y_{10},x_{11},y_9).\cr
\eea
\eex

Il reste maintenant \`{a} d\'eterminer les briques de bases $C_{Id}^{(k)}$. Elles sont donn\'ees par
\bl
Les $C_{(Id)}$ sont d\'efinis de mani\`{e}re unique par la relation de r\'ecurrence
\bea\label{recF}
C_{(Id)}^{(1)}(x_1,y_1) &:=& 1 ,\cr
C_{(Id)}^{(k)}(x_1,\dots,y_k) &:=&  \sum_{j=1}^{k-1} {C_{(Id)}^{(j)}(x_1,y_1,\dots,x_j,y_j) C_{(Id)}^{(k-j)}(x_{j+1},y_{j+1},\dots,x_k,y_k) \over (x_k-x_1)(y_k-y_j)}. \cr
\eea
\el

Il existe une mani\`{e}re efficace et simple de d\'ecrire les solutions de cette \'equation \`{a} l'aide d'une repr\'esentation
sous forme d'arbres. La proc\'edure \`{a} mettre en oeuvre est d\'ecrite en appendice de \EOBethe.

\section{D\'eveloppements topologiques et g\'eom\'etrie alg\'ebrique.}

Cette partie repr\'esente le coeur de cette th\`{e}se. Apr\`{e}s avoir rappel\'e les d\'efinitions et propri\'et\'es de g\'eom\'etrie alg\'ebrique
n\'ecessaires \`{a} la compr\'ehension des r\'esultats principaux de ce chapitre, nous montrerons comment calculer les expressions explicites de tout le
d\'eveloppement topologique d'une grande famille d'observables du syst\`{e}me ainsi que de l'\'energie libre comme fonctions
sur la courbe spectrale classique.

\subsection{Rappels de g\'eom\'etrie alg\'ebrique.}\label{geoalg}

Le propos de cette th\`{e}se n'\'etant pas la g\'eom\'etrie alg\'ebrique \`{a} proprement parler, nous ne pr\'esenterons ici
que les notions n\'ecessaires \`{a} la lecture des paragraphes suivants. Le lecteur int\'eress\'e pourra se reporter aux ouvrages
\cite{Fay} et \cite{Farkas} pour les approfondir.

\subsubsection{Equation alg\'ebrique et surface de Riemann compacte associ\'{e}e.}

Soit une \'equation alg\'ebrique
\beq\label{eqalg}
\CE(x,y) = 0
\eeq
o\`{u} $\CE$ est un polyn\^{o}me de degr\'es $d_1+1$ et $d_2+1$ en ses deux variables $x$ et $y$. On appelle $\overline{\Sigma}$
la surface de Riemann compacte associ\'ee, c'est-\`{a}-dire qu'il existe deux fonctions meromorphes $x: \overline{\Sigma} \to \mathbb{C}$
et $y: \overline{\Sigma} \to \mathbb{C}$ telles que
\beq
\CE(x,y) = 0
\Leftrightarrow
\exists p \in \overline{\Sigma} \; \; \hbox{tel que} \;\;
\left\{\begin{array}{l}
x=x(p) \cr
y = y(p) \cr
\end{array} \right. .
\eeq
On associe ainsi \`{a} une courbe alg\'ebrique $\CE$ une surface de Riemann compact $\overline{\Sigma}$. Dans toute la suite,
on parlera donc indistinctement de surface de Riemann ou de courbe alg\'ebrique.

\br
Lorsque l'\'equation alg\'ebrique est de la forme
\beq
y^2 = \prod_{i=0}^{nbp} (x-a_{2i}) ( x - a_{2i+1}),
\eeq
on parle de courbe hyperelliptique. Ces courbes ont des propri\'et\'es particuli\`{e}res que nous d\'ecrirons dans la partie \ref{part1MM} du chapitre suivant.
\er

\subsubsection{Polytope de Newton.}
La courbe alg\'ebrique $\CE$ peut \^etre caract\'eris\'ee par les coefficients de son d\'evelop- pement
autour de $x,y \to \infty$:
\beq \label{eqalg2}
\CE(x,y) = \sum_{i=0}^{d_1+1} \sum_{j=0}^{d_2+1} E_{ij} \; x^i \, y^j .
\eeq
Un outil tr\`es utile \`{a} l'\'etude des propri\'et\'es de $\overline{\Sigma}$ consiste \`{a} repr\'esenter graphiquement
ces coefficients: on associe \`{a} tout $E_{ij}\neq 0$ un point \`{a} la position $(i,j)$ du plan. Le polygone ainsi obtenu
est appel\'e {\bf polytope de Newton} de $\CE$.

On peut directement lire sur ce polytope certaines propri\'et\'es de $\overline{\Sigma}$. Par exemple le genre maximal
de la surface $\overline{\Sigma}$ associ\'ee
est \'egal au nombre de points strictement contenus \`{a} l'int\'erieur du polytope.

On peut \'egalement y lire les propri\'et\'es des points \`{a} l'infini de $\overline{\Sigma}$, c'est-\`a-dire des p\^oles de $x$ et de $y$. Pour ce faire, il faut regarder
l'enveloppe externe du polytope, i.e. la fronti\`{e}re du plus petit domaine convexe contenant tous les points du
polytope. Cette fronti\`{e}re est compos\'ee des deux segments $[(0,0),(d_1+1,0)]$ et $[(0,0),(0,d_2+1)]$ d'une
part mais aussi d'un ensemble de segments $\{Seg_i = [u_i,u_{i+1}]\}_{i=1 \dots \CP-1}$ ordonn\'es tels que $u_1 = (0,d_2+1)$
et $u_\CP = (d_1+1,0)$. On peut alors montrer que chaque segment $Seg_i$ correspond \`{a} un p\^{o}le de la forme
$ydx$ dont le degr\'e est donn\'e par la pente du segment correspondant.

\bex
Consid\'erons la courbe
\beq
\CE(x,y) = x + x^3 + x^2 y^2 + y^3 + x y^2 .
\eeq
Son polytope est
\beq
\begin{array}{r}
{\rm \includegraphics[width=5cm]{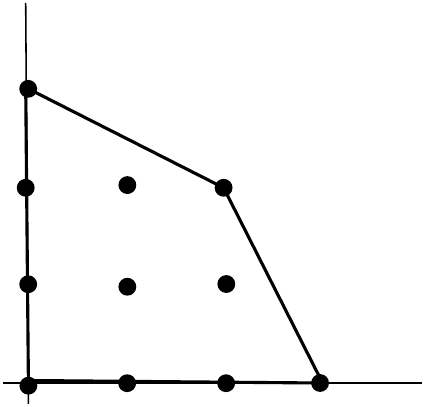}}
\end{array}
\eeq

On peut y voir par exemple que $\CE$ a un genre $g<3$ et deux points \`{a} l'infini.
\eex

\subsubsection{Structure en feuillets.}
Consid\'erons l'\'equation alg\'ebrique \eq{eqalg} comme une \'equation en $y$ \`{a} $x$ fix\'e. Elle a $d_2 + 1$ solutions,
i.e. pour tout $x$ (ou presque)
\beq
\exists (p^0,p^1,\dots,p^{d_2}) \in \overline{\Sigma}^{d_2+1} \; \hbox{tels que} \; \forall i \; x(p^i) = x.
\eeq
On a donc une structure en $d_2+1$ feuillets en $x$, c'est-\`{a}-dire que l'on a, de mani\`{e}re g\'en\'erique, $d_2+1$ copies de $\mathbb{C}$ superpos\'ees
se projetant toutes sur la base $x$.

De m\^eme, on notera les diff\'erents feuillets en $y$ par:
\beq
\exists (\tilde{p}^0,\tilde{p}^1,\dots,\tilde{p}^{d_2}) \in \overline{\Sigma}^{d_1+1} \; \hbox{tels que} \; \forall i \; y(\tilde{p}^i) = y.
\eeq

On peut exprimer $\CE$ de mani\`{e}re \`{a} faire appara\^{i}tre explicitement ces structures en feuillets:
\beq
\CE(x(p),y(q)) = E_{0,d_2} \prod_{i = 0}^{d_2} ( y(q) - y(p^i)) = E_{d_1,0} \prod_{i = 0}^{d_1} ( x(p) - x(\td{q}^i)).
\eeq

\subsubsection{Points de branchement:}
On nomme {\bf point de branchement} en $x$ tout point $a_i$ o\`{u} la diff\'erentielle $dx$ s'annule:
\beq
dx(a_i) = 0.
\eeq
Dans cette th\`ese, nous supposerons toujours que les points de branchements sont des z\'eros simples de $dx$, c'est-\`{a}-dire qu'au voisinage
d'un point de branchement $a_i$, $x$ se comporte comme le carr\'e de $y$:
\beq
y(p) {\displaystyle \sim_{p \to a_i}} \sqrt{x(p) -x(a_i)}.
\eeq

\begin{figure}
\hspace{4cm}  \includegraphics[width=8cm]{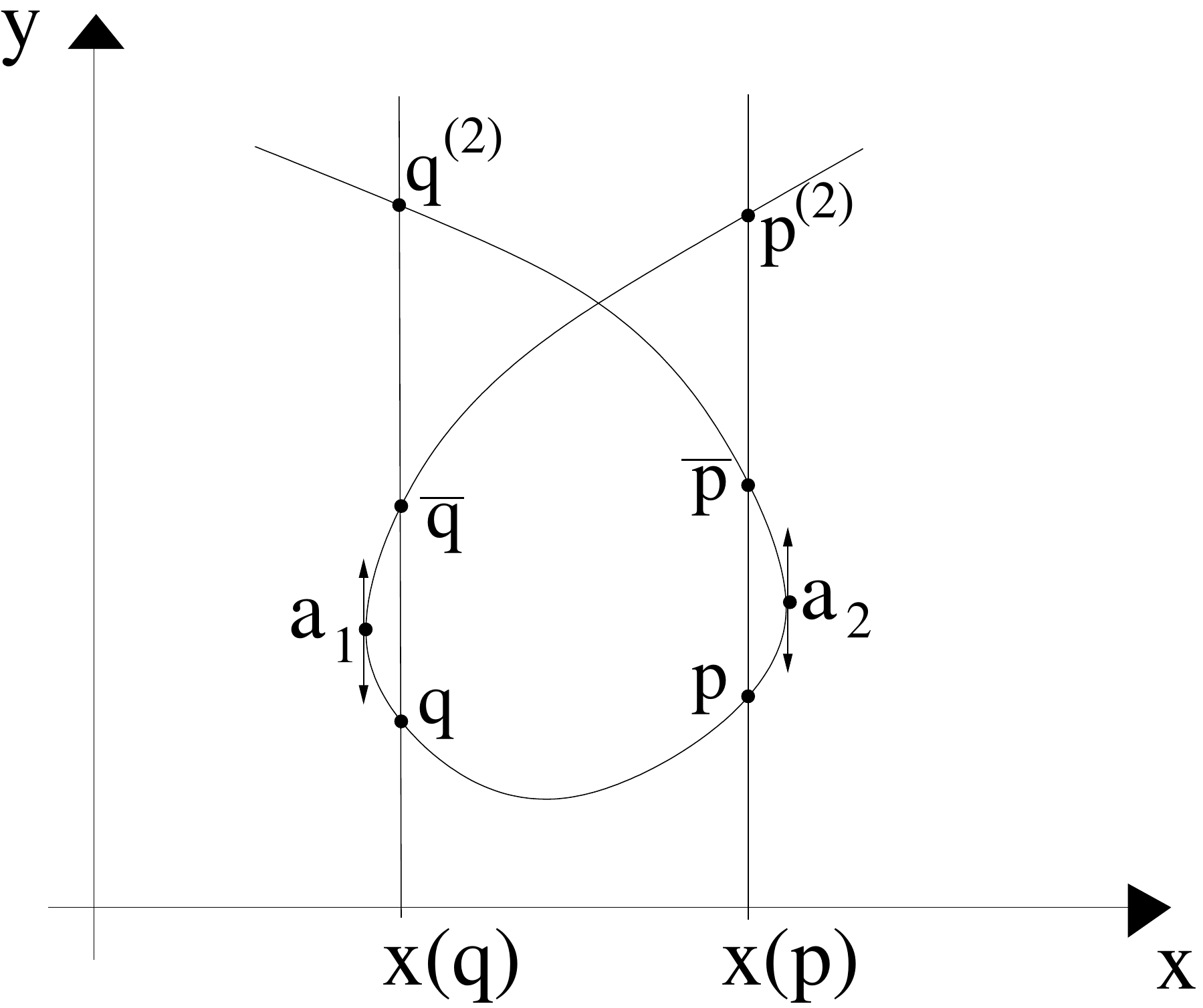}\\
  \caption{Exemple de courbe $\CE(x,y)$ pr\'esentant deux points de branchements en $x$. On peut voir que le point conjugu\'e n'est pas
  globalement d\'efini.}\label{branch1}
\end{figure}

Ces points correspondent aux points o\`{u} deux feuillets se croisent: deux points $p^{i}$ correspondant au m\^eme $x$
se rencontrent. En effet, pour tout point $p$ au voisinage d'un point de branchement $a_i$, il existe un unique point $\pbar$ tel que $x(p) = x(\pbar)$ et $\pbar$ approche $a_i$ quand
$p \to a_i$ (voir fig. \ref{branch1}). La courbe alg\'ebrique $\CE$ peut alors \^etre vue comme $d_2 +1$ sph\`{e}res de Riemann
recoll\'ees entre elles par des segments reliant les points de branchement en $x$. On passe de l'une \`{a} l'autre en traversant
ces segments.

\begin{figure}
\hspace{4.5cm}  \includegraphics[width=8cm]{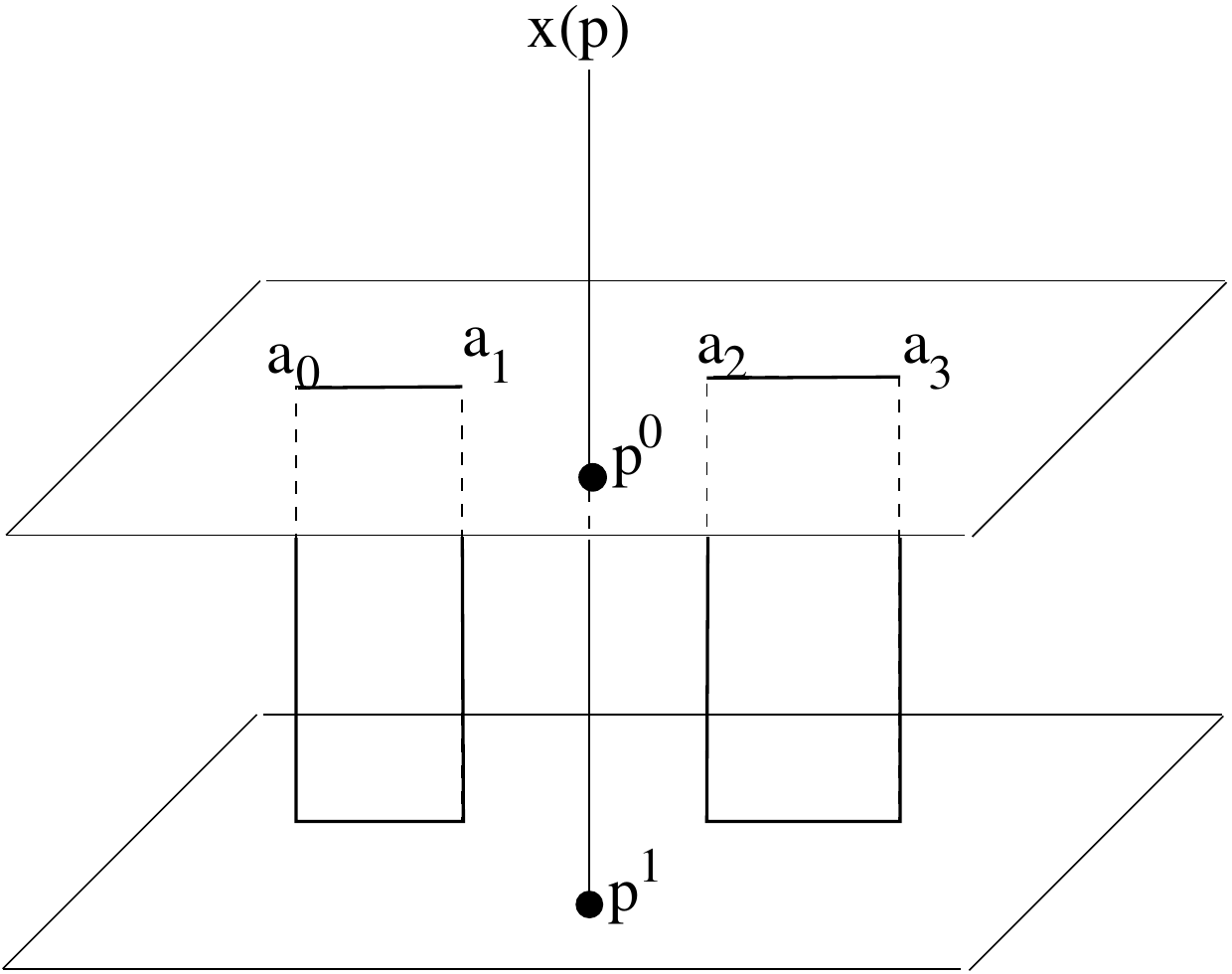}\\
  \caption{Repr\'esentation du tore sous la forme de deux feuillets en $x$ reli\'es par deux coupures, c'est-\`{a}-dire deux sph\`{e}res
  de Riemann recoll\'ees le long de deux segments.}\label{torussheet}
\end{figure}

\br
La d\'efinition du {\bf point conjugu\'e} $\pbar$ de $p$ d\'epend en g\'en\'eral du point de branchement $a_i$ consid\'er\'e
(voir par exemple la figure \ref{branch1}). Cependant, dans la mesure o\`{u} il sera toujours clair \`{a} quel point
de branchement on se r\'ef\`{e}re, nous omettrons de pr\'eciser celui-ci dans la suite.
\er

\br
La repr\'esentation de la surface de Riemann sous forme de feuillets reli\'es par des coupures semble indiquer qu'il existe
une relation entre le nombre de points de branchements et le genre de la surface. Ce lien est obtenu par le th\'eor\`{e}me de
Riemann-Hurwitz qui dit que le nombre de points de branchements $\# b.p$ et le genre $g$ sont reli\'es par l'\'equation:
\beq
g = - d_2 + {\# b.p \over 2}.
\eeq
\er

\subsubsection{Structure modulaire.}

$\bullet$ {\bf Base de cycles.}

\begin{figure}
  \includegraphics[width=15cm]{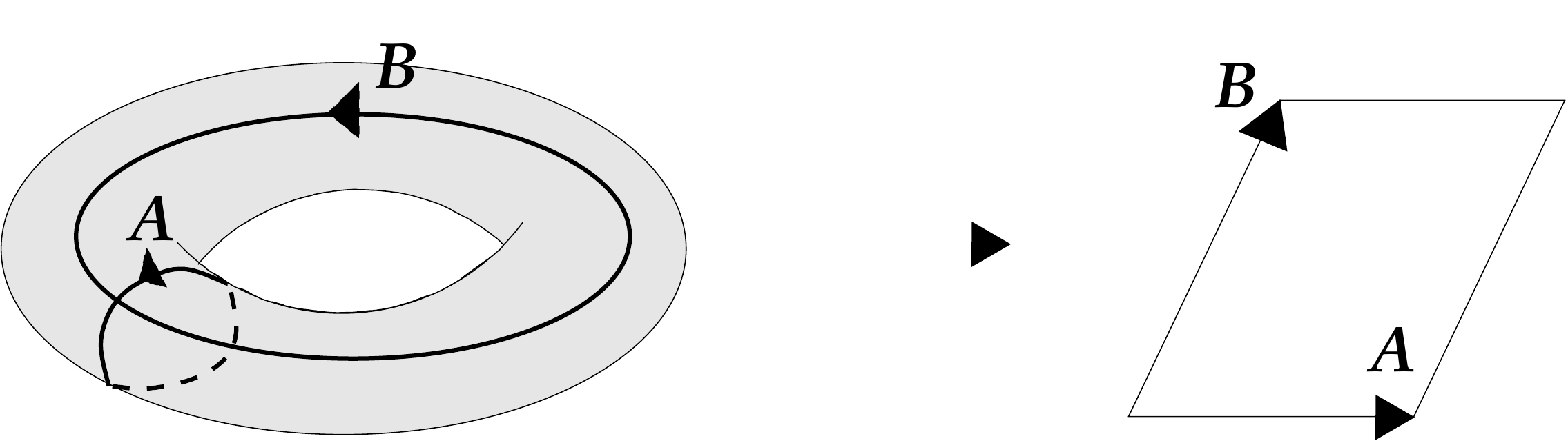}\\
  \caption{Le domaine fondamental du tore est un losange obtenu par d\'ecoupage de la surface de d\'epart suivant les cycles $\CA$
  et $\CB$.}\label{fundamental}
\end{figure}

Lorsque la courbe $\curve$ a pour genre $g$, il existe $2g$ cycles non triviaux homologiquement ind\'ependants. On peut alors
choisir une base symplectique $\{(\underline\CA_i,\underline\CB_i)\}_{i=1 \dots g}$ telle que
\beq
\underline\acycle_i\cap \underline\bcycle_j=\delta_{ij}
\virg
\underline\acycle_i\cap \underline\acycle_j=0
\virg
\underline\bcycle_i\cap \underline\bcycle_j=0
.
\eeq
En d\'ecoupant la surface le long des cycles $\CA$ et $\CB$ ainsi choisis, on obtient un domaine simplement connexe comme
d\'ecrit sur la figure \ref{fundamental}: le {\bf domaine fondamental}.

Sur cette courbe, il existe $g$ formes holomorphes lin\'eairement ind\'ependantes \eol
$du_1,\dots, du_g$, que l'on choisit normalis\'ees sur les cycles $\CA$:
\beq
\oint_{\underline\acycle_j} du_i = \delta_{ij}.
\eeq

Ces diff\'erentielles permettent de d\'efinir la {\bf matrice des p\'eriodes de Riemann} caract\'eristique de la surface $\overline{\Sigma}$.
C'est une matrice $\tau$ de taille $g \times g$ dont les \'el\'ements sont les int\'egrales des diff\'erentielles holomorphes sur les cycles $\underline\CB$:
\beq
\tau_{ij} = \oint_{\underline\bcycle_j} du_i.
\eeq
Cette matrice $\tau$ est sym\'etrique et satisfait
\beq
\tau_{ij}=\tau_{ji} \virg \Im\,\tau>0.
\eeq

D\'es lors, consid\'erant un point $p \in \overline{\Sigma}$ n'appartenant \`{a} aucun cycle $\underline\CA$ ni $\underline\CB$, on d\'efinit
l'application d'Abel
\beq
u_i(p) = \int_{p_0}^{p} du_i
\eeq
o\`{u} le contour d'int\'egration est dans le domaine fondamental.

Le vecteur \`{a} $g$ composantes ${\bf u}(p) = (u_1(p),\dots,u_g(p))$ transporte la courbe dans sa jacobienne.

Notons que l'application d'Abel d\'epend du point de base $p_0$ mais celui-ci
n'aura aucune influence dans toute la suite.

\subsubsection{Fonctions theta et formes premi\`{e}res}

On appelle {\bf caract\'eristique} tout vecteur ${\bf z} \in \mathbb{C}^g$ s'il existe deux vecteurs \`{a} coefficients entiers ${\bf a}\in \mathbb{Z}^g$
et ${\bf b}\in \mathbb{Z}^g$ tels que:
\beq
{\bf z}={{\bf a}+\tau.{\bf b}\over 2}.
\eeq
On dit que ${\bf z}$ et une caract\'eristique impaire si
\beq
\sum_{i=1}^g a_i b_i = {\rm impair}.
\eeq
Etant donn\'ees une caract\'eristique ${\bf z}={{\bf a}+\tau {\bf b}\over 2}$, une matrice sym\'etrique $\tau_{ij}=\tau_{ji}$
telle que $\Im\tau$ est d\'efinie positive et un vecteur ${\bf v} \in \mathbb{C}^g$, on d\'efinit la {\bf fonction th\'eta}
\beq
\theta_{\bf z}({\bf v},\tau) = \sum_{{\bf n}\in Z^g} \ee{i\pi ({\bf n}-{\bf b}/2)^t \tau ({\bf n}-{\bf b}/2)}\,\ee{2i\pi ({\bf v}+{\bf a}/2)^t.({\bf n}+{\bf b}/2)}.
\eeq
Si ${\bf z}$ est une caract\'eristique impaire, $\theta_{\bf z}$ est une fonctions impaire de ${\bf v}$\footnote{En particulier, on a alors $\theta_z(0,\tau)=0$}
et on d\'efinit la forme holomorphe
\beq
dh_{\bf z}(p) = \sum_{i=1}^g du_i(p)\,\,. \,\left.{\partial \theta_{\bf z}({\bf v})\over \partial v_i}\right|_{{\bf v}=0}.
\eeq
Elle a $g-1$ z\'eros qui sont des z\'eros doubles. On peut donc consid\'erer leurs racines carr\'ees d\'efinies dans la domaine fondamental
et la {\bf forme premi\`{e}re}:
\beq
\primef(p,q) = {\theta_{\bf z}(u(p)-u(q))\over \sqrt{dh_{\bf z}(p)\,dh_{\bf z}(q)}}.
\eeq
On peut montrer qu'elle est {\bf ind\'ependante de ${\bf z}$} et qu'elle n'a qu'un z\'ero sur la diagonale $p=q$ et aucun p\^ole.

\subsubsection{Structure complexe.}
En plus de la structure modulaire de $\overline{\Sigma}$, on doit \'egalement d\'ecrire sa structure complexe, c'est-\`{a}-dire
les propri\'et\'es des fonctions m\'eromorphes $x(p)$ et $y(p)$. Plus pr\'ecis\'ement, il nous suffit ici de pr\'eciser les modules
caract\'erisant la forme diff\'erentielles $ydx$: sa structure de p\^{o}les, son comportment au voisinage de ceux-ci ainsi que
son comportement le long des cycles $\underline\CA$.

On note $\{\alpha_i\}_{i=1\dots {\CP}}$ les p\^{o}les de $ydx$\footnote{Rappelons que le nombre de points \`a l'infini, ${\cal P}$,
peut \^etre connu en repr\'esentant le polytope de Newton de $\CE$.} et $z_{\alpha_i}$ une variable locale au voisinage de $\alpha_i$.
Pour d\'efinir cette derni\`{e}re, consid\'erons trois cas:
\begin{itemize}
\item soit $\alpha_i$ est un p\^{o}le de degr\'es $d_i$ de $x$, auquel cas on d\'efinit:
\beq
z_{\alpha_i}(p) := x(p)^{1 \over d_i};
\eeq

\item soit $\alpha_i$ n'est ni un p\^ole de $x$ ni un point de branchement en $x$ mais un p\^ole de $y$. Alors, on d\'efinit:
\beq
z_{\alpha_i}(p) := {1 \over x(p) - x(\alpha_i)};
\eeq

\item soit $\alpha_i$ est un point de branchement en $x$ (i.e. un z\'ero simple de $dx$) et donc un p\^ole de $y$. Dans ce cas
\beq
z_{\alpha_i}(p) := {1 \over \sqrt{x(p)-x(\alpha_i)}}.
\eeq

\end{itemize}

Dans tous les cas, $z_{\alpha_i}(p)$ a un p\^ole simple quand $p \to \alpha_i$ et donc $\zeta_{\alpha_i}(p):={1 \over z_{\alpha_i}}(p)$ a
un z\'ero simple, d\'efinissant ainsi une bonne variable locale au voisinage du point $\alpha_i$.

\vspace{0.5cm}

{\bf Les temp\'eratures.}

Un premier ensemble de modules est simplement donn\'e par les r\'esidus de $ydx$ en ses p\^{o}les:
\beq
t_{0,i}:= \Res_{p \to \alpha_i} y(p) \, dx(p).
\eeq

\br
Les temp\'eratures ne sont pas des modules ind\'ependants. En effet, il est facile de voir que
\beq
\sum_{i=1}^{\CP} t_{0,i} = 0.
\eeq
\er

\vspace{0.3cm}

{\bf Les modules aux p\^{o}les.}

Les ordres sous-dominants dans le d\'eveloppement de taylor de $ydx$ donnent les autres modules de cette forme li\'es aux points
\`{a} l'infini sur la surface. Pour les d\'ecrire, on introduit un "potentiel" associ\'e \`{a} chacune des singularit\'es
de la surface:
\beq
V_{i}(p) := \Res_{q \to \alpha_i} y(q) dx(q)\; \ln \left( 1 - {z_{\alpha_i}(p) \over z_{\alpha_i}(q)} \right).
\eeq
C'est un polyn\^{o}me en $z_{\alpha_i}(p)$. On peut donc l'\'ecrire:
\beq
V_{i}(p) := \sum_{j =1}^{\hbox{deg} V_i} t_{j,i} \; z_{\alpha_i}^j(p).
\eeq

\br
La donn\'ee de ces modules caract\'erise bien le comportement de $ydx$ aux voisinage des p\^{o}les car
\beq
ydx(p) {\displaystyle \sim_{p \to \alpha_i}} dV_i(p) - t_{0,i} {dz_{\alpha_i}(p) \over z_{\alpha_i}(p)} + O \left({dz_{\alpha_i}(p) \over z_{\alpha_i}^2(p)} \right)
\eeq
o\`{u} la diff\'erentielle
\beq
dV_i(p) = \Res_{q \to \alpha_i} y(q) dx(q) {dz_{\alpha_i}(p) \over z_{\alpha_i}(p) - z_{\alpha_i}(q)}
\eeq
satisfait
\beq
\Res_{p \to \alpha_i} dV_i(p) = 0.
\eeq
\er

\vspace{0.3cm}

{\bf Les fractions de remplissage\footnote{Cette d\'enomination vient directement de l'interpr\'etation des mod\`{e}les
de matrices comme un gaz de particules fermioniques (voir \cite{Mehta,courseynard}).}.}

On d\'efinit les fractions de remplissage $\epsilon_i$ pour $i = 1 \dots g$ comme les int\'egrales sur les cycles $\underline\CA$ de
la diff\'erentielle $ydx$:
\beq
\epsilon_i = {1 \over 2 i \pi} \oint_{\underline\CA_i} ydx.
\eeq

\vspace{0.3cm}

\subsubsection{Fonctions et diff\'erentielles fondamentales sur la courbe.}

Les fonctions de corr\'elation et les termes du d\'eveloppement topologique de l'\'energie libre peuvent \^{e}tre exprim\'es
uniquement \`{a} l'aide de deux briques \'el\'ementaires, deux diff\'erentielles d\'efinies sur la courbe spectrale:
le noyau de Bergmann et la diff\'erentielle Abelienne de troisi\`{e}me esp\`{e}ce.

{\bf Noyau de Bergmann.}

Sur la surface $\overline{\Sigma}$, il existe une unique diff\'erentielle bilin\'eaire $\underline{B}(p,q)$ ayant un unique
p\^{o}le double sur la diagonale $p=q$ sans r\'esidu et avec des cycles $\underline\CA$ nuls. C'est-\`{a}-dire qu'il est donn\'e par
les contraintes:
\beq
\underline{B}(p,q) = {dz(p) dz(q) \over (z(p) -z(q))^2} + \; \hbox{terme fini}
\;\; \hbox{et} \;\; \oint_{\underline\CA} \underline{B}(p,q) = 0
\eeq
o\`{u} $z$ est n'importe quelle variable locale au voisinage de $q$.

\vspace{0.5cm}

{\bf Diff\'erentielle Abelienne de troisi\`{e}me esp\`{e}ce.}

De m\^{e}me, pour tout couple de points $(q_1,q_2)$ de $\overline{\Sigma}$, il existe une unique forme diff\'erentielle $\underline{dS}_{q_1,q_2}(p)$ sur la surface avec deux p\^{o}les simples en $p \to q_1$ et $p \to q_2$
avec r\'esidus respectifs $1$ et $-1$ et normalis\'ee par l'annulation de ses int\'egrales sur les cycles $\underline\CA$:
\beq
\Res_{q_1} \underline{dS}_{q_1,q_2} = 1 = - \Res_{q_2} \underline{dS}_{q_1,q_2}
\virg
\oint_{\acycle_i} \underline{dS}_{q_1,q_2} = 0
.
\eeq

\vspace{0.5cm}

{\bf Propri\'et\'es.}

On peut tout d'abord remarquer que la diff\'erentielle Abelienne de troisi\`{e}me esp\`{e}ce peut \^{e}tre obtenue par int\'egration
du noyau de Bergmann sur un chemin du domaine fondamental allant de $q_2$ \`{a} $q_1$:
\beq
\underline{dS}_{q_1,q_2}(p) = \int_{q_2}^{q_1} \underline{B}(p,q)
\eeq
ou bien dans sa version d\'eriv\'ee
\beq
d_{q_1} \left(\underline{dS}_{q_1,q_2}(p)\right) =  \underline{B}(q_1,p).
\eeq

Le noyau de Bergmann, en tant que d\'eriv\'ee seconde d'une foncion $\theta$, est sym\'etrique:
\beq
\underline{B}(p,q) = d_p d_q \ln{(\theta_{\bf z}(u(p)-u(q)))} = \underline{B}(q,p)
\eeq
o\`{u} ${\bf z}$ est une caract\'eristique impaire quelconque. Par int\'egration, on obtient:
\beq
\underline{dS}_{q_1,q_2}(p) = d_p \ln{\left(\theta_{\bf z}({\bf u}(p)-{\bf u}(q_1))\over \theta_{\bf z}({\bf u}(p)-{\bf u}(q_2))\right)}
\eeq
et
\beq
\underline{dS}_{q_1,q_2} = - \underline{dS}_{q_2,q_1}.
\eeq

Les int\'egrales sur les cycles $\underline\CB$ sont donn\'ees par
\beq
\oint_{q\in\underline\bcycle_i} \Bergmann(p,q) = 2i\pi\, du_i(p)
\qquad
\hbox{et}
\qquad
\oint_{\bcycle_i} \underline{dS}_{q_1,q_2} = 2i\pi (u_i(q_1)-u_i(q_2)).
\eeq

\vspace{0.3cm}

{\bf Formules de Cauchy.}

Pour une fonction meromorphe $f(p)$, les formules de Cauchy ainsi que les propri\'et\'es de $\underline{dS}$ et $\underline{B}$ permettent d'\'ecrire
\beq
f(p) = - \Res_{q_1\to p} \underline{dS}_{q_1,q_2}(p) f(q_1)
\eeq
ainsi qu'une expression pour la diff\'erentielle de $f$ :
\beq\label{Bergmanndiff}
df(p) = \Res_{q\to p} \Bergmann(p,q) f(q).
\eeq

\subsubsection{Identit\'e bilin\'eaire de Riemann.}

Tout au long de ce chapitre, nous aurons besoin de d\'eplacer des contours d'int\'egra- tion sur une surface de Riemann.
Si l'int\'egrand a des int\'egrales non nulles le long de cycles non-triviaux, celles-ci doivent \^etre prises en compte
lorsque l'on d\'eplace le contour d'int\'egration. L'identit\'e bilin\'eaire de Riemann \cite{Farkas} explicite cette d\'ependance.

Soient $\om_1$ et $\om_2$ deux formes m\'eromorphes sur la surface $\overline\Sigma$
et un point $p_0$ arbitraire. Consid\'erons la fonction $\Phi_1$ d\'efinie sur le domaine fondamental par
\beq
\Phi_1(p) = \int_{p_0}^p \om_1
\eeq
o\`u le chemin d'int\'egration reste \`a l'int\'erieur du domaine fondamental.

On a alors l'{\bf identit\'e bilin\'eaire de Riemann}:
\beq\label{Riemannbilinear}
\Res_{p\to {\rm tous \; les \; poles}} \Phi_1(p)\om_2(p) = {1\over 2i\pi}\, \sum_{i=1}^g \oint_{\underline\acycle_i} \om_1 \oint_{\underline\bcycle_i} \om_2 - \oint_{\underline\bcycle_i} \om_1 \oint_{\underline\acycle_i} \om_2 .
\eeq
En particulier, pour $\om_1(p)=B(p,q)$, on obtient:
\beq
\Res_{p\to {\rm tous \; les \; poles}} dS_{p,p_0}(q)\, \om (p) = - \sum_{i=1}^g du_i(q)\,\oint_{\acycle_i} \om
\eeq
ainsi que:
\beq\label{blineardSCauchy}
\om(q) = \Res_{p\to {\rm poles \, de}\,\om} dS_{p,p_0}(q)\, \om (p) + \sum_{i=1}^g du_i(q)\,\oint_{\acycle_i} \om
.\eeq

\subsubsection{D\'ecomposition de $ydx$ sur les modules de la structure complexe.}

En utilisant l'\'equation bilin\'eaire de Riemann \eq{blineardSCauchy}, on peut r\'e\'ecrire $ydx$ de mani\`{e}re \`{a} faire appara\^{i}tre
explicitement les modules de la courbe $\CE$:
\beq\label{decompoydx2MM}
ydx = \sum_{i,j} t_{j,i} B_{j,i} + \sum_i t_{0,i} \underline{dS}_{\alpha_i,o} + 2 i \pi  \sum_i \epsilon_i du_i
\eeq
avec
\beq
B_{j,i}(p) = - \Res_{q \to \alpha_i} \underline{B}(p,q) z_{\alpha_i}(q)^j.
\eeq
Cette d\'ecomposition est tr\`{e}s utile puisqu'elle nous permet de reporter toute variation des modules introduits jusqu'ici en une variation de
la forme diff\'erentielle $ydx$ seule.

\subsubsection{Briques \'el\'ementaires.}

A partir de ces fonctions, on d\'efinit deux \'el\'ements qui seront la base de la r\'esolution des \'equations de boucles.
Pour un point $p\in \overline\Sigma$ quelconque et un point $q$ proche d'un  point de branchement $a_i$ (i.e. tel que l'on puisse d\'efinir
un point conjugu\'e $\qbar$.), on d\'efinit les formes diff\'erentielles:
\beq
\omega(q) := (y(q)-y(\qbar))dx(q)
\eeq
et
\beq
\underline{dE}_q(p) = {1 \over 2} \int_{q}^{\qbar} B(.,p)
\eeq
o\`{u} le chemin d'int\'egration reste dans un voisinage du point de branchement $a_i$.

\subsubsection{Fonction tau de Bergmann.}

La {\bf fonction tau de Bergmann}, $\tau_{Bx}$, a \'et\'e introduite dans \cite{KoKo,KoKo2,EKK} pour l'\'etude des espaces de Hurwitz et
de la premi\`ere correction de l'\'energie libre des mod\`eles de matrices. Elle est d\'efinie de mani\`ere unique par les contraintes:
\beq\label{deftauBx}
{\d \ln{(\tau_{Bx})}\over \d x(a_i)} =  \Res_{p\to a_i} {B(p,\pbar)\over dx(p)}
\eeq
pour tout point de branchement $a_i$. La formule variationnelle de Rauch \cite{Rauch} assure que le membre de droite
est une forme ferm\'ee et donc que cette d\'efinition n'est pas caduque. Cependant, notons que cette fonction est d\'efinie \`a
une constante multiplicative pr\`es qui ne jouera aucun r\^{o}le dans la suite.

\subsection{Propri\'et\'es de la courbe spectrale du mod\`{e}le \`{a} deux matrices.}

Dans ce paragraphe nous allons nous attarder sur les propri\'et\'es particuli\`{e}res de la courbe spectrale classique d 'un mod\`{e}le
\`{a} deux matrices dont nous rappelons la forme:
\beq
\CE_{2MM}(x,y) = (V_1'(x) - y)(V_2'(y)-x) + P(x,y)
\eeq
o\`{u} $V_1$ et $V_2$ sont des polyn\^{o}mes de degr\'es respectifs $d_1+1$ et $d_2+1$ et $P(x,y)$ est un polyn\^{o}me
en ses deux variables de degr\'e $d_1-1$ en $x$ et $d_2-1$ en $y$.

\subsubsection{Polytope de Newton.}

\begin{figure}
\hspace{3cm}  \includegraphics[width=10cm]{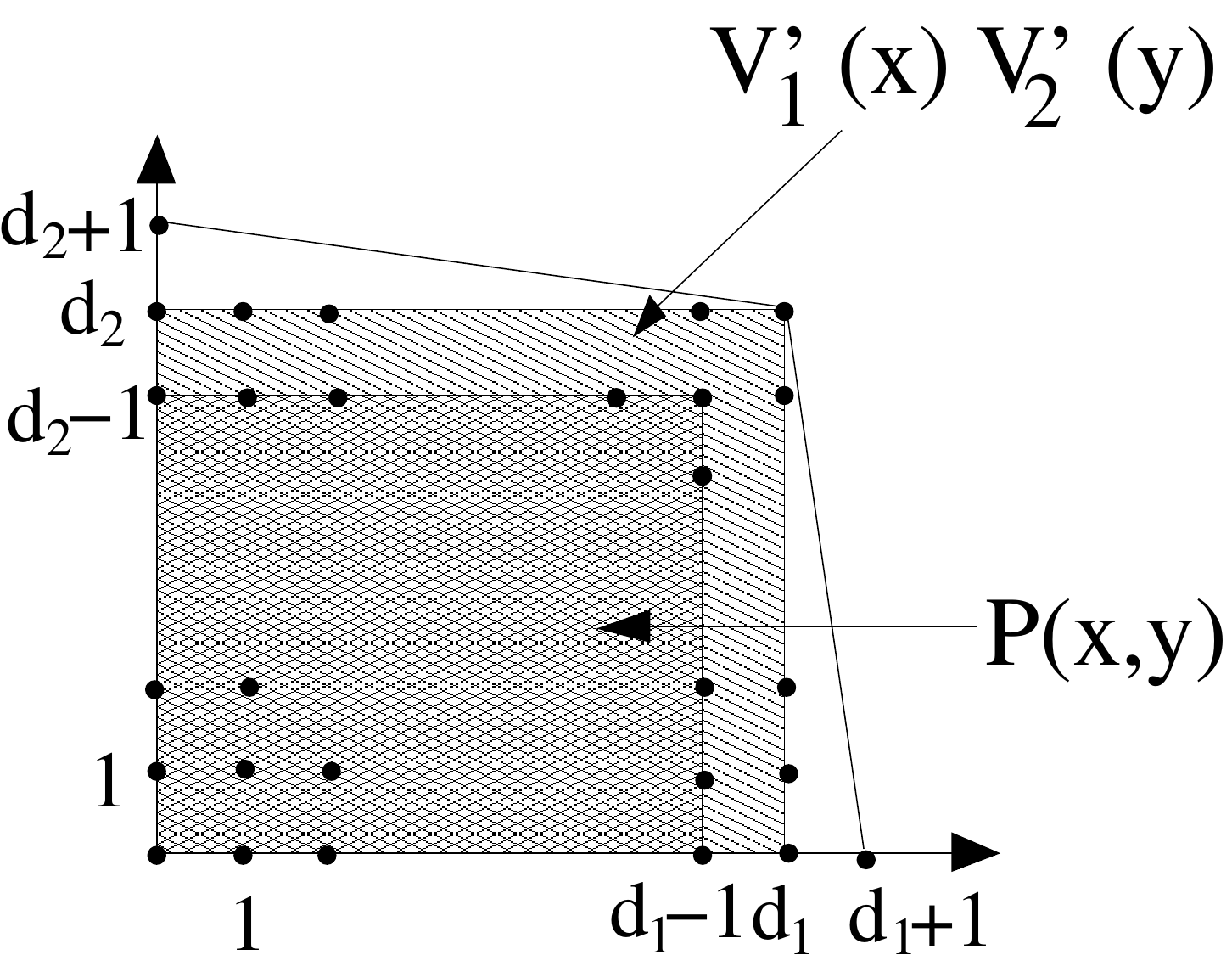}\\
  \caption{Polytope de Newton de $\CE_{2MM}$. Les parties hachur\'ees mettent en valeur les contributions des diff\'erents
  termes, $P(x,y)$ et $V_1'(x)V_2'(y)$ respectivement. La partie non hachur\'ee correspond \`{a} la contribution
  exclusive de $x V_1'(x) + y V_2'(y)$.}\label{pol2MM}
\end{figure}

Le d\'eveloppement de $\CE_{2MM}$ peut s'\'ecrire:
\beq
\CE_{2MM} = t_{d_1} x^{d_1 +1} + \tilde{t}_{d_2} y^{d_2+1} + x^{d_1} (t_{d_1} V_2'(y) + t_{d_1-1}) + y^{d_2} (\tilde{t}_{d_2} V_1'(x) + \tilde{t}_{d_2-1})
+ P(x,y).
\eeq
Il est alors clair que $P(x,y)$ est contenu strictement \`{a} l'int\'erieur du polytope et que ce dernier prend la forme repr\'esent\'ee
dans la figure (\ref{pol2MM}).

La simple lecture de ce polytope nous permet de conclure que la surface de Riemann correspondante a un genre $g$ major\'e
par
\beq
g \leq d_1 d_2 -1 .
\eeq
On peut \'egalement voir que sa fronti\`{e}re est compos\'ee de deux segments, donc la forme $ydx$ a deux p\^{o}les dont
nous \'etudions les propri\'et\'es dans le paragraphe suivant.

\subsubsection{Points \`{a} l'infini.}

Il y a deux points \`{a} l'infini sur la surface $\overline{\Sigma}$, c'est-\`{a}-dire deux points qui sont des p\^{o}les
de la fonction meromorphe $x(p)$ ou de $y(p)$:

\begin{itemize}
\item L'un, not\'e $\infty_x$, est un p\^{o}le simple pour $x(p)$ et de degr\'es $d_1$ pour $y(p)$ (on peut le voir d'apr\`{e}s le facteur
$y-V_1'(x)$ dans $\CE_{2MM}$ par exemple);

\item L'autre, not\'e $\infty_y$, est un p\^{o}le simple pour $y(p)$ et de degr\'es $d_2$ pour $x(p)$ (d'apr\`{e}s le facteur
$x-V_2'(y)$ dans $\CE_{2MM}$).

\end{itemize}

Si l'on s'int\'eresse maintenant \`{a} la structure de p\^{o}les de la forme diff\'erentielle $ydx$, on voit qu'elle a un p\^{o}le de
degr\'es $d_1+2$ en $\infty_x$ et de degr\'es $d_2+2$ en $\infty_y$.

\subsubsection{Structure en feuillets.}

Nous nous int\'eressons  dans ce paragraphe \`{a} la structure en feuillets en $x$, c'est \`{a} dire que nous fixons la variable
complexe $x$ dans l'\'equation alg\'ebrique $\CE_{2MM}(x,y)=0$ et regardons les diff\'erentes solutions correspondantes en $y$.

De mani\`{e}re g\'en\'erique $\CE_{2MM}(x,y)=0$ a $d_2+1$ solutions en $y$. Il existe donc $d_2+1$ points $\{p^{i}\}_{i=0 \dots d_2}$ de la surface alg\'ebrique
compacte $\overline\Sigma$ se projetant sur un m\^{e}me point $x= x(p^{i})$ dans le plan complexe: il y a $d_2+1$ feuillets en $x$,
chacun \'etant associ\'e \`{a} un $p^{i}$ diff\'erent. Chaque exposant se r\'ef\`{e}rera donc \`{a} partir de maintenant \`{a} un feuillet.
Ces feuillets  ne sont cependant pas tous \'equivalents. En effet, l'un de ces feuillets est caract\'eris\'e par le fait
qu'il contient le point $\infty_x$ alors que tous les autres feuillets se rejoignent en $\infty_y$. On appelle {\bf feuillet physique} le feuillet contenant
$\infty_x$ et on lui associe l'exposant $0$. Nous expliquerons cette d\'enomination dans la partie \ref{partfeuillet}.

\begin{figure}
\hspace{4cm}  \includegraphics[width=8cm]{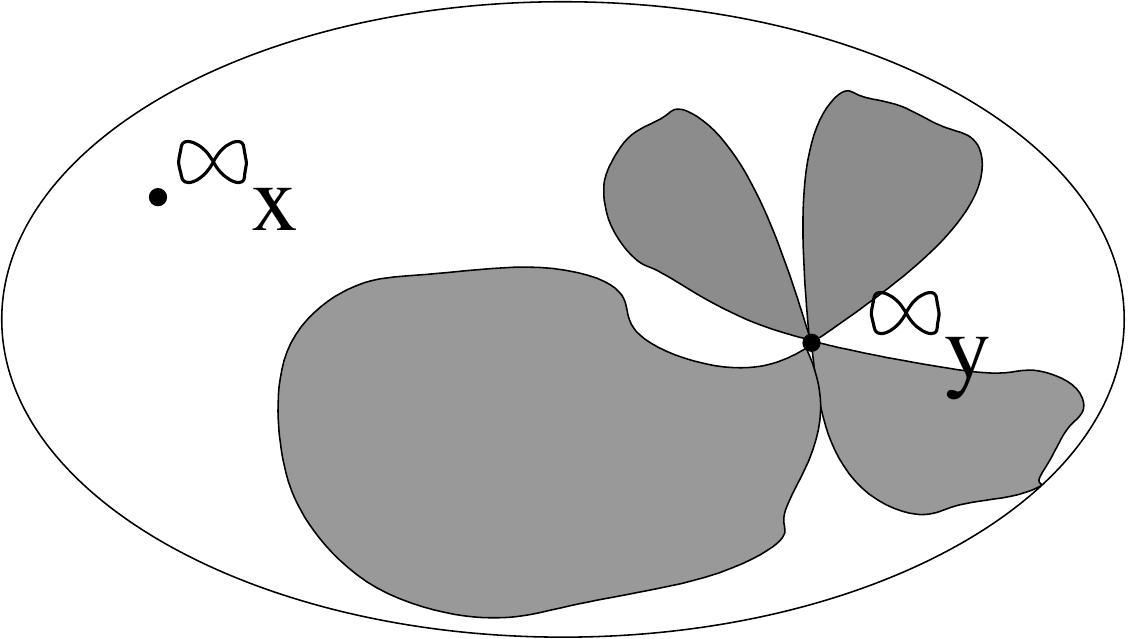}\\
  \caption{La structure en feuillets en $x$ de $\CE_{2MM}$ fait appara\^{i}tre un unique
  feuillet contenant $\infty_x$ alors que les quatre autres (en gris\'e) contiennent $\infty_y$. On a repr\'esent\'e ici
  l'exemple d'une surface de genre 0 par souci de simplicit\'e, mais le m\^eme type de description s'applique bien s\^ur \`{a} des surfaces de genre
  plus \'elev\'e.}\label{feuil}
\end{figure}

\subsubsection{Cas particulier: Courbe de genre 0.}
Si l'on est int\'eress\'e par la combinatoire des cartes portant une structure de spin, on se place dans le cadre d\'ecrit dans la partie \ref{partcombi}
et la courbe $\CE_{2MM}$ est alors de genre nul. Il n'y
a alors plus de fraction de remplissage $\epsilon_i$ ni de base de cycles $(\underline\CA,\underline\CB)$. Ce cas correspond vraiment
 \`{a} traiter un mod\`{e}le d'Ising sur surface al\'eatoire.

La condition de genre 0 assure \'egalement que l'on peut trouver une param\'etrisation rationnelle de la courbe $\CE_{2MM}(x,y)$.
Cela signifie que l'on peut trouver deux fonctions rationnelles $P(z)$ et $Q(z)$ d\'efinies dans le plan complexe telles que:
\beq
\forall \,  z \in \mathbb{C} \, , \; \CE_{2MM}(P(z),Q(z)) = 0.
\eeq
On repr\'esente donc $\CE_{2MM}$ par cette param\'etrisation:
\beq
\CE_{2MM} := \left\{ \begin{array}{l}
x= P(z) \cr
y= Q(z) \cr
\end{array}
\right.
\eeq
D\'es lors, toute fonction et diff\'erentielle sur $\overline{\Sigma}$ s'exprime simplement en termes de ces fonctions rationnelles.
En particulier le noyau de Bergmann est donn\'e par:
\beq
\underline{B}(z_1,z_2) = {P'(z_1) P'(z_2) dz_1 dz_2 \over (P(z_1) - P(z_2))^2}
\eeq
o\`{u} $P'(z)$ est la d\'eriv\'ee de la fonction rationnelle $P(z)$ par rapport \`{a} sa variable complexe $z$.

\section{D\'eveloppement topologique du mod\`{e}le \`{a} deux matrices.}\label{partdevtopo}

Dans cette partie, nous allons montrer comment on peut calculer tous les termes du d\'eveloppement topologique de n'importe quelle
fonction de corr\'elation du mod\`{e}le \`{a} deux matrices hermitiennes, y compris l'\'energie libre. Nous ne pr\'esenterons ici
que les r\'esultats avec parfois des id\'ees des d\'emonstrations, les preuves compl\`{e}tes \'etant assez techniques et \'ecrites
dans nos articles avec L. Chekhov et B. Eynard \eyno\CEO.

Nous allons pr\'esenter les r\'esultats par ordre de complexit\'e croissante:

\begin{itemize}
\item Dans un premier temps, nous montrons comment d\'efinir une courbe spectrale semi-classique \`{a} partir des corrections
en ${1 \over N^2}$ de la courbe spectrale classique;

\item Nous verrons ensuite comment celle-ci nous permet de calculer tout le \eol
d\'eveloppement topologique des fonctions de correlation
simples;

\item L'introduction d'un nouvel op\'erateur nous permet alors de remonter jusqu'au d\'eveloppement topologique de l'\'energie libre;

\item Enfin, nous pr\'esenterons une formule permettant le calcul de n'importe quelle fonction de corr\'elation \`{a} n'importe quel
ordre de correction en ${1 \over N^2}$.

\end{itemize}

\subsection{D\'efinitions, solutions des \'equations de boucles et feuillet physique.}\label{partfeuillet}

Comme nous l'avons vu plus t\^ot dans la partie \ref{masterloop} consacr\'ee \`{a} l'\'equation de boucles ma\^itresse, la r\'esolution
des \'equations de boucles donne g\'en\'eralement plusieurs solutions pour les fonctions de corr\'elation: ce sont donc
des fonctions multivalu\'ees du plan complexe. Dans toute la suite, nous allons r\'esoudre les \'equations de boucles
en \'etudiant les fonctions de corr\'elations comme des diff\'erentielles sur la courbe spectrale classique o\`{u} elles sont
monovalu\'ees.
Le fait que ces derni\`{e}res donnent lieu \`{a} plusieurs solutions diff\'erentes dans le plan complexe vient alors
tout simplement de la structure en feuillets (que ce soit en $x$ ou en $y$) de la surface de Riemann. En effet, pour une valeur complexe de $x(p)$ ou de $y(p)$,
il existe plusieurs points $p$ correspondant sur la courbe $\CE_{2MM}$, chacun d'eux induisant une valeur diff\'erente pour les
fonctions de corr\'elation.

Cependant, en revenant \`{a} la d\'efinition m\^eme de ces objets, on peut s\'electionner la solution
physiquement (ou combinatoirement) int\'eressante. En effet, les fonctions de corr\'elation sont des s\'eries g\'en\'eratrices
valables uniquement comme s\'eries formelles  quand leur param\`{e}tre de d\'eveloppement $x$ (resp. y) tend vers l'infini
avec un pole simple\footnote{Cette condition est souvent r\'esum\'ee par le comportement asymptotique de la r\'esolvante:
$W_{1,0}(x) \sim {1 \over x} + O(x^{-2})$ quand $x \to \infty$.}. Or, parmi les diff\'erents ant\'ec\'edents $p^i$ de $x(p)$
(resp. $\td{p}^j$ de $y(p)$), un seul satisfait cette condition: c'est l'ant\'ec\'edent se trouvant dans le feuillet physique
en $x$, $p^0$ (resp. le feuillet physique en $y$, $\td{p}^0$), expliquant ainsi cette d\'enomination.

D\'es \`{a} pr\'esent, et ce tout au long de ce chapitre, le terme fonction de corr\'elation ne se rapportera plus aux
fonctions complexes $\overline{H}_{{\bf k};l;m}^{(h)}$ d\'efinies plus haut mais aux diff\'erentielles
\beq
H_{{\bf k};l;m}^{(h)}(S_{k_1}, \dots, S_{k_i} ; {\bf p_L}; {\bf q_M})
\eeq
dont les arguments $\{p_{i,j},q_{i,j}\}_{j=1}^{k_i}$ pour $i = 1, \dots, k$, $\{p_i\}_{i=1}^l$ et $\{q_i\}_{i=1}^m$ sont
des points de la surface de Riemann $\overline{\Sigma}$ et satisfaisant la condition
\bea
&&H_{{\bf k};l;m}^{(h)}(S_{k_1}, \dots, S_{k_i} ; {\bf p_L}; {\bf q_M}) = \cr
 &&\overline{H}_{{\bf k};l;m}^{(h)}(S_{k_1}, \dots, S_{k_i} ; {\bf x(p_L)}; {\bf y(q_M)})
\prod dx(p_{i,j}) \, dy(q_{i,j})\,dx(p_i) \, dy(q_j)\cr
&& + \delta_{h,0} \delta_{l,0} \delta_{m,0} \delta_{i,1} \delta_{k_1,1} dx(p_{1,1}) dy(q_{1,1})\cr
&& + \delta_{h,0} \delta_{i,0} \delta_{k,2} \delta_{l,0} {dx(p_1) dx(p_2) \over (x(p_1)-x(p_2))^2}\cr
&& + \delta_{h,0} \delta_{i,0} \delta_{k,0} \delta_{l,2} {dy(q_1) dy(q_2) \over (y(q_1)-y(q_2))^2}\cr
\eea
pour des arguments $p$ dans le feuillet physique en $x$ et $q$ dans le feuillet physique en $y$.

On notera \'egalement les fonctions de corr\'elation simples:
\beq
W_{k,l}^{(g)}:= H_{0;k;l}^{(g)}.
\eeq

Dans la suite, nous allons donc calculer ces fonctions de corr\'elation en n'importe quel point de la surface de Riemann,
mais il ne faudra pas oublier que l'on n'obtient de valeur physique que pour des arguments dans leurs feuillets physiques respectif.

On introduit finalement les fonctions auxiliaires suivantes:
\beq
\widetilde{U}_{{\bf k};l;m}^{(h)}(S_{k_1}, \dots, S_{k_i} ; {\bf p_L}; {\bf q_M}):=Pol_{x(p_{1,1})} V_1'(x(p_{1,1})) H_{{\bf k};l;m}^{(h)}(S_{k_1}, \dots, S_{k_i} ; {\bf p_L}; {\bf q_M}),
\eeq
\beq
{U}_{{\bf k};l;m}^{(h)}(S_{k_1}, \dots, S_{k_i} ; {\bf p_L}; {\bf q_M}):=Pol_{y(q_{1,k})} V_2'(y(q_{1,k})) H_{{\bf k};l;m}^{(h)}(S_{k_1}, \dots, S_{k_i} ; {\bf p_L}; {\bf q_M})
\eeq
et
\beq
{P}_{{\bf k};l;m}^{(h)}(S_{k_1}, \dots, S_{k_i} ; {\bf p_L}; {\bf q_M}):=Pol_{x(p_{1,1})} V_1'(x(p_{1,1})) U_{{\bf k};l;m}^{(h)}(S_{k_1}, \dots, S_{k_i} ; {\bf p_L}; {\bf q_M})
\eeq
qui sont des polyn\^omes respectivement en $x(p_{1,1})$, $y(q_{1,k})$ et $x(p_{1,1})$ et $y(q_{1,k})$.

\subsection{Quelques propri\'et\'es des fonctions de corr\'elation.}\label{propcorr}

Dans ce paragraphe nous r\'esumons quelques propri\'et\'es importantes des fonctions de corr\'elation.

\bl\label{intWA}
Les fonctions de corr\'elation simples ont toute des int\'egrales nulles autour des cycles $\underline{\CA}$:
\beq
\int_{p \in \underline\CA} W_{k+1;l}^{(h)}(p,{\bf p_K};{\bf q_L}) =0
\eeq
pour tout $k+l+h>0$.
\el
Cette propri\'et\'e vient directement du fait que les fractions de remplissage $\epsilon_i$ sont les int\'egrales
de $ydx$ et donc de $W_{1;0}^{(0)}$ sur ces cycles et qu'elles sont ind\'ependantes des coefficients des potentiels
$V_1$ et $V_2$.

\bl\label{polW}
Les fonctions de corr\'elation simples $W_{k+1;l}^{(h)}(p,{\bf p_K};{\bf q_L})$ ont des p\^{o}les seulement aux points de branchement $p \to a_i$.
\el
Ce r\'esultat s'obtient par r\'ecurrence gr\^ace aux \'equations de boucles.

Un autre r\'esultat classique donne la fonction \`a deux points de genre 0 comme une fonction fondamentale sur
la courbe spectrale clasique:
\bl\label{W2Berg}
La fonction \`a deux points de genre 0 est le noyau de Bergmann sur la courbe spectrale $\CE$:
\beq\encadremath{
W_{2;0}^{(0)}(p,q) = W_{0;2}^{(0)}(p,q) = - W_{1;1}^{(0)}(p;q) = \underline{B}(p,q).}
\eeq
\el
Ce r\'esultat est fondamental puisqu'il nous servira de base dans toute la suite. Nous nous servirons de cette fonction \`a deux points
comme donn\'ee initiale pour une r\'esolution par r\'ecurrence.

\subsection{Courbe spectrale compl\`ete.}

Reprenons l'\'equation de boucle maitresse \eq{eqmaitresse1}, mais conservons le terme correctif en ${1 \over N^2}$:
\beq\label{masteq}
E(x,Y(x)) = {1 \over N^2} U(x,Y(x);x)
\eeq
avec
\beq
E(x,y) = {\cal{E}}_{2MM}(x,y) + \sum_{g=1}^\infty N^{-2g} E^{(g)}(x,y).
\eeq

On peut montrer que les deux fonctions $E$ et $U$ apparaissant dans cette \'equation peuvent s'\'ecrire:
\bt\label{completecurve}
\beq\label{expE}
\encadremath{
E(x(p),y) = -\td{t}_{d_2+1}\,"\left<\prod_{i=0}^{d_2} (y-V'_1(x(p^i))+\hbar \Tr{1\over x(p^i)-M_1}) \right>"
}\eeq
et
\beq\label{expU}
U_{0}(p,y) = -\td{t}_{d_2+1}\,"\left<\prod_{i=1}^{d_2} (y-V'_1(x(p^i))+ \hbar \Tr{1\over x(p^i)-M_1})  \right>" .
\eeq
o\`{u} les guillemets $"\left<.\right>"$ signifient que chaque fois que l'on rencontre une fonction \`{a} deux points en d\'eveloppant en cumulants,
on la remplace par
\beq
W_{2,0}(x,x'):= \left< \Tr{1 \over x - M_1} \Tr{1 \over x' - M_1}\right> +{1\over (x-x')^2}.
\eeq
\et

Ce r\'esultat est obtenu simplement  en montrant que \eq{masteq} a une unique solution, \'etant donn\'ees les propri\'et\'es de $U$ et $E$, et que les formules donn\'ees dans le th\'eor\`{e}me sont
effectivement solutions ( voir \CEO pour plus de d\'etails).

A priori cette formule d\'efinit une fonction $E(x(p),y)$ polyn\^omiale en $y$ mais dont le comportement en son autre variable $p$
est moins \'evident: sa sym\'etrie sur tous les feuillets $p^{i}$ permet juste de conclure qu'il s'agit d'une fonction rationnelle de $x(p)$.
N\'eanmoins, une \'etude approfondie de la structure de ses p\^{o}les permet de v\'erifier qu'il s'agit bien d'un polyn\^{o}me
en $x(p)$ \CEO. On peut m\^{e}me aller plus loin et d\'emontrer que
\bl
$E(x,y)$ est sym\'etrique dans le r\^{o}le de $x$ et $y$:
\bea
E(x(p),y(q)) &=& -\td{t}_{d_2+1}\,"\left<\prod_{i=0}^{d_2} (y(q)-V'_1(x(p))+\hbar \Tr{1\over x(p^i)-M_1}) \right>" \cr
&=& - t_{d_1+1}\,"\left<\prod_{i=0}^{d_1} (x(p)-V'_2(y(q))+\hbar \Tr{1\over y(\tilde{q}^i)-M_2}) \right>".\cr
\eea
\el

Cette propri\'et\'e, tr\`{e}s forte \`a premi\`ere vue, est tout \`{a} fait naturelle. En effet, la courbe spectrale, qu'elle soit classique ou semi-classique,
est obtenue par la d\'erivation d'\'equations de boucles par changement de variable dans l'int\'egrale matricielle. Nous avons
ici choisi d'effectuer le changement $\delta M_1 :=  {1 \over x-M_1} {V_2'(y)-V_2'(y) \over y-M_2}$ brisant la sym\'etrie naturelle
du mod\`{e}le en $M_1$ et $M_2$. Mais nous aurions tout aussi bien pu consid\'erer le changement de variable
$\delta M_2 := {V_1'(x) - V_1'(M_1) \over x-M_1} {1 \over y-M_2}$ pour obtenir la m\^{e}me information \`{a} travers une autre
courbe spectrale correspondant \`a la premi\`{e}re par l'\'echange des r\^{o}les respectifs de $x$ et de $y$. D\'ej\`{a} au niveau classique la courbe est explicitement
sym\'etrique:
\beq
\CE_{2MM}(x,y) = (V_1'(x) - y)(V_2'(y)-x) + P^{(0)}(x,y).
\eeq
Il n'est donc pas surprenant que cette sym\'etrie se retrouve dans tout le d\'eveloppement topologique de cette courbe spectrale
qui contient toute l'information n\'ec\'essaire \`{a} la r\'esolution du mod\`{e}le (mis \`{a} part un petit nombre de param\`{e}tres:
les fractions de remplissage).

\subsection{Fonctions de corr\'elation simples et repr\'esentation diagrammatique.}
\label{partdiag2MM}

\subsubsection{Premi\`{e}re relation de r\'ecurrence.}

Sachant que $E(x,y)$ est un polyn\^ome en ses deux variables, \'etudions un \`{a} un les coefficients des diff\'erentes puissances
de $y$ et en rappelant que
\beq
E(x,y) = (V_1'(x) - y)(V_2'(y)-x) + P(x,y)
\eeq
o\`{u} $P(x,y)$ est une s\'erie formelle en ${1 \over N^2}$ dont chacun des termes est un polyn\^ome en $x$ et en $y$. A l'ordre $h$ dans le d\'eveloppement
topologique de cette \'equation, le d\'eveloppement en $y\to \infty$
des membres de gauche et de droite de cette \'equation donnent les \'egalit\'es suivantes:

\begin{itemize}
\item {\bf Coefficient de $y^{d_2+1}$:} Le r\'esultat est une \'egalit\'e triviale du type "$0 = 0$".

\item {\bf Coefficient de $y^{d_2}$:} On obtient une \'equation liant fortement les diff\'erents feuillets et tr\`{e}s utile
dans toute la suite:
\beq\label{somfeuillet}
\forall \,  h\geq 1, \; \;  \sum_{i=0}^{d_2} W_{1,0}^{(h)}(p^{i}) = 0.
\eeq

\item {\bf Coefficient de $y^{d_2-1}$:} On obtient cette fois-ci une \'equation bilin\'eaire:
\beq
\begin{array}{l}
2 {\displaystyle \sum_{i=0}^{d_2}} y(p^{i}) W_{1,0}^{(h)}(p^{i}) dx(p) =\cr
 = {\displaystyle \sum_{i=0}^{d_2} \sum_{m=1}^{h-1}} W_{1,0}^{(m)}(p^{i}) W_{1,0}^{(h-m)}(p^{i}) + {\displaystyle \sum_{i=0}^{d_2}} \overline{W}_{2,0}^{(h-1)}(p^{i},p^{i})
+ 2 Q^{(h)}(x(p)) \left(dx(p)\right)^2\cr
\end{array}
\eeq
o\`{u} $Q(x) = {\displaystyle \sum_h} N^{-2h} Q^{(h)}(x):= {1 \over N} \left< \Tr {V_1'(x)-V_1'(M_1) \over x-M_1} \right> $ est un polyn\^{o}me
en $x$ de degr\'e $d_1 -1$.

Etant donn\'ees les int\'egrales sur les cycles $\underline\CA_i$ des fonctions de corr\'elation:
\beq
\forall \, (i,k,h), \;\; \oint_{p\in \underline\CA_i} W_{k+1,0}^{(h)}(p,{\bf p_K}) = 0
\eeq
et la structure de p\^{o}les de ces derni\`{e}res d\'ecrite par le lemme \ref{polW}, on peut en d\'eduire
\bea\label{rec1}
W_{1,0}^{(h)}(q) &=& - \sum_{i} \Res_{p \to a_i} { \underline{dE}_{p}(q)  ( \ovl{W}_{2,0}^{(h-1)}(p,p) + \sum_{m=1}^{h-1} W_{1,0}^{(m)}(p) W_{1,0}^{(h-m)}(p) ) \over(y(p)-y(\overline{p})) dx(p)}\cr
&=& \sum_{i} \Res_{p \to a_i} { \underline{dE}_{p}(q)  (W_{2,0}^{(h-1)}(p,\pbar) + \sum_{m=1}^{h-1} W_{1,0}^{(m)}(p) W_{1,0}^{(h-m)}(\pbar) ) \over(y(p)-y(\overline{p})) dx(p)} .\cr
\eea
\end{itemize}

On a ainsi exprim\'e la fonction de corr\'elation \`{a} un point de genre $h$ en termes de fonctions de corr\'elation de genres
moins \'elev\'es. Cependant, ce syst\`{e}me d'\'equations n'est \'evidemment pas ferm\'e puisqu'il ne donne pas acc\`{e}s aux fonctions
\`{a} plusieur points n\'ecessaires \`a la construction d'une r\'ecurrence. Pour fermer ce syst\`{e}me, il faut donc \^etre capable de passer de $W_{k,0}^{(h)}$ \`{a} $W_{k+1,0}^{(h)}$.

\subsubsection{Op\'erateur d'insertion de boucle.}

Ces op\'erateurs faisant passer de $W_{k,l}$ \`{a} $W_{k+1,l}$ et $W_{k,l+1}$ sont connus et \'etudi\'es depuis longtemps:
ce sont les op\'erateurs d'insertion de boucle introduits dans \eq{loopinsertion} permettant d'obtenir toutes les fonctions de
corr\'elation simples lorsqu'ils sont appliqu\'es successivement sur l'\'energie libre (voir \eq{succinsert}).

Ainsi pour obtenir toutes les fonctions de corr\'elation \`a partir de la fonction \`a un point, il suffit de savoir comment
ces op\'erateurs agissent sur les diff\'erents \'el\'ements composant les r\`egles de r\'ecurence.
Nous n'aurons donc besoin que de deux propri\'et\'es d\'ej\`{a} bien connues de l'op\'erateur ${\partial \over \partial V_1}$, \`{a} savoir son action
sur les fonctions fondamentales:
\beq\label{insert1}
{\partial y(p) \over \d V_1(x(r))}\,dx(r) = - {\underline{B}(p,r)\over dx(p)}
\eeq
et
\bea\label{difB}
{\d \underline{B}(p,q) \over \d V_1(x(r))}\,dx(r) &=& \sum_\alpha
\Res_{\xi\rightarrow \mu_\alpha} {2 \underline{dE}_{\xi}(q)
\underline{B}(p,\overline{\xi}) \underline{B}(\xi,r) \over \left( y(\xi)-y(\overline{\xi})
\right) dx(\xi)}\cr
&=& \sum_\alpha \Res_{\xi\rightarrow
\mu_\alpha} {\underline{dE}_{\xi}(q)
\left[\underline{B}(p,\overline{\xi}) \underline{B}(\xi,r) + \underline{B}(r,\overline{\xi}) \underline{B}(\xi,p)
\right] \over \left( y(\xi)-y(\overline{\xi}) \right) dx(\xi)}.\cr
\eea

En appliquant cet op\'erateur plusieurs fois sur \eq{rec1}, on obtient un ensemble de relations de r\'ecurrence ferm\'e:

\bt
\beq\label{conjecture} \encadremath{
\begin{array}{rcl}
 W_{k+1,0}^{(h)}(q,p_K)
&=&    \sum_{\alpha} \Res_{p \to \mu_\alpha} {\underline{dE}_{p,\pbar}(q)\over (y(p)-y(\pbar))\,dx(p)}\left(
W_{k+1,0}^{(h-1)}(p,\overline{p},p_K) + \right. \cr && \;\;\; +
\left. \sum_{j,m} W_{j+1,0}^{(m)}(p,p_J) \,
W_{k+1-j,0}^{(h-m)}(\overline{p},p_{K-J}) \right) , \cr
\end{array}}\eeq
\et
La d\'emonstration tient au fait que cette r\`{e}gle est stable sous l'application de ${\partial \over \partial V_1}$.

\subsubsection{Repr\'esentation de la r\`{e}gle de r\'{e}currence.}

Ces relations de r\'ecurrence peuvent \^{e}tre repr\'esent\'ees de mani\`{e}re diagrammatique permettant une meilleure
compr\'ehension du r\'esultat. Pour ce faire, on reprend les repr\'esentations des fonctions de corr\'elations sous forme diagrammatique
introduites dans la partie \ref{partcombi}.

Commen\c{c}ons par repr\'esenter la fonction de corr\'elation $W_{k,0}^{(g)}({\bf p_K})$ par une surface de genre $g$ avec
$k$ pattes marqu\'ees par les points $\{p_i\}_{i=1 \dots k}$ par analogie avec les surfaces qu'elle g\'en\`{e}re:
\beq
W_{k+1,0}^{(g)}(p,{\bf p_K}) :=\begin{array}{l}
{\rm \includegraphics[width=6cm]{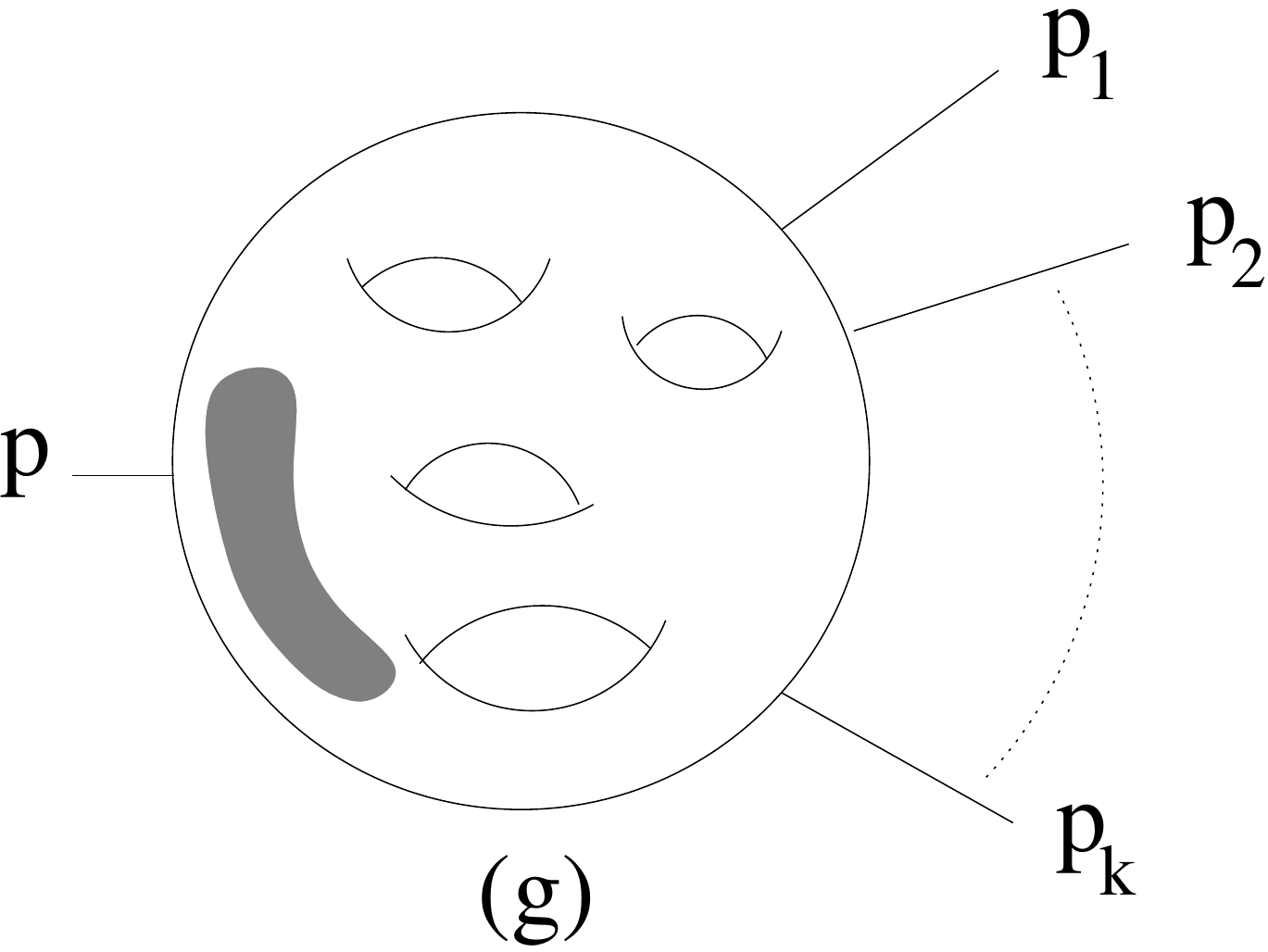}}
\end{array} .
\eeq
La fonction \`{a} deux points de genre 0, $W_{2,0}^{(0)}(p,q)$, est alors une sph\`{e}re \`{a} deux pattes: on la repr\'esente plus simplement
par une ar\^ete non-orient\'ee joignant $p$ et $q$:
\beq
W_{2,0}^{(0)}(p,q):=\begin{array}{l}
{\rm \includegraphics[width=2cm]{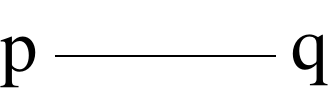}}
\end{array}= \underline{B}(p,q).
\eeq

Introduisons un dernier \'el\'ement graphique compos\'e d'une ar\^ete orient\'ee allant d'un point $p$ vers un
vertex trivalent dont les deux autres pattes sont associ\'ees \`{a} $q$ et $\qbar$ respectivement:
\beq
\sum_i \Res_{q \to a_i} {\underline{dE}_q(p) \over \omega(q)}:=\begin{array}{r}
{\rm \includegraphics[width=3cm]{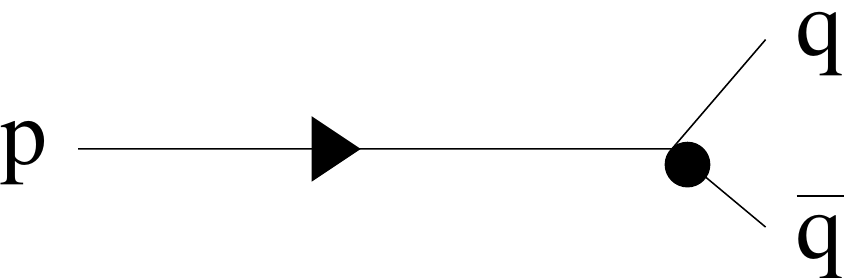}}
\end{array}.
\eeq
On rep\`{e}re par un point la patte correspondant \`{a} $\qbar$ et on la d\'enommera enfant droit dans toute la suite pour la diff\'erencier
de l'enfant gauche associ\'e \`a $q$.

La relation de r\'ecurrence \eq{conjecture} est alors repr\'esent\'ee par
\beq\label{recdiag}
\begin{array}{r}
{\rm \includegraphics[width=10cm]{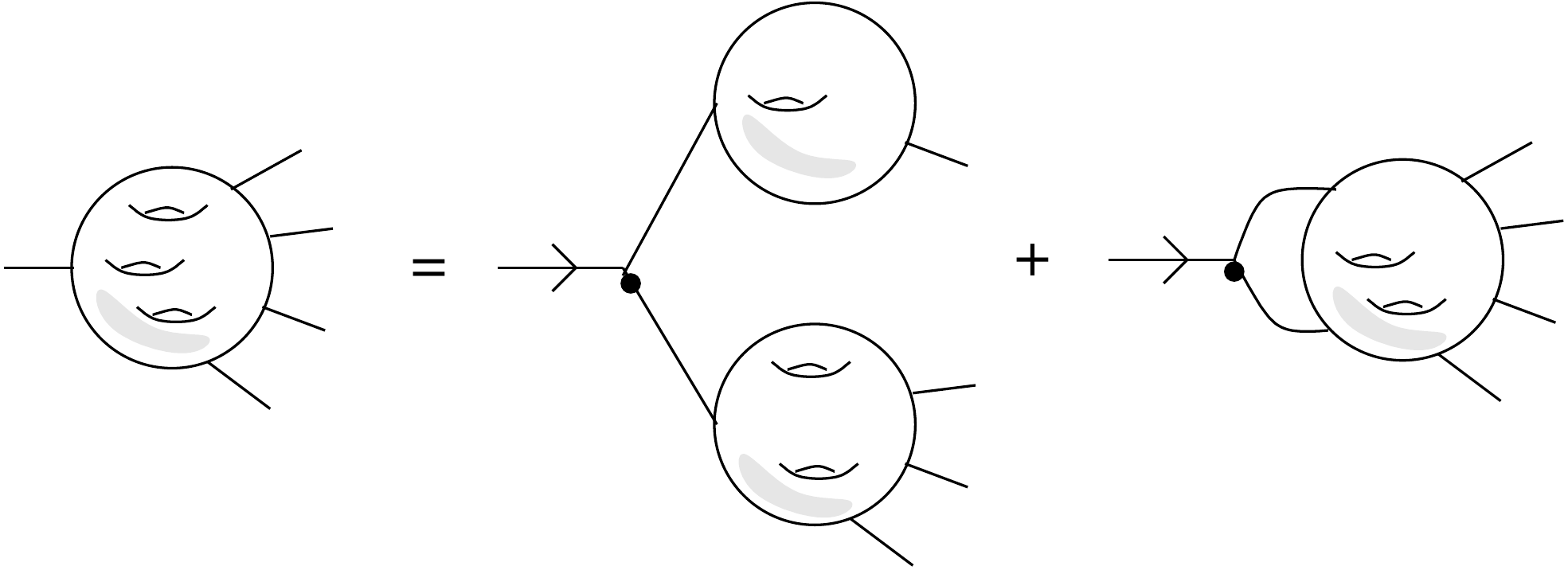}}
\end{array}.
\eeq
Sous cette forme, il est clair que l'ensemble des relations de r\'ecurrence \eq{conjecture} forme un syst\`{e}me triangulaire puisqu'\`{a}
chaque \'etape on r\'eduit soit le nombre de "trous" dans la surface ou le nombre de pattes (ou les deux). Ainsi, apr\`{e}s un
nombre fini d'\'etapes, on sera r\'eduit au calcul de la fonction \`{a} deux points $W_{2,0}^{(0)}$ qui est d\'ej\`{a} connue.

\br
Cette repr\'esentation diagrammatique est \'egalement pratique pour expliquer comment passer de la relation pour la fonction \`{a}
un point \`{a} la r\`{e}gle de r\'ecurrence g\'en\'erale. En effet, il suffit de voir comment l'op\'erateur d'insertion de
boucles agit sur les diff\'erents \'el\'ements de cette m\'ethode diagrammatique.

Par d\'efinition, l'op\'erateur d'insertion de boucle ajoute une patte \`{a} n'importe quel objet \`{a} $k$ pattes
et $g$ trous:
\beq
{\partial \over \partial V_1(q)}\begin{array}{r}
{\rm \includegraphics[width=3cm]{Wk}}
\end{array}
=
\begin{array}{r}
{\rm \includegraphics[width=3cm]{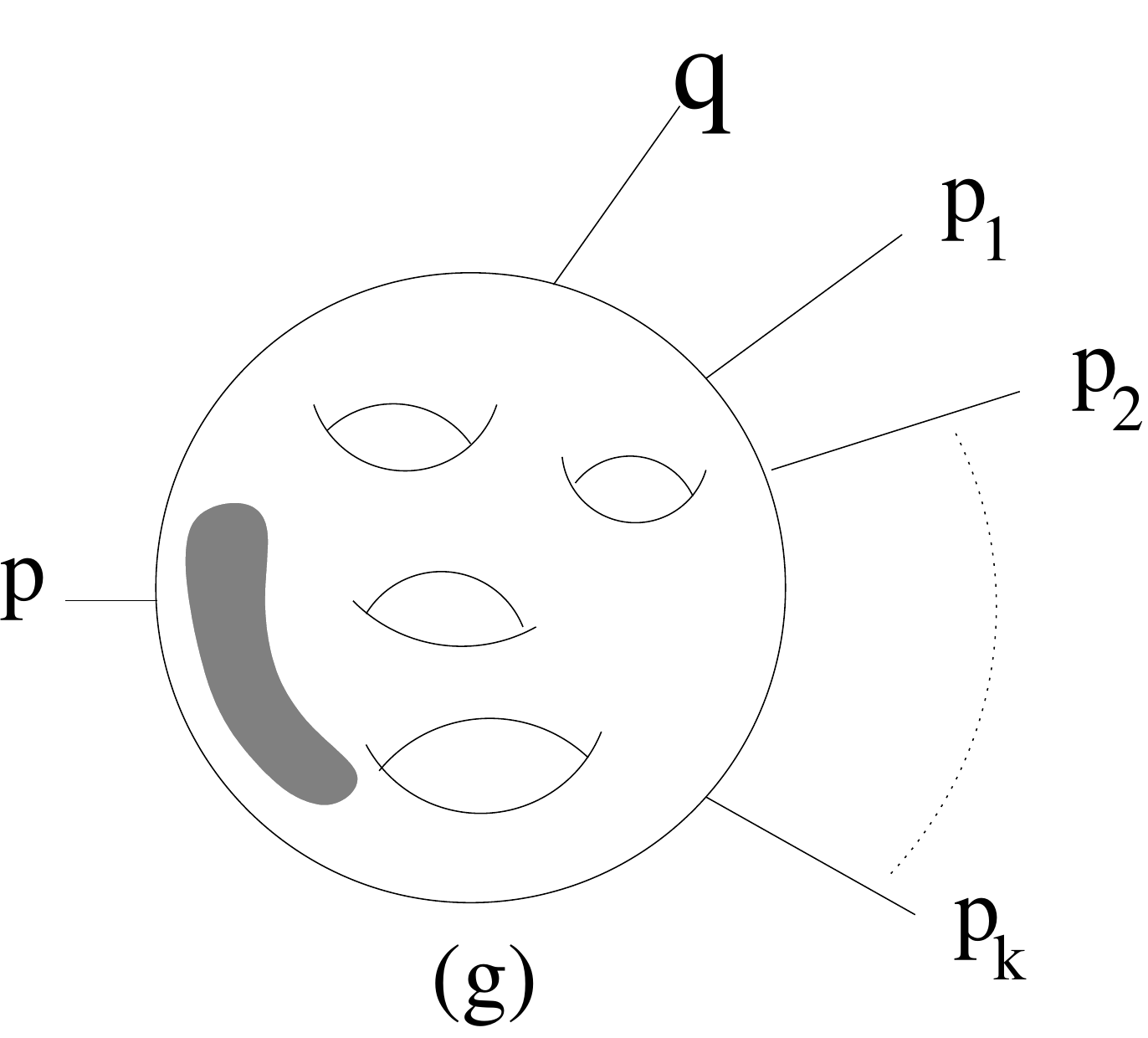}}
\end{array}.
\eeq

Les propri\'et\'es \eq{insert1} et \eq{difB} permettent quant \`{a} elles de d\'eterminer comment l'op\'erateur d'insertion de boucles
agit sur le vertex:
\beq
dx(r)\,{\partial \over \partial V_1(x(r))}\,
\begin{array}{l} {\rm \includegraphics[width=2.4cm]{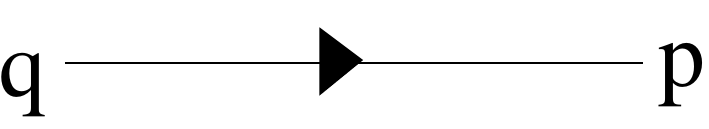}}
\end{array}
=
\begin{array}{l}  {\rm \includegraphics[width=8cm]{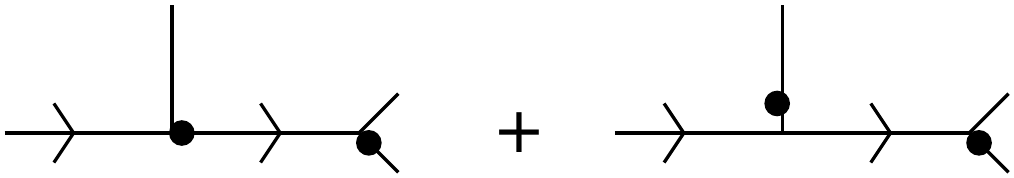}}
\end{array}.
\eeq
Il est alors facile de v\'erifier que si l'on applique ${\partial \over \partial V_1}$ \`{a} l'\'equation \eq{recdiag} d\'efinissant
$W_{k+1,0}^{(g)}$, on obtient bien l'\'equation donnant $W_{k+2,0}^{(g)}$.

\er

\subsubsection{R\'esultat de la r\'eccurence.}
En it\'erant $2g+k-2$ fois la relation de r\'ecurrence pour le calcul de $W_{k,0}^{(g)}$, on obtient une expression
exacte de $W_{k,0}^{(g)}$ comme somme sur un ensemble de diagrammes trivalents \`{a} $g$ boucles et $k$ pattes dont le poids est
donn\'e par les r\`egles pr\'ec\'edentes:
\bt\label{thdiag2MM}
Pour tout $k\geq 0$ et tout $g\geq 0$ tels que $2g+k-2 \geq 1$ et tout ensemble de points $\{p,p_1, \dots ,p_k \} \in \overline\Sigma^{k+1}$,
la fonction de corr\'elation \`{a} $k+1$ points de genre $g$ est donn\'ee par:
\beq
\encadremath{
W_{k+1,0}^{(g)}(p,{\bf p_K}) = \sum_{G \in \CG_k^{g}(p, {\bf p_K})} \CP(G)
}
\eeq
o\`{u} $\CG_k^{g}(p, {\bf p_K})$ est l'ensemble des graphes connexes trivalents d\'efinis par les contraintes:

\vspace{0.3cm}

1. ils comportent $2g+k-1$ vertex trivalents que l'on appelle {\bf vertex};

\vspace{0.3cm}

2. ils comportent un unique vertex monovalent marqu\'e par le point $p$ appel\'e {\bf racine};

\vspace{0.3cm}

3. ils comportent $k$ vertex monovalents marqu\'es par $p_1,p_2, \dots, p_k$ appel\'es {\bf feuilles} du graph;

\vspace{0.3cm}

4. il sont constitu\'es de $3g+2k-1$ ar\^{e}tes;

\vspace{0.3cm}

5. les ar\^{e}tes peuvent \^{e}tre soit orient\'ees par une fl\`{e}che soit non orient\'ees. Il y a $k+g$ ar\^{e}tes
non orient\'ees;

\vspace{0.3cm}

6. l'ar\^{e}te partant de la racine $p$ est orient\'ee par une fl\`{e}che partant de $p$;

\vspace{0.3cm}

7. les ar\^etes partant des feuilles $p_1, \dots, p_k$ ne sont pas orient\'ees;

\vspace{0.3cm}

8. les ar\^{e}tes orient\'ees forment un "squelette d'arbre binaire planaire couvrant\footnote{Un squelette d'arbre binaire
est un arbre binaire dont on a ot\'e toutes les feuilles, i.e. un arbre dont les vertex ont pour valence 1, 2 ou 3. La condition
de planarit\'e signifie que les enfants droit (marqu\'e par un point) et gauche (non marqu\'e)  d'un vertex ne sont pas \'equivalents.
Enfin, le fait que l'arbre soit couvrant signifie qu'il passe par tous les vertex.}". L'orientation des ar\^etes va de la racine
vers les feuilles munissant le graphe d'un ordre partiel sur les vertex;

\vspace{0.3cm}

9. parmi les $k+g$ ar\^etes non orient\'ees, $k$ d'entre elles lient un vertex \`{a} une feuille et les $g$ autres lient deux vertex trivalents entre eux.
Deux vertex trivalents peuvent \^{e}tre li\'es entre eux seulement si ils sont comparable selon l'ordre fournit par les fl\`eches\footnote{
C'est-\`{a}-dire si on peut aller de l'un \`{a} l'autre en suivant des ar\^etes orient\'es suivant leur orientation. Dans ce cas
le vertex de d\'epart est appel\'e parent et le vertex d'arriv\'ee descendant.};

\vspace{0.3cm}

10. si les enfants d'un vertex sont une ar\^ete orient\'ee et une ar\^ete non orient\'ee, alors l'ar\^ete orient\'ee est l'enfant de gauche.
Cette pr\'escription ne s'applique que si l'ar\^ete non orient\'ee lie ce vertex \`{a} l'un de ses descendants (et non \`{a}
l'un de ses parents).

\vspace{0.5cm}

et le poids $\CP(G)$ d'un graphe $G$ est donn\'e par les r\`{e}gles suivantes:
\begin{itemize}
\item On marque tout vertex trivalent du graphe par un point courant $r_i$ de $\overline{\Sigma}$ et on associe $r_i$ \`{a}
son enfant de gauche et $\overline{r}_i$ \`{a} son enfant de droite. Toute ar\^ete relie alors deux points de la surface
$\overline{\Sigma}$;

\item A une ar\^ete non orient\'ee liant $r$ et $r'$, on associe le facteur $\underline{B}(r,r')$;

\item A toute ar\^ete orient\'ee allant  de $r$ vers $r'$, on associe le facteur ${\underline{dE}_{r'}(r) \over (y(r')-y(\overline{r}')) dx(r')}$;

\item Suivant les fl\`eches en sens inverse (des feuilles vers la racine), \`{a} chaque vertex $r$, on calcule la somme sur tous
les points de branchement $a_i$ des r\'esidus quand $r \to a_i$: ${\displaystyle \sum_i \Res_{r \to a_i}}$\footnote{Notons que ceci donne un
sens au poids donn\'e aux ar\^etes fl\'ech\'ees ainsi qu'au choix d'associer $\overline{r}$ \`{a} l'un des enfants de chaque vetex.
En effet, l'application $p \to \overline{p}$ n'est d\'efinie localement qu'au voisinage des points de branchement. On peut v\'erifier
qu'ici cette application ne sera invoqu\'ee que pour des points marquant un vertex et donc proches d'un point de branchement.}.

\item Apr\`{e}s avoir calcul\'e l'ensemble de ces r\'esidus, on obtient le poids du graphe.

\end{itemize}
\et

Gr\^{a}ce  \`{a} ce th\'eor\`{e}me, on a une m\'ethode rapide, simple et ais\'ee \`{a} se souvenir pour calculer n'importe quel
terme dans le d\'eveloppement topologique des fonctions de corr\'elation simples $W_{k,0}$, c'est-\`{a}-dire les fonctions
g\'en\'eratrices des surfaces discr\'etis\'ees avec $k$ bords sans op\'erateurs de bords de genre quelconque. Cette technique
est simple car elle ne n\'ecessite qu'une fonction de base et le calcul de r\'esidus, i.e. le d\'eveloppement de
Taylor du noyau de Bergmann au voisinage des points de branchement. Elle est \'egalement rapide car elle n\'ecessite le calcul d'un nombre
tr\`es restreint de termes et est facile \`{a} programmer pour une r\'esolution informatique.

\br
Par construction les quantit\'es ainsi d\'efinies sont bien sym\'etrique en toutes leurs variables sauf la premi\`{e}re.
Il est cependant possible de montrer qu'elles sont en fait heureusement bien sym\'etriques en toutes leurs variables
comme attendu.

Ces graphes ainsi que les poids qui leur sont associ\'es ont d'autres propri\'et\'es peu \'evidentes \`{a} premi\`{e}re vue
et nous les discuterons dans le chapitre 5.
\er

\br
Comment construire de mani\`{e}re pratique les graphes correspondant \`{a} la fonction \`{a} $k+1$ points de genre $g$?
\begin{itemize}
\item
On commence par tracer tous les arbres enracin\'es en $p$ possibles compos\'es de $2g+k-1$ ar\^etes et $2g+k$ vertex de valence 1, 2 ou 3
avec la contrainte que la racine soit un vertex de valence 1.

\item On oriente ensuite toutes les ar\^{e}tes en partant de la racine.

\item On branche de toutes les mani\`eres possibles les $k$ ar\^etes correspondant aux feuilles $p_1, \dots, p_k$ sur les vertex
de mani\`{e}re \`{a} ce que leur valence ne d\'epasse toujours pas 3.

\item On compl\`{e}te le graph en branchant $g$ ar\^{e}tes non orient\'ees de mani\`{e}re \`{a} ce que tous les vertex aient valence 3
en interdisant aux ar\^etes de relier deux vertex non comparables.

\item Il ne reste plus qu'\`{a} marquer de toutes les mani\`{e}res non \'equivalentes possibles les enfants droits et gauches
en suivant la pr\'escription 10.

\end{itemize}

\er

Etudions explicitement les premiers exemples:

\bex
Commen\c{c}ons par le calcul de $W_{3,0}^{(0)}(p,p_1,p_2)$ et donc la construction des \'el\'ements de $\CG_3^0(p,p_1,p_2)$.
On repr\'esente tous les arbres enracin\'es \`{a} 2 vertex et 1 ar\^ete, orient\'es depuis la racine $p$:
\beq
\begin{array}{l} {\rm \includegraphics[width=2.4cm]{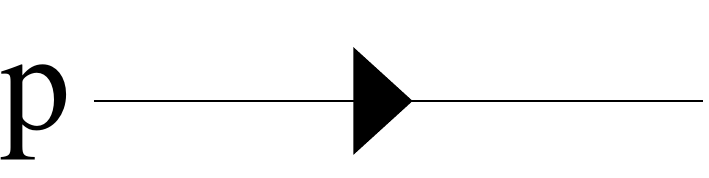}}
\end{array}
\eeq
On ajoute ensuite les deux feuilles $p_1$ et $p_2$:
\beq
\begin{array}{l} {\rm \includegraphics[width=2.4cm]{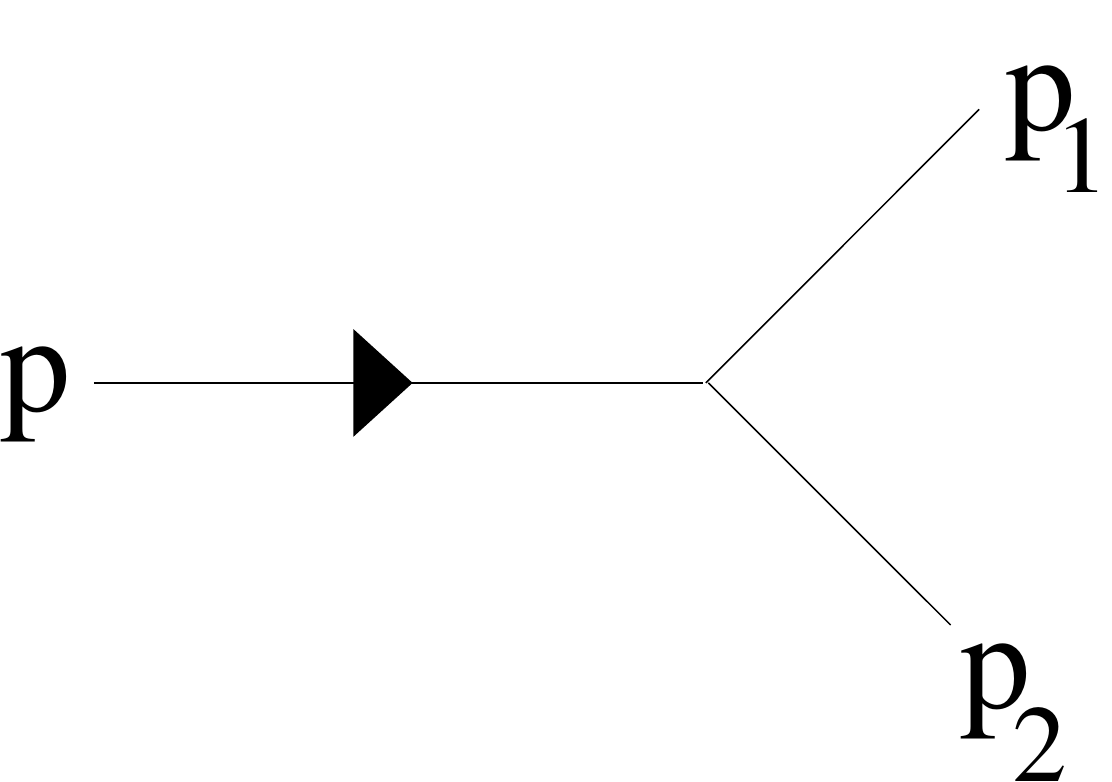}}
\end{array}
\eeq
On marque finalement l'enfant droit et l'enfant gauche:
\beq
W_{3,0}^{(0)}(p,p_1,p_2) = \CP \left(\begin{array}{l} {\rm \includegraphics[width=2.4cm]{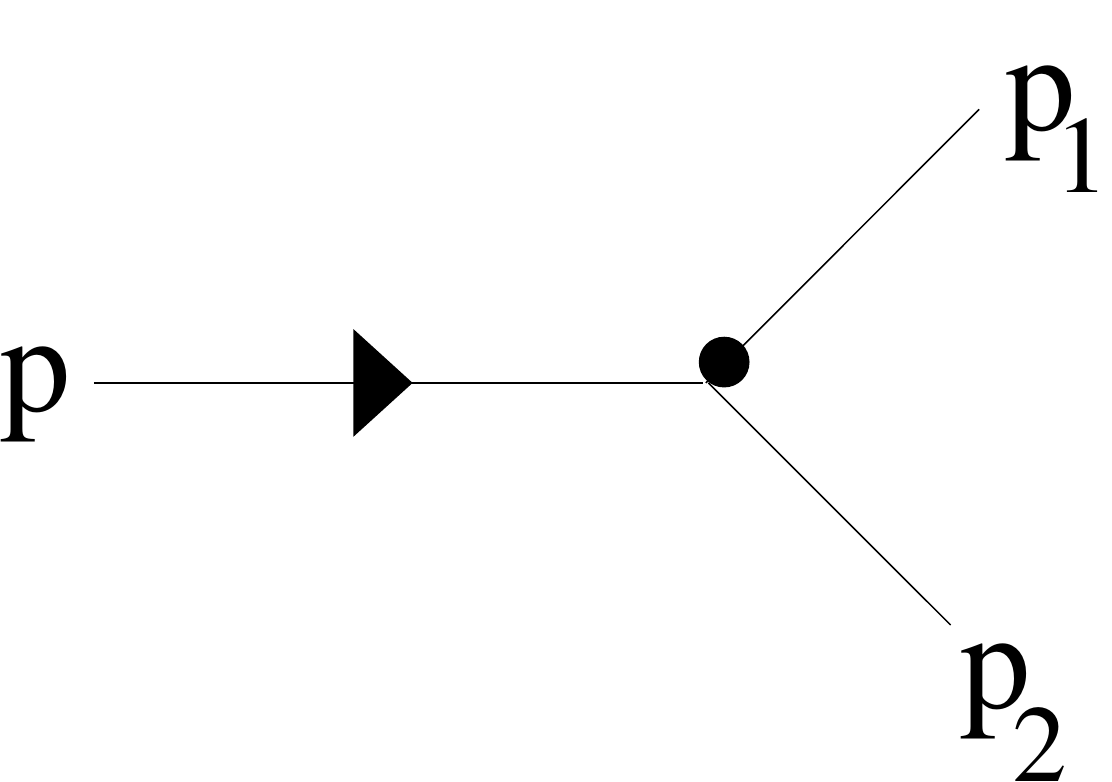}}
\end{array}\right)
+ \CP \left(\begin{array}{l} {\rm \includegraphics[width=2.4cm]{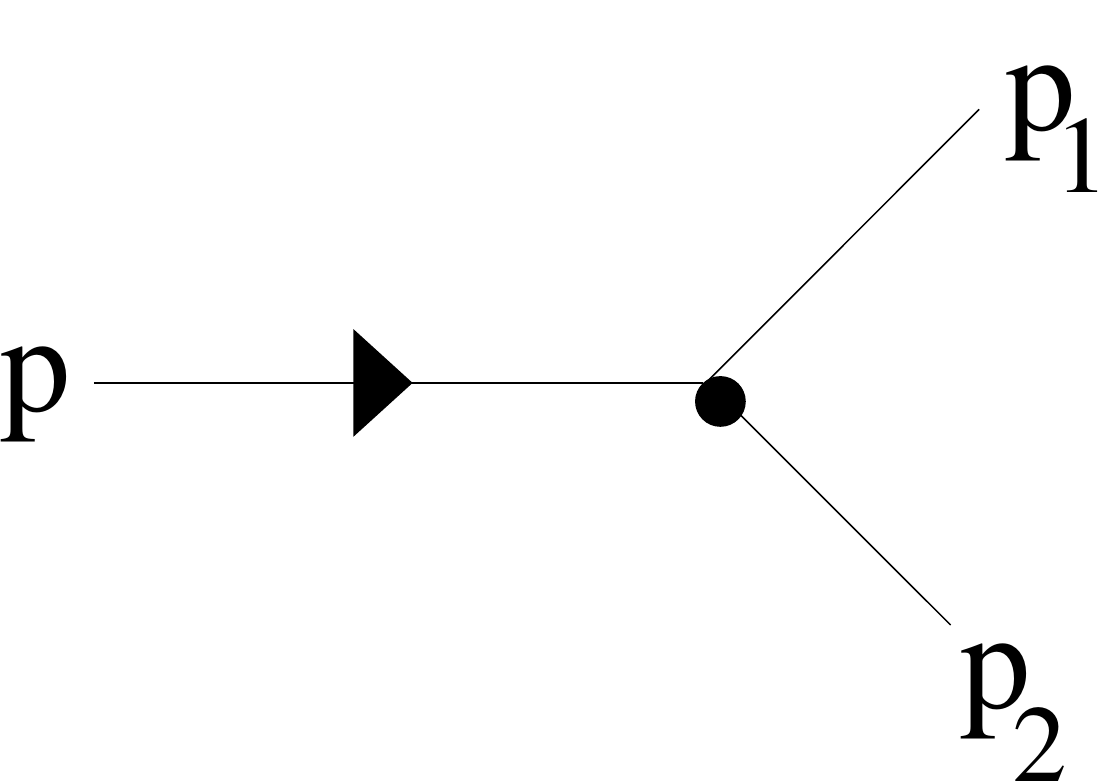}}
\end{array}\right).
\eeq
Calculons le poids du premier graphe. Notons $r$ la variable d'int\'egration associ\'ee \`{a} l'unique vertex trivalent de ce graphe.
Les trois ar\^etes le composant ont alors pour poids respectifs $\underline{B}(\overline{r},p_1)$, $\underline{B}(r,p_2)$ et
${\underline{dE}_{r}(p) \over (y(r)-y(\overline{r})) dx(r)}$. Donc
\beq
\CP \left(\begin{array}{l} {\rm \includegraphics[width=2.4cm]{arrow4}}
\end{array}\right)
= \sum_i \Res_{r \to a_i} {\underline{dE}_{r}(p) \over (y(r)-y(\overline{r})) dx(r)}\underline{B}(\overline{r},p_1) \underline{B}(r,p_2),
\eeq
ce qui donne finalement:
\beq
W_{3,0}^{(0)}(p,p_1,p_2) =  \sum_i \Res_{r \to a_i} {\underline{dE}_{r}(p) \over (y(r)-y(\overline{r})) dx(r)}\left[\underline{B}(\overline{r},p_1) \underline{B}(r,p_2)
+ \underline{B}(\overline{r},p_2) \underline{B}(r,p_1) \right].
\eeq

De la m\^{e}me mani\`{e}re, il est facile de construire le seul \'el\'ement de $\CG_{1}^{1}$ et d'obtenir:
\bea
W_{1}^{(1)}(p) &=& \begin{array}{r}
{\rm \includegraphics[width=4cm]{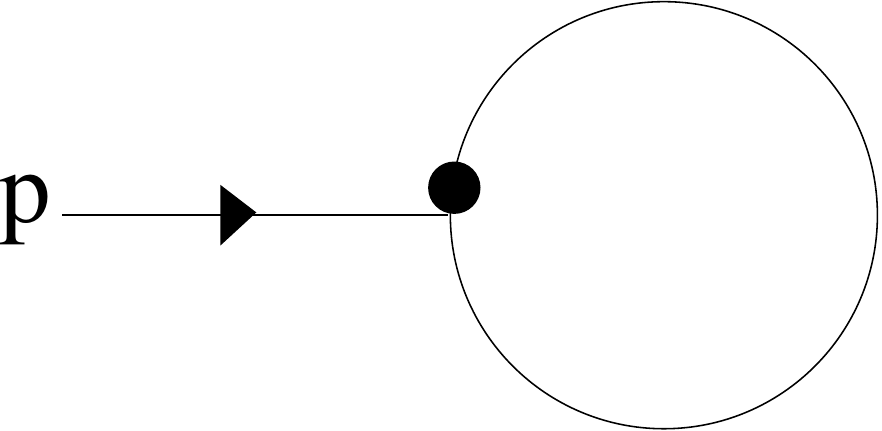}}
\end{array} \cr
&=& \Res_{q \to {\bf a}} {\underline{dE}_q(p) \over \omega(q)} \underline{B}(q, \qbar).\cr
\eea

\vs

Etudions un dernier exemple: $W_{1,0}^{(2)}(p)$.

La premi\`{e}re \'etape de la construction donne:
\beq
\begin{array}{r}
{\rm \includegraphics[width=2cm]{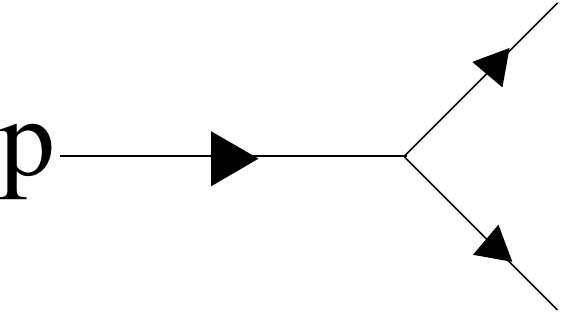}}
\end{array}
\virg
\begin{array}{r}
{\rm \includegraphics[width=2.5cm]{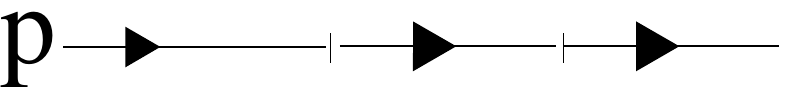}}
\end{array}.
\eeq
On compl\`{e}te alors les vertex par des ar\^etes non orient\'ees de mani\`{e}re \`{a} ce qu'ils soient tous trivalents.
D\'ej\`{a} ici, la prescription 9) interdit une possibilit\'e.
\beq
\begin{array}{r}
\begin{array}{r}
{\rm \includegraphics[width=2.5cm]{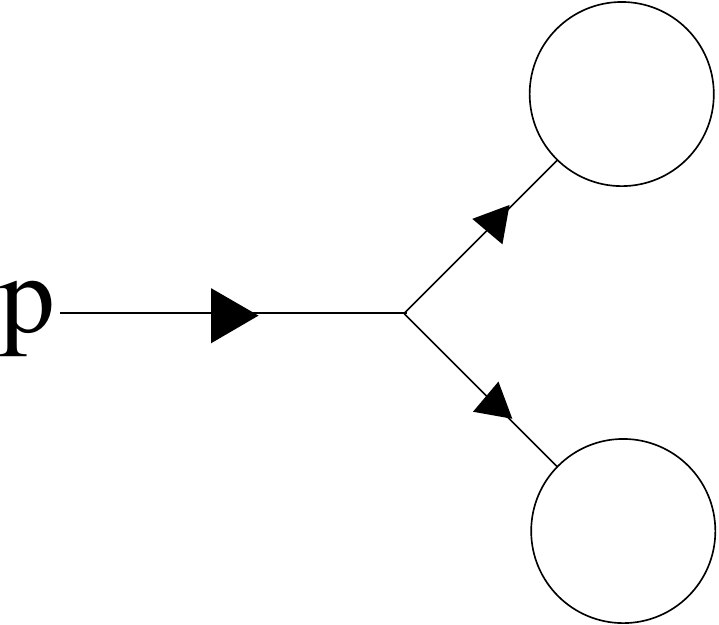}}
\end{array}
\virg
{\rm \includegraphics[width=2.2cm]{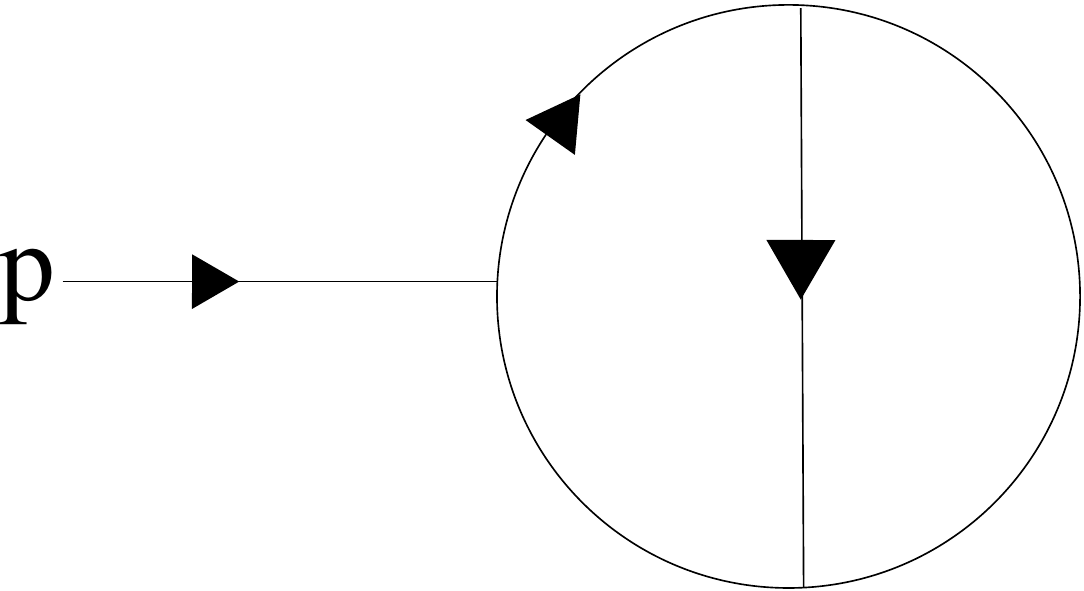}}
\end{array}
\virg
\begin{array}{r}
{\rm \includegraphics[width=3cm]{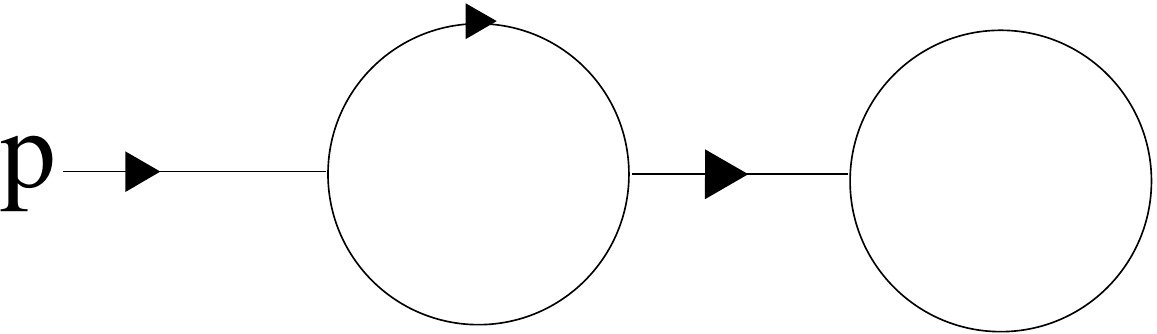}}
\end{array}.
\eeq

Il reste alors \`{a} sp\'ecifier les enfants droits et gauches en respectant la contrainte 10):
\bea
\begin{array}{r}
{\rm \includegraphics[width=2.2cm]{y123}}
\end{array}
\virg
\begin{array}{r}
{\rm \includegraphics[width=2.5cm]{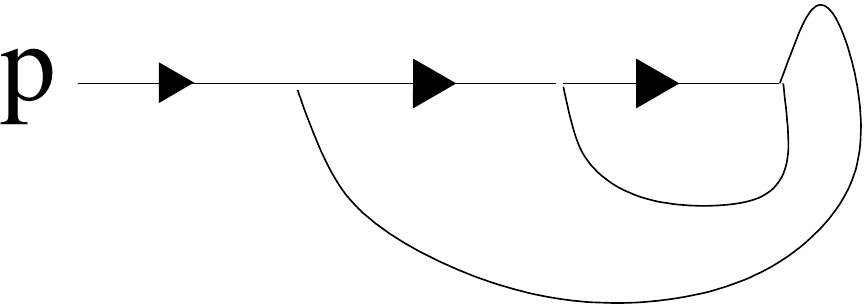}}
\end{array}
\virg
\begin{array}{r}
{\rm \includegraphics[width=3cm]{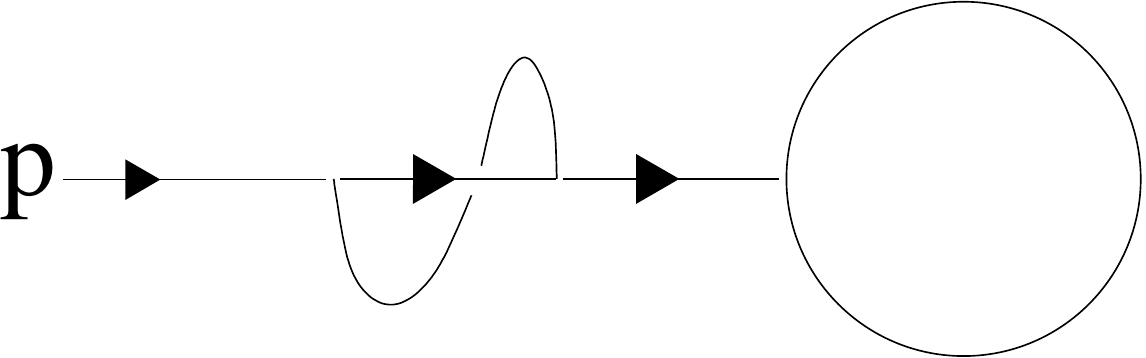}}
\end{array} \cr
\begin{array}{r}
{\rm \includegraphics[width=2.5cm]{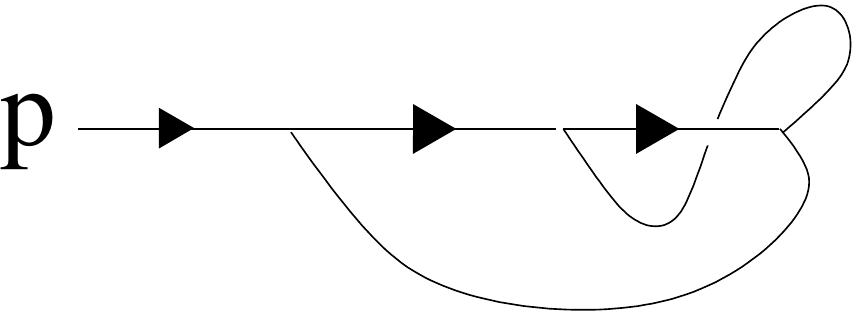}}
\end{array}
\virg
\begin{array}{r}
{\rm \includegraphics[width=3cm]{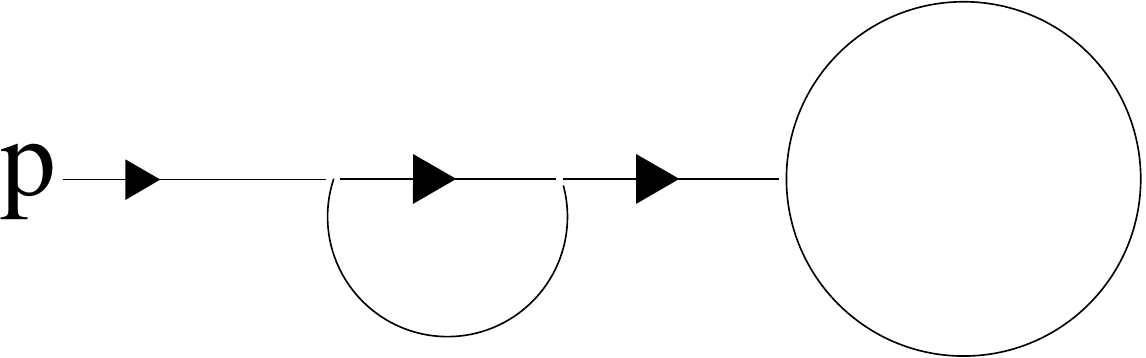}}
\end{array} .\cr
\eea

Ce qui s'\'ecrit en indiquant l'enfant droit par un point:
\bea
\begin{array}{r}
{\rm \includegraphics[width=2.2cm]{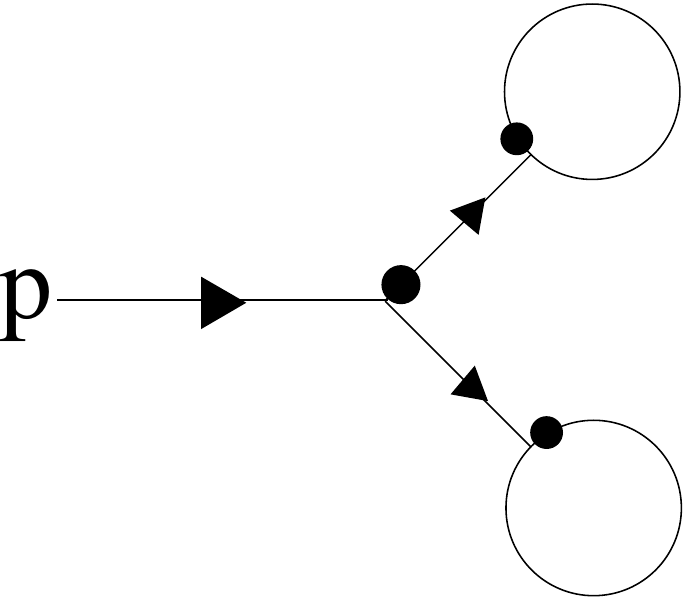}}
\end{array}
\virg
\begin{array}{r}
{\rm \includegraphics[width=2.5cm]{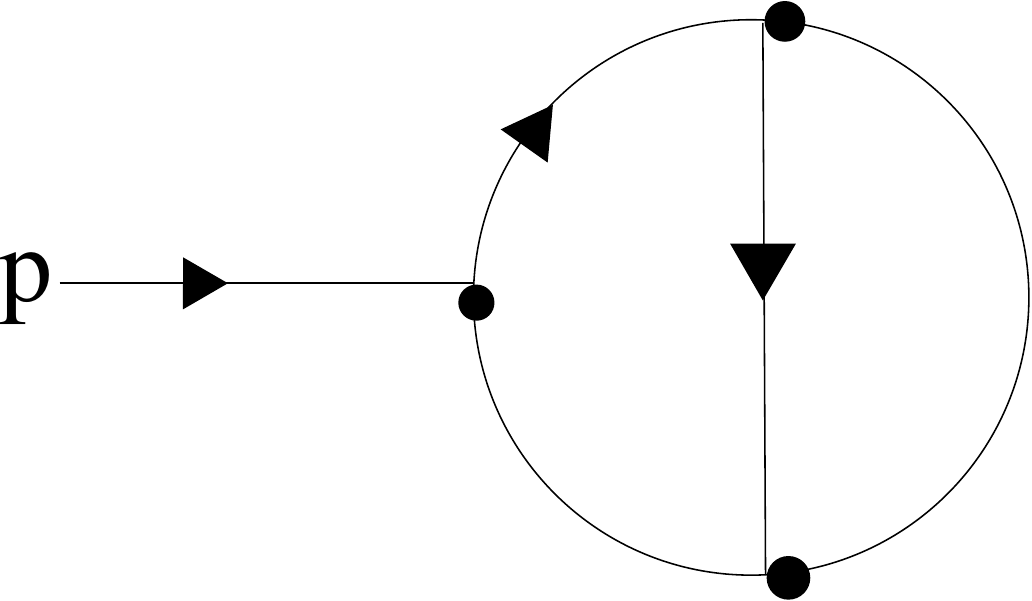}}
\end{array}
\virg
\begin{array}{r}
{\rm \includegraphics[width=3cm]{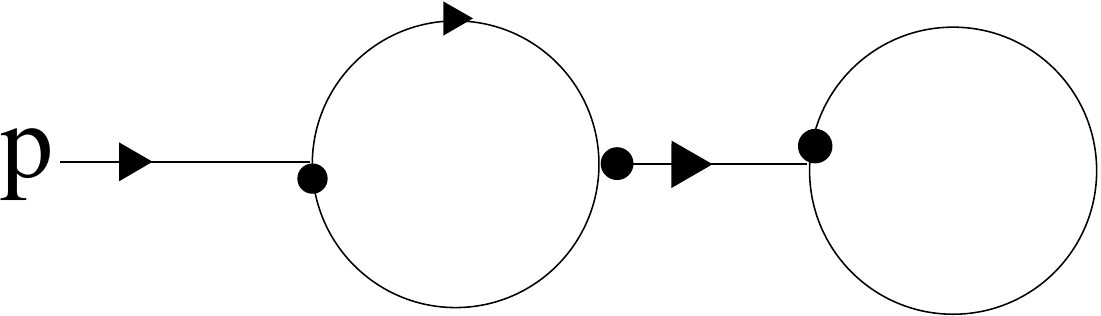}}
\end{array} \cr
\begin{array}{r}
{\rm \includegraphics[width=2.5cm]{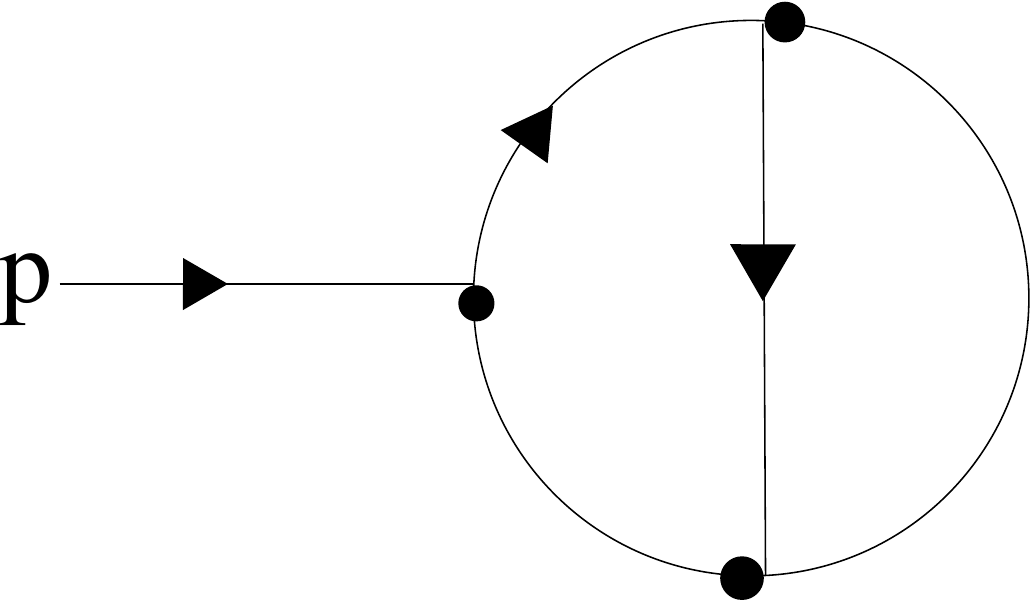}}
\end{array}
\virg
\begin{array}{r}
{\rm \includegraphics[width=3cm]{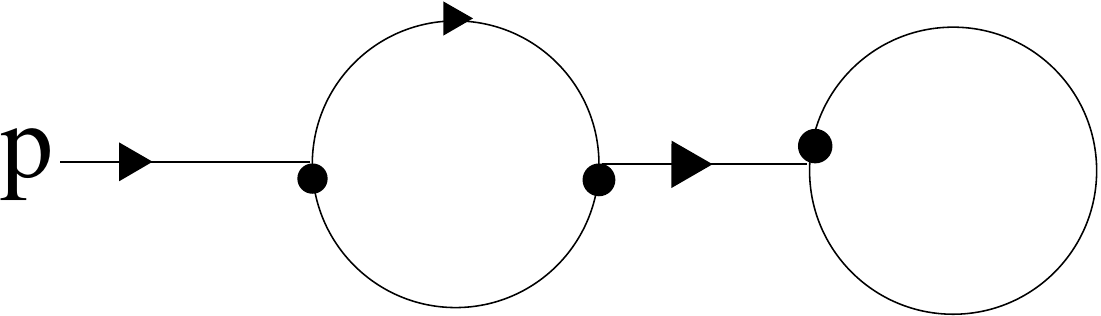}}
\end{array} \cr
\eea
On peut remarquer que sans les contraintes 9) et 10), on aurait obtenu 13 graphes diff\'erents au lieu de 5.

Ceci nous donne la valeur de la fonction \`{a} un point de genre 2:
\bea
W_1^{(2)}(p)
&=&  \Res_{q \to {\bf a}} \Res_{r \to {\bf a}} \Res_{s \to {\bf a}}  {\underline{dE}_q(p) \over \omega(q)}
{\underline{dE}_r(q) \over \omega(r)} {\underline{dE}_s(\qbar) \over \omega(s)} \underline{B}(r,\overline{r}) \underline{B}(s, \overline{s})\cr
&& + \Res_{q \to {\bf a}} \Res_{r \to {\bf a}} \Res_{s \to {\bf a}}  {\underline{dE}_q(p) \over \omega(q)}
{\underline{dE}_r(q) \over \omega(r)} {\underline{dE}_s(\overline{r}) \over \omega(s)} \underline{B}(r,\qbar) \underline{B}(s,\overline{s})\cr
&& + \Res_{q \to {\bf a}} \Res_{r \to {\bf a}} \Res_{s \to {\bf a}}  {\underline{dE}_q(p) \over \omega(q)}
{\underline{dE}_r(q) \over \omega(r)} {\underline{dE}_s(r) \over \omega(s)} \left[ \underline{B}(\qbar,\overline{r}) \underline{B}(s,\overline{s})
 \right. \cr
&& \left. + \underline{B}(\overline{s},\qbar) \underline{B}(s,\overline{r}) + \underline{B}(s,\qbar) \underline{B}(\overline{s},\overline{r}) \right]\cr
\eea

\eex

\subsection{D\'eveloppement topologique de l'\'energie libre.}

On a maintenant une expression exacte pour tous les $W_{k,0}^{(g)}$ avec $k>0$ \`{a} partir des $W_{1,0}^{(g)}$
gr\^ace \`{a} l'op\'erateur d'insertion de boucles ${\partial \over \partial V_1}$. Pour obtenir les termes du d\'eveloppement
topologique de l'\'energie libre $F^{(g)}$, pouvant \^etre vus comme les $W_{0,0}^{(g)}$, il faut \^etre capable de passer
de $W_{k+1,0}^{(g)}$ \`{a} $W_{k,0}^{(g)}$, c'est-\`{a}-dire trouver l'op\'erateur inverse de ${\partial \over \partial V_1}$
dans la base des $W_{k,0}^{(g)}$.

\subsubsection{Inversion de l'op\'erateur d'insertion de boucles.}

\bd
Soit l'op\'erateur $H_x$ tel que pour toute diff\'erentielle m\'eromorphe $\phi$:
\beq\encadremath{
H_x. \phi :=
\Res_{\infty_x} V_1(x)\,\phi -\Res_{\infty_y} (V_2(y)-xy)\,\phi +
\int_{\infty_x}^{\infty_y} \phi + \sum_i \epsilon_i \oint_{{\cal
B}_i} \phi.}
\eeq
\ed

Son action sur les fonctions de corr\'elation est d\'ecrite par
\bt
Pour tout $h\geq0$ et $k\geq 0$:
\beq
H_x.W_{k+1,0}^{(h)}(.,{\bf p_K}) = (2-k-2h) W_{k,0}^{(h)}({\bf p_K}).
\eeq
o\`{u} l'on a adopt\'e la notation:
\beq
W_{0,0}^{(h)}:= - F^{(h)}.
\eeq
\et

\bc
Pour tout $h\geq0$ et $k\geq 0$ et une primitive quelconque de $ydx$:
\beq
\phi:= \int ydx,
\eeq
on a
\beq
\Res_{q \to {\bf a}} \phi(q) W_{k+1}^{(h)}(q,{\bf p_K}) = (2-2h-k)  W_{k}^{(h)}({\bf p_K}).
\eeq
\ec

\proof{
La d\'emonstration compl\`{e}te de ce th\'eor\`{e}me est donn\'ee dans \CEO. Je ne vais pr\'esenter
ici qu'une id\'ee de celle-ci en \'evitant les d\'etails techniques.

On peut tout d'abord remarquer que si les graphes composant $W_{k+1,0}^{(h)}$ ont plusieurs pattes externes, on peut
choisir d'agir sur l'une des feuilles plut\^ot que sur la racine.

Etudions alors comment $H_x$ agit sur les diff\'erents \'el\'ements composant les diagrammes. On peut montrer que l'on a:
\beq
\begin{array}{l}
{\rm \includegraphics[width=3cm]{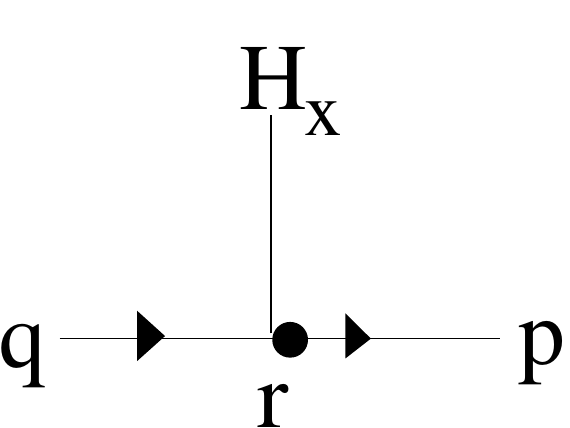}}
\end{array}
+ \begin{array}{l} {\rm \includegraphics[width=3cm]{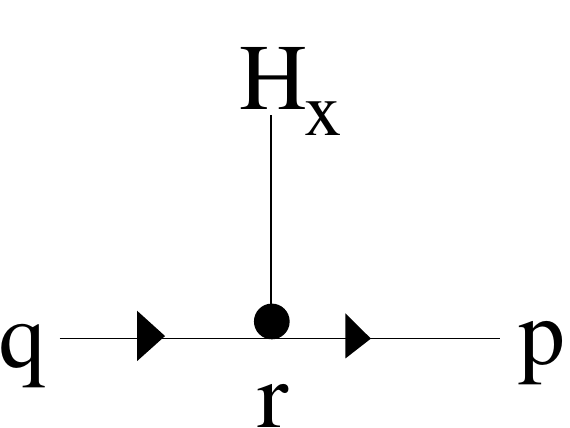}}
\end{array}= -
\begin{array}{l}
{\rm \includegraphics[width=3cm]{arrow}}
\end{array}
\eeq

et

\beq
\begin{array}{l}
{\rm \includegraphics[width=3cm]{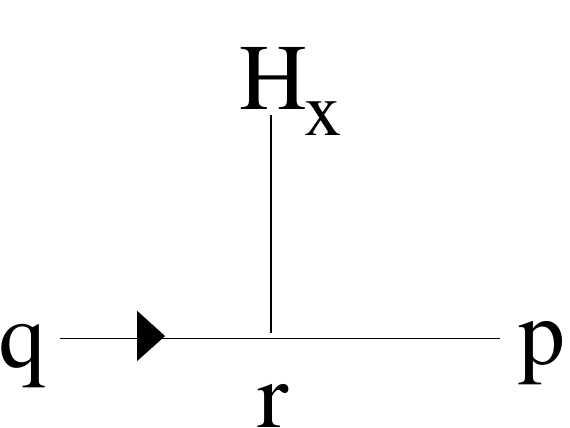}}
\end{array}
= 0. \eeq

On peut alors en d\'eduire que
\beq
H_x.W_{k+1,0}^{(h)}(.,{\bf p_K}) =  H_x. {\partial \over \partial V_1} W_{k,0}^{(h)}({\bf p_K}) =(2-k-2h) W_{k,0}^{(h)}({\bf p_K}).
\eeq
o\`{u} le facteur combinatoire $2h+k-2$ vient du fait qu'il y a $2h+k-2$ ar\^{e}tes orient\'ees composant les diagrammes
de $W_{k,0}^{(h)}$ et donc autant de possibilit\'es de faire appara\^itre une nouvelle patte externe par l'action de
${\partial \over \partial V_1}$ que l'on fait ensuite disparaitre par $H_x$.

La formule pour \'energie libre
\beq
(2-2h) \CF^{(h)} = - H_x.W_{1,0}^{(h)}
\eeq
 est obtenue en montrant qu'elle redonne bien la fonction \`{a} un point sous l'action de ${\partial \over \partial V_1}$.

Le corrolaire est obtenus par des combinaisons d'int\'egrations par partie et de changement de contours d'int\'egration.
}

\subsubsection{Expression de $\CF^{(g)}$ pour $g>1$.}

Ce th\'eor\`{e}me nous donne donc acc\`{e}s \`{a} une formule explicite pour n'importe quel terme du d\'eveloppement
topologique de l'\'energie libre sauf le terme sous-dominant, c'est-\`{a}-dire  la premi\`{e}re correction en ${1 \over N^2}$:
\beq\encadremath{
(2-2h) \CF^{(h)} = - H_x.W_{1,0}^{(h)}.}
\eeq
En effet, pour $h=1$, le facteur combinatoire s'annule et cette \'equation ne nous donne aucune information quant \`{a} la
forme de $\CF^{(1)}$.

\subsubsection{Expression de $\CF^{(0)}$.}

Pour ce qui est de $\CF^{(0)}$, sa valeur est connue depuis longtemps (voir par exemple \cite{MarcoF}) sous la forme:
\beq\encadremath{
2{\cal F}^{(0)} = \Res_{\infty_x}
(V_1(x)+V_2(y)-xy)\,ydx + T \Pint_{\infty_x}^{\infty_y} ydx +
\sum_i \epsilon_i \oint_{{\cal B}_i} ydx}
\eeq
i.e
\beq
2{\cal F}^{(0)} = H_x.y dx.
\eeq
Cette formulation correspond \`{a} la d\'efinition:
\beq
W_{1,0}^{(0)}:= -ydx
\eeq
naturelle lorsque l'on consid\`{e}re l'ordre dominant de la d\'efinition:
\beq
Ydx = V'_1 - W_{1,0}.
\eeq

\subsubsection{Expression de $\CF^{(1)}$.}

L'expression de l'\'energie libre de genre 1 ne peut \'evidemment pas \^etre obtenue \`a partir de l'action de $H_x$ puisque
le pr\'efacteur ${1 \over 2 -2g}$ diverge dans ce cas. Elle a en effet \'et\'e obtenue par une m\'ethode bien diff\'erente
dans \cite{EKK} et peut \^etre mise sous la forme:
\beq\encadremath{
\CF^{(1)} = {- 1 \over 2} \ln (\tau_{Bx}) - {1 \over 24} \ln \left(\prod_i {dy\over dz_i}(a_i) \right)}
\eeq
o\`u $\tau_{Bx}$ est la fonction $\tau$ de Bergmann associ\'ee \`a $x$ d\'efinie par \eq{deftauBx}.

\subsection{D\'eveloppement topologique des fonctions de corr\'elation mixtes.}

Dans cette partie, nous allons r\'esoudre l'\'equation de boucle la plus g\'en\'erale donnant acc\`es au d\'eveloppement
topologique de n'importe quelle fonction de corr\'elation $H_{{\bf k_L};m;n}^{(h)}$. Nous allons une nouvelle fois pr\'esenter le r\'esultat
sous forme d'un syst\`{e}me triangulaire que nous repr\'esenterons de mani\`{e}re diagrammatique par souci de simplicit\'e.

\subsubsection{Equations de boucles.}

Le changement de variables
\beq
\begin{array}{rcl}
\delta M_1 &:=&{1 \over x(p_{1,1})-M_1}{1 \over y(p_{1,1})-M_2} {1 \over x(p_{1,2})-M_1}{1 \over y(p_{1,2})-M_2}
\dots {1 \over x(p_{1,k_1})-M_1}{1 \over y(p_{1,k_1})-M_2} \cr
&&
\prod_{i=2}^l \Tr \left({1 \over x(p_{i,1})-M_1}{1 \over y(p_{i,1})-M_2} {1 \over x(p_{i,2})-M_1}{1 \over y(p_{i,2})-M_2}
\dots {1 \over x(p_{i,k_i})-M_1}{1 \over y(p_{i,k_i})-M_2}\right) \cr
&& \qquad \; \; \prod_{j=1}^m \Tr { 1 \over x(p_j)-M_1} \prod_{s=1}^n \Tr {1 \over y(q_s)-M_2}\cr
\end{array}
\eeq
donne l'\'equation de boucles g\'en\'erale:
\beq
\begin{array}{l}
{\displaystyle (Y(p_{1,1})-y(q_{1,k_1})-  \hbox{Pol}_{x(p_{1,1})} V_1'(x(p_{1,1})) ) H_{k_1,\dots,k_l;m;n}({\bf S_L};{\bf p_M};{\bf q_N}) = }\cr
{\displaystyle \sum_{A\bigcup B=\{2, \dots,l\}}  \sum_{I,J} H_{k_1,{\bf k_A};\left|I\right|;\left|J\right|}(S_1,{\bf S_A}; {\bf p_{I}};{\bf q_J})
H_{{\bf k_B};m-\left|I\right|+1;n-\left|J\right|}({\bf S_B}; p_{1,1} {\bf p_{M/I}};{\bf q_N/J})}\cr
{\displaystyle + \sum_{A\bigcup B=\{2, \dots,l\}}\sum_{\alpha =2}^{k_1} \sum_{I,J} H_{k_1-\alpha+1,{\bf k_B};m-\left|I\right|;n-\left|J\right|}(\{p_{1,\alpha},q_{1,\alpha},\dots p_{1,k_1},q_{1,k_1}\},{\bf S_B}; {\bf p_{M/I}};{\bf q_{N/J}})}\cr
 \times { H_{\alpha-1,{\bf k_A};\left|I\right|;\left|J\right|}(\{p_{1,1},q_{1,1},\dots p_{1,\alpha-1},q_{1,\alpha-1}\},{\bf S_A}; {\bf p_{I}};{\bf q_J})
-  H_{\alpha-1,{\bf k_A};\left|I\right|;\left|J\right|}(\{p_{1,\alpha},q_{1,1},\dots p_{1,\alpha-1},q_{1,\alpha-1}\},{\bf S_A}; {\bf p_{I}};{\bf q_J}) \over x(p_{1,\alpha})-x(p_{1,1})} \cr
{\displaystyle \sum_{i=2}^{l} \sum_{\alpha=1}^{k_i} {1 \over x(p_{i,\alpha}) - x(p_{1,1})} }\cr
\left[ H_{k_1+k_i,{\bf k_{L/\{1,i\}}};m;n}( \{S_1,p_{i,\alpha},q_{i,\alpha},p_{i,\alpha+1}, \dots ,q_{i,k_i},p_{i,1},\dots ,p_{i,\alpha-1},q_{i,\alpha-1}\},{\bf S_{L/\{1,i\}}};{\bf p_M};{\bf q_N}) \right. \cr
\left. \left. \qquad \qquad \qquad -  \hbox{Idem} \right|_{p_{1,1}:=p_{i,\alpha}} \right] \cr
- \sum_{i=1}^{m} \partial_{p_i} \left[{H_{{\bf k_L};m-1;n}({\bf S_L};{\bf p_{M/\{i\}}};{\bf q_N}) - \left. H_{{\bf k_L};m-1;n}({\bf S_L};{\bf p_{M/\{i\}}};{\bf q_N})\right|_{p_{1,1}:=p_i} \over x(p_i) - x(p_{1,1})} \right] \cr
+ {1 \over N^2} \sum_{\alpha =2}^{k_1} { 1 \over x(p_{1,\alpha})-x(p_{1,1})} \times \cr
\left[ H_{\alpha-1, k_1-\alpha+1, { \bf k_{L/\{1\}}}; m;n}(S_{1|\alpha}, {\bf S_{L/\{1\}}};{\bf p_M};{\bf q_N}) \right. \cr
\left.\left. - H_{\alpha-1, k_1-\alpha+1, { \bf k_{L/\{1\}}}; m;n}(S_{1|\alpha}, {\bf S_{L/\{1\}}};{\bf p_M};{\bf q_N}) \right]\right|_{p_{1,1}:=p_{1,\alpha}} \cr
+ {1 \over N^2} H_{{\bf k_L};m+1;n}({\bf S_K};p_{1,1},{\bf p_M};{\bf q_N}).\cr
\end{array}
\eeq
o\`u $S_{1|\alpha}$ est le couple de cycles obtenu en coupant le cycle $S_1$ avant $p_{1,\alpha}$ et avant $p_{1,1}$:
\beq
S_{1,\alpha}:=\{p_{1,1},q_{1,1},\dots p_{1,\alpha-1},q_{1,\alpha-1}\},\{p_{1,\alpha},q_{1,\alpha},\dots p_{1,k_1},q_{1,k_1}\}.
\eeq

Le coefficient de ${1 \over N^{2g}}$ s'\'ecrit alors:
\beq \label{loopeq}
\begin{array}{l}
(y(p_{1,1})-y(q_{1,k_1})-  Pol_{x(p_{1,1})} V_1'(x(p_{1,1})) ) H_{k_1,\dots,k_l;m;n}^{(g)}(S_1,S_2, \dots, S_l;p_1 ,\dots, p_m;q_1,\dots,q_n) = \cr
{\displaystyle \sum_{h=1}^{g} H_{0;1;0}^{(h)}(p_{1,1}) H_{k_1,\dots,k_l;m;n}^{(g-h)}(S_1,S_2, \dots, S_l;p_1 ,\dots, p_m;q_1,\dots,q_n) }\cr
{\displaystyle +\sum_h \sum_{A\bigcup B=\{2, \dots,l\}}  \sum_{I,J} H_{k_1,{\bf k_A};\left|I\right|;\left|J\right|}^{(h)}(S_1,{\bf S_A}; {\bf p_{I}};{\bf q_J})
H_{{\bf k_B};m-\left|I\right|+1;n-\left|J\right|}^{(g-h)}({\bf S_B}; p_{1,1} {\bf p_{M/I}};{\bf q_N/J})}\cr
{\displaystyle + \sum_h \sum_{A\bigcup B=\{2, \dots,l\}}\sum_{\alpha =2}^{k_1} \sum_{I,J} H_{k_1-\alpha+1,{\bf k_B};m-\left|I\right|;n-\left|J\right|}^{(h)}(\{p_{1,\alpha},q_{1,\alpha},\dots p_{1,k_1},q_{1,k_1}\},{\bf S_B}; {\bf p_{M/I}};{\bf q_{N/J}})}\cr
\qquad  \times { H_{\alpha-1,{\bf k_A};\left|I\right|;\left|J\right|}^{(g-h)}(\{p_{1,1},q_{1,1},\dots p_{1,\alpha-1},q_{1,\alpha-1}\},{\bf S_A}; {\bf p_{I}};{\bf q_J})
-  H_{\alpha-1,{\bf k_A};\left|I\right|;\left|J\right|}^{(g-h)}(\{p_{1,\alpha},q_{1,1},\dots p_{1,\alpha-1},q_{1,\alpha-1}\},{\bf S_A}; {\bf p_{I}};{\bf q_J}) \over x(p_{1,\alpha})-x(p_{1,1})} \cr
+{\displaystyle  \sum_{i=2}^{l} \sum_{\alpha=1}^{k_i} {1 \over x(p_{i,\alpha}) - x(p_{1,1})} }\cr
\left[ H_{k_1+k_i,{\bf k_{L/\{1,i\}}};m;n}^{(g)}( \{S_1,p_{i,\alpha},q_{i,\alpha},p_{i,\alpha+1}, \dots ,q_{i,k_i},p_{i,1},\dots ,p_{i,\alpha-1},q_{i,\alpha-1}\},{\bf S_{L/\{1,i\}}};{\bf p_M};{\bf q_N}) \right. \cr
\left.  \left. \qquad \qquad \qquad - \hbox{Idem} \right|_{p_{1,1}:=p_{i,\alpha}} \right] \cr
+ \sum_{i=1}^{m} \partial_{p_i} \left[{\left. H_{{\bf k_L};m-1;n}({\bf S_L};{\bf p_{M/\{i\}}};{\bf q_N})\right|_{p_{1,1}:=p_i} \over x(p_i) - x(p_{1,1})} \right] \cr
+ \sum_{\alpha =2}^{k_1} { 1 \over x(p_{1,\alpha})-x(p_{1,1})} \times \cr
\left[ H_{\alpha-1, k_1-\alpha+1, { \bf k_{L/\{1\}}}; m;n}^{(g-1)}(S_{1|\alpha}, {\bf S_{L/\{1\}}};{\bf p_M};{\bf q_N}) \right. \cr
\left. - H_{\alpha-1, k_1-\alpha+1, { \bf k_{L/\{1\}}}; m;n}^{(g-1)}(S_{1|\alpha}, {\bf S_{L/\{1\}}};{\bf p_M};{\bf q_N}) \right] \cr
+ H_{{\bf k_L};m+1;n}^{(g-1)}({\bf S_K};p_{1,1},{\bf p_M};{\bf q_N}).\cr
\end{array}
\eeq

\section{Solution g\'en\'erale des \'equations de boucles.}

Pour r\'esoudre l'\'equation de boucle \eq{loopeq}, remarquons tout d'abord que le membre de droite ne contient que des fonctions de
corr\'elation d'ordre inf\'erieur \`{a} $H_{{\bf k_L};m;n}^{(g)}$ dans le sens o\`{u} elles correspondent \`{a} un genre strictement
inf\'erieur \`{a} $g$ ou bien d\'ecrivent des surfaces avec moins de bords, i.e. elles contiennent moins de facteurs
de la forme $\Tr(.)$.
Nous allons donc proc\'eder par r\'ecurrence sur le genre et le nombre de traces, plus pr\'ecis\'ement sur $2g+l+m+n$.

Consid\'erons donc tous les termes du membre de droite de \eq{loopeq} connus et calculons les termes du membre de gauche.
Par d\'efinition le dernier terme du membre de gauche est un polyn\^ome en $x(p_{1,1})$ de degr\'e $d_1-1$. Or, $H_{{\bf k_L};m;n}^{(g)}$
n'a pas de p\^ole en $p_{1,1}:=\tilde{q}_{1,k_1}^{j}$ pour $j = 1, \dots, d_1$, on peut donc \'evaluer cette \'equation en ces
points pour obtenir la valeur du polyn\^ome
\beq
U_{k_1,\dots,k_l;m;n}^{(g)}(x(p_{1,1})):=
Pol_{x(p_{1,1})} V_1'(x(p_{1,1}))  H_{k_1,\dots,k_l;m;n}^{(g)}({\bf S_L};{\bf p_M};{\bf q_N}).
\eeq
en ces points ind\'ependament de $H_{{\bf k_L};m;n}^{(g)}$ puisque sa contribution est annul\'ee par son pr\'efacteur $y(p_{1,1})-y(q_{1,k_1})$.
La formule d'interpolation de Lagrange s'\'ecrit alors:
\beq
U_{k_1,\dots,k_l;m;n}^{(g)}(x(p_{1,1}))=
\sum_{j=1}^{d_1} \Res_{r \to \tilde{q}^{j}}{ H_{1;0;0}^{(0)}(p_{1,1},q_{1,k_1})U_{k_1,\dots,k_l;m;n}^{(g)}(x(r))(y(p_{1,1})-y(q)) dx(r) \over
(x(p_{1,1})-x(r))(y(r)-y(q))H_{1;0;0}^{(0)}(r,q_{1,k_1})},
\eeq
pour ce polyn\^ome de degr\'e $d_1-1$ dont la valeur est connue en $d_1$ points.

En ins\'erant cette solution dans \eq{loopeq}, on obtient:
\beq
H_{{\bf k_L} ;m;n}^{(g)}({\bf S_L};p_1 ,\dots, p_m;q_1,\dots,q_n) =
\Res_{r \to p_{1,1} , \tilde{q}^{j}} { H_{1;0;0}^{(0)}(p_{1,1},q_{1,k_1})\; \hbox{RHS}|_{p_{1,1}:=r} \over
(x(p_{1,1})-x(r))(y(q)-y(r))H_{1;0;0}^{(0)}(r,q_{1,k_1})}
\eeq
o\`{u} $\hbox{RHS}$ est le terme de droite de \eq{loopeq}.

On peut \'eliminer certains termes en d\'epla\c{c}ant le contour d'int\'egration et en utilisant la structure de p\^oles des
fonctions de corr\'elation ainsi que le caract\`{e}re rationnel de certaines combinaisons de celles-ci pour finalement obtenir:
\beq\label{gensoluce}\encadremath{
\begin{array}{l}
H_{{\bf k_L} ;m;n}^{(g)}({\bf S_L};p_1 ,\dots, p_m;q_1,\dots,q_n) =\cr
{\displaystyle \Res_{r \to p_{i,\alpha},p_j, \tilde{q}_{1,k_1}^{j}} { H_{1;0;0}^{(0)}(p_{1,1},q_{1,k_1}) \over
(x(p_{1,1})-x(r))(y(q_{1,k_1})-y(r))H_{1;0;0}^{(0)}(r,q_{1,k_1})}}\cr
\Big\{
{\displaystyle \sum_{h=1}^{g} H_{0;1;0}^{(h)}(r) H_{k_1,\dots,k_l;m;n}^{(g-h)}(S_1(r),S_2, \dots, S_l;p_1 ,\dots, p_m;q_1,\dots,q_n) }\cr
{\displaystyle +\sum_{A\bigcup B=\{2, \dots,l\}}  \sum_{h,I,J} H_{k_1,{\bf k_A};\left|I\right|;\left|J\right|}^{(h)}(S_1(r),{\bf S_A}; {\bf p_{I}};{\bf q_J})
H_{{\bf k_B};m-\left|I\right|+1;n-\left|J\right|}^{(g-h)}({\bf S_B}; r, {\bf p_{M/I}};{\bf q_N/J})}\cr
{\displaystyle + \sum_{A\bigcup B=\{2, \dots,l\}}\sum_{\alpha =2}^{k_1} \sum_{h,I,J} H_{k_1-\alpha+1,{\bf k_B};m-\left|I\right|;n-\left|J\right|}^{(h)}(\{p_{1,\alpha},q_{1,\alpha},\dots p_{1,k_1},q_{1,k_1}\},{\bf S_B}; {\bf p_{M/I}};{\bf q_{N/J}})}\cr
\qquad  \times { H_{\alpha-1,{\bf k_A};\left|I\right|;\left|J\right|}^{(g-h)}(\{r,q_{1,1},\dots p_{1,\alpha-1},q_{1,\alpha-1}\},{\bf S_A}; {\bf p_{I}};{\bf q_J})
 \over x(p_{1,\alpha})-x(r)} \cr
{\displaystyle + \sum_{i=2}^{l} \sum_{\alpha=1}^{k_i} {1 \over x(p_{i,\alpha}) - x(r)}  }
 H_{k_1+k_i,{\bf k_{L/\{1,i\}}};m;n}^{(g)}( S_{1;i\alpha}(r),{\bf S_{L/\{1,i\}}};{\bf p_M};{\bf q_N})  \cr
{\displaystyle + \sum_{\alpha =2}^{k_1} { 1 \over x(p_{1,\alpha})-x(r)} }
 H_{\alpha-1, k_1-\alpha+1, { \bf k_{L/\{1\}}}; m;n}^{(g-1)}(S_{1|\alpha}(r), {\bf S_{L/\{1\}}};{\bf p_M};{\bf q_N}) \cr
+ H_{{\bf k_L};m+1;n}^{(g-1)}({\bf S_K}(r);r,{\bf p_M};{\bf q_N})
\Big\} \cr
\end{array}}
\eeq
o\`u l'on a not\'e $S(r)$ lorsque l'on remplace $p_{1,1}$ par $r$ dans $S$ et $S_{1;i\alpha}$ pour le cycle obtenu
en recollant $S_1$ et $S_i$ \`a la position $\alpha$ de $S_i$:
\beq
S_{1;i\alpha}:= \{S_1,p_{i,\alpha},q_{i,\alpha},p_{i,\alpha+1}, \dots ,q_{i,k_i},p_{i,1},\dots ,p_{i,\alpha-1},q_{i,\alpha-1}\}.
\eeq

Comme nous l'avons vu plus haut, ce syst\`{e}me est triangulaire et admet une solution unique \'etant donn\'es $H_{1;0;0}^{(0)}$,
$H_{0;1;0}^{(0)}$, $H_{0;0;2}^{(0)}$ et $H_{0;1;1}^{(0)}$ d\'ej\`{a} d\'etermin\'es dans la litt\'erature de nombreuses fois (voir par exmple \cite{eynm2m}).

\subsection{Repr\'esentation diagrammatique}

Nous allons \`{a} nouveau repr\'esenter ce r\'esultat sous forme de diagramme en utilisant la repr\'esentation introduite dans
la partie \ref{partcombi}. Cette repr\'esenation a plusieurs avantages. Tout d'abord, elle permet de mettre en avant la structure de ce
r\'esultat et de bien voir que ce syst\`{e}me d'\'equations est triangulaire. Elle est \'egalement tr\`{e}s utile pour se
souvenir du r\'esultat qui semble a priori compliqu\'e alors que sa structure est finalement assez simple. Enfin, cette
repr\'esentation appelle une interpr\'etation combinatoire et fait le lien entre les r\'esultat pr\'ec\'edement obtenus sur
l'ordre dominant des traces mixtes et le d\'eveloppement topologique des traces simples.

\subsubsection{Blocs de base: deux nouveaux vertex}

Repr\'esentons les fonctions de corr\'elation par des surfaces de genre $g$ avec $l+m+n$ bords comme dans \eq{diagmix}:
\beq
H_{{\bf k_L} ;m;n}^{(g)}({\bf S_L};p_1 ,\dots, p_m;q_1,\dots,q_n):=
\begin{array}{r}
{\rm \includegraphics[width=5cm]{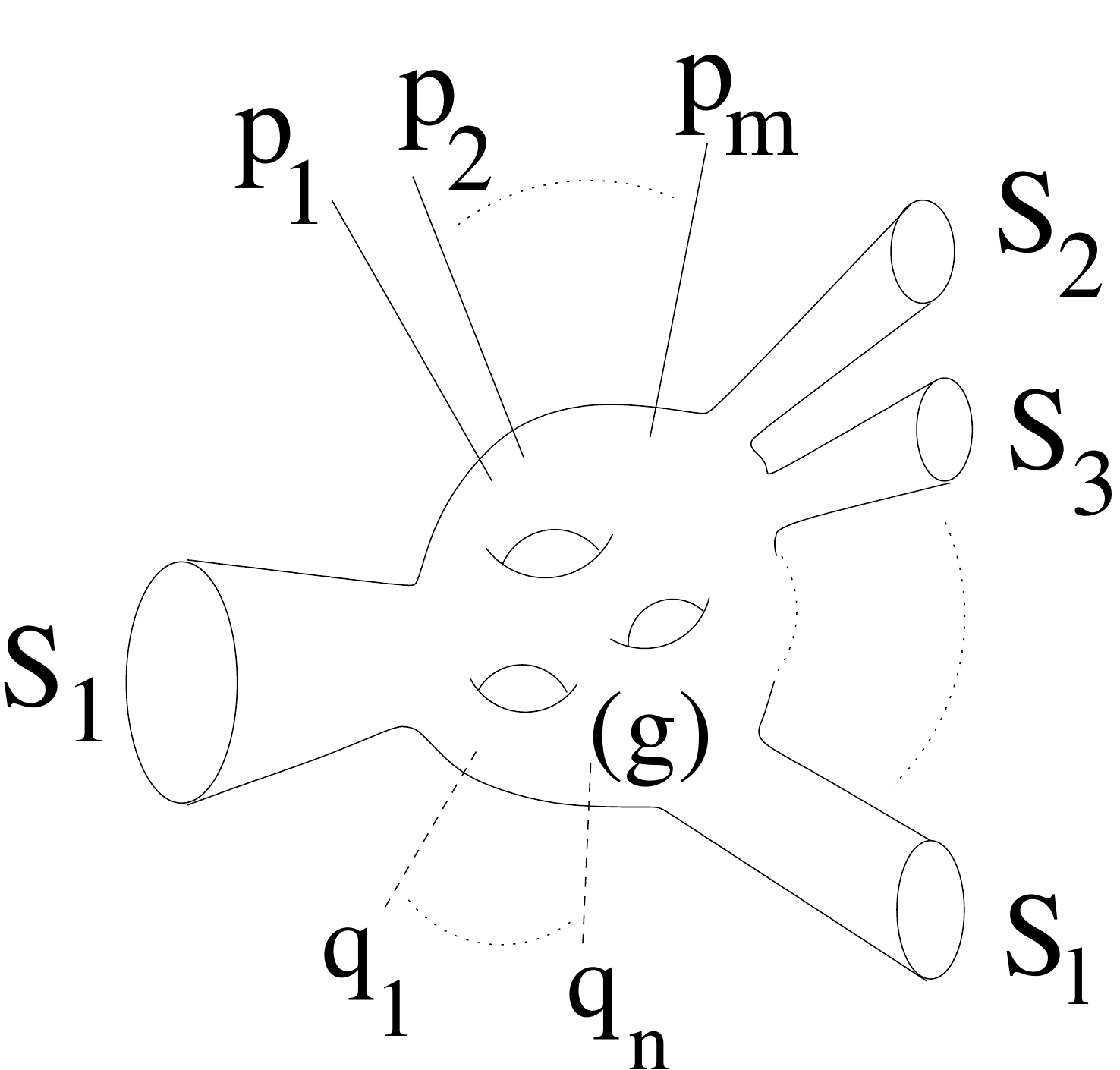}}
\end{array}.
\eeq

Introduisons \'egalement deux nouveaux vertex

$\bullet$ {\bf Vertex fin:}
\beq
\Res_{r \to p, \tilde{q}^j}
{ H_{1;0;0}^{(0)}(p,q) \over
(x(p)-x(r))(y(q)-y(r))H_{1;0;0}^{(0)}(r,q)}:=
\begin{array}{r}
{\rm \includegraphics[width=4cm]{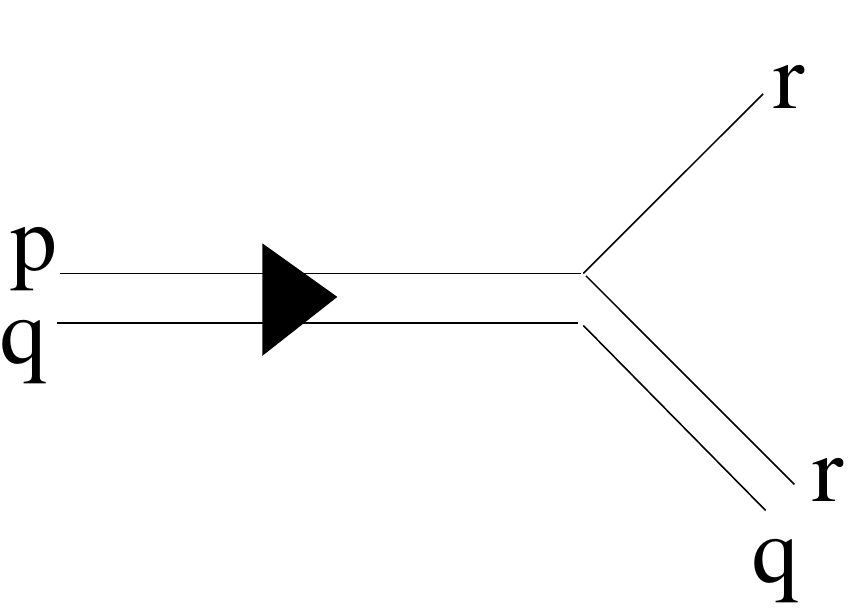}}
\end{array}
\eeq

$\bullet$ {\bf Vertex \'epais:}
\beq
\Res_{r \to p, \tilde{q}^j,p'}
{ H_{1;0;0}^{(0)}(p,q) \over
(x(p)-x(r))(y(q)-y(r))(x(p')-x(r)) H_{1;0;0}^{(0)}(r,q)}:=
\begin{array}{r}
{\rm \includegraphics[width=4cm]{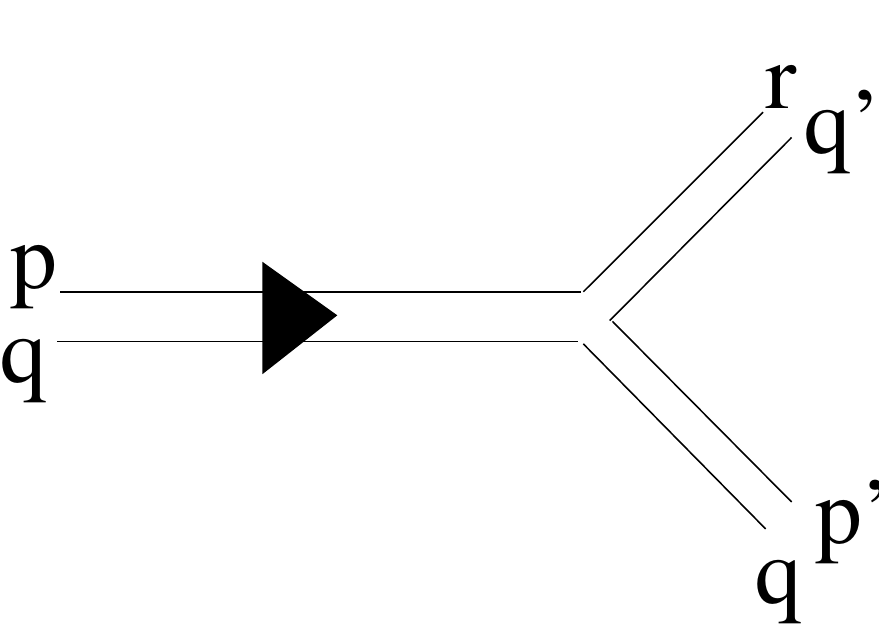}}
\end{array}
\eeq

Ces deux nouveaux vertex vont nous permettre de g\'en\'eraliser la repr\'esentation diagrammatique introduite pr\'ec\'edemment.
Nous aurons ainsi acc\`es \`a une repr\'esentation graphique de toutes les fonctions de corr\'elation.

\eop

\subsubsection{Repr\'esentation de la relation de r\'ecurrence.}

La relation de r\'ecurrence \eq{gensoluce} prend alors simplement la forme:
\beq
\begin{array}{l}
\begin{array}{r}
{\rm \includegraphics[width=4.5cm]{blobmix1}}
\end{array}
=
\begin{array}{r}
{\rm \includegraphics[width=7cm]{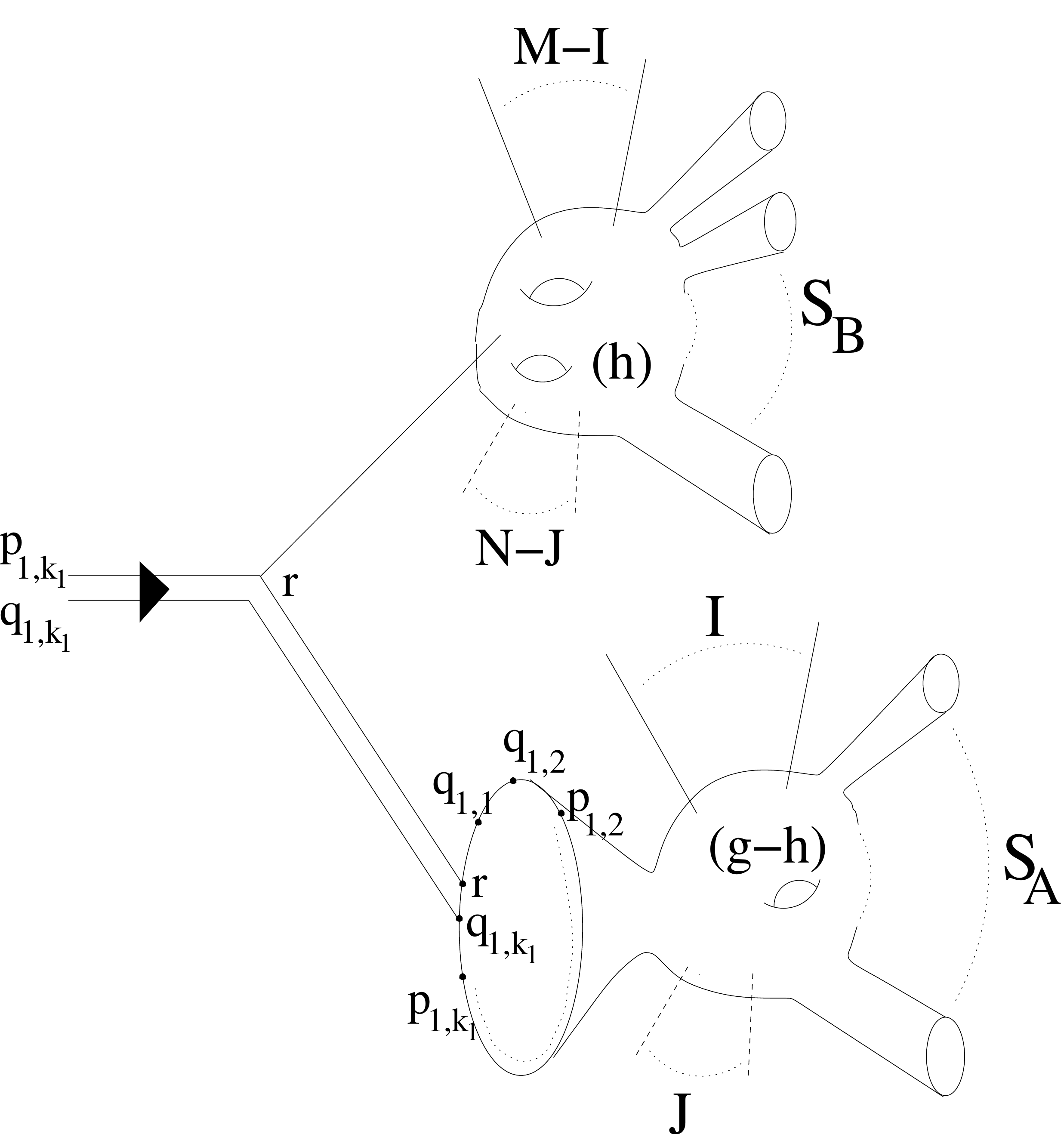}}
\end{array} \cr
+\begin{array}{r}
{\rm \includegraphics[width=7.5cm]{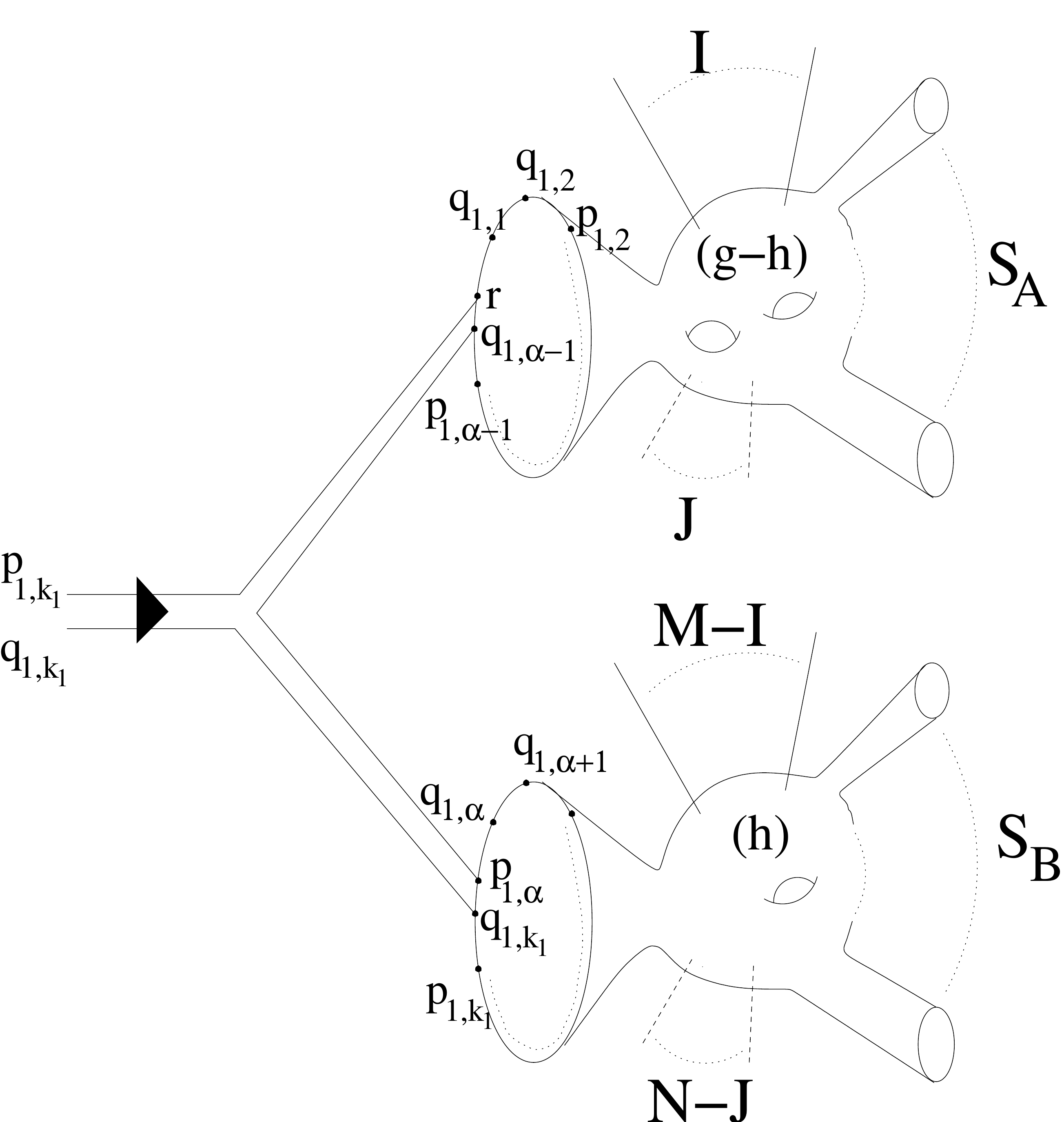}}
\end{array}
+\begin{array}{r}
{\rm \includegraphics[width=7.5cm]{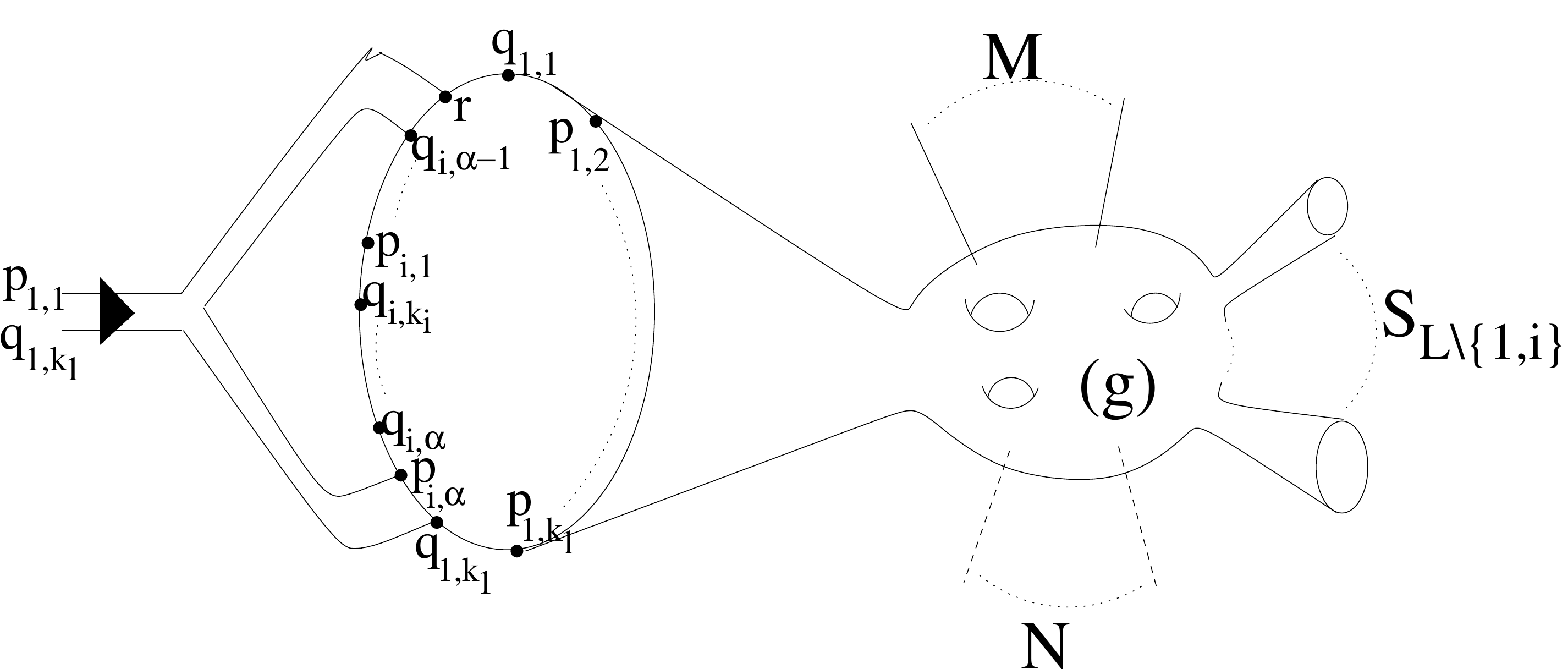}}
\end{array}\cr
+\begin{array}{r}
{\rm \includegraphics[width=7.5cm]{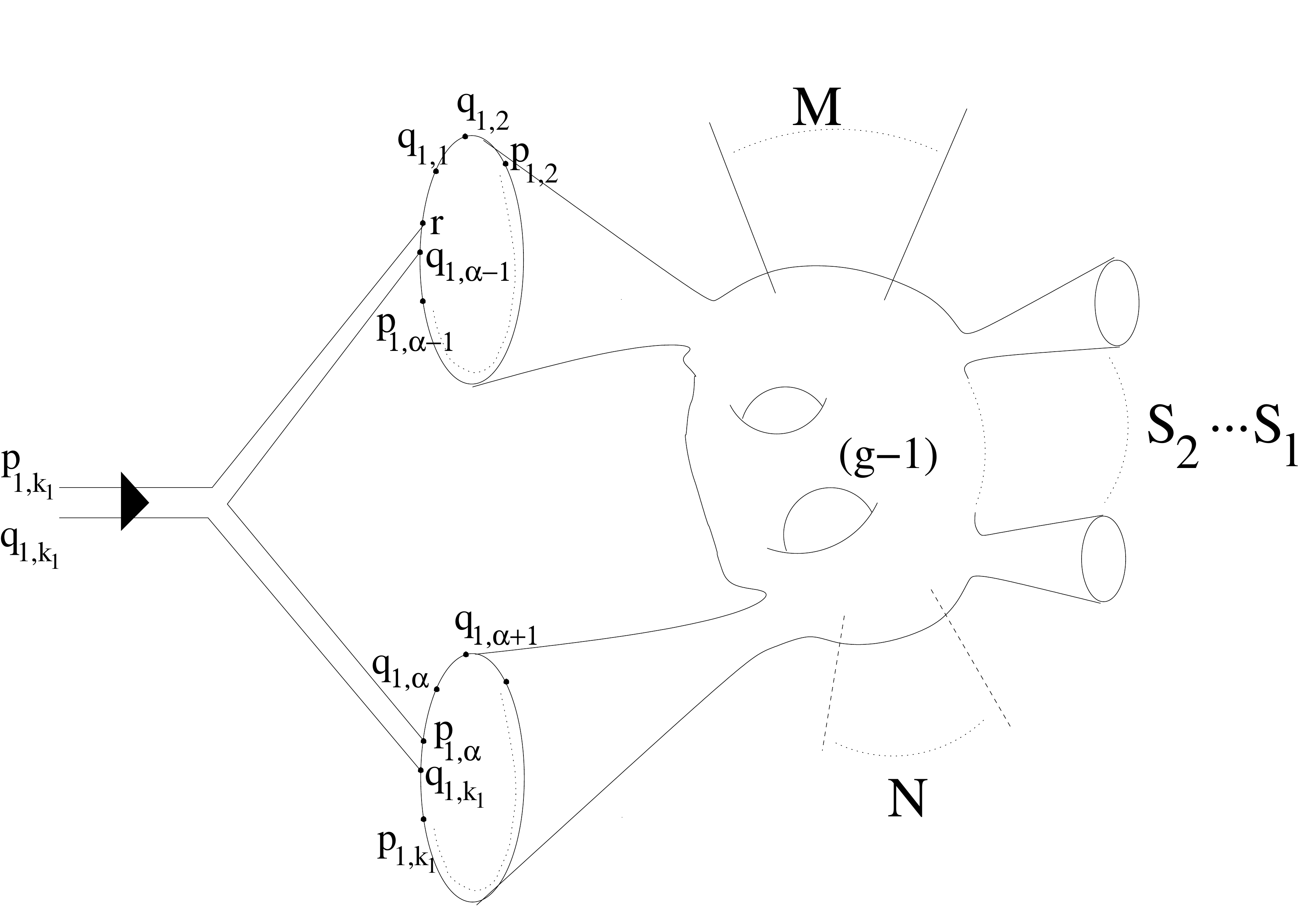}}
\end{array}
+\begin{array}{r}
{\rm \includegraphics[width=7.5cm]{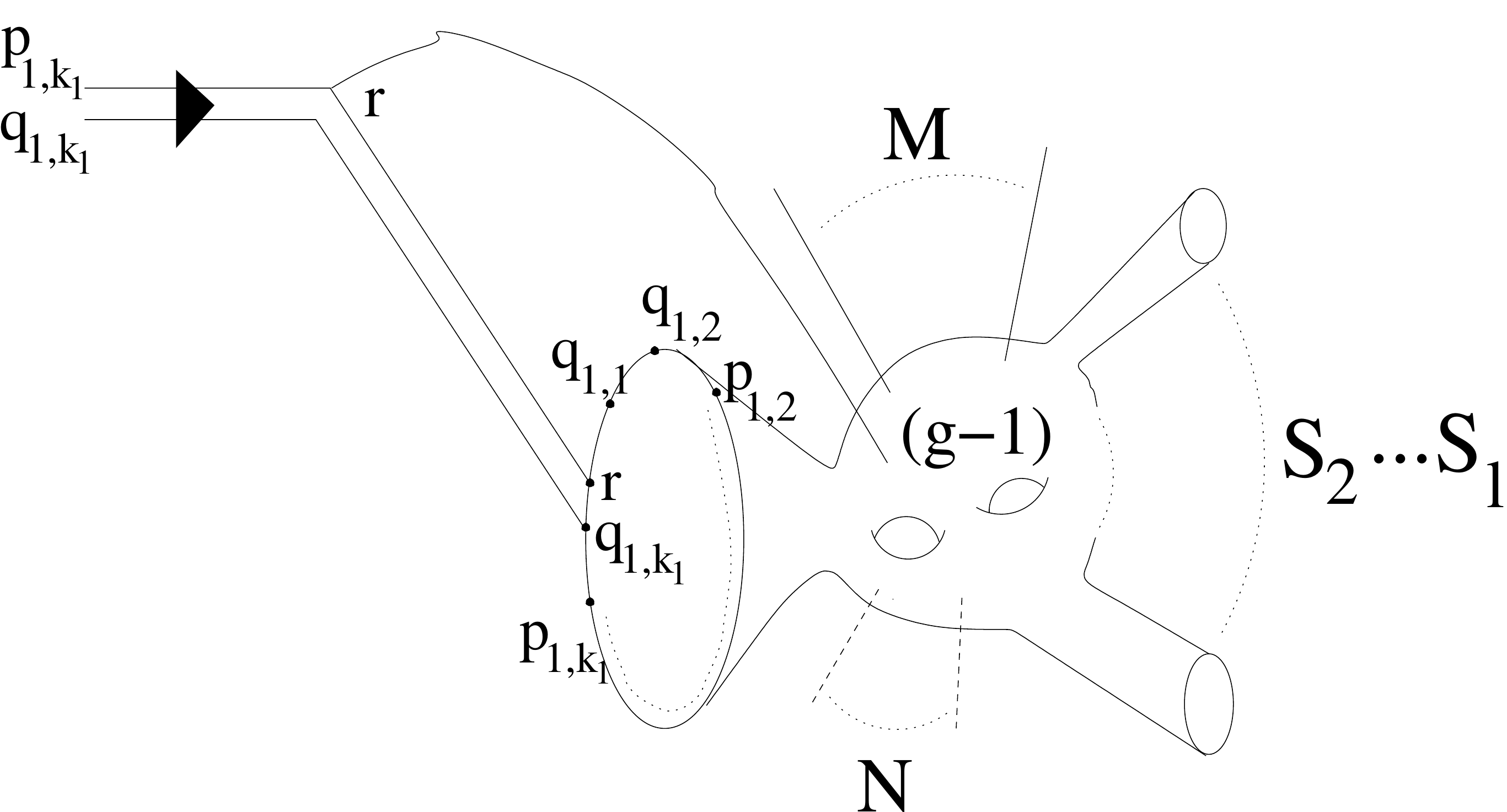}}
\end{array}.\cr
\end{array}
\eeq

\subsection{Retour sur le d\'eveloppement topologique des traces simples.}

Ces relations g\'en\'erales doivent contenir les r\`egles de r\'ecurrence pour les traces simples \eq{conjecture} puisque les
traces simples peuvent \^etre obtenues en \'etudiant le comportement des traces mixtes lorsque l'une des variables tend vers l'infini:
\beq
\left<\Tr {1 \over x(p)-M_1} {1 \over y(q)- M_2} \right> \sim_{q \to \infty_y} {1 \over y(q)} \left< \Tr {1 \over x(p) -M_1} \right>.
\eeq
Plus pr\'ecis\'ement, les fonctions de corr\'elation simples et mixtes sont li\'ees par le lemme
\bl\label{leee}
Les fonctions de corr\'elations simples peuvent \^etre obtenues \`a partir des fonctions mixtes par:
\beq
W_{k+1,0}^{(h)}(p,{\bf p_K}) = \Res_{q \to \infty_y} {H_{1;k;0}^{(h)}(p,q;{\bf p_K}) dy(q) \over H_{1;0;0}^{(0)}(p,q)}.
\eeq
\el

La fonction de corr\'elation n'ayant que des traces simples sauf une trace mixte est
donn\'ee par la relation de r\'ecurrence:
\bea
H_{1;k;0}^{(g)}(p,q;{\bf p_K}) &=& \Res_{r \to p, \tilde{q}^j, {\bf p_K}} {H_{1;0;0}^{(0)}(p,q) \over (x(p)-x(r))(y(r)-y(q))H_{1;0;0}^{(0)}(r,q)}
\Big( H_{1;k+1;0}^{(g-1)}(r,q;r,{\bf p_K}) \cr
&& + \sum_{h,l} H_{0;l+1;0}^{(h)}(r, {\bf p_L}) H_{1;k-l;0}^{(g-h)}(r,q; {\bf p_{K/L}}) \Big).
\eea

Par d\'efinition, l'application du lemme \ref{leee} \`a cette relation redonne les relations de r\'ecurrence \eq{recdiag}
qui n'en sont donc qu'un cas particulier. Cette propri\'et\'e signifie que le passage \`a la limite
$y \to \infty$ consiste \`a "\'ecraser" le propagateur \'epais associ\'e \`a $y$ pour le rendre fin comme pour les fonctions
de corr\'elation simples.

\subsection{Retour sur la limite planaire des traces mixtes.}

Lorsque l'on s'int\'eresse seulement au calcul de l'ordre dominant des traces mixtes la relation de r\'ecurrence se
referme sur les $H_{k}:=H_{k;0;0}^{(0)}$ et devient
\bea
&&H_{k}(p_1,q_1,p_2,q_2,\dots,p_k,q_k) = \cr
&& = \Res_{r \to \tilde{q}_k^j,p_i} {H_{1}(p_1,q_k) \over (x(p_1)- x(r)) (y(q_k)-y(r)) H_1(r,q_k)} \times \cr
&& \qquad \times \sum_{l=2}^k {H_{l-1}(r,q_1,p_2,q_2,\dots,q_{l-1})H_{k-l+1}(p_l,q_l,\dots,p_k,q_k) \over x(p_l)-x(r)} .\cr
\eea
Cette formule semble faire explicitement appel \`a la structure de la courbe spectrale classique. Cependant nous allons montrer
qu'elle peut simplement \^etre ramen\'ee \`a la relation de r\'ecurrence \eq{Ansatz}.

En effet, on peut mettre cette \'equation sous la forme:
\bea
H_k &=& \sum_{l=2}^k \Res_{r \to \tilde{q}_k^j,p_l} {H_{1}(p_1,q_k) H_{l-1}(r,q_1,\dots,q_{l-1})H_{k-l+1}(p_l,\dots,q_k)\over (x(p_1)- x(r)) (y(q_k)-y(r))(x(p_l)-x(r)) H_1(r,q_k)}\cr
&=&\sum_{l=2}^k \Res_{r \to \tilde{q}_k^j,p_l} {H_{1}(p_1,q_k) H_{l-1}(r,q_1,\dots,q_{l-1})H_{k-l+1}(p_l,\dots,q_k)\over (x(p_1)- x(r)) (x(p_l)-x(r)) H_1(r,q_k)}\cr
&& \qquad \times \left( {1 \over (y(q_k)-y(r))} - {1 \over (y(q_k)-y(q_1))} + {1 \over (y(q_k)-y(q_1))} \right) \cr
&=&\sum_{l=2}^k \Big[ \Res_{r \to \tilde{q}_k^j,p_l,p_1} {H_{1}(p_1,q_k) H_{l-1}(r,q_1,\dots,q_{l-1})H_{k-l+1}(p_l,\dots,q_k)\over (x(p_1)- x(r)) (x(p_l)-x(r))(y(q_k)-y(q_1)) H_1(r,q_k)}\cr
&+& \Res_{r \to \tilde{q}_k^j,p_l,p_1} {H_{1}(p_1,q_k) H_{k-l+1}(p_l,\dots,q_k) (y(r)-y(q_1))H_{l-1}(r,q_1,\dots,q_{l-1})\over (x(p_1)- x(r)) (x(p_l)-x(r))(y(q_k)-y(q_1))(y(q_k)-y(r)) H_1(r,q_k)} \Big]\cr
\eea
Or, $(y(r)-y(q_1))H_{l-1}(r,q_1,\dots,q_{l-1})$ et $(y(q_k)-y(r)) H_1(r,q_k)$ sont des polyn\^omes en $x(r)$. L'int\'egrant dans le
second terme est donc une fraction rationnelle en $x(r)$. On peut facilement v\'erifier que le contour d'int\'egration
entoure tous ses p\^oles dans le plan complexe et ce second terme est donc nul. Le premier terme ne pr\'esente quant \`a lui
que des p\^oles simples et on retrouve donc
\beq
H_k = \sum_{l=2}^k \Res_{r \to p_l,p_1}{H_{1}(p_1,q_k) H_{l-1}(r,q_1,\dots,q_{l-1})H_{k-l+1}(p_l,\dots,q_k)\over (x(p_1)- x(r)) (x(p_l)-x(r))(y(q_k)-y(q_1)) H_1(r,q_k)}
\eeq
qui n'est rien d'autre que l'\'equation \eq{Ansatz2} avec les r\^oles de $x$ et $y$ invers\'es.

\subsection{Interpr\'etation des relations de r\'ecurrence.}

Cette repr\'esentation sous forme de diagrammes des fonctions de corr\'elation, c'est-\`{a}-dire des fonction g\'en\'eratrices
de surfaces discr\'etis\'ees, pose la question de l'existence d'une th\'eorie des champs sous-jacente dont les graphes
pr\'esent\'es ici seraient les diagrammes de Feynman. En fait, une telle th\'eorie n'existe probablement pas.
 En effet, le nombre de diagrammes ainsi cr\'e\'es croit beaucoup moins vite avec le nombre de boucles
que ce que l'on devrait obtenir avec une th\'eorie des champs standard. Ceci est li\'e aux
pr\'escriptions fortement non-locales dans la construction des diagrammes participant aux traces simples.

Par contre, il existe une fa\c{c}on de se rappeler simplement les diff\'erents termes apparaissant dans le terme de droite
des relations de r\'ecurrences \eq{gensoluce}. Le terme de gauche $H_{{\bf k_L} ;m;n}^{(g)}({\bf S_L};p_1 ,\dots, p_m;q_1,\dots,q_n)$
repr\'esentant la fonction g\'en\'eratrices de toutes les surfaces $\CS_{{\bf k_L} ;m;n}^{(g)}$
du type
\beq
\CS_{{\bf k_L} ;m;n}^{(g)} = \begin{array}{r}
{\rm \includegraphics[width=5cm]{blobmix1}},
\end{array}
\eeq
on peut le voir comme une int\'egrale sur l'espace des modules $\CM_{{\bf k_L} ;m;n}^{(g)}$ de telles surfaces.
Le terme de droite peut alors \^etre vu comme les termes de bords venant de cette int\'egrale, c'est-\`a-dire
la somme sur toutes les surfaces d\'eg\'en\'er\'ees de cet ensemble. Ils sont donc obtenus de mani\`ere pratique
en "pin\c{c}ant" tous les chemins non triviaux possibles sur la surface $\CS_{{\bf k_L} ;m;n}^{(g)}$.
On voit alors qu'il existe plusieurs chemins ayant des r\^oles non \'equivalents:
\begin{itemize}
\item On peut tout d'abord pincer un cycle non trivial sur la surface. Deux cas de figure se pr\'esentent: soit l'on d\'ecoupe
ainsi la surface en deux composantes :
\beq
\begin{array}{r}
{\rm \includegraphics[width=12cm]{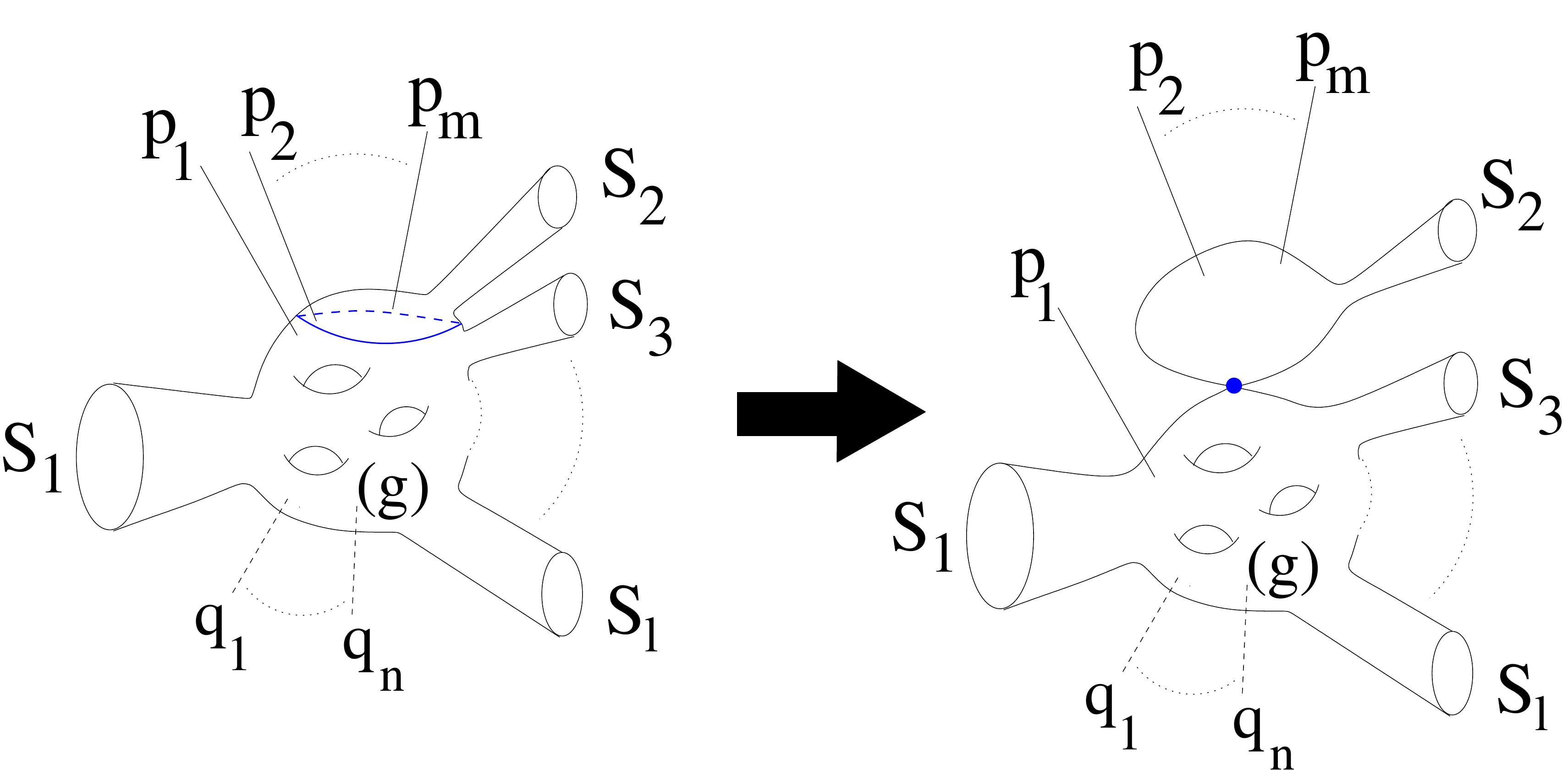}}
\end{array}
\eeq
et l'on obtient le premier terme de \eq{gensoluce}; soit l'on pince juste une jambe interne:
\beq
\begin{array}{r}
{\rm \includegraphics[width=12cm]{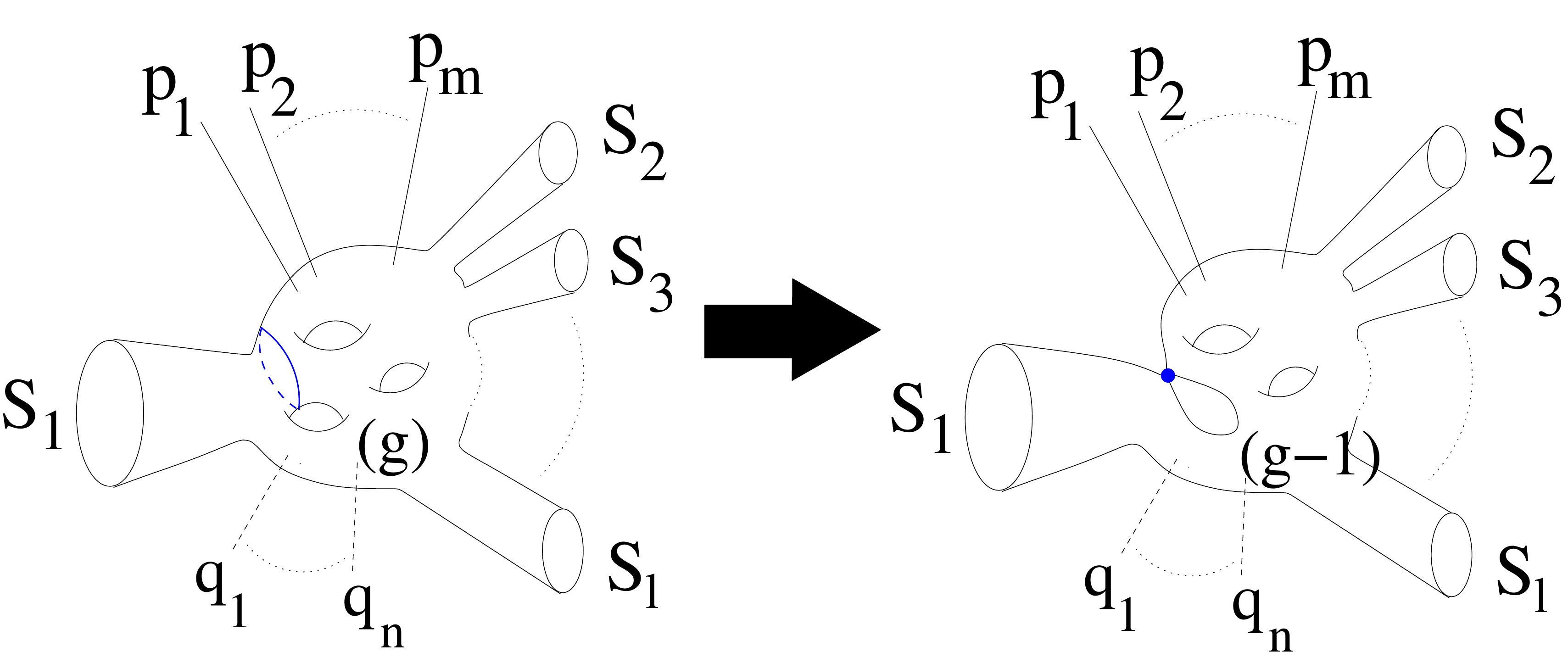}}
\end{array}
\eeq
et l'on obtient le cinqui\`eme terme de \eq{gensoluce}.

\item On peut \'egalement pincer un chemin allant d'un cycle $S_i$ a un autre cycle $S_j$:
\beq
\begin{array}{r}
{\rm \includegraphics[width=12cm]{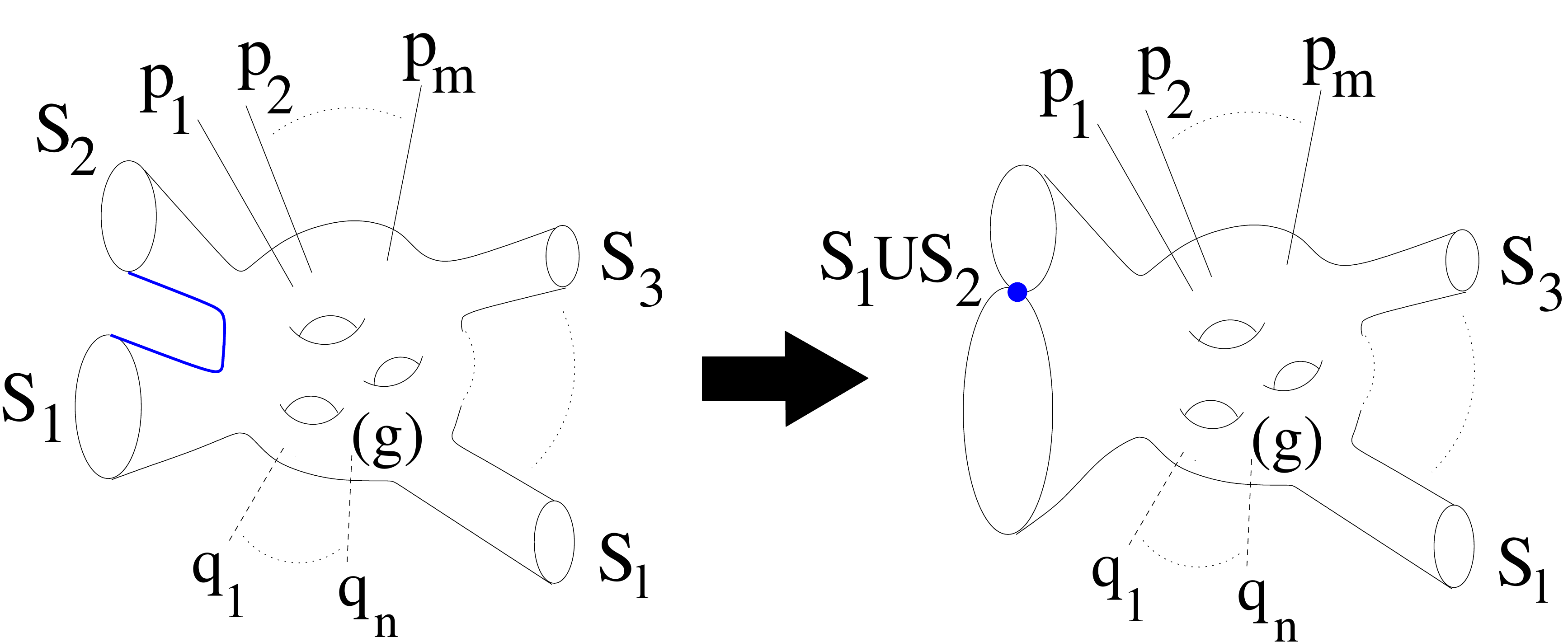}}
\end{array}
\eeq
pour obtenir le troisi\`eme terme de \eq{gensoluce}\footnote{Notons que ce cas est le seul correpondant \`a la r\`egle merge des \'equations
de boucles.}.

\item Enfin, on peut pincer un cycle $S_i$ lui-m\^eme
\beq
\begin{array}{r}
{\rm \includegraphics[width=12cm]{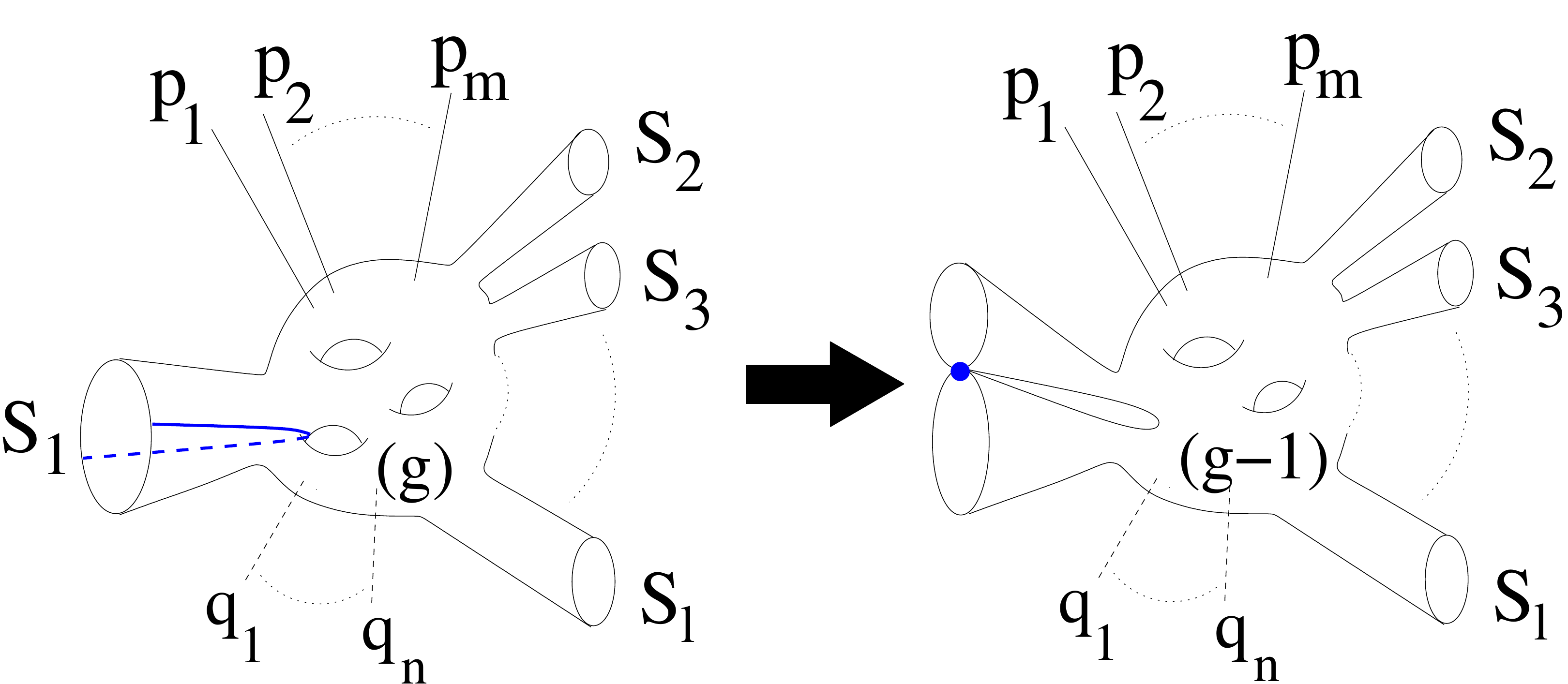}}
\end{array}
\eeq
pour obtenir les deuxi\`emes et quatri\`emes termes de \eq{gensoluce} selon si cette d\'egenerescence coupe la surface
en deux parties ou non.

\end{itemize}

\section{Conclusion du chapitre.}

Dans ce chapitre, nous avons montr\'e comment r\'esoudre enti\`erement le mod\`ele \`a deux matrices hermitiennes,
c'est-\`a-dire comment calculer tous les termes du d\'evelop- pement topologique de toutes les fonctions de corr\'elation
du mod\`ele. Cette m\'ethode de r\'esolution, utilisant avantageusement le langage de la g\'eom\'etrie alg\'ebrique,
pr\'esente plusieurs avantages:
\begin{itemize}

\item Elle est syst\'ematique et montre bien que toutes les observables du mod\`ele \`a deux matrices hermitiennes  ne d\'ependent
que de la courbe spectrale et plus particuli\`erement de ses points de branchements. Le r\'esultat prend la m\^eme forme
que l'on soit dans un cas \`a une ou plusieurs coupures.

\item Contrairement \`a d'autres m\'ethodes de r\'esolutions comme celle d\'evelopp\'ee originellement dans \cite{ACKM}, celle-ci
 est facilement programmable pour une r\'esolution automatique. En effet, une fois la courbe spectrale connue, il suffit de
 calculer le noyau de Bergmann et effectuer des d\'eveloppement de Taylor au voisinage des points de branchement: il n'est
 pas n\'ecessaire de reconstruire une base dans laquelle travailler \`a chaque \'etape du calcul.

\item Comme on peut repr\'esenter le r\'esultat de cette m\'ethode sous forme de graphes, il est assez facile \`a retrouver
avec tr\`es peu d'\'el\'ements (il suffit de connaitre la valeur des vertex et propagateurs). En effet, on peut retrouver
la forme des \'equations de r\'ecurrence gr\^ace \`a une analogie avec les surfaces g\'en\'er\'ees par les fonctions de corr\'elation et
leurs d\'egenerescences possibles.

\end{itemize}

Si le mod\`ele \`a deux matrices formel est r\'esolu en termes de sa courbe spectrale ce chapitre appelle \'egalement des questions.
D'une part, il serait int\'eressant de pouvoir utiliser ces r\'esultats pour faire le lien avec ceux obtenus
dans l'\'etude de l'int\'egrale convergente en s'int\'eressant aux effets d'instantons interdits dans le cas formel.
D'autre part, si ces r\`egles diagrammatiques sont efficaces, leur interpr\'etation est plus d\'elicate et nous ne sommes pas
en mesure pour le moment d'expliquer leur origine combinatoire (ou g\'eom\'etrique).
Enfin, la simplicit\'e du r\'esultat en terme de la courbe spectrale classique laisse supposer que cette m\'ethode doit pouvoir
s'adapter \`a d'autres mod\`eles de matrices: c'est ce que nous montrons dans le chapitre suivant.


\chapter{Autres mod\`eles de matrices.}

Le mod\`{e}le \`{a} deux matrices hermitiennes est un mod\`{e}le de matrices al\'eatoires particulier parmi bien d'autres.
Cependant, la structure alg\'ebrique pr\'esent\'ee dans la chapitre pr\'ec\'edent est pr\'esente dans d'autres mod\`{e}les
de matrices hermitiennes.
Dans ce chapitre, nous \'etudions d'autres exemples de mod\`{e}les de matrices classiques reposant sur une m\^eme structure
alg\'ebrique: chacun de ces mod\`{e}les a une interpr\'etation combinatoire et peut \^{e}tre associ\'e et totalement
caract\'eris\'e par une courbe alg\'ebrique et des fractions de remplissage.
On peut alors montrer qu'ils peuvent \^{e}tre r\'esolus par la m\^{e}me m\'ethode que le mod\`{e}le \`{a} deux matrices.

Si ces mod\`{e}les pr\'esentent les propri\'et\'es n\'ec\'essaires pour une r\'esolution alg\'ebrique,
la raison n'en est pas encore claire. Bien \'evidemment de tels miracles sont li\'es aux propri\'et\'es d'int\'egrabilit\'e des mod\`{e}les consid\'er\'es
mais le lien profond entre une int\'egrale de matrice donn\'ee et l'existence d'une \'eventuelle structure alg\'ebrique associ\'ee
n'est pour l'instant pas claire. On peut, par exemple, se demander si de tels r\'esultats peuvent \^{e}tre \'etendus \`{a} des
mod\`{e}les \'equivalents avec des matrices appartenant \`{a} un autre groupe: par exemple \`a des matrices quaternioniques ou r\'eelles symm\'etriques.
Nous ne discuterons pas cette question ici puisque la r\'eponse n'en est pas encore connue en g\'en\'eral: on a seulement
observ\'e, au cas par cas, que l'on retrouvait une structure similaire pour plusieurs mod\`eles diff\'erents.

\section{Mod\`ele \`a une matrice.}\label{part1MM}

Le mod\`{e}le \`{a} une matrice hermitienne est tr\`{e}s proche du mod\`{e}le \`{a} deux matrices\footnote{En effet, comme
nous le verrons dans la partie \ref{partlien2MM}, ce mod\`{e}le est un cas particulier du pr\'ec\'edent.}. Pour le d\'efinir, consid\'erons
la fonction de partition $\CZ_{1MM}$ d\'efinie par l'int\'egrale de matrices formelle
\beq
\CZ_{1MM}:= \int dM \, e^{-{N \over T} Tr \, V(M)},
\eeq
o\`{u} $V(x)$ est un potentiel dont la d\'eriv\'ee est une fonction rationnelle donn\'ee par ses z\'eros $\{X_i\}_{i=1 \dots d}$ et
son d\'enominateur $D(x)$:
\beq
V'(x) ={{\displaystyle \prod_{i=1}^d} (x-X_i) \over D(x)},
\eeq
et l'int\'egrale sur les matrices
hermitiennes de taille $N\times N$ est \`{a} comprendre au sens formel de mani\`{e}re similaire \`{a} l'int\'egrale
\`{a} deux matrices \eq{defZ} du chapitre pr\'ec\'edent.

Pour cela, choisissons une partition de $N$ \`{a} $d$ \'el\'ements:
\beq
\vec{n} = \{ n_1, n_2, \dots ,n_d \}
\eeq
telle que
\beq
N = \sum_{i=1}^d n_i ,
\eeq
\`{a} laquelle on associe une matrice diagonale
\beq
M_{\vec{n}} = (\overbrace{X_1, \dots ,X_1}^{n_1}, \dots ,\overbrace{X_d, \dots ,X_d}^{n_d} )
\eeq
solution de $V'(M_{\vec{n}})=0$. D\'ecomposons \'egalement le potentiel $V(M_{\vec{n}}+M)$ en mettant \`{a} part
sa partie quadratique en $M$ dans son d\'eveloppement de Taylor pour un petit $M$:
\beq
\Tr V(M) = \Tr \left( V(M_{\vec{n}}) + V_{quad}(M) + \delta V(M)\right)
\eeq
o\`{u} $V_{quad}(M)$ contient tous les termes quadratiques en $M$.
Le d\'eveloppement en s\'erie de la partie non quadratique s'\'ecrit comme une s\'erie formelle
\beq\label{devexp}
\exp{-{N \over T} \delta V(M)} = \sum_{j=0}^\infty {1 \over j!}  T^{-j} N^j (\delta V(M))^j.
\eeq
On d\'efinit alors la fonction de partition du mod\`{e}le \`{a} une matrice comme la s\'erie formelle:
\beq
\CZ_{1MM}:= \sum_{k=0}^\infty T^k \sum_{j=0}^{2k} A_{k,j}
\eeq
o\`{u} les coefficients $A_{k,j}$ sont des polyn\^{o}mes de Laurent en $N^2$ d\'efinis par les valeurs moyennes
par rapport \`{a} la mesure gaussienne $\exp{\left(-{N \over T} V_{quad}(M)\right)} dM$
des termes du d\'eveloppement en s\'erie de \eq{devexp}:
\beq
{\int dM e^{-{N \over T} V_{quad}(M)} {1 \over j!}  T^{-j} N^j (\delta V(M))^j \over \int dM e^{-{N \over T} V_{quad}(M)}}
: = \sum_{k={j\over 2}}^{{dj \over 2}-1} A_{k,j} T^k.
\eeq

Ce mod\`{e}le, \`{a} premi\`{e}re vue bien plus simple que le pr\'ec\'edent, pr\'esente d\'ej\`{a} beaucoup des ingr\'edients
n\'ecessaires \`{a} la r\'esolution de ce dernier. Il a donc souvent servi de premier test pour des nouvelles m\'ethodes de r\'esolutions.
Ainsi, depuis l'introduction du d\'eveloppement topologique (et donc du mod\`ele formel) dans \cite{BIPZ}, le mod\`ele a \'et\'e
r\'esolu par \'etapes successives. Il a d'abord \'et\'e montr\'e dans le cas \`a une coupure dans \cite{staudacher,ACKM,Chekh} que les observables du mod\`ele
sont plus facilement d\'ecrites en termes des points de branchements d'une courbe alg\'ebrique associ\'ee: la courbe spectrale. Cette m\'ethode
a \'et\'e r\'ecemment am\'elior\'ee par la technique diagrammatique d\'evelopp\'ee par Eynard dans \cite{eynloop1mat} permettant
de calculer le d\'eveloppement topologique de toutes les fonctions de corr\'elation quel que soit le nombre de coupures. C'est cette m\'ethode qui a servi de base \`a tous les travaux
pr\'esent\'es dans cette th\`ese. Elle a ensuite \'et\'e \'etendue dans \cite{ec1loopF} pour donner acc\`es au d\'eveloppement
topologique de l'\'energie libre.

Dans cette partie, nous pr\'esentons donc les r\'esultats obtenus dans ces deux articles \cite{eynloop1mat,ec1loopF} par Chekhov et Eynard.

\subsection{Interpr\'etation combinatoire.}

Tout comme le mod\`{e}le \`{a} deux matrices, l'int\'egrale formelle $\CZ_{1MM}$ a une interpr\'eta- tion combinatoire en termes de
comptage de cartes, i.e. de surfaces compos\'ees de polygones. En effet, de mani\`{e}re analogue on peut voir cette
int\'egrale comme le d\'eveloppement en diagrammes de Feynman \'epais autour d'un point col gaussien. Alors, $\CZ_{1MM}$ est la fonction
g\'en\'eratrice de toutes les surfaces ferm\'ees compos\'ees de polygones avec au plus $d$ c\^ot\'es. Comme il n'y a plus
qu'une matrice, les polygones ne portent plus ici de spins.

On peut ainsi d\'efinir l'{\bf\'energie libre} comme la fonction g\'en\'eratrice des surfaces ferm\'ees connexes:
\beq\label{F1MM}
\CF_{1MM}:= - {1 \over N^2} \ln (\CZ_{1MM}),
\eeq
ainsi que les {\bf fonctions de corr\'elation} \`{a} $k$ points, $W_{k}({\bf x_K})$, comme les fonctions g\'en\'eratrices
des surfaces avec $k$ bords:
\beq
W_{k}({\bf x_K}) := N^{k-2} \left< \prod_{i=1}^{k} \Tr {1 \over x_i-M} \right>_c .
\eeq
On peut \'egalement s\'el\'ectionner le genre $h$ des surfaces g\'en\'er\'ees en \'ecrivant les d\'evelop- pements topologiques
respectifs de ces fonctions:
\beq
\CF_{1MM} = \sum_{h=0}^{\infty} N^{2-2h} F^{(h)}
\;\;\; \hbox{et} \;\;\; W_{k}({\bf x_K}) = \sum_{h=0}^{\infty} N^{2-2h-k} W_{k}^{(h)}.
\eeq

\subsection{Courbe spectrale classique.}

\subsubsection{Equations de boucles.}

Consid\'erons l'\'equation de boucles associ\'ee au changement de variable $\delta M = {1 \over x-M}$. Elle peut s'\'ecrire sous
la forme:
\beq
Y(x)^2 -{1\over N^2} W_{2}(x,x) = {V'(x)^2 \over 4} - {1 \over N} \left< \Tr {V'(x) -V(M) \over x-M} \right>
\eeq
o\`{u} l'on a d\'efini la fonction $Y(x) = {V'(x) \over 2} - {1 \over N} \left< \Tr {1 \over x-M} \right>$.

En oubliant le terme en ${1 \over N^2}$, on construit la courbe spectrale classique associ\'ee au mod\`{e}le
\`{a} une matrice:
\beq\label{E1MM}\encadremath{
\CE_{1MM}(x,y) = D(x)^2 (y^2(x) - {V'(x)^2 \over 4} + P(x) )}
\eeq
o\`{u} $D(x) P(x)$ est un polyn\^{o}me de degr\'e au plus $d-1$. Il est compl\`{e}tement fix\'e par le choix des fractions de
remplissage $\vec{n}$ par la prescription:
\beq
\forall i \, , \; {1 \over 2 i \pi} \oint_{\CA_i} Y^{(0)}(x) dx = {n_i T \over N}
\eeq
pour un choix de base de cycles canonique $\{\CA_i,\CB_i \}$ sur $\CE_{1MM}(x,y)$ et le comportement \`{a} petit $T$:
\beq
Y^{(0)}(x) \sim {V'(x) \over 2} - {1 \over N} \sum_{i=1}^d {n_i \over x - X_i} + O(T^2).
\eeq

\subsubsection{Propri\'et\'es de la courbe spectrale.}

La courbe spectrale $\CE_{1MM}(x,y)$ est une courbe hyperelliptique puisque quadratique en l'une de ses variables. Cette
simple observation a plusieurs implications directes qui permettent de simplifier son \'etude et la description des fonctions
qu'elle engendre.

\begin{figure}
\hspace{5cm} \includegraphics[width=5cm]{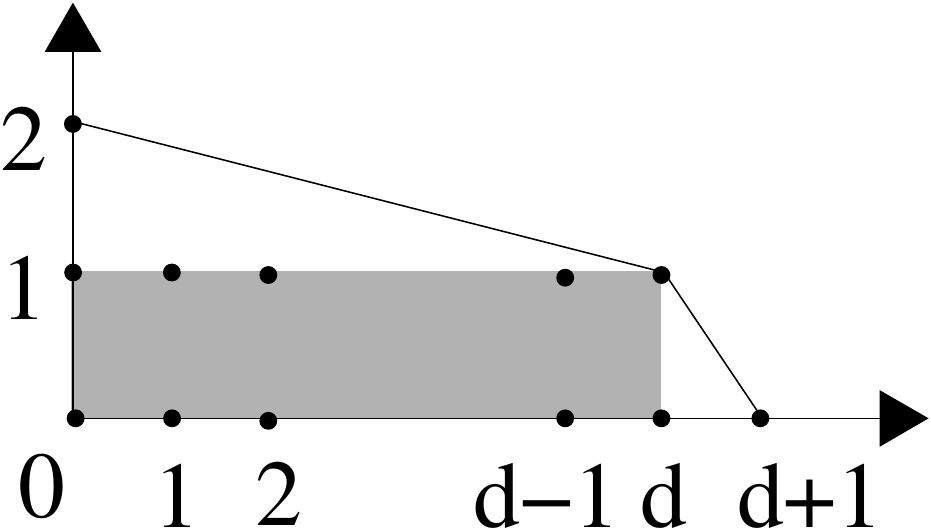}\\
  \caption{Le polytope de la courbe spectrale classique du mod\`ele \`a une matrice.}\label{poly1mm}
\end{figure}

Comme dans le cas du mod\`{e}le \`{a} deux matrices, $\CE_{1MM}$ peut \^etre vue comme une surface de Riemann compacte
$\overline{\Sigma}_{1MM}$ munie de deux fonctions m\'eromorphes $x$ et $y$ de $\overline{\Sigma}_{1MM}$ dans $\mathbb{C}$
telles que:
\beq
\forall \, p \in \overline{\Sigma}_{1MM} , \; \; \CE_{1MM}(x(p),y(p))=0 .
\eeq

La simple lecture du polytope de $\CE_{1MM}$ de la figure \ref{poly1mm} nous permet de voir que le genre maximal de $\CE_{1MM}$ est $d-1$ et qu'il existe deux points \`{a} l'infini sur cette
courbe comme dans le cas du mod\`{e}le \`{a} deux matrices: $\infty_x$ et $\infty_y$. Ces deux points sont caract\'eris\'es
par le comportement de la fonction $y(x)$ en leur voisinage:
\beq
ydx \sim_{\infty_x} V'(x) dx \;\;\; \hbox{et} \;\;\; y \; \hbox{a un p\^ole simple en $\infty_y$} .
\eeq

\begin{figure}
  \hspace{5cm} \includegraphics[width=5cm]{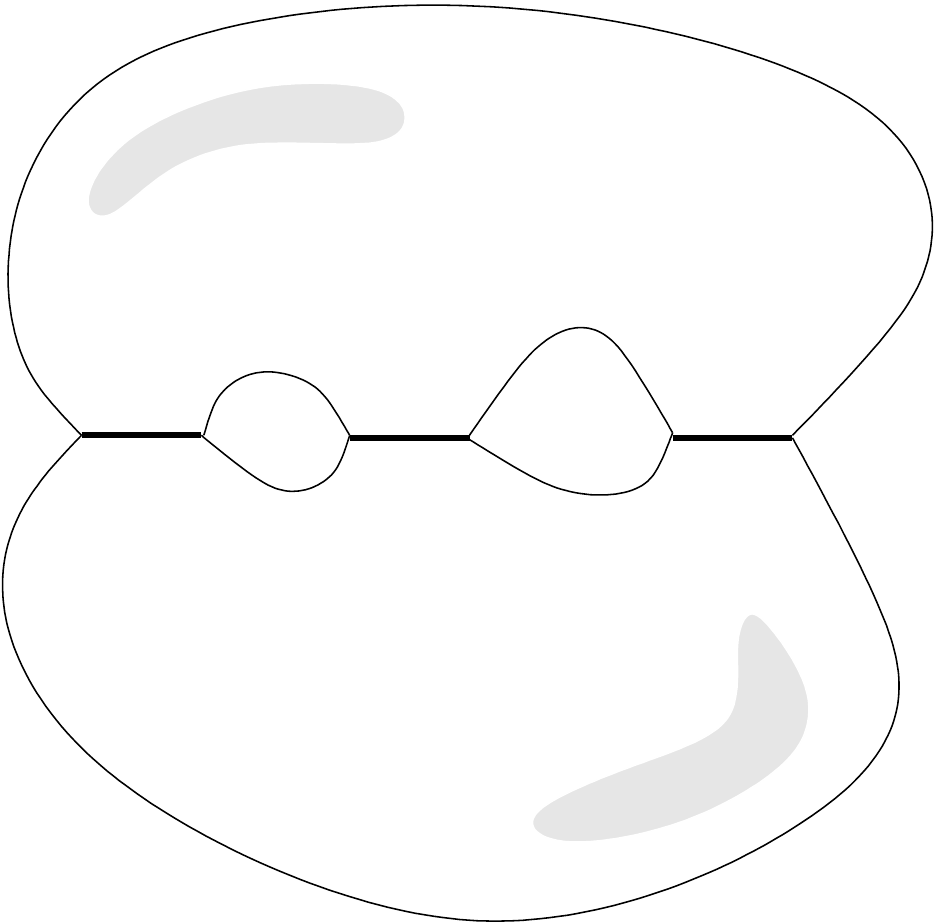}\\
  \caption{Exemple de courbe hyperelliptique avec 6 points de branchement. Elle forme une surface de genre 2.}\label{ellipt}
\end{figure}

La propri\'et\'e d'hyperellipticit\'e de la courbe spectrale en elle m\^{e}me signifie qu'il n'y a que deux feuillets en $x$.
En effet, pour un $x$ g\'en\'erique fix\'e, $\CE_{1MM}(x,y)$ est un polyn\^{o}me de degr\'es 2 en $y$ et l'\'equation $\CE_{1MM}(x,y)=0$
a deux solutions en $y$. Cependant, pour certaines valeurs particuli\`{e}res de $x$, cette \'equation peut avoir un unique
z\'ero double en $y$: les deux feuillets se rejoignent donc en ces points $x(a_i)$, les {\bf points de branchements}.
Ils peuvent \^etre caract\'eris\'es par l'annulation de la diff\'erentielle $dx(a_i)=0$ en leur rel\`evement sur
la surface $\overline{\Sigma}_{1MM}$ par la fonction $x$. La courbe $\CE_{1MM}$ peut donc \^etre vue comme deux copies de la
sph\`{e}re de Riemann coll\'ees selon des coupures joignant les diff\'erents points de branchements (voir figure \ref{ellipt}).

\br
Le nombre de points de branchements est directement li\'e au genre de $\overline{\Sigma}_{1MM}$ par la formule
de Riemann-Hurwitz:
\beq
\hbox{Genre} = - 1 + {\hbox{Nombre de points de branchments}\over 2}.
\eeq
\er

Pour tout point $p$ de $\overline{\Sigma}_{1MM}$, il existe donc un unique autre point $\overline{p}$ avec la m\^{e}me projection
selon $x$:
\beq
\forall p \in \overline{\Sigma}_{1MM} , \; \exists ! \overline{p} \;\; \hbox{tel que} \;\; x(p) = x(\overline{p}).
\eeq
La forme de la courbe $y^2 = f(x)$, permet m\^eme de dire que les valeurs de $y$ dans les deux feuillets sont oppos\'ees
\beq
y(p) = - y(\overline{p}),
\eeq
le choix d'un feuillet consistant \`{a} choisir l'une des deux racine de $f(x)$. Ainsi, les points de branchements co\"{i}ncident
avec les z\'eros de la fonction $y$.

La notation de $\overline{p}$ pour repr\'esenter la deuxi\`{e}me solution de $\CE_{1MM}(x(p),y) = 0$ n'a pas \'et\'e prise au
hasard puisqu'elle fait r\'ef\'erence au point conjugu\'e d\'efini dans la partie \ref{geoalg} du chapitre 2. En effet, dans le cas d'une courbe hyperelliptique,
puisqu'il n'existe que deux feuillets, l'application qui, \`a un point $p$ associe son point conjugu\'e $\overline{p}$,
est globalement d\'efinie et elle coincide avec la d\'efinition donn\'ee ici.

%
%
%
%
%
%
%
%
%
%
%

\subsection{Solution des \'equations de boucles.}
On peut montrer que la m\'ethode utilis\'ee pour r\'esoudre le mod\`{e}le \`{a} deux matrices fonctionne encore pour le mod\`{e}le
\`{a} une matrice et que les d\'eveloppements topologiques des fonctions de corr\'elation et de l'\'energie libre sont donn\'es
par le syst\`{e}me triangulaire de relations de r\'ecurrence:

\bt
Les termes du d\'eveloppement topologique des fonctions de corr\'elation sont r\'ecursivement d\'efinis par
\beq\label{conjecturerecW1MM} \encadremath{
\begin{array}{rcl}
 W_{k+1}^{(h)}(q,p_K)
&=&    \sum_{\alpha} \Res_{p \to \mu_\alpha} {{1\over
2}dE_{p,\pbar}(q)\over (y(p)-y(\pbar))\,dx(p)}\left(
W_{k+1}^{(h-1)}(p,\overline{p},p_K) + \right. \cr && \;\;\; +
\left. \sum_{j,m} W_{j+1}^{(m)}(p,p_J) \,
W_{k+1-j}^{(h-m)}(\overline{p},p_{K-J}) \right) , \cr
\end{array}}\eeq
et ceux de l'\'energie libre par
\beq
\forall \, h >1 , \; (2h-2) F^{(h)} = H_x. W_1^{(h)},
\eeq
avec l'op\'erateur $H_x$ d\'efini par son action sur une forme diff\'erentielle $\phi$:
\beq
H_x. \phi :=
\Res_{\infty_x} V_1(x)\,\phi +\Res_{\infty_y} xy\,\phi +
\int_{\infty_x}^{\infty_y} \phi + \sum_i \epsilon_i \oint_{{\cal
B}_i} \phi.
\eeq
\et

On peut noter une l\'eg\`{e}re simplification de ces r\`{e}gles gr\^ace \`{a} l'hyperellipticit\'e de la courbe qui assure
une d\'efinition globale de $\overline{p}$ satisfaisant:
\beq\encadremath{
y(p)-y(\overline{p}) = 2 y(p).}
\eeq

Les premiers termes du d\'eveloppement topologique de l'\'energie libre sont connus depuis longtemps (voir par exemple \cite{Chekh,EKK,MarcoF}) et s'\'ecrivent:
\bt
\beq
F^{(0)} = -{1 \over 2} H_x. ydx
\eeq
et
\beq
F^{(1)} = {- 1 \over 2} \ln (\tau_{Bx}) - {1 \over 24} \ln \left(\prod_i {dy\over dz_i}(a_i) \right)
\eeq
o\`u $\tau_{Bx}$ est la fonction $\tau$ de Bergmann associ\'ee \`a $x$ d\'efinie par \eq{deftauBx} du chapitre 2.
\et

\subsection{Lien avec le mod\`{e}le \`{a} deux matrices.}\label{partlien2MM}

Revenons au mod\`{e}le \`{a} deux matrices d\'efini dans le chapitre pr\'ec\'edent en consid\'erant le cas particulier
o\`{u} l'un des potentiels est quadratique, par exemple:
\beq
V_2(y):= y^2 .
\eeq
On peut alors r\'e\'ecrire l'int\'egrale formelle \eq{defZ} du chapitre 2 sous la forme:
\bea
\CZ &=&
\int dM_1 e^{-{1 \over \hbar} \Tr V_1(M_1)} \int dM_2 e^{-{1 \over \hbar} \Tr \left[ \left(M_2-{M_1 \over 2} \right)^2 - {M_1^2 \over 4} \right]} \cr
&=& \int dM_1 e^{-{1 \over \hbar} \Tr \left( V_1(M_1) - {M_1^2 \over 4} \right) } \int dM e^{-{1 \over \hbar} \Tr M^2 } \cr
& \propto & \int dM_1 e^{-{1 \over \hbar} \Tr \left( V_1(M_1) - {M_1^2 \over 4} \right)} \cr
\eea
o\`{u} l'on est pass\'e de la premi\`ere ligne \`{a} la deuxi\`{e}me par un simple changement de variables dans la seconde int\'egrale
et le coefficient de proportionnalit\'e de la derni\`{e}re ligne est simplement l'int\'egrale gaussienne sur $M$ de la seconde ligne.

On voit ainsi que le mod\`{e}le \`{a} une matrice est \'equivalent \`{a} un mod\`{e}le \`{a} deux matrices dont l'un
des potentiels est gaussien. Il est facile de v\'erifier que tous les \'el\'ements d\'ecrits dans cette partie \`{a} propos du
mod\`{e}le \`{a} une matrice co\"{i}ncident bien avec les r\'esultats d'un mod\`{e}le \`{a} deux matrices avec potentiel quadratique.
On peut par exemple v\'erifier que la courbe spectrale classique du mod\`{e}le \`{a} deux matrice se r\'eduit \`{a} celle
du mod\`{e}le \`{a} une matrice si $V_2$ est quadratique. Toutes les autres propri\'et\'es de cette partie peuvent en \^etre
d\'eriv\'ees.

\section{Mod\`ele de matrices en champ ext\'erieur.}\label{partMext}

\subsection{D\'efinition et interpr\'etation combinatoire.}

Dans cette partie, nous abordons le mod\`{e}le \`{a} une matrice en champ ext\'erieur d\'efini par l'int\'egrale formelle \cite{PZinnmatext}:
\beq
Z_{\rm Mext}(\hat\L) = \int_{H_n} dM \, e^{-N
Tr(V(M) - M\widehat{\L} )}
\eeq
o\`{u} ${\displaystyle V'(x) = {\sum_{k=0}^{d} g_k x^k \over D(x)}}$ est une fraction rationnelle de d\'enominateur $D(x)$, et $\hat\L$ est une matrice ext\'erieure de taille $N \times N$ suppos\'ee
diagonale:
\beq
\widehat{\L}={\rm diag}\,(\,\mathop{\overbrace{\widehat{\lambda}_1,\dots,\widehat{\lambda}_1}}^{n_1},\mathop{\overbrace{\widehat{\lambda}_2,\dots,\widehat{\lambda}_2}}^{n_2},\dots,\mathop{\overbrace{\widehat{\lambda}_s,\dots,\widehat{\lambda}_s}}^{n_s}\,).
\eeq

Ce mod\`{e}le a une importance particuli\`{e}re du point de vue math\'ematique. En effet, le cas particulier o\`u $V(x) = x^3$
et $\widehat{\L} = \L^2$ a \'et\'e introduit par Kontsevich \cite{kontsevich} dans un contexte tout \`a fait diff\'erent pour calculer
les nombres d'intersections de surfaces de Riemann. Ce rapprochement entre deux probl\`emes \`a priori tr\`es \'eloign\'es
est principalement due au fait que l'on peut construire une bijection entre l'ensemble des classes de Chern des surfaces de Riemann
et les cartes avec des bords marqu\'es et que l'int\'egrale de Kontsevich est une r\'ealisation explicite de la fonction $\tau$ de
la hierarchie int\'egrable de $KdV$ \cite{kontsevich}.

De plus, ce mod\`ele a \'et\'e li\'e au pr\'ec\'edent. En effet, en se basant sur une comparaison des \'equations
de boucles, il a \'et\'e montr\'e que l'on pouvait obtenir cette int\'egrale comme le carr\'e d'une double
limite d'\'echelle d'un mod\`ele \`a une matrice \cite{MMM}, avant de montrer qu'elle
peut \^etre obtenue \'egalement directement comme le r\'esultat d'une autre double limite d'\'echelle
du mod\`ele \`a une matrice \cite{AK}.

En fait, en consid\'erant un mod\`ele \`a une matrice en champ ext\'erieur, on pourra avoir acc\`es aux fonctions $\tau$
de n'importe quelle r\'eduction de la hierarchie $KP$ et l'on ne sera pas limit\'e \`a $KdV$. Ceci semble indiquer que
le r\'esultat des doubles limites d'\'echelles du mod\`ele \`a deux matrices est contenu dans l'ensemble des mod\`eles
\`a une matrice en champ ext\'erieur.

Dans cette partie, nous montrons comment on peut appliquer exactement la m\^eme m\'ethode que pr\'ec\'edemment pour r\'esoudre
ce mod\`ele.

\subsection{R\'esolution du mod\`{e}le.}

Comme dans les mod\`{e}les \`{a} une et deux matrices, on d\'efinit de mani\`{e}re analogue les d\'eveloppements topologiques
de l'\'energie libre et des fonctions de corr\'elation par:
\beq
\overline{w}_{k}(x_1,\dots,x_k) := N^{k-2}\left< \prod_{i=1}^k \tr{1\over x_{i}-M}\right>_c + {\delta_{k,2}  \over (x_1-x_2)^2},
\eeq
\beq\label{GKdefwklh}
\overline{w}_{k}(\bfx_K) = \sum_{h=0}^\infty {1\over N^{2h}}\,\overline{w}_{k}^{(h)}(\bfx_K)
\eeq
et
\beq\label{FMext}
\CF_{\rm Mext} := -{1 \over N^2} \ln Z_{\rm Mext} = \sum_{h=0}^\infty N^{2-2h} F_{\rm Mext}^{(h)}.
\eeq

Il est aussi important d'introduire le polyn\^ome minimal de $\hat\L$:
\beq
S(y):=\prod_{i=1}^s (y-\widehat{\lambda}_i),
\eeq
ainsi que les deux polyn\^omes
\beq\label{GKdefubarkl}
u_{k}(x,y;\bfx_K) := N^{k-1} \left< \tr {1\over x-M} {S(y)-S(\L)\over y-\L}\,\,
\prod_{r=1}^{k} \tr{1\over x_{i_r}-M}\right>_c -\delta_{k,0}S(y)
\eeq
et
\beq\label{GKdefpkl}
P_{k}(x,y;\bfx_K) := N^{k-1} \left< \tr {V'(x)-V'(M_1)\over x-M} {S(y)-S(\L)\over y-\L}\,\,
\prod_{r=1}^{k} \tr{1\over x_{i_r}-M}\right>_c.
\eeq
On voit que $u_{k}(x,y;\bfx_K)$ est un polynome en $y$ de degr\'e $s-1$ et que
$D(x) P_{k}(x,y;\bfx_K)$ est un polyn\^ome en $x$ de degr\'e $d-1$ et en $y$ de degr\'e $s-1$.

\subsubsection{Equation de boucles ma\^itresse et courbe spectrale classique.}

Le changement de variable
\beq
\delta M = {1\over x-M}{S(y)-S(\L)\over y-\L}
\eeq
donne l'\'equation de boucles
\beq\label{ext1}
(y+\overline{w}_1(x)-V'(x))({u}_0(x,y)-S(y))+{1\over N^2}{u}_1(x,y;x)
= (V'(x)-y)S(y) - P_0(x,y).
\eeq

On d\'efinit la {\bf courbe spectrale} du mod\`{e}le de matrice en champ ext\'erieur par
\beq\label{EMext}\encadremath{
E_{\rm Mext}(x,y):= \left((V'(x)-y)S(y) - P_0(x,y)\right) D(x)}
\eeq
et la fonction
\beq
Y(x):=V'(x)-\overline{w}_1(x).
\eeq
L'\'equation de boucles \eq{ext1} s'\'ecrit alors
\beq\label{GKloop1}
E_{\rm Mext}(x,Y(x))={1\over N^2}u_1(x,Y(x);x)D(x)
\eeq
et devient une \'equation alg\'ebrique \`{a} l'ordre dominant dans le d\'eveloppement en ${1 \over N^2}$:
\beq
E_{\rm Mext}^{(0)}(x,Y(x))
= 0
\eeq
d\'efinissant ainsi la {\bf courbe spectrale classique} de ce mod\`ele.
\vs

{\bf Propri\'et\'es de la courbe spectrale classique.}

Etudions les propri\'et\'es de la courbe alg\'ebrique $\curve_{\rm Mext}(x,y)=E_{\rm Mext}^{(0)}(x,y)=0$.

Puisque $y$ est solution d'une \'equation de degr\'e $s+1$ \`{a} $x$ fix\'e, la courbe $\curve_{\rm Mext}(x,y)$  pr\'esente $s+1$ feuillets en $x$.
Ceux-ci peuvent \^etre discrimin\'es par le comportement de $x(p)$ lorsque $p$ approche un p\^ole:

$\bullet$ dans le feuillet physique, $y(p)\sim V'(x(p))-1/x+O(1/x^2(p))$ quand $x(p) \to \infty$;

$\bullet$ Dans les autres feuillets, $y(p)\sim \widehat{\lambda}_i + {n_i\over N}\,{1\over x(p)}+O(1/x^2(p))$.

\vs

La courbe $\curve_{\rm Mext}$ est de genre maximal $g \leq s-1$ et on d\'efinit les fractions de remplissages comme
des param\`{e}tres du mod\`{e}le:
\beq
\epsilon_i:={1\over 2i\pi}\oint_{{\cal A}_i} y dx.
\eeq

\subsubsection{Equations de boucles et d\'eveloppement topologique.}

Comme pr\'ec\'edemmment, on pr\'ef\`ere travailler avec des fonctions de corr\'elations d\'efinies comme des diff\'erentielles
sur la courbe spectrale classique plut\^ot qu'avec des fonctions multivalu\'ees dans le plan complexe. On d\'efinit donc:
\beq
W_k({\bf p_K}) :=\overline{w}_k({\bf x(p_K)})\, dx(p_1)\dots dx(p_k)
\eeq
et
\beq
U_k(p,y;{\bf p_K}) :=u_k(x(p),y;{\bf x(p_K)})\, dx(p)dx(p_1)\dots dx(p_k)
\eeq
de m\^eme que leur d\'eveloppements toplogiques:
\beq
W_k({\bf p_K}) = \sum_{h=0}^\infty N^{2-2h}\,W_k^{(h)}({\bf p_K})
\;\;\; \hbox{and} \;\;\;
U_k(p,y;{\bf p_K}) = \sum_{h=0}^\infty N^{2-2h}\,U_k^{(h)}(p,y;{\bf p_K}).
\eeq

Le d\'eveloppement topologique de \eq{GKloop1} donne pour $h>1$:
\bea\label{GKmastloop2} E^{(h)}(x,y) &=&
D(x) (y-Y(x)) u_0^{(h)}(x,y) + D(x) w_{1,0}^{(h)}(x) u_0^{(0)}(x,y) \cr && +
D(x) \sum_{m=1}^{h-1} w_{1,0}^{(m)}(x) u_0^{(h-m)}(x,y) + D(x) u_1^{(h-1)}(x,y;x), \cr \eea
o\`{u} $E^{(h)}(x,y)$ est le d\'eveloppement topologique du polyn\^ome $E_{\rm Mext}$.

Cette \'equation est suffisante pour d\'emontrer le th\'eor\`eme:
\bt
\beq \label{GKdefE}
\begin{array}{l}
E_{\rm Mext}(x,y) = \cr
-D(x)``\left<\prod_{i=0}^s (y-V'(x(p))+{1 \over N} \Tr{1 \over x(p^{(i)})-M}) \right>'' = D(x)\left[(V'(x)-y)S(y) - P_0(x,y)\right]\cr
\end{array}
\eeq
et
\beq
U_0(p,y) = - ``\left<\prod_{i=1}^s (y-V'(x(p))+{1 \over N} \Tr{1 \over x(p^{(i)})-M}) \right>''
\eeq
o\`{u} $``<.>''$ est d\'efini comme dans le th\'eor\`{e}me \ref{completecurve}.
\et

\proof{La preuve est compl\`etement similaire \`{a} celle du th\'eor\`{e}me \ref{completecurve} du chapitre 2.}

Comme dans les cas pr\'ec\'edents, cette propri\'et\'e nous permet de montrer que les \'equations de boucles sont r\'esolues par
le syst\`eme triangulaire de d\'efinitions:
\beq\label{GKconjecturerecW} \encadremath{
\begin{array}{rcl}
 W_{k+1}^{(h)}(q,p_K)
&=&   \sum_{\alpha} \Res_{p \to \mu_\alpha} {{1\over
2}dE_{p,\pbar}(q)\over (y(p)-y(\pbar))\,dx(p)}\left(
W_{k+1}^{(h-1)}(p,\overline{p},p_K) + \right. \cr && \;\;\; +
\left. \sum_{j,m} W_{j+1}^{(m)}(p,p_J) \,
W_{k+1-j}^{(h-m)}(\overline{p},p_{K-J}) \right) , \cr
\end{array}}\eeq
avec la donn\'ee initiale de la fonction \`{a} deux points:
\beq
W_2^{(0)}(p_1,p_2) = \underline{B}(p_1,p_2),
\eeq
o\`{u} les notations sont les m\^emes que dans les parties pr\'ec\'edentes pour des objets d\'efinis cette fois ci sur la courbe
alg\'ebrique $\curve_{\rm Mext}$.

On obtient \'egalement ainsi le d\'eveloppement topologique de l'\'energie libre:
\beq\encadremath{
\forall h >1 \, , \; F_{Mext}^{(h)} = {1 \over 2 - 2h} \Res_{q \to {\bf a}} \phi(q) W_1^{(h)}(q),
}\eeq
o\`{u} $\phi(q)$ est n'importe quelle primitive de $ydx(q)$.

\section{Conclusion du chapitre.}

Nous avons montr\'e dans ce chapitre comment il est possible d'appliquer la m\^eme proc\'edure pour r\'esoudre le mod\`ele
\`a une matrice et le mod\`ele en champ ext\'erieur que pour r\'esoudre le mod\`ele \`a deux matrices. Pour cela il suffit
de {\bf garder les m\^emes r\'esultats en changeant seulement la courbe spectrale}. On a donc obtenu une formule g\'en\'erale
unique pour r\'esoudre ces diff\'erents mod\`eles de matrices formels.

Une premi\`ere question se pose: \`a quelle cat\'egorie de mod\`eles de matrices peut-on appliquer la m\^eme proc\'edure?
Pour le moment, la r\'eponse \`a cette question n'est pas connue en g\'en\'eral m\^eme si l'on conna\^it des mod\`eles
pour lesquels cette m\'ethode semble fonctionner comme le mod\`ele \`a plusieurs matrices coupl\'ees en chaine.

Une autre question se pose alors: il para\^it \'evident que l'on peut g\'en\'eraliser cette proc\'edure au del\`a des mod\`eles
de matrices. A toute courbe alg\'ebrique, on peut associer des familles infinies d'\'energies libres et de fonctions de corr\'elation
par la m\^eme m\'ethode. Que repr\'esentent alors ces objets lorsque la courbe alg\'ebrique n'est pas issue d'un mod\`ele
de matrice? Ont-ils des propri\'et\'es particuli\`eres? Nous tenterons de r\'epondre autant que possible
\`a ces questions dans le chapitre suivant.


\chapter{Invariants alg\'ebriques.}

Dans les chapitres pr\'ec\'edents, nous avons \'etudi\'e diff\'erents mod\`{e}les de matrices hermitiennes. Nous avons ainsi pu
calculer les d\'eveloppements topologiques de l'\'energie libre et des fonctions de corr\'elation respectifs de ces mod\`{e}les.
Etant donn\'e un mod\`{e}le, il \'etait suppos\'e depuis longtemps que les coefficients de ces d\'eveloppements
pouvaient s'exprimer uniquement en termes des modules d'une courbe alg\'ebrique: la courbe spectrale classique associ\'ee au
mod\`{e}le consid\'er\'e. Cependant il est plus surprenant que les r\'esultats pr\'esent\'es pr\'ec\'edemment montrent
qu'il existe une formule unique pour r\'esoudre ces diff\'erents mod\`{e}les en changeant simplement la courbe alg\'ebrique sur
laquelle travailler: c'est-\`{a}-dire que l'on est capable de d\'efinir des fonctions $F^{(g)}(\CE_{\rm Modele})$ qui coincident
avec le d\'eveloppement de 't Hooft de l'\'energie libre $F^{(g)}_{\rm Modele}$ du mod\`{e}le dont la courbe spectrale est
$\CE_{\rm Modele}$.

Il est alors naturel de se poser la question d'une g\'en\'eralisation de cette proc\'edure au del\`{a} des mod\`{e}les de matrices.
Etant donn\'ee une courbe alg\'ebrique $\CE$ quelconque, nous d\'efinissons dans ce chapitre des nombres $F^{(g)}(\CE)$ par la m\^eme proc\'edure
que pour le calul de l'\'energie libre des mod\`eles de matrices et \'etudions leurs propri\'et\'es. Tous les r\'esultats
pr\'esent\'es ici peuvent \^etre trouv\'es dans \EOinvariants avec les d\'emonstrations associ\'ees.

Dans un premier temps, nous allons bri\`evement rappeler les diff\'erents param\`{e}tres d\'efinissant une courbe alg\'ebrique $\CE$,
c'est-\`{a}-dire les modules de notre probl\`eme. Nous d\'efinirons ensuite une famille de formes diff\'erentielles
sur la courbe $\CE$ en imitant la r\'esolution des \'equations de boucles des mod\`{e}les de matrices hermitiennes.
Nous nous attacherons ensuite \`{a} l'\'etude des variations de ces formes diff\'erentielles sous les d\'eformations de la
courbe alg\'ebrique. Nous montrerons enfin comment la connaissance de ces lois de d\'eformation ainsi que des propri\'et\'es
d'invariance en d\'ecoulant, sont un outil tr\`{e}s puissant, en particulier pour d\'emontrer l'\'equivalence de diff\'erents mod\`{e}les.

\section{D\'efinition des modules de la courbe.}

Consid\'erons une {\bf courbe alg\'ebrique quelconque} d\'efinie comme dans la partie \ref{geoalg} du chapitre 2\footnote{Dans tout ce chapitre,
nous reprendrons les notations de la partie \ref{geoalg} du chapitre 2 sauf indication contraire. On pourra s'y reporter directement sans avoir lu au pr\'ealable le reste du chapitre
ayant trait au mod\`ele \`a deux matrices.} par l'\'equation:
\beq
\CE(x,y) = 0
\eeq
et qui peut \^etre vue comme une surface de Riemann compacte $\Sigma$ de genre donn\'e $\CG$ muni de deux fonctions m\'eromorphes
$x,y$ telles que:
\beq
\forall p \in \Sigma \, , \; \CE(x(p),y(p)) = 0
\eeq
ainsi que d'une base symplectique de cycles $\{(\underline\CA_i, \underline\CB_i)\}_{i = 1}^{\CG}$.

Ceci fait, les p\^oles $\alpha_i$ de la forme diff\'erentielle $ydx$ et les points de branchements $a_i$ annulant
$dx(a_i) = 0$ sont fix\'es.
On peut alors d\'ecrire les modules de la courbe:
\begin{itemize}

\item Les fractions de remplissage:
\beq
\underline{\epsilon}_i := {1 \over 2 i \pi} \oint_{\underline\CA_i} ydx;
\eeq

\item Les temp\'eratures:
\beq
t_{0,i}:= \Res_{\alpha_i} ydx;
\eeq

\item Les autres modules aux p\^oles: $t_{j,i}$ d\'efinis par les polyn\^omes
\beq
V_i(z_{\alpha_i}(p)):=\sum_j t_{j,i} z_{\alpha_i}^j(p) := \Res_{q \to \alpha_i} y(q) dx(q) \ln \left( 1 - {z_{\alpha_i} \over z_{\alpha_i}(q)} \right)
\eeq
o\`{u} ${1 \over z_{\alpha_i}}$ est une bonne variable locale au voisinage du p\^ole $\alpha_i$ ind\'ependante de $y$.

\end{itemize}

\section{D\'eformation par $\kappa$.}

En plus de ces modules propres \`{a} la courbe alg\'ebrique elle-m\^eme, on introduit un param\`{e}tre suppl\'ementaire
pemettant de changer la normalisation des fonctions de base.

Pour une {\bf matrice sym\'etrique quelconque} $\kappa$, on d\'eforme les briques de bases de la construction des fonctions de
corr\'elation du mod\`{e}le \`{a} deux matrices en d\'efinissant:

\bd
Pour toute matrice sym\'etrique $\kappa$, on d\'efinit le noyau de Bergmann $\kappa$-d\'eform\'e:
\beq
B(p,q)[\kappa] := \underline{B}(p,q) + 2 i \pi \sum_{i,j} du_i(p) \kappa_{ij} du_j(q)
\eeq
ainsi que la d\'eform\'ee de la diff\'erentielle Abelienne de troisi\`{e}me esp\`{e}ce:
\beq
{dS}_{q_1,q_2}(p) = \int_{q_2}^{q_1} B(p,q)
\eeq
o\`{u} le chemin d'int\'egration se situe enti\`{e}rement dans le domaine fondamental, et, au voisinage d'un point de branchement,
\beq
dE_q(p)[\kappa] := \underline{dE}_q(p) + 2 i \pi \sum_{i,j} du_i(p) \kappa_{ij} (u_j(\overline{q})-u_j(q)).
\eeq
\ed

On d\'efinit \'egalement un ensemble de cycles d\'ependant de $\kappa$ m\'elangeant les \'el\'ements de la base
$\{(\underline{\CA}_i,\underline{\CB}_i)\}$:
\bd
Pour une matrice sym\'etrique $\kappa$ quelconque, on d\'efinit l'ensemble de cycles $\kappa$-d\'eform\'es $\{(\CA_i,\CB_i)\}_{i=^1}^\CG$ par:
\beq
\left\{\begin{array}{l}
\CA := (1 + \kappa \tau) \underline{\CA} - \kappa \underline{\CB} \cr
\CB := \underline{\CB} - \tau \underline{\CA} \cr
\end{array} \right.
\eeq
o\`{u} l'on a utilis\'e une notation vectorielle pour d\'ecrire les cycles: $\CA:= (\CA_i)$.
\ed

Ceci nous permet de d\'efinir des fractions de remplissage associ\'ees \`a chaque valeur de $\kappa$:
\bd
On d\'efinit les fractions de remplissage $\kappa$ d\'eform\'ees par:
\beq
\epsilon_i:= {1 \over 2 i \pi} \oint_{\CA_i} ydx .
\eeq
\ed

Ces cycles permettent \'egalement de mieux d\'ecrire les propri\'et\'es de normalisation des diff\'erentielles $\kappa$-d\'eform\'ees.
Ces diff\'erentielles satisfont les propri\'et\'es suivantes:

\begin{itemize}
\item Sym\'etrie et normalistion du noyau de Bergmann d\'eform\'e:
\beq
B(p,q)[\kappa]=B(q,p)[\kappa]
\virg
\oint_{\acycle_i} B[\kappa] =0
\virg
\oint_{q\in\bcycle_i} B(p,q)[\kappa] = 2i\pi\, du_i(p).
\eeq

\item Formule de Cauchy:
pour toute fonction m\'eromorphe $f(p)$, sa diff\'erentielle est donn\'ee par
\beq
df(p) = \Res_{q\to p} B(p,q) f(q)
.
\eeq

\item $B(p,q)[0] = \underline{B}(p,q)$.

\item Structure de p\^oles et normalisation de $dS_{q_1,q_2}$: C'est l'unique forme m\'eromorphe
avec seulement deux p\^oles simples en $q_1$ et $q_2$, telle que
\beq
\Res_{q_1} dS_{q_1,q_2} = 1 = - \Res_{q_2} dS_{q_1,q_2}
\virg
\oint_{\acycle_i} dS_{q_1,q_2} = 0
\eeq
et elle satisfait:
\beq
\oint_{\bcycle_i} dS_{q_1,q_2} = 2i\pi (u_i(q_1)-u_i(q_2)).
\eeq

\item Sym\'etrie de $dS$:
\beq
dS_{q_1,q_2} = - dS_{q_2,q_1},
\eeq

\item Expression de $dS$ en termes des fonctions $\theta$:
\beq
dS_{q_1,q_2}(p) = d_p \ln{\left(\theta_{\bf z}({\bf u}(p)-{\bf u}(q_1))\over \theta_{\bf z}({\bf u}(p)-{\bf u}(q_2))\right)} + 2i\pi\, \sum_{i,j}\, du_i(p) \kappa_{ij} (u_j(q_1)-u_j(q_2)).
\eeq

\item D\'erivation et int\'egration de $dS$:
\beq
d_{q_1} \left(dS_{q_1,q_2}(p)\right) =  B(q_1,p)
\eeq
et
\beq
\int_{p_1}^{p_2} dS_{q_1,q_2} = \int_{q_1}^{q_2} dS_{p_1,p_2}.
\eeq

\item
Formule de Cauchy: pour toute fonction m\'eromorphe $f$:
\beq
f(p) = - \Res_{q_1\to p} dS_{q_1,q_2}(p) f(q_1).
\eeq

\item
L'identit\'e bilin\'eaire de Riemann reste v\'erifi\'ee avec ces nouveaux cycles d\'eform\'es.

Consid\'erant $\om_1$ et $\om_2$ deux formes m\'eromorphes sur la surface $\Sigma$
et un point $p_0$ arbitraire,on d\'efinit $\Phi_1$ sur le domaine fondamental par
\beq
\Phi_1(p) = \int_{p_0}^p \om_1
\eeq
o\`u le chemin d'int\'egration reste \`a l'int\'erieur du domaine fondamental.

On a alors l'{\bf identit\'e bilin\'eaire de Riemann}:
\bea\label{Riemannbilinear2}
\Res_{p\to {\rm all\, poles}} \Phi_1(p)\om_2(p) &=& {1\over 2i\pi}\, \sum_{i=1}^g \oint_{\underline\acycle_i} \om_1 \oint_{\underline\bcycle_i} \om_2 - \oint_{\underline\bcycle_i} \om_1 \oint_{\underline\acycle_i} \om_2 \cr
&=& {1\over 2i\pi}\, \sum_{i=1}^g \oint_{\acycle_i} \om_1 \oint_{\bcycle_i} \om_2 - \oint_{\bcycle_i} \om_1 \oint_{\acycle_i} \om_2 .
\eea

\item
On peut \'egalement d\'ecomposer $ydx$ (comme dans \eq{decompoydx2MM} du chapitre 2) sur les modules en tenant compte des $\kappa$ d\'eformations:
\beq\label{decompoydx}
ydx = \sum_{i,j} t_{j,i} B_{j,i} + \sum_i t_{0,i} dS_{\alpha_i,o} + 2 i \pi  \sum_i \epsilon_i du_i
\eeq
avec
\beq
B_{j,i}(p) = - \Res_{q \to \alpha_i} \underline{B}(p,q) z_{\alpha_i}(q)^j.
\eeq

\end{itemize}

\section{Fonctions de corr\'elation et \'energies libres.}

\subsection{D\'efinitions.}
Dans les chapitres pr\'ec\'edents, on a vu comment, \'etant donn\'ee la courbe spectrale classique $\CE_{MM}$ d'un mod\`{e}le de matrices
al\'eatoires, on peut calculer les d\'eveloppements topologiques de ses fonctions de corr\'elations $W_{k,MM}$ et son \'energie libre $F_{MM}$
gr\^{a}ce \`{a} un ensemble d'applications $F^{(g)}$ (resp. $W_k^{(g)}$) allant de l'ensemble des courbes alg\'ebriques $\CE$
dans l'ensemble des nombres complexes (resp. des $k$-formes sur $\CE$):
\beq
W_{k,MM} = \sum_{g=0}^{\infty} N^{-2g} W_{k}^{(g)}(\CE_{MM}) \virg F_{MM}:= \sum_{g=0}^{\infty} N^{-2g} F^{(g)}(\CE_{MM}).
\eeq
Si ces applications donnent des r\'esultats convaincants lorsqu'elles sont utilis\'ees dans le cadre des mod\`{e}les de matrices, il
est int\'eressant de les \'etendre au del\`a de ces derniers. Dans cette partie nous allons donc \'etudier leurs propri\'et\'es
en toute g\'en\'eralit\'e. Nous allons commencer par d\'efinir ces objets {\bf sans supposer l'existence d'un quelconque mod\`{e}le
de matrices sous jacent}.

Par imitation du d\'eveloppement topologique des fonctions de corr\'elation et de l'\'energie libre du mod\`{e}le \`{a}
deux matrices, on d\'efinit ainsi:\bd\label{defloopfctions}
{\bf Fonctions de corr\'elation:}
\beq
W_k^{(g)}=0 \quad {\rm if}\,\, g<0
\eeq
\beq
W_1^{(0)}(p) = 0
\eeq
\beq
W_2^{(0)}(p_1,p_2) = B(p_1,p_2)
\eeq
et on d\'efinit par r\'ecurence sur $k$ et $g$ les formes multilin\'eaires m\'eromorphes:
\beq\label{defWkgrecursive}
\encadremath{
\begin{array}{l}
W_{k+1}^{(g)}(p,{\bf p_K}) = \cr
= {\displaystyle \sum_i \Res_{q\to a_i}} {dE_{q}(p)\over \om(q)}\,\Big(
{\displaystyle \sum_{m=0}^g} \sum_{J\subset K} W_{j+1}^{(m)}(q,{\bf p_J})W_{k-j+1}^{(g-m)}(\qbar,{\bf p_{K/J}})
+ W_{k+2}^{(g-1)}(q,\qbar,{\bf p_K}) \Big).\cr
\end{array}}
\eeq

{\bf Energies libres:}

\beq \encadremath{
\forall g>1 \, , \; F^{(g)} = {1\over 2-2g}\,\sum_i \Res_{q\to a_i} \Phi(q) W_{1}^{(g)}(q)}
\eeq
o\`u $\Phi$ est une primitive quelconque de $ydx$,
et pour $g=1$
\beq\encadremath{
F^{(1)}=-{1\over 2}\ln{(\tau_{Bx})}\,\,-{1\over 24}\ln{\left(\prod_i y'(a_i) \right)}+ \ln\left(\det\left({1 \over \kappa}\right)\right)}
\eeq
o\`{u}
\beq
y'(a_i) = {dy(a_i)\over dz_i(a_i)}
\virg
z_i(p) = \sqrt{x(p)-x(a_i)}
\eeq
et $\tau_{Bx}$ fonction $\tau$ de Bergmann d\'efinie par \eq{deftauBx}.

L'ordre dominant est quant \`{a} lui d\'efini par:
\beq\encadremath{
F^{(0)}
= {1\over 2}\sum_{i} \Res_{\alpha_i} V_i ydx + {1\over 2} \sum_{i} t_{0,i} \mu_{\alpha_i}
- {1\over 4i\pi}\sum_i \oint_{\acycle_i} ydx \oint_{\bcycle_i} ydx}
\eeq
o\`{u}
\beq
\mu_{\alpha_i}
= \int_{\alpha_i}^{o} (ydx - dV_i + t_{0,i} {dz_{\alpha_i}\over z_{\alpha_i}}) + V_i(o) - t_{0,i} \ln{(z_{\alpha_i}(o))}.
\eeq
\ed

Dans la suite, on notera:
\beq
\forall i=1 \dots \CG \, , \; \Gamma_i  := \oint_{\bcycle_i} ydx.
\eeq

On d\'efinit \'egalement les {\bf fonctions sp\'eciales} correspondant au cas o\`u $\kappa \to 0$:
\bd
{\bf Fonctions de corr\'elation sp\'eciales:}
\beq
\underline{W}_k^{(g)}=0 \quad {\rm if}\,\, g<0
\eeq
\beq
\underline{W}_1^{(0)}(p) = 0
\eeq
\beq
\underline{W}_2^{(0)}(p_1,p_2) = \underline{B}(p_1,p_2)
\eeq
o\`u $\underline{B}$ est le noyau de Bergmann non d\'eform\'e,
et on d\'efinit par r\'ecurrence sur $k$ et $g$ les fomes multilin\'eaires m\'eromorphes:
\beq\label{defWkgrecursivespecial}
\encadremath{
\begin{array}{l}
\underline{W}_{k+1}^{(g)}(p,{\bf p_K}) = \cr
= {\displaystyle \sum_i \Res_{q\to a_i}} {\underline{dE}_{q}(p)\over \om(q)}\,\Big(
{\displaystyle \sum_{m=0}^g} \sum_{J\subset K} \underline{W}_{j+1}^{(m)}(q,{\bf p_J})\underline{W}_{k-j+1}^{(g-m)}(\qbar,{\bf p_{K/J}})
+ \underline{W}_{k+2}^{(g-1)}(q,\qbar,{\bf p_K}) \Big).\cr
\end{array}}
\eeq

{\bf Energies libres sp\'eciales:}

\beq \encadremath{
\forall g>1 \, , \; \underline{F}^{(g)} = {1\over 2-2g}\,\sum_i \Res_{q\to a_i} \Phi(q) \underline{W}_{1}^{(g)}(q)}
\eeq
et, pour $g=1$,
\beq\encadremath{
\underline{F}^{(1)}=-{1\over 2}\ln{(\tau_{Bx})}\,\,-{1\over 24}\ln{\left(\prod_i y'(a_i) \right)}}.
\eeq
L'ordre dominant est quant \`{a} lui d\'efini par:
\beq\encadremath{
F^{(0)}
= {1\over 2}\sum_{i} \Res_{\alpha_i} V_i ydx + {1\over 2} \sum_{i} t_{0,i} \mu_{\alpha_i}
- {1\over 4i\pi}\sum_i \oint_{\underline\acycle_i} ydx \oint_{\underline\bcycle_i} ydx.}
\eeq
\ed

\subsection{Repr\'esentation diagrammatique.}\label{partdiag}

On reprend la repr\'esentation diagrammatique introduite dans la partie \ref{partdiag2MM} du chapitre 2 en changeant l\'eg\`{e}rement
les valeurs des propagateurs et vertex: on change partout les noyaux de Bergmann par leurs $\kappa$ d\'eform\'ees, y compris
dans les diff\'erentielles ab\'elienes de troisi\`{e}me esp\`{e}ce.

Ainsi, les fonctions de corr\'elation et \'energies libres sont obtenues en sommant sur le m\^eme ensemble de graphes
que celui d\'efini par le th\'eor\`eme \ref{thdiag2MM} du chapitre 2 mais
avec les poids d\'efinis par:
\bd
Le poids $\CP(G)$ d'un graphe $G$ est donn\'e par les r\`{e}gles suivantes:
\begin{itemize}
\item On marque tout vertex trivalent du graphe par un point courant $r_i$ de ${\Sigma}$ et on associe $r_i$ \`{a}
son enfant de gauche et $\overline{r}_i$ \`{a} son enfant de droite. Toute ar\^ete relie alors deux points de la surface
${\Sigma}$;

\item A une ar\^ete non orient\'ee liant $r$ et $r'$, on associe le facteur ${B}(r,r')$;

\item A toute ar\^ete orient\'ee allant  de $r$ vers $r'$, on associe le facteur ${{dE}_{r'}(r) \over (y(r')-y(\overline{r}')) dx(r')}$;

\item Suivant les fl\`eches en sens inverse (des feuilles vers la racine), \`{a} chaque vertex $r$, on calcule la somme sur tous
les points de branchement $a_i$ des r\'esidus quand $r \to a_i$: ${\displaystyle \sum_i \Res_{q \to a_i}}$.

\item Apr\`{e}s avoir calcul\'e l'ensemble de ces r\'esidus, on obtient le poids du graphe.

\end{itemize}
\ed

Cela peut simplement s'\'ecrire:
\beq
\CP\left(\begin{array}{l}
  \includegraphics[width=2cm]{Berg}
\end{array}\right):=B(p,q),
\eeq
et
\beq
\CP\left(\begin{array}{r}
\includegraphics[width=3cm]{vertex}
\end{array}\right):= \sum_i \Res_{q \to a_i} {{dE}_q(p) \over \omega(q)}.
\eeq

Les fonctions de corr\'elations sont alors donn\'ees par
\bt
\beq
\encadremath{
W_{k+1,0}^{(g)}(p,{\bf p_K}) = \sum_{G \in \CG_k^{g}(p, {\bf p_K})} \CP(G).
}
\eeq
\et

\subsection{Propri\'et\'es.}

Ces nouveaux objets poss\`edent de nombreuses propri\'et\'es dont certaines sont largement inspir\'ees des mod\`{e}les de matrices.
Nous pr\'esentons ici un ensemble de caract\'eristiques utiles \`{a} la manipulation et \`{a} la compr\'ehension de ces objets.
Toutes les d\'emonstrations peuvent \^etre trouv\'ees dans \EOinvariants.

Tout d'abord, et ceci est presque une d\'efinition, les chapitres pr\'ec\'edents nous permettent d'\'ecrire le th\'eor\`{e}me:
\bt
Les d\'eveloppements topologiques des \'energies libres des mod\`{e}les \`{a} une matrice, deux matrices et une matrice
en champ ext\'erieur d\'efinies respectivement par \eq{F1MM} du chapitre 3, \eq{F2MM} du chapitre 2 et \eq{FMext} du chapitre 3 sont donn\'ees par
\beq
\CF_{1MM} = \sum_{g=0}^\infty N^{-2g} \underline{F}^{(g)}(\CE_{1MM})
\virg
\CF_{2MM} = \sum_{g=0}^\infty N^{-2g} \underline{F}^{(g)}(\CE_{2MM})
\eeq
et
\beq
\CF_{\rm Mext} = \sum_{g=0}^\infty N^{-2g} \underline{F}^{(g)}(\CE_{\rm Mext})
\eeq
o\`{u} $\CE_{1MM}$, $\CE_{2MM}$ et $\CE_{\rm Mext}$ sont les courbes spectrales classiques des mod\`{e}les \`{a} une matrice,
deux matrices et une matrice en champ ext\'erieur respectivement, d\'efinies par \eq{E1MM} du chapitre 3, \eq{E2MM} du chapitre 2 et \eq{EMext} du chapitre 3 et $\kappa$ a \'et\'e
prise \'egale \`{a} 0.
\et

Les fonctions de corr\'elation ainsi d\'efinies sont \'evidement sym\'etriques en tous leurs arguments sauf le premier.
Cependant, on peut voir que la fonction \`{a} trois points de genre $0$ peut \^{e}tre mise sous une forme explicitement sym\'etrique
en tous ses arguments:
\bt\label{thW30}
La $3$-forme $W_3^{(0)}$ s'\'ecrit:
\beq
W_3^{(0)}(p,p_1,p_2) = \Res_{q\to \bfa} {B(q,p)B(q,p_1)B(q,p_2)\over dx(q) dy(q)}.
\eeq
\et
Cette formule est la g\'en\'eralisation de la formule donn\'ee par Krichever dans le cadre des mod\`{e}les de matrices \cite{Kri}.

Il est tr\`{e}s important de conna\^itre la structure de p\^oles des fonctions de corr\'elation pour pouvoir les manipuler.
On peut montrer qu'elles sont donn\'ees par:
\bt\label{thpolesWkgbp}
Pour tout couple $(k,g)\neq (1,0)$, la fonction de corr\'elation $W_{k+1}^{(g)}$ n'a de p\^{o}le en chacune de ses variables
qu'aux points de branchements en $x$, $a_i$.
\et
Ce r\'esultat qui est ici un th\'eor\`{e}me est une condition n\'ec\'essaire pour que ces formes diff\'erentielles soient bien
les observables voulues dans le cadre des mod\`{e}les de matrices.

De m\^{e}me, dans le contexte des int\'egrales matricielles, ces observables doivent \^{e}tre bien d\'efinies sur chaque feuillet
et donc avoir des int\'egrales nulles sur les cycles $\CA_i$. Dans ce cadre plus g\'en\'eral, on montre que:
\bt\label{thWkcycle}
Pour tout couple $(k,g)$, on a:
\beq
\forall i=1,\dots,\CG \qquad \oint_{p\in{\cal A}_i} W_{k+1}^{(g)}(p,p_1,\dots,p_k) = 0
,
\eeq
\beq
\forall i=1,\dots,\CG , \;  \forall m=1,\dots,k \qquad \oint_{p_m\in{\cal A}_i} W_{k+1}^{(g)}(p,p_1,\dots,p_k) = 0
.
\eeq
\et
Si dans le cadre particulier des mod\`{e}les de matrices o\`{u}, rappelons le, $\kappa = 0$, on consid\'erait les int\'egrales
sur les cycles $\underline{\CA}$ de la base choisie au d\'epart, on voit qu'il est important de consid\'erer ici les cycles $\CA[\kappa]$ pour
$\kappa \neq0$.

Deux propri\'et\'es \'egalement tr\`{e}s utiles des mod\`{e}les de matrices concernant la somme sur l'ensemble des feuillets
sont conserv\'ees:
\bt\label{thsumWk}
Pour tout couple $(k,g)$, on a:
\beq\label{thsumWk1}
\sum_i W_{k+1}^{(g)}(p^i,p_1,\dots,p_k)= \delta_{k,1}\delta_{g,0}\,{dx(p)dx(p_1)\over (x(p)-x(p_1))^2}
\eeq
et $k\geq 1$:
\beq\label{thsumWk2}
\sum_i W_{k+1}^{(g)}(p_1,p^i,p_2,\dots,p_k)= \delta_{k,1}\delta_{g,0}\,{dx(p)dx(p_1)\over (x(p)-x(p_1))^2}
\eeq
o\`{u} la somme porte sur tous les feuillets, c'est-\`{a}-dire tous les points $p^i$ tels que $x(p^i)=x(p)$.
\et

\bt\label{thPkpol}
Pour $(k,g)\neq (0,1)$,
\bea
P_k^{(g)}(x(p),p_K)
&=& {1\over dx(p)^2}\,\sum_i \Big[-2y(p^i)dx(p) W_{k+1}^{(g)}(p^i,p_K) \cr
&& + \sum_{m=0}^g \sum_{J\subset K} W_{j+1}^{(m)}(p^i,p_J)W_{k-j+1}^{(g-m)}(p^i,p_{K/J})+ W_{k+2}^{(g-1)}(p^i,p^i,p_K) \Big] \cr
\eea
est une fonction rationnelle de $x(p)$ sans p\^ole aux points de branchements en $x$.
\et

Ces propri\'et\'es nous permettent finalement de montrer la sym\'etrie des fonctions de corr\'elation
en tous leurs arguments:
\bt\label{thsymWk}
$W_k^{(g)}(p_1, \dots ,p_k)$ est une fonction sym\'etrique de toutes ses variables $p_i$.
\et
Cette propri\'et\'e est encore une fois naturelle dans le cadre des mod\`{e}les de matrices mais je tiens \`{a} souligner \`{a}
nouveau que rien n'assurait \`{a} priori qu'elle soit vraie dans un cadre plus g\'en\'eral.

Ceci nous permet d'affiner la d\'escription du comportement des fonctions de cor- r\'elations en leurs p\^oles:
\bc\label{corResxyWk}
\beq
\forall i,\qquad \Res_{a_i} W^{(g)}_{k+1}(p,p_1,\dots,p_k) =0
,
\eeq
\beq
\forall i,\qquad \Res_{a_i} x(p) W^{(g)}_{k+1}(p,p_1,\dots,p_k) =0
,
\eeq
\beq
\sum_i \Res_{a_i} y(p) W^{(g)}_{k+1}(p,p_1,\dots,p_k) =0
,
\eeq
\beq
\sum_i \Res_{a_i} x(p)y(p) W^{(g)}_{k+1}(p,p_1,\dots,p_k) =0
.
\eeq
\ec

\subsection{Cas particuliers: genres 0 et 1.}

Dans cette partie nous montrons comment ces relations de r\'ecurrences s'\'ecrivent explicitement dans les cas o\`{u} la
courbe alg\'ebrique $\CE$ est de genre 0 ou 1. En effet, dans ces cas l\`{a}, on peut les exprimer dans un langage peut-\^etre plus familier.

\subsubsection{Genre 0.}

Comme nous l'avons vu dans le chapitre 2, si l'on veut faire de la combinatoire de cartes par l'interm\'ediaire des
mod\`{e}les de matrices, il faut imposer que la courbe spectrale classique associ\'ee soit de genre 0 en imposant toutes les
fractions de remplissage nulles sauf l'une d'elles.
Alors, la surface de Riemann compacte $\Sigma$ est \'equivalente \`{a} la sph\`{e}re de Riemann sous les transformations
conformes. Il existe donc une param\'etrisation rationnelle de $\CE$, c'est-\`{a}-dire que l'on peut trouver deux fonctions rationnelles
$X(p)$ et $Y(p)$ telles que:
\beq
\curve(x,y) = 0 \quad \Leftrightarrow \quad \exists p\in {\bf C}\, , \,\, x=X(p)\, , \, y=Y(p).
\eeq

Dans ce cas, le noyau de Bergmann est d\'efini comme le noyau de Bergmann de la sph\`{e}re par:
\beq
B(p,q)=\Bergmann(p,q) = {dp dq \over (p-q)^2} = d_p\,d_q\,\ln{(p-q)}
\eeq
et la forme premi\`ere est simplement:
\beq
\primef(p,q) =\underline\primef(p,q)= {p-q\over \sqrt{dp\,dq}} .
\eeq
Les points de branchement sont les z\'eros de la fonction $X(p)$:
\beq
X(a_i) = 0.
\eeq
Le vertex est alors donn\'e par:
\beq
\Res_{q \to {\bf a}} {dp \over 2 (Y(q)-Y(\overline{q})) X'(q) dq} \left[ {1 \over q-p} - {1 \over \qbar -p} \right]
\eeq
et les diff\'erentes fonctions de corr\'elation sont facilement obtenues par d\'evelopppement de Taylor des fonctions
$X$ et $Y$ au voisinage des z\'eros de $X'$.

\subsubsection{Genre 1.}

Si la courbe est de genre $\CG=1$, alors on peut la param\'etrer sur le losange correspondant au domaine fondamental d'un
tore (cf. fig.\ref{fundamental} du chapitre 2) i.e. on peut trouver une param\'etrisation elliptique de $\CE$.
Ainsi il existe deux fonctions elliptiques $X(p)$ et $Y(p)$ \cite{ellipticalf} telles que
\bea
\curve(x,y) = 0 \quad \Leftrightarrow \quad \exists p\in {\bf C}\, , \,\, x=X(p)\, , \, y=Y(p) \cr
X(p+1)=X(p+\tau)=X(p) \virg
Y(p+1)=Y(p+\tau)=Y(p).
\eea

Alors le noyau de Bergmann est quasiment la fonction de Weierstrass correspondante \cite{ellipticalf}:
\beq
\Bergmann(p,q) = \left(\wp(p-q,\tau)+{\pi\over \Im\tau}\right)\,\,dp dq .
\eeq
La forme premi\`{e}re est donn\'ee par:
\beq
\underline\primef(p,q) = {\theta_1(p-q,\tau)\over \theta'_1(0,\tau) \sqrt{dp\,dq}} .
\eeq

Notons que si $\kappa={-1\over 2i \Im\tau}$, le noyau de Bergmann d\'eform\'e se r\'eduit \`{a} la fonction de
Weierstrass
\beq
B(p,q) = \wp(p-q,\tau)\,\,dp dq
\eeq
et est donc invariant modulaire. Nous verrons dans la prochaine partie que cette propri\'et\'e s'\'etend \`{a} toutes les courbes
alg\'ebriques et n'est pas un artefact du genre 1.

\section{D\'eformations de la courbe.}
Nous avons  d\'efini les fonctions de corr\'elation et \'energies libres comme des fonctionnelles d'une courbe alg\'ebrique $\CE$,
on peut alors se poser la question de leur d\'ependance en les modules de cette courbe: comment varient ces quantit\'es lorsque l'on
change la param\'etrisation de la courbe ou bien la courbe elle m\^eme par l'interm\'ediaire de ses modules?

\subsection{Invariance symplectique.}

Le th\'eor\`{e}me suivant montre que sous certaines transformations de la courbe alg\'ebri- que $\CE$, les \'energies libres restent inchang\'ees.

\bt\label{thinvsymp}
Pour tout $g$, $F^{(g)}$ reste inchang\'ee lorsque l'on change $\CE$ par l'une des transformations suivantes:
\begin{itemize}
\item $x \to {a x +b \over cx +d}$ et $y \to {(cx+d)^2 \over ad-bc} y$ avec $(a,b,c,d) \in \mathbb{C}^4$;

\item $x \to x$ et $y \to y + R(x)$ pour toute fonction rationnelle $R(x)$;

\item $x \to -x$ et $y \to y$;

\item $x \to y$ et $y \to x$.

\end{itemize}
\et
La d\'emonstration des trois premi\`{e}res propri\'et\'es d\'ecoule directement des d\'efinitions par l'interm\'ediaire de
la repr\'esentation diagrammatique pr\'esent\'ee dans la partie \ref{partdiag}.
La d\'emonstration de la derni\`{e}re sym\'etrie est beaucoup plus difficile. Dans le cadre du mod\`{e}le \`{a} deux matrices, elle
correspond \`{a} dire que le calcul de l'\'energie libre donne le m\^{e}me r\'esultat que l'on r\'esolve les \'equations de
boucles obtenues par variation de la matrice $M_1$ ou par variation de $M_2$, c'est-\`{a}-dire que le r\'esultat ne d\'epend
pas du mode de calcul. La d\'emonstration passe ici par la g\'en\'eralisation des fonctions de corr\'elation mixtes
$H_{1;k;l}^{(g)}$ dans ce cadre plus g\'en\'eral et leur \'etude pr\'ecise. Cette d\'erivation tr\`{e}s technique peut \^etre trouv\'ee
dans \EOSym.

Ce th\'eor\`{e}me montre que les \'energies libres sont des invariants d'une certaine classe de courbes alg\'ebriques, c'est
pourquoi nous nommerons dor\'enavant invariants alg\'ebriques les $F^{(g)}$\footnote{En effet, le terme \'energie libre
inspir\'e des mod\`{e}les de matrices n'a plus de sens ici tant qu'aucun syst\`{e}me physique sous-jacent n'a \'et\'e identifi\'e.
}.
Il nous fournit un outil tr\`{e}s puissant pour comparer diff\'erents mod\`{e}les caract\'eris\'es par une courbe alg\'ebrique.
Nous montrerons par exemple son utilit\'e dans la comparaison de plusieurs mod\`{e}les de matrices dans la partie \ref{partapplic} de ce
chapitre.

\br
On parle ici d'invariance symplectique parce que toutes ces transformations conservent la forme symplectique
\beq
|dx \wedge dy |.
\eeq
Cependant, il n'est pas clair pour le moment si $F^{(g)}$ est conserv\'ee par toute transformation laissant cette forme
inchang\'ee. Plus pr\'ecis\'ement, on ne conna\^it pas de caract\'eristaion de l'ensemble des transformations conservant les
$F^{(g)}$.
\er

\subsection{Variation par rapport aux modules de la courbe.}\label{variatmod}

Consid\'erons une variation infinit\'esimale de la courbe $\curve\to \curve+\delta \curve$,
obtenue par une variation de $y(x)$ \`{a} $x$ fix\'e:
\beq
\delta_\Omega\,\, y|_x \,dx = - \Omega.
\eeq

\br
On peut travailler \`{a} n'importe quelle coordonn\'ee locale $z$ fix\'ee plut\^ot que $x$ fix\'e.
On a alors une structure de Poisson:
\beq\label{deltaydxxdy}
\delta_\Omega\, y|_z\,dx - \delta_\Omega\, x|_z\, dy = - \Omega
\eeq
o\`{u} $\Omega$ est une forme diff\'erentielle quelconque dont nous restreignons les propri\'et\'es dans la suite.
\er

Nous allons \'etudier comment les diff\'erentes formes diff\'erentielles et invariants introduits plus haut varient sous
cette transformation. Pour ce faire, nous allons d'abord \'etudier comment varie l'\'el\'ement de base de cette construction:
le noyau de Bergmann.

\subsubsection{Formule variationnelle de Rauch et variation du noyau de Bergmann.}

Sous cette transformation, \eq{deltaydxxdy} montre que la variation de la projection des points de branchements dans le
plan $x$ est donn\'ee par:
\beq
\delta_\Omega\, x(a_i) =  {\Omega(a_i)\over dy(a_i)}.
\eeq
en supposant ici que ${\Omega\over dy}$ n'a pas de p\^ole aux points de branchements.

La formule variationnelle de Rauch \cite{Rauch, Fay} nous dit alors que la variation du noyau de Bergmann s'\'ecrit
\bea
\left. \delta_\Omega \underline{B}(p,q)\right|_{x(p),x(q)}
&=&  \sum_i \Res_{r\to a_i} {\Omega(r)\underline{B}(r,p)\underline{B}(r,q)\over dx(r)dy(r)}. \cr
\eea

En int\'egrant sur un cycle $\underline\bcycle$, on obtient la variation des diff\'erentielles holomorphes:
\bea
\left. \delta_\Omega du(p)\right|_{x(p)}
&=&  \sum_i \Res_{r\to a_i} {\Omega(r)\underline{B}(r,p){\bf du}(r)\over dx(r)dy(r)}, \cr
\eea
puis, par une seconde int\'egration, la variation de la matrice des p\'eriodes de Riemann:
\bea
 \delta_\Omega \tau
&=&  2i\pi \sum_i \Res_{r\to a_i} {\Omega(r) {\bf du}(r) {\bf du}^t(r)\over dx(r)dy(r)} .\cr
\eea

Etudions de m\^eme la variation du noyau de Bergmann $\kappa$ d\'eform\'e. Il y a maintenant deux termes \`{a} faire varier:
d'une part le noyau de Bergmann nu et d'autre part la d\'eformation dont la variation se d\'ecompose elle-m\^eme en la variation
des diff\'erentielles holomorphes et de $\kappa$ lui m\^eme. En effet, aucune contrainte ne vient s'opposer \`{a} un choix de
$\kappa$ d\'ependant de la courbe alg\'ebrique elle m\^eme. La formule variationnelle de Rauch peut alors s'\'ecrire:
\bea
\left. \delta_\Omega B(p,q)\right|_{x(p),x(q)}
&=&  \Res_{r\to \bfa} {\Omega(r) B(r,p)B(r,q)\over dx(r)dy(r)}
-  2i\pi \, {\bf du}^t(p) ( \kappa \, \delta_\Omega\tau\, \kappa - \delta_\Omega \kappa )  {\bf du}(q) \cr
&=& 2 \Res_{r\to \bfa} {\Omega(r) dE_r(p)B(r,q)\over \om(r)}  -  2i\pi \, {\bf du}^t(p) ( \kappa \, \delta_\Omega\tau\, \kappa - \delta_\Omega \kappa ) {\bf du}(q) \cr
\eea
que l'on peut r\'e\'ecrire sous la forme
\bea
 \left( \delta_\Omega + \tr \left[\kappa \,\delta_\Omega\tau \,\kappa - \delta_\Omega \kappa\right]{\partial \over \partial \kappa}\right)_{x(p),x(q)}\,\,  B(p,q)
&=&  \Res_{r\to \bfa} {\Omega(r) B(r,p)B(r,q)\over dx(r)dy(r)}  \cr
&=& - 2 \Res_{r\to \bfa} {\Omega(r) dE_r(p)B(r,q)\over \om(r)}   \cr
\eea
En d\'efinissant la {\bf d\'eriv\'ee covariante} par:
\beq\label{devcov}
\encadremath{
D_\Omega = \delta_\Omega + \tr\left[(\kappa \,\delta_\Omega\tau \,\kappa -\delta_\Omega \kappa) {\partial \over \partial \kappa}\right]
}
\eeq
on peut r\'esumer la variation du noyau de Bergmann d\'eform\'e par:
\bea\label{variatB}
D_\Omega B(p,q) &=& - 2 \Res_{r\to \bfa} {\Omega(r) dE_r(p)B(r,q)\over \om(r)}   \cr
&=& \Res_{r\to \bfa} {dE_r(p)\over \om(r)} \left[ \Omega(r) B(\overline{r},q) + \Omega(\overline{r}) B(r,q) \right].\cr
\eea

En int\'egrant au voisinage d'un point de branchement $a_j$, on obtient \'egalement:
\bea\label{variatdE}
\left. D_\Omega\, dE_{q}(p)\right|_{x(p),x(q)}  &=&  - 2 \Res_{r\to \bfa} { dE_{r}(p) \over \om(r)}\, \Omega(r) dE_{q}(r) \cr
&=& \Res_{r\to \bfa} { dE_{r}(p) \over \om(r)}\, \left[ \Omega(r) dE_{q}(\overline{r}) + \Omega(\overline{r}) dE_{q}(r) \right]. \cr
\eea

Le sens de ces deux \'equations peut \^etre \'eclairci en utilisant la repr\'esentation diagrammatique des fonctions de corr\'elation.
En effet, le noyau de Bergmann et $dE$ sont les deux types d'ar\^{e}tes composant les diagrammes et les deux seules quantit\'es variant sous l'action de $\delta_\Omega$.
Les \'equations \eq{variatB} et \eq{variatdE} expliquent alors comment les diagrammes changent sous la variation $\Omega$ de la courbe alg\'ebrique
$\CE$:
$$ \mbox{\includegraphics[width=12cm]{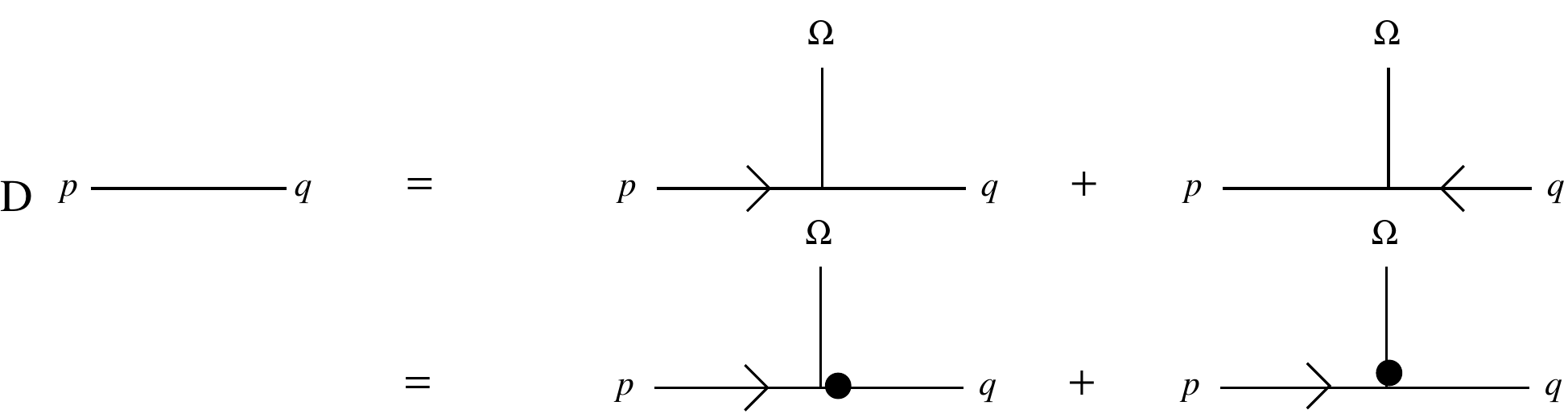}}$$
et
$$ \mbox{\includegraphics[width=12cm]{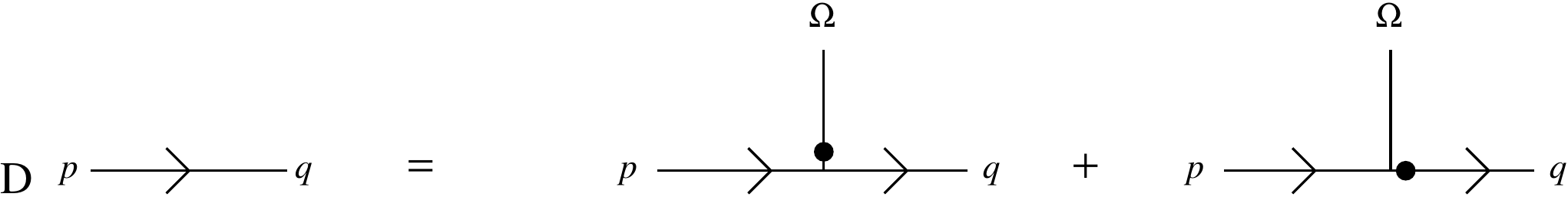}}.$$

On reconstruit alors la d\'eriv\'ee covariante des fonctions de corr\'elations ainsi que des invariants alg\'ebriques par
le lemme suivant:
\bl \label{lemDOmega} Pour toute forme bilin\'eaire sym\'etrique $f(q,p)=f(p,q)$:
\bea
D_\Omega \left( \sum_j \Res_{q\to a_j} {dE_{q}(p)\over \om(q)}\, f(q,\qbar) \right)_{x(p)}
&=& 2    \sum_{i,j} \Res_{r\to a_i} \Res_{q\to a_j} { dE_{r}(p) \over \om(r)}\, \Omega(r)\, {dE_{q}(r) \over \om(q)}\, f(q,\qbar) \cr
 && + \sum_j \Res_{q\to a_j} {dE_{q}(p)\over \om(q)} \,\, D_\Omega \left(f(q,\qbar)\right)_{x(q)} .\cr
\eea
\el
Ce lemme exprime en fait l'action de $D_\Omega$ sur le vertex:
$$ \mbox{\includegraphics[width=12cm]{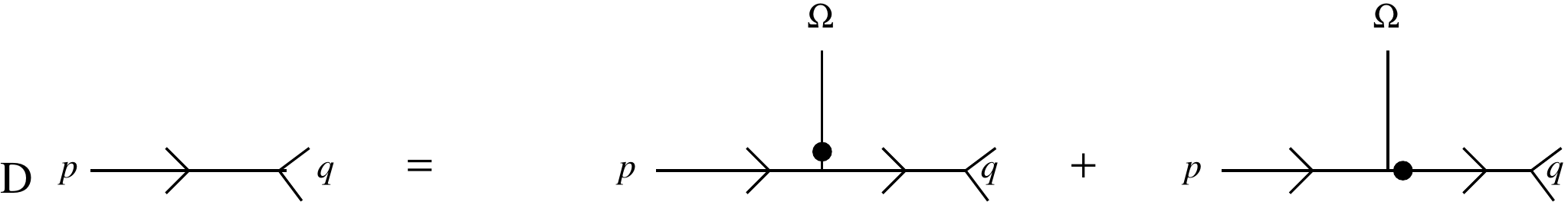}}.$$
Graphiquement, cela signifie simplement que la variation d'un diagramme est obtenue en ajoutant une patte $\Omega$ \`{a}
toutes les ar\^{e}tes possibles.

En particulier, si la variation de la courbe peut se mettre sous la forme
\beq
\Omega(p) = \int_{\partial\Omega} B(p,q) \Lambda(q)
\eeq
o\`{u} ${\partial\Omega}$ est un contour d'int\'egration assez \'eloign\'e des points de branchements
\footnote{Ceci exclut entre autres le cas o\`{u} $\Omega$ correspond \`{a} la variation d'une "ar\^{e}te dure", cf \cite{eynhardedges,  chekhovhe, Marco2}.},
on peut d\'ecrire les variations correspondantes par:
\bt \label{variat} Variations des fonctions de correlation et \'energies libres:
pour $g+k>1$, elles sont donn\'ees par:
\beq\encadremath{
\left. D_\Omega W_k^{(g)}(p_1,\dots,p_k)\right|_{x(p_i)} =  \int_{\partial\Omega} W_{k+1}^{(g)}(p_1,\dots,p_k,q) \Lambda(q)
}\eeq
et,
pour $g \geq 1$,
\beq\label{deltaFgintW1L}\encadremath{
D_\Omega F^{(g)} = - \int_{\partial\Omega} W_1^{(g)}(p) \L(p).
}\eeq

\et

Nous allons utiliser cette description pour \'etudier les variations des invariants alg\'ebriques par rapport aux modules
de la courbe $\CE$. En effet, \eq{decompoydx} nous permet d'identifier les $\Omega$ correspondant \`{a} la variation de chacun
des modules.

\subsubsection{Variation des fractions de remplissage.}

Consid\'erons la variation de la courbe donn\'ee par
\beq
\Omega(p) = -2i\pi du_i(p)  = -\oint_{\bcycle_i} B(p,q).
\eeq
i.e. $\partial\Omega=\bcycle_i$ et $\L=-1$.
Elle correspond \`{a}:
\beq
\delta_\Omega \epsilon_j =  \delta_{ij}
\virg
\delta_\Omega t_{0,j}= 0
\virg
\delta_\Omega V_j= 0.
\eeq
Cette variation est donc \'equivalente \`{a} faire varier uniquement l'une des fractions de remplissage
$\epsilon_i={1\over 2i\pi}\oint_{\acycle_i} y dx$:
\beq
D_{-2i\pi du_i} = {\partial \over \partial \epsilon_i}.
\eeq

Le th\'eor\`eme \ref{variat} implique
\beq
{\partial \over \partial \epsilon_i} \, W_{k}^{(g)}(p_1,\dots,p_k) = - \oint_{\bcycle_i} W_{k+1}^{(g)}(p_1,\dots,p_k,q),
\eeq
\beq
{\partial \over \partial \epsilon_i} \, F^{(g)} =   \oint_{\bcycle_i} W_{1}^{(g)}(q),
\eeq
et
\beq
 {\partial \over \partial \epsilon_i} \, F^{(0)}= - \oint_{\bcycle_j} ydx + {1 \over 4 i \pi} \left( \kappa  \oint_{\bcycle} ydx\right)^t \; \delta_{- 2 i \pi du_i} (\tau) \;  \kappa  \oint_{\bcycle} ydx.
\eeq

\subsubsection{Variation des temp\'eratures.}

Soient $\alpha_i$ et $\alpha_j$ deux p\^oles distincts de $ydx$.
Consid\'erons la variation de la courbe donn\'ee par
\beq
\Omega(p) = - dS_{\alpha_i,\alpha_j}(p)  = \int_{\alpha_i}^{\alpha_j} B(p,q)
\virg{\rm i.e.}\,\,\,
\partial\Omega = [\alpha_i,\alpha_j] \,\,\, , \,\, \L=1
\eeq
Elle correspond \`{a}:
\beq
\delta_\Omega \epsilon_j =0
\virg
\delta_\Omega t_{0,k}= \delta_{i,k} - \delta_{j,k}
\virg
\delta_\Omega V_k= 0.
\eeq
Cette variation est donc \'equivalente \`{a} faire varier uniquement les temp\'eratures $t_{0,i}$ et $t_{0,j}$:
\beq
D_{-dS_{\alpha_i,\alpha_j}} = {\partial \over \partial t_{0,i}} - {\partial \over \partial t_{0,j}}.
\eeq

\br
Toutes les temp\'eratures sont li\'ees par la condition:
\beq
\sum_i t_{0,i} = 0.
\eeq
On ne peut donc pas faire varier une seule d'elles ind\'ependamment. C'est pourquoi nous avons d\^u en changer deux \`{a} la fois
ici.
\er

Le th\'eor\`eme \ref{variat} implique
\beq
\left( {\partial \over \partial t_{0,i}} - {\partial \over \partial t_{0,j}} \right) \, W_{k}^{(g)}(p_1,\dots,p_k) = \int_{\alpha_i}^{\alpha_j} W_{k+1}^{(g)}(p_1,\dots,p_k,q),
\eeq
\beq
\left({\partial \over \partial t_{0,i}} - {\partial \over \partial t_{0,j}}\right) \, F^{(g)} =  \int_{\alpha_j}^{\alpha_i} W_{1}^{(g)}(q)
\eeq
et
\beq
\left({\partial \over \partial t_{0,i}} - {\partial \over \partial t_{0,j}}\right)  F^{(o)} =  \mu_{\alpha_i}-\mu_{\alpha_j} + {1 \over 4 i \pi} \left( \kappa  \oint_{\bcycle} ydx\right)^t \; \delta_{-dS_{\alpha_i,\alpha_j}} (\tau) \;  \kappa  \oint_{\bcycle} ydx.
\eeq

\subsubsection{Variation des modules aux p\^oles}

Soit $\alpha_i$ un p\^ole de $y dx$.
Consid\'erons la variation de la courbe donn\'ee par
\beq
\Omega(p) = - B_{k,i}  =  \Res_{\alpha_i} B(p,q) z_{\alpha_i}^k(q),
\eeq
i.e. $\partial\Omega$ est un petit cercle entourant $\alpha_i$ et $\L={1\over 2i\pi}\, z_{\alpha_i}^k$.
On a alors:
\beq
\delta_\Omega \epsilon_j = 0
\virg
\delta_\Omega t_{0,j}= 0
\virg
\delta_\Omega t_{k',j}= \delta_{i,j} \delta_{k,k'}
\eeq
Cette variation \'equivaut donc \`{a} ne changer que le coefficient $t_{k,i}$:
\beq
D_{-B_{k,i}} = {\partial \over \partial t_{k,i}}.
\eeq

Le th\'eor\`eme \ref{variat} donne
\beq
{\partial \over \partial t_{k,i}} \, W_{k}^{(g)}(p_1,\dots,p_k) =  \Res_{\alpha_i} z_{\alpha_i}^k(q) W_{k+1}^{(g)}(p_1,\dots,p_k,q),
\eeq
\beq
{\partial \over \partial t_{k,i}} \, F^{(g)} = - \Res_{\alpha_i} z_{\alpha_i}^k(q) W_{1}^{(g)}(q)
\eeq
et
\beq
{\partial \over \partial t_{k,i}} F^{(o)} =   \Res_{\alpha_i} ydx z_{\alpha_i}^k + {1 \over 4 i \pi} \left( \kappa  \oint_{\bcycle} ydx\right)^t \; \delta_{B_{k,i}} (\tau) \;  \kappa  \oint_{\bcycle} ydx.
\eeq

Il peut \^etre int\'eressant de conna\^itre non seulement les d\'eriv\'ees premi\`{e}res mais \'egalement les d\'eriv\'ees
secondes de $F^{(0)}$ par rapport aux modules de la courbe. En effet, dans le cadre des mod\`{e}les de matrices, ce sont
ces derni\`{e}res qui sont "universelles". Nous les avons donc regroup\'ees dans l'appendice \ref{appF0} sous la forme d'un formulaire.

\subsubsection{Homog\'en\'eit\'e}

On peut utiliser ces formules pour montrer une propri\'et\'e d'homog\'en\'eit\'e permettant de d\'ecomposer
les invariants sur la base de ces modules:
\bt \label{homogeneity} Pour tout $g>1$, $F^{(g)}$ satisfait la propri\'et\'e d'homog\'en\'eit\'e:
\beq
(2-2g)F^{(g)}=
\sum_{i,k} t_{k,i} {\partial\over \partial t_{k,i}} F^{(g)}
+ \sum_{i} t_{0,i} {\partial\over t_{0,i}} F^{(g)}
+ \sum_{i} \epsilon_i {\partial\over \partial \epsilon_i} F^{(g)}
\eeq
i.e. $F^{(g)}$ est homog\`{e}ne de degr\'e $2-2g$.
\et

\subsection{Op\'erateur d'insertion de boucle et son inverse.}
On peut \'egalement utiliser ces d\'eformations de la courbe alg\'ebrique pour reconstruire les fonctions de corr\'elation $W_k^{(g)}$ \`{a}
partir des invariants alg\'ebriques $F^{(g)}$.

En effet, pour tout point $q$ situ\'e loin de tout point de branchement, consid\'erons la variation:
\beq
\Omega(p) = B(p,q).
\eeq
Elle induit une variation des fonctions de corr\'elation donn\'ee par:
\bt
\beq
D_B \, W_{k}^{(g)}(p_1,\dots,p_k) = W_{k+1}^{(g)}(p_1,\dots,p_k,q)
\eeq
\beq
D_B \, F^{(g)} = - W_{1}^{(g)}(q)
\eeq
et
\beq
D_B \, F^{(0)} = y(q)dx(q) +{1 \over 4 i \pi} \left( \kappa \oint_{\bcycle}ydx \right)^t \oint_{\bcycle} \oint_{\bcycle} W_{3,0}
\kappa \oint_{\bcycle}ydx
.
\eeq
\et
C'est \`{a} dire que l'op\'erateur $D_{B(.,q)}$ agit sur les fonctions de corr\'elation en ajoutant une patte externe au point $q$.
C'est exactement le r\^ole jou\'e par l'op\'erateur ${\partial \over \partial V_1}$ d\'efini par \eq{loopinsertion} du chapitre 2.
On a donc \'etendu la notion d'{\bf op\'erateur d'insertion de boucle} \`{a} ce cadre plus g\'en\'eral d'invariants alg\'ebriques:
il consiste simplement \`a ajouter un p\^ole double \`a la forme $ydx$ en un point marqu\'e de la courbe spectrale.

Dans le chapitre 2, nous avons introduit l'op\'erateur $H_x$ pour retirer une patte aux fonctions de corr\'elations, i.e.
inverser l'op\'erateur d'insertion de boucles\footnote{Ce n'est pas \`{a} proprement parler l'inverse de ${\partial \over \partial V_1}$
puisque cette propri\'et\'e n'est valable que dans l'espace des fonctions de corr\'elations et qu'il appara\^it un facteur
d\'ependant de la fonction consid\'er\'ee.}.

Nous pouvons d\'efinir ici le m\^eme type d'op\'erateur par:
\bt\label{thintPhi}
Pour $k\geq 1$, on a:
\bea
\Res_{p_{k+1}\to \bfa,p_1,\dots,p_k} \Phi(p_{k+1}) W^{(g)}_{k+1}(p_1,\dots,p_k,p_{k+1}) &=& (2g+k-2) W^{(g)}_k(p_1,\dots,p_k) \cr
&& + \delta_{g,0}\delta_{k,1} y(p_1)dx(p_1)
\cr\eea
o\`u $\Phi$ est n'importe quelle primitive de $ydx$:
\beq
d\Phi = ydx.
\eeq
\et
Il est \`{a} noter que pour $k=0$ et $g\geq 2$, ceci d\'efinit l'\'energie libre comme la fonction de corr\'elation \`{a} 0 points $W^{(g)}_0=F^{(g)}$.

\subsection{Transformations modulaire.}\label{sectmodular}

Nous allons maintenant \'etudier comment les diff\`erentes fonctions de corr\'elation d\'ependent du choix d'une base de cycles
$(\underline\CA,\underline\CB)$.

Consid\'erons une transformation modulaire quelconque des cycles:
\beq
\pmatrix{\underline\acycle \cr \underline\bcycle} = \pmatrix{\delta_{\acycle \acycle'} & \delta_{\acycle \bcycle'}\cr \delta_{\bcycle \acycle'} & \delta_{\bcycle \bcycle'}} \pmatrix{\underline\acycle' \cr \underline\bcycle'}
\virg
\pmatrix{\underline\acycle' \cr \underline\bcycle'} = \pmatrix{\delta_{\acycle' \acycle} & \delta_{\acycle' \bcycle}\cr \delta_{\bcycle' \acycle} & \delta_{\bcycle' \bcycle}} \pmatrix{\underline\acycle \cr \underline\bcycle}
\eeq
o\`{u} $\delta_{\acycle' \acycle} = \delta_{\bcycle \bcycle'}^t$, $\delta_{\acycle' \bcycle} = - \delta_{\acycle \bcycle'}^t$,
$\delta_{\bcycle' \bcycle} = \delta_{\acycle \acycle'}^t$, $\delta_{\bcycle' \acycle} = - \delta_{\bcycle \acycle'}^t$ et les matrices $\delta_{\acycle \acycle'}$, $\delta_{\acycle \bcycle'}$, $\delta_{\bcycle \acycle'}$ et $\delta_{\bcycle \bcycle'}$ ont des coefficients entiers
qui satisfont $\delta_{\acycle \acycle'} \delta_{\bcycle \bcycle'}^t - \delta_{\bcycle \acycle'} \delta_{\acycle \bcycle'}^t = Id$.

Sous ce changement de base de cycles d'homologie, les diff\'erentiels holomorphes et la matrice des p\'eriodes de Riemann
sont chang\'ees suivant
\beq
{\bf du}' = {\cal{J}} {\bf du} \virg {\bf du} = {\cal{J}}^{-1} {\bf du}'
\eeq
o\`{u}
\beq
{\cal{J}} =
(\delta_{\acycle \acycle'}^t+\tau' \delta_{\acycle \bcycle'}^t) = (\delta_{\bcycle \bcycle'}-\tau \delta_{\acycle \bcycle'})^{-1}
\eeq
et
\beq
\tau' = (\delta_{\bcycle \bcycle'}-\tau \delta_{\acycle \bcycle'})^{-1} (-\delta_{\bcycle \acycle'}+\tau \delta_{\acycle \acycle'})
\virg
\tau = (\delta_{\acycle \acycle'}^t+\tau' \delta_{\acycle \bcycle'}^t)^{-1}\,(\delta_{\bcycle \acycle'}^t+\tau' \delta_{\bcycle \bcycle'}^t).
\eeq

Nous commen\c{c}ons une nouvelle fois par \'etudier les variations induites sur les blocs de bases.
Ainsi le noyau de Bergmann est chang\'e en:
\beq\label{modularchB}
\underline{B}' = \underline{B} + 2i\pi\, {\bf du}^t \widehat{\kappa} {\bf du} .
\eeq
o\`{u} l'on a utilis\'e la matrice sym\'etrique de taille $\CG \times \CG$:
\beq
\widehat{\kappa}=\widehat{\kappa}^t=(\delta_{\bcycle \bcycle'} \delta_{\acycle \bcycle'}^{-1}-\tau )^{-1}
=
\delta_{\acycle \bcycle'} {\cal{J}}.
\eeq

Le noyau de Bergmann $\kappa$-d\'eform\'e change lui suivant:
\bea
B'(p,q) &=& \underline{B}(p,q) + 2 i \pi \left[ {\bf du}'^t(p) \kappa {\bf du}'(q) + {\bf du}^t(p) \widehat{\kappa} {\bf du}(q) \right] \cr
&=& \underline{B}(p,q) + 2 i \pi {\bf du}^t(p) \left( \widehat{\kappa} + {\cal{J}}^t
\kappa {\cal{J}} \right) {\bf du}(q).
\eea

En d'autres termes, une transformation modulaire est \'equivalente \`{a} un changement du param\`{e}tre $\kappa$ par
$\kappa\to \widehat{\kappa} + {\cal{J}}^t \kappa {\cal{J}} $ dans la d\'efinition de la d\'eformation du noyau de Bergmann $B(p,q)$.
Ceci nous permet de conna\^{i}tre les variations modulaires des \'energies libres:
\bt
\label{thmodular}
Pour $g\geq 2$ la transformation modulaire de l'invariant $F^{(g)}$ consiste \`{a} changer $\kappa$ en
$\widehat{\kappa} + {\cal{J}}^t \kappa {\cal{J}}$ dans les d\'efinitions des noyaux de Bergmann et diff\'erentielles Abeliennes
d\'eform\'es.

Pour $g=1$, $F^{(1)}$ est chang\'ee en:
\beq
F^{(1)'} = F^{(1)} - {1\over 2} \ln{(\delta_{\bcycle \bcycle'} -\tau \delta_{\acycle \bcycle'})}.
\eeq
\et

\br
La transformation de $F^{(0)}$ sous ce changement de base de cycles d'homologie est bien plus compliqu\'ee
puisque le r\'esultat final d\'epend explicitement de la position des p\^oles de $ydx$ dans le domaine fondamental
et donc de tous les param\`{e}tres de la transformation modulaire.
\er

Comme les transformations modulaires des \'energies libres d\'ependent directement de $\kappa$ par l'application:
\beq
\kappa \to \widehat{\kappa} + {\cal{J}}^t \kappa {\cal{J}},
\eeq
on peut se demander si cette application a un point fixe, i.e. si il existe un choix de $\kappa$ pour lequel
les $F^{(g)}$ sont des invariants modulaires. La r\'eponse est donn\'ee par le th\'eor\`{e}me suivant:
\bt
Pour le choix $\kappa={i\over 2\, \Im\tau}$,  les $F^{(g)}[\kappa]$ sont des {\bf invariants modulaires}.
\et

En effet, pour ce choix de $\kappa$, le noyau de Bergmann d\'eform\'e est le noyau de Schiffer dont l'invariance modulaire est connue \cite{BergSchif}.
Comme la seule d\'ependance modulaire des $F^{(g)}$ se trouve dans la d\'eformation du noyau de Bergmann, on obtient simplement
le r\'esultat.

Nous verrons dans le chapitre suivant que cette valeur de $\kappa$ permet de faire le lien avec les fonctions de partition
de th\'eorie des cordes topologiques et nous reviendrons donc sur ce cas particulier dans la partie correspondante.

\subsection{Variation par rapport \`{a} $\kappa$.}\label{sectvarikappa}

Comme nous l'avons vu dans la partie pr\'ec\'edente, la matrice $\kappa$ peut \^etre utilis\'ee pour d\'ecrire les variations
modulaires des \'energies libres. On peut ainsi voir les variations de $\kappa$ comme des transformations modulaires infinit\'esimales.
Nous allons donc calculer la d\'ependance en $\kappa$ des fonctions de corr\'elation et \'energies libres.

Tout d'abord, en comptant le nombre d'ar\^{e}tes dans leur repr\'esentation diagrammatique, il est facile de voir que $W^{(g)}_k$ est un polyn\^ome en $\kappa$ de degr\'e $3g+2k-3$ tandis que $F^{(g)}$
est un polyn\^ome de degr\'e $3g-3$ pour $g>1$.

Leurs d\'eriv\'ees par rapport aux \'el\'ements de $\kappa$ sont quant \`{a} elles donn\'ees par:
\bt\label{thdWdkappa}
\beq\encadremath{
\begin{array}{lcl}
2i\pi {\partial \over \partial \kappa_{ij}}\,W^{(g)}_{k}({\bf p_K})
&=&  {1\over 2}\,\oint_{r\in\bcycle_j}\oint_{s\in\bcycle_i} W^{(g-1)}_{k+2}({\bf p_K},r,s) \cr
&& + {1\over 2}\,\sum_h \sum_{L\subset K} \oint_{r\in\bcycle_i} W^{(h)}_{|L|+1}({\bf p_L},r) \oint_{s\in\bcycle_j} W^{(g-h)}_{k-|L|+1}({\bf p_{K/L}},s) \cr
\end{array}
}\eeq
et, en particulier pour $g \geq 2$:
\beq\encadremath{
- 2i\pi {\partial \over \partial \kappa_{ij}}\,F^{(g)}
=  {1\over 2}\,\oint_{r\in\bcycle_j}\oint_{s\in\bcycle_i} W^{(g-1)}_{2}(r,s) + {1\over 2}\,\sum_{h=1}^{g-1} \oint_{r\in\bcycle_i} W^{(h)}_{1}(r) \oint_{s\in\bcycle_j} W^{(g-h)}_{1}(s)
}\eeq
alors que pour $g=1$, on a
\beq\encadremath{
 {\partial \over \partial \kappa_{ij}}\,F^{(1)} =  \left(\kappa^{-1}\right)_{ij}.}
\eeq

\et

On peut noter une forte similitude entre ces \'equations et les \'equations d'anomalie holomorphe de la th\'eorie de
Kodaira--Spencer \cite{ABK,ADKMV,BCOV}\EMO. En fait, nous montrerons dans le chapitre suivant qu'elles coincident effectivement pour
la valeur de $\kappa$ rendant les \'energies libres invariantes modulaires.

\section{Limite singuli\`ere.}\label{partsing}

Nous allons \'etudier dans cette partie comment les objets introduits jusqu'ici se comportent lorsque la courbe $\CE$ devient singuli\`ere.

Pour ce faire, consid\'erons une famille de courbes alg\'ebriques param\'etr\'ee par une variable $t$ permettant d'approcher une singularit\'e:
\beq
\curve(x,y,t)
\eeq
telle que la courbe pour $t=0$ a un point de branchement singulier $a$ avec une singularit\'e
de type $p/q$, c'est-\`{a}-dire, d\'ecrite en termes d'une variable locale $z$ au voisinage de $a$, par:
\beq
\left\{\begin{array}{l}
t=0\cr
x(z) \sim x(a) + (z-a)^q \cr
y(z) \sim y(a) + (z-a)^p \cr
\end{array}\right. .
\eeq

Pour $t\neq 0$, la singularit\'e dispara\^it et l'on a une param\'etrisation en terme de la variable locale
$\zeta = z\, t^{-\nu}$:
\beq
\left\{\begin{array}{l}
x(z,t) \sim x(a) + t^{q\nu}\,\,Q(\zeta) + o(t^{q\nu}) \cr
y(z,t) \sim y(a) + t^{p\nu}\,\,P(\zeta) +o(t^{p\nu}) \cr
\end{array}\right.
\eeq
o\`{u} $Q$ et $P$ sont des polyn\^omes de degr\'es respectifs $q$ et $p$ et o\`{u} $\nu$ est un exposant critique d\'ependant explicitement du choix du param\`{e}tre $t$.

La {\bf courbe spectrale singuli\`{e}re}
\beq
\curve_{\rm sing}(\xi,\eta)=\left\{\begin{array}{l}
\xi(\zeta) = Q(\zeta)  \cr
\eta(\zeta) = P(\zeta)  \cr
\end{array}\right.
= {\rm Resultant}(Q-\xi,P-\eta)
\eeq
est d\'efinie comme la partie la plus divergente de la courbe lorsque $t \to 0$.

Si la courbe $\CE(t)$ elle m\^{e}me pr\'esente une singularit\'e, les \'energies libres $F^{(g)}(\curve(t))$
pr\'esentent elles aussi une singularit\'e pour $t \to 0$. Elles se comportent suivant:
\beq
F^{(g)}(\curve(t)) \sim t^{\gamma_g} F^{(g)}_{\rm sing} + o(t^{\gamma_g}) \qquad ,\,\, {\rm for}\, g\geq 2
\eeq
\beq
F^{(1)}(\curve(t)) \sim -{1\over 24}\,(p-1)(q-1)\nu\,\ln{(t)}  + O(1) \qquad ,\,\, {\rm for}\, g=1 .
\eeq
On appelle {\bf double limte d'\'echelle} de $F^{(g)}$ la partie singuli\`{e}re $F^{(g)}_{\rm sing}$ en r\'eference aux
mod\`{e}les de matrices.
Elle est d\'etermin\'ee ainsi que l'exposant $\gamma_g$ par le th\'eor\`{e}me suivant:

\bt\label{thsinglimit} La limite singuli\`{e}re de l'\'energie libre est donn\'ee par:
\beq\encadremath{
F^{(g)}_{\rm sing}(\curve) = F^{(g)}(\curve_{\rm sing}) \qquad ,\,\, {\rm for}\, g\geq 2
}\eeq
et l'exposant critique s'\'ecrit
\beq
\gamma_g = (2-2g)(p+q)\nu .
\eeq
En d'autres termes, par une normalisation appropri\'ee, notre construction des $F^{(g)}$ commute avec les limites singuli\`eres.
\et

Cette propri\'et\'e de passage \`{a} la limite a de nombreuses implications comme nous le verrons plus loin. Elle permet, entre
autres choses, d'appliquer notre m\'ethode \`{a} l'\'etude des th\'eories conformes et permet de faire le lien entre diff\'erents mod\`{e}les.

\section{Applications.} \label{partapplic}

Nous montrons dans cette partie l'efficacit\'e de notre m\'ethode dans deux cas particuliers. D'une part, nous montrons comment
elle permet d'avoir acc\`es de mani\`{e}re directe et syst\'ematique aux mod\`{e}les minimaux $(p,q)$ et nous
montrons d'autre part comment elles permettent de retrouver tr\`es facilement des r\'esultats classiques relatifs \`a l'int\'egrale
de Kontsevich \cite{kontsevich}.

\subsection{Double limite d'\'echelle et mod\`{e}les minimaux des th\'eories conformes.}
On sait depuis longtemps que les limites singuli\`{e}res des mod\`{e}les \`{a} une et deux matrices hermitiennes,
connues sous le nom de doubles limites d'\'echelle, sont fortement
li\'ees aux mod\`{e}les minimaux de type $(p,q)$ obtenus en th\'eorie des champs conforme \cite{Kos, ZJDFG, DKK,Douglas1}.
Dans cette partie, nous explicitons ce lien en montrant de mani\`{e}re pr\'ecise comment traiter la singularit\'e
due \`{a} la double limite d'\'echelle gr\^ace aux invariants d\'efinis plus haut. Nous montrons, entre autres, comment les
exposants critiques correspondants aux mod\`{e}les $(p,q)$ apparaissent et quelles sont les courbes spectrales correspondantes.

Nous avons vu dans le chapitre 2 que, tant que la courbe spectrale classique associ\'ee \`{a} un mod\`{e}le \`a deux matrices
est r\'eguli\`ere, tous les coefficients $F^{(g)}$ du d\'eveloppement topologique de l'\'energie libre peuvent \^etre calcul\'es.
Ainsi, le rayon de convergence en $T$ de $F^{(g)}(T)$ est atteint pour certaines courbes singuli\`{e}res. Nous nous restreignons
ici, comme dans la partie \ref{partsing}, aux singularit\'es rationnelles.

Consid\'erons donc le cas o\`{u} les potentiels $V_1$ et $V_2$ sont choisis de mani\`{e}re \`{a} ce que la courbe spectrale
classique $\curve_{\rm 2MM}$ ait une singularit\'e de type $p/q$ lorsque $T$ est \'egale \`{a} la valeur critique $T_c$.
\beq
\left\{\begin{array}{l}
T=T_c\cr
x(z) \mathop{\sim}_{p \to a}\, x(a) + (z-z(a))^q \cr
y(z) \mathop{\sim}_{p \to a}\, y(a) + (z-z(a))^p \cr
\end{array}\right.
\eeq
Notons que les valeurs de $p$ et $q$ sont major\'ees par les degr\'es des potentiels $V_1$ et $V_2$. Ainsi, le mod\`ele
\`{a} une matrice donne seulement acc\`{e}s \`a $q=2$.

On peut se ramener \`a la limite singuli\`ere d\'ecrite dans la partie pr\'ec\'edente en utilisant la notation $t = T - T_c$.
Ainsi, \`{a} $t\neq0$, il n'y a pas de singularit\'e et l'on d\'ecrit la courbe $\CE_{\rm 2MM}$  en termes de la variable
 locale $\zeta = z t^{-\nu}$:
\beq\label{dsllimitcurve2MM}
\left\{\begin{array}{l}
x(z,t) \sim x(a) + t^{q\nu}\,\,Q(\zeta) + o(t^{q\nu}) \cr
y(z,t) \sim y(a) + t^{p\nu}\,\,P(\zeta) +o(t^{p\nu}) \cr
\zeta = z t^{-\nu}
\end{array}\right.
\eeq
o\`{u} $Q$ (resp. $P$) est un polyn\^ome de degr\'es $q$ (resp. $p$).

On d\'efinit alors la courbe spectrale singuli\`{e}re par
\beq
\curve_{\rm sing}(\xi,\eta)=\left\{\begin{array}{l}
\xi = Q(\zeta)  \cr
\eta = P(\zeta)  \cr
\end{array}\right.
= {\rm Resultant}(Q-\xi,P-\eta).
\eeq
Or, la d\'ependance en $T$ du mod\`ele \`{a} deux matrices se caract\'erise par
\beq
\left. {d\over dT} ydx\right|_{x} = dS_{\infty_x,\infty_y}
\eeq
o\`u $\infty_x$ et  $\infty_y$ sont les deux p\^oles de $ydx$ d\'ecrits dans la partie \ref{geoalg} du chapitre 2.
Notons qu'ils sont loins des points de branchement, et en particulier du point de branchement devenant singulier.
Au voisinage de la singularit\'e, on a donc
\beq
\left. {d\over dT} ydx\right|_{x} \sim C\, t^\nu \,d\zeta + O(t^{2\nu})
\eeq
o\`u $C$ est une constante d'ordre 1.
En utilisant la param\'etrisation \eq{dsllimitcurve2MM}, on obtient la relation de Poisson \cite{DKK,ZJDFG} entre les polyn\^omes:
\beq\label{Poissonpq1}
pP(\zeta)Q'(\zeta) - q Q(\zeta)P'(\zeta) = {C\over\nu} \, t^{1-(p+q-1)\nu}
\eeq
qui fixe
\beq
\nu = {1\over p+q-1}.
\eeq

Alors le th\'eor\`eme \ref{thsinglimit} montre que le terme dominant des \'energies libres est donn\'e par
\beq
F_{\rm 2MM}^{(g)}(T) \mathop{\sim}_{T\to T_c} (T-T_c)^{(2-2g)(p+q)/(p+q-1)} F^{(g)}(\curve_{\rm sing})
 \qquad ,\,\, {\rm pour}\, g\geq 2
 \eeq
 lorsque l'on s'approche de la singularit\'e \`a $T = T_c$.
Les deux premier termes du d\'eveloppement topologique sont aussi donn\'es par:
\beq
F_{\rm 2MM}^{(0)}(T) \mathop{\sim}_{T\to T_c} {C^2\over 2} {(p+q-1)^2\over (p+q)(p+q+1)}\,(T-T_c)^{2+2\nu} + {\rm reg} \qquad ,\,\, {\rm pour}\, g=0
\eeq
et
\beq
F_{\rm 2MM}^{(1)}(T) \mathop{\sim}_{T\to T_c}  -{1\over 24}\,(p-1)(q-1)\nu\,\ln{(T-T_c)}  + O(1) \qquad ,\,\, {\rm pour}\, g=1.
\eeq
On a ainsi un moyen de calculer explicitement la double limite d'\'echelle des termes du d\'eveloppement topologique de l'\'energie
libre sans faire de passage \`{a} la limite: il suffit de calculer les quantit\'es correspondantes directement sur la courbe
spectrale singuli\`ere.

\subsubsection{Mod\`eles minimaux (p,q).} \label{sectpq}

Etudions plus pr\'ecis\'ement la courbe singuli\`ere $\curve_{(p,q)}$ correspondant \`{a} un mod\`{e}le minimal $(p,q)$ (cf \cite{DKK})
donn\'ee par
\beq\label{curvepq}
\curve_{(p,q)}(x,y)=
\left\{\begin{array}{l}
x = Q(\zeta) \cr
y = P(\zeta)  \cr
\end{array}\right.
= {\rm Resultant}(Q-x,P-y).
\eeq
o\`u $P$ et $Q$ sont deux polyn\^omes de degr\'es respectifs $p$ et $q$ satisfaisant la relation de Poisson:
\beq\label{Poissonpq}
 p PQ'-q QP'  = {t_1\over \nu}
\eeq
dont la solution peut \^etre \'ecrite \cite{ZJDFG}:
\beq
P= (Q^{p/q})_+
\eeq
et
\beq
(Q^{p/q})_- = {t_1\over q} \zeta^{1-q} + O(\zeta^{-q}),
\eeq
o\`u nous avons utilis\'e les notations $f= f_+ + f_-$ avec $f_+$ et $f_-$ d\'esignant respectivement les parties positives et n\'egatives
du d\'eveloppement en s\'erie de Laurent de $f$, i.e. la partie polyn\^omiale et la partie p\^olaire.
On voit ais\'ement que cette derni\`{e}re \'equation implique $q-2$ contraintes pour les \'el\'ements de $Q$.

La courbe $\curve_{(p,q)}$ ainsi d\'efinie est de genre nul et telle que les fonctions rationnelles $x$ et $y$ la param\'etrisant
n'ont qu'un seul p\^ole situ\'e \`{a} l'infini: $\alpha := \infty$.
Le noyau de Bergmann est alors (cf partie \ref{geoalg} du chapitre 2):
\beq
B(\zeta_1,\zeta_2) = {d\zeta_1 \, d\zeta_2\over (\zeta_1-\zeta_2)^2}.
\eeq
Les modules au p\^ole de cette courbe sont alors les coefficients $Q_k$ et $P_k$ donn\'es par
\beq
Q(\zeta) = \sum_{k=0}^q Q_k \zeta^k
\virg
P(\zeta) = \sum_{k=0}^p P_k \zeta^k.
\eeq
Par translation sur la variable locale $\zeta$, on peut supposer que $Q_{q-1}=0$ et, par changement d'\'echelle, que $Q_{q-2}=-q Q_q$
sans perte de g\'en\'eralit\'e. La relation de Poisson \eq{Poissonpq} dit alors que $P_{p-1}=0$ et $P_{p-2}=-p P_p$, et donc que:
\beq
Q_{q-1}=P_{p-1} = 0 \virg {Q_{q-2}\over Q_q}=-q \virg {P_{p-2}\over P_p}=-p.
\eeq

On retrouve alors:
\beq
F^{(0)}(\curve_{(p,q)})= 0 .
\eeq

\br
L'invariance symplectique du th\'eor\`eme \ref{thinvsymp} autorisant l'\'echange des r\^oles respectifs de $x$ et de $y$ permet de montrer
tr\`es simplement la propri\'et\'e bien connue d'\'equiva- lence des mod\`{e}les $(p,q)$ et $(q,p)$.
\er

\subsubsection{Autres modules.}

Dans un contexte plus g\'en\'eral \cite{ZJDFG}, on peut d\'eformer les mod\`{e}les minimaux $(p,q)$ avec $p+q-2$ autres
modules $t_1,\dots, t_{p+q-2}$.
Pour ce faire, consid\'erons un polyn\^ome $Q$ monique de degr\'e $q$:
\beq
Q(\zeta) = \zeta^q + \sum_{j=0}^{q-2} u_{q-j} \zeta^j
\eeq
dont les coefficients $u_2,\dots, u_{q}$ sont des fonctions de $q-1$ modules $t_1,\dots, t_{q-1}$ d\'etermi- n\'ees par
la contrainte
\beq
(Q^{p/q})_- = \sum_{j=1}^{q-2} {q-j\over q} t_{q-j} Q^{-j/q} + {t_1\over q} \zeta^{1-q} + O(\zeta^{-q}).
\eeq
D\'efinissons alors un polyn\^ome $P$ monic de degr\'e $p$ par:
\beq
P(\zeta) = \zeta^p + \sum_{j=0}^{p-2} v_{p-j} \zeta^j =   Q^{p/q}_+ - \sum_{j=1}^{p-1} {j+q\over q} t_{q+j-1} Q^{j/q}_+
\eeq
d\'ependant de $p-1$ autres param\`etres $t_q, \dots , t_{q+p-2}$.

La courbe spectrale classique correspondante est alors:
\beq
\curve_{(p,q)}(x,y) = {\rm Resultant}(x-Q,y-P).
\eeq
Il est bien connu que cette courbe d\'ependant des param\`{e}tres $t_1,\dots, t_{p+q-2}$ est la courbe spectrale de la hierarchie
int\'egrable de Witham sans dispersion \cite{krichwitham} et l'on a ainsi acc\`es aux observables de cette hierarchie.

On peut v\'erifier que $t_2=t_3=\dots=t_{p+q-1}=0$ redonne bien s\^ur le mod\`{e}le minimal $(p,q)$.

\subsubsection{Exemples de Mod\`{e}les minimaux.}
D\'ecrivons les premiers exemples de mod\`{e}les $(p,q)$ en donnant \`{a} chaque fois la courbe spectrale correspondante ainsi
que les polyn\^omes permettant une param\'etrisation rationnelle.

$\bullet$ {\bf Courbe d'Airy $(2,1)$.}

La courbe spectrale classique pour le mod\`ele minimal $(2,1)$ est
\beq
\curve_{(2,1)}(x,y) = y^2 +t_1 - x ,
\eeq
i.e.
\beq
Q(\zeta)=\zeta^2+t_1 \virg P(\zeta)=\zeta.
\eeq
On l'\'etudiera plus en d\'etail dans la partie \ref{sectairy} puisqu'elle d\'ecrit le comportement g\'en\'erique de n'importe quelle courbe au voisinage d'un point de branchement et coincide avec la loi bien connue
de Tracy-Widom \cite{TWlaw}.

\vs

\noindent $\bullet$ {\bf Gravit\'e pure $(3,2)$.}

La param\'etrisation
\beq
Q(\zeta) = \zeta^2 - 2 v
\virg
P(\zeta) = \zeta^3 - 3v \zeta
\virg
t_1=3 v^2
\eeq
donne la courbe spectrale classique
\beq
\curve_{(3,2)}(x,y) = x^3-3v^2 x - y^2 + 2 v^3
\eeq
que nous \'etudierons dans la partie \ref{sectpure} .

\vs

\noindent $\bullet$ {\bf Mod\`ele d'Ising $(4,3)$.}

Si l'on habille la gravit\'e pure d'une structure de spin, c'est-\`{a}-dire que l'on ajoute de la mati\`ere
de la mani\`{e}re la plus simple possible, on obtient la param\'etrisation
\beq
Q(\zeta) = \zeta^3 - 3 v \zeta - 3 w
\virg
P(\zeta) = \zeta^4 - 4v \zeta^2 - 4w\zeta +2v^2 - {5\over 3}t_5(\zeta^2-2v)
\eeq
avec
\beq
t_1 = 4v^3+6w^2  \virg t_2=6vw.
\eeq
La courbe spectrale classique est alors
\bea
\curve_{(4,3)}(x,y)
&=& x^4 - y^3 - 4 v^3 x^2 + 3 v^4 y +   2 v^6 \cr
&& +12 w v ( - x y + v^2  x)
+6 w^2 (-   x^2 + 2 v  y - 4 v^3 )
+ 8 w^3 x
- 3 w^4
 \cr
&& + 5 t_5(-   x^2 y-    v^2 x^2+ 2   v^3 y+ 2 v^5
- 2   w y x+ 2   v^2 w x+ 3   w^2 y-  17 v^2 w^2) \cr
&& + {25\over 3} t_5^2 ( v^2 y+ 2 v^4- 4 v w x- 12 v w^2) \cr
&& + {125\over 27} t_5^3 (-x^2+ 2 v^3- 6  w x- 9 w^2).
\eea
Les variations par rapport aux modules $t_5$, $t_2$ et $t_1$ correspondent \`{a} des variations de la forme $ydx$ donn\'ees respectivement par
\beq
\Omega_5 = -d\L_5 = d(\zeta^5 - 5 v \zeta^3 + 10 v^2 \zeta),
\eeq
\beq
\Omega_1 = -d\L_1 = d(\zeta +{5\over 12}\,{t_5\over v^3-w^2}\,(2v^2\zeta + 2 v w - w \zeta^2)),
\eeq
et
\beq
\Omega_2 = -d\L_2 = d(\zeta^2 +{5\over 6}\,{t_5\over v^3-w^2}\,(v^2\zeta^2 - 2 v w \zeta  -2 w^2))
\eeq
en utilisant les notations de \ref{variatmod}.

\vs

\noindent $\bullet$ {\bf Mod\`eles unitaires $(q+1,q)$.}

Lorsque $p:=q+1$, la courbe spectrale classique prend la forme:
\beq
\curve_{(q+1,q)}(x,y)=T_{q+1}(x)-T_q(y)
\eeq
o\`{u} $T_p$ est le $p$-i\`{e}me polyn\^ome de Tchebychev.
Elle est alors param\'etr\'ee par
\beq
Q(\zeta) = T_q(\zeta)
\virg
P(\zeta) = T_{q+1}(\zeta).
\eeq

\subsection{Application \`{a} l'int\'egrale de Kontsevich.}
\label{sectkontse}
Dans ce paragraphe nous \'etudions un cas particulier tr\`es simple de mod\`{e}le \`{a} une matrice en champ ext\'erieur
introduit par Kontsevich dans le cadre de l'\'etude des nombres d'intersection des surfaces de Riemann \cite{kontsevich}. Il serait int\'eressant
de d\'evelopper le profond lien entre les invariants introduits dans cette th\`ese et le calcul de ces nombres d'intersection
et la conjecture de Witten-Kontsevich \cite{Witten}, mais ce n'est pas le propos de ce paragraphe et nous discuterons brievement ce point
au cours de la conclusion.

L'int\'egrale de Kontsevich est la fonction g\'en\'eratrice permettant de calculer les nombres d'intersection de l'espace des
modules des surfaces de Riemann (cf. \cite{kontsevich, DReview}). Elle est d\'efinie par:
\beq
Z_{\rm Kontsevich} = \int dM\,\,\ee{-N\Tr ({M^3\over 3}-M (\L^2+t_1))} \virg t_1={1\over N}\Tr {1\over \L}
\eeq
o\`{u} $\L$ a pour valeurs propres $\l_1,\dots,\l_N$.
Cette int\'egrale correspond donc au mod\`ele \`a une matrice en champ ext\'erieur $Z_{\rm Mext}$ introduit dans la partie \ref{partMext} du
chapitre 3 avec $V(x)=x^3/3$ et $\hat\L=\L^2 + t_1$.
La courbe spectrale classique correspondante est donc
\beq
\curve_{\rm Kontsevich}(x,y) = (x^2-y)S(y)-xS_1(y)-S_2(y)
\eeq
o\`{u} $S_1(y)$ et $S_2(y)$ sont des polyn\^omes de degr\'es inf\'erieurs strictement \`{a} $s$ qui est le nombre de valeurs propres
distinctes de $\L$.

Pour que $F_{\rm Kontsevich}^{(g)}$ soit bien la fonction g\'en\'eratrice des nombres d'intersection, cette courbe spectrale
doit \^etre de genre nul n\'ecessairement. Alors, on peut \'ecrire explicitement une param\'etrisation rationnelle:
\beq
\curve_{\rm Kontsevich}(x,y) =
\left\{
\begin{array}{l}
x(z) = z + {1\over 2N}\,\Tr {1\over \L}{1\over z-\L}\cr
y(z) = z^2 + {1\over N}\,\Tr{1\over \L}
\end{array}
\right. .
\eeq
Sous cette forme, on peut voir que les points de branchements en $x$ sont nombreux et compliqu\'es \`{a} d\'ecrire.
Heureusement, l'invariance symplectique du th.\ref{thinvsymp} nous permet de travailler avec les r\^oles de $x$ et $y$
invers\'es. Nous allons donc \'etudier la forme $xdy$ et les points de branchement en $y$.
En fait, il y a un unique point de branchement en $y$ solution de $y'(z)=0$: il est situ\'e \`{a} l'origine $z=0$.
Or, les formules d\'efinissant les $F^{(g)}$ consistent en le calcul de r\'esidus en ce point de branchement, on a donc
uniquement besoin de conna\^itre le d\'eveloppement de Taylor de $x(z)$ autour de $z=0$ pour calculer tous les $F^{(g)}$,
i.e.
\beq
\curve_{\rm Kontsevich}(x,y) =
\left\{
\begin{array}{l}
x(z) = z - {1\over 2}\sum_{k=0}^\infty t_{k+2} z^k  \cr
y(z) = z^2 + t_1\cr
\end{array}
\right.
\eeq
o\`{u} l'on a d\'efini les temps de Kontsevich:
\beq
t_k = {1\over N}\Tr \L^{-k}
\eeq
de mani\`ere \`a ce que:
\beq
\encadremath{
F^{(g)}_{\rm Kontsevich} = F^{(g)}(\curve_{\rm Kontsevich}).
}\eeq

On peut une nouvelle fois utiliser l'invariance symplectique du th\'eor\`eme \ref{thinvsymp} et ainsi ajouter \`{a} $x(z)$ n'importe
quelle fonction rationnelle de $y(z)$. On obtient ainsi le th\'eor\`eme suivant:
\bt
$F_{\rm Kontsevich}^{(g)}$ d\'epend uniquement des temps impairs $t_{2k+1}$, with $k\leq 3g-2$:
\beq
\encadremath{
F_{\rm Kontsevich}^{(g)}=F_{\rm Kontsevich}^{(g)}(t_1,t_3,t_5,\dots,t_{6g-3})
}\eeq
\et

Supposons maintenant que $t_k=0$ pour tout $k\geq p+2$. La courbe spectrale classique devient alors:
\beq
\left\{
\begin{array}{l}
x(z) = z - {1\over 2}\sum_{k=0}^p  t_{k+2} z^k  \cr
y(z) = z^2 +t_1 \cr
\end{array}
\right.
\eeq
qui n'est autre que la courbe du mod\`ele minimal $(p,2)$ introduite dans le paragraphe pr\'ec\'edent.
D'une part cela d\'emontre ais\'ement le r\'esultat obtenu par Ambjorn et Kristjansen \cite{AK} liant la double limite d'\'echelle du mod\`ele
\`a une matrice et l'int\'egrale de Kontsevich. D'autre part, cela d\'emontre en une ligne le th\'eor\`eme classique

\bt
$Z_{\rm Kontsevich}$ est la fonction $\tau$ de la hierarchie de KdV.
\et

\subsubsection{Quelques fonctions de corr\'elations et \'energies libres.}

Pour la courbe de Kontsevich, les \'el\'ements de base sont donn\'es par:
\beq
B(z,z')={dz \, dz'\over (z-z')^2}
\virg
dE_z(z') = {z \,dz'\over z^2-z'^2}
\;
, \;
\om(z)=2z^2 dz \,(2 - \sum_j t_{2j+3}\, z^{2j})
\eeq
et l'unique point de branchement est situ\'e en $z=0$.

Suivant la d\'efinition \ref{defloopfctions}, on calcule les premi\`eres fonctions de corr\'elation:
\bea
W_1^{(1)}(z)
&=& - {dz\over 8 (2-t_3)} \,\left({1\over z^4}+{t_5\over (2-t_3)z^2}\right), \cr
\eea

\bea
W_3^{(0)}(z_1,z_2,z_3)
&=& - {1 \over 2-t_3}\, {dz_1\,dz_2\,dz_3\over z_1^2 z_2^2 z_3^2},\cr
\eea

\bea
 {W_2^{(1)}(z_1,z_2) }
&=& {dz_1 \, dz_2 \over 8 (2-t_3)^4 z_1^6 z_2^6} \Big[ (2-t_3)^2 ( 5 z_1^4 + 5 z_2^4+3z_1^2 z_2^2) \cr
&& \qquad + 6 t_5^2 z_1^4 z_2^4 + (2-t_3) \big(6 t_5 z_1^4 z_2^2 +6 t_5 z_1^2 z_2^4 + 5 t_7 z_1^4 z_2^4\big) \Big] \cr
\eea
et
\bea
{W_1^{(2)}(z) } &=& - { dz \over 128 (2-t_3)^7 z^{10}} \Big[ 252\, t_5^4 z^8 + 12\, t_5^2 z^6 (2-t_3) (50\, t_7 z^2 + 21\, t_5) \cr
&& \quad + z^4 (2-t_3)^2 ( 252\, t_5^2 + 348\, t_5 t_7 z^2 + 145\, t_7^2 z^4 + 308\, t_5 t_9 z^4) \cr
&& \qquad + z^2 (2-t_3) (203\, t_5 +145\, z^2 t_7 + 105\, z^4 t_9 +105\, z^6 t_{11}) \cr
&& \qquad \quad + 105\, (2 -t_3)^4 \Big] .\cr
\eea

\vs

Les premiers termes du d\'eveloppement topologique de l'\'energie libre sont alors donn\'es par
\beq
F_{\rm Kontsevich}^{(1)} = - {1 \over 24} \ln\left(1-{t_3\over 2} \right)
\eeq
et
\beq
F_{\rm Kontsevich}^{(2)} = {1 \over 1920}\, {252\, t_5^3 + 435\, t_5 t_7 (2-t_3) + 175\, t_9 (2-t_3)^2 \over (2 -t_3)^5 }.
\eeq
Ces expressions coincident bien avec les r\'esultats d\'ej\`a connus dans la litt\'erature \cite{IZK}.

Il est ici facile de calculer les termes suivants du d\'eveloppement topologique en utilisant la m\^eme m\'ethode (on peut facilement
\'ecrire un programme faisant le travail).

\subsection{Exemple: courbe d'Airy.}
\label{sectairy}

La courbe $y=\sqrt{x}$ est d'un int\'er\^et particulier. En effet, non seulement elle correspond au mod\`{e}le minimal le plus
simple (cf \cite{ZJDFG, BookPDF}), i.e. $(2,1)$, appel\'e \'egalement loi de Tracy-Widom \cite{TWlaw}, mais elle
est le prototype du comportement de n'importe quelle courbe au voisinage de l'un de ses points de branchements.

\bigskip

Consid\'erons la courbe
\beq
\curve(x,y)=y^2-x
\eeq
et l'uniformisation $p=y$:
\beq
\left\{
\begin{array}{lll}
x(p)&=& p^2 \cr
y(p)&=& p
\end{array}.
\right.
\eeq
Il n'y a ici qu'un p\^ole situ\'e en $p = \infty$, et un seul point de branchement en $p=0$. Le point conjugu\'e
est alors simplement $\overline{p}=-p$ et le noyau de Bergmann:
\beq
B(p,q)={dp \, dq\over (p-q)^2}
\virg
dE_q(p) = {q \,dp\over q^2-p^2}
\virg
\om(q)=4q^2 dq.
\eeq

On peut ais\'ement voir que toutes les fonctions de corr\'elations avec $2g+k\geq 3$ sont de la forme:
\beq
W_k^{(g)}(p_1,\dots,p_k) = \om_k^{(g)}(p_1^2,\dots,p_k^2)\,dp_1\dots dp_k
\eeq
On peut m\^eme observer que les r\`egles diagrammatiques sont homog\`enes et donc les fonctions $W_1^{(g)}(p)$ doivent \^etre
homog\`enes en $p$. On peut montrer que:
\beq
W_1^{(g)}(p) = {c_g\, dp\over p^{6g-2}}
\eeq
et que la fonction de corr\'elation \`{a} un point resomm\'ee est
\beq\label{Adefiry1ptfc}
W_1(p,N) := - N ydx + \sum_{g=1}^\infty N^{1-2g} W^{(g)}_1(p) = W_1(N^{1\over 3}p,1).
\eeq
De m\^eme, la fonction \`{a} deux points est donn\'ee par
\beq\label{Adefiry2ptfc}
W_2(p,q,N) := \sum_{g=0}^\infty N^{-2g} W^{(g)}_2(p,q) = W_2(N^{1\over 3}p,N^{1\over 3}q,1)
\eeq
et plus g\'en\'eralement
\beq\label{Adefirykptfc}
W_k(p_1,\dots,p_k,N) := \sum_{g=0}^\infty N^{2-2g-k} W^{(g)}_k(p_1,\dots,p_k) = W_k(N^{1\over 3}p_1,\dots,N^{1\over 3}p_k,1).
\eeq

\medskip

En fait, la solution de la relation de r\'ecurrence \eq{defWkgrecursive} peut s'exprimer explicitement en termes de la fonction
d'Airy.
Pour cela, consid\'erons $g(x)=Ai'(x)/Ai(x)$ o\`u $Ai(x)$ est la fonction d'Airy, i.e. $g'(x)+g^2(x)=x=p^2$.
En termes de la variable $p$, on traduit cela par:
\beq
f(p)=g(p^2)
\virg
f^2 + {f'\over 2p} = p^2
\eeq
dont le d\'eveloppement \`{a} grand $p$ donne:
\beq
f(p) = \sum_{k=0}^\infty f_k p^{1-3k}
= p - {1\over 4 p^2} - {9\over 32 p^5} + \dots
\eeq
o\`u les coefficients du d\'eveloppement satisfont:
\beq
{4-3k\over 2}f_{k-1}+\sum_{j=0}^k f_j f_{k-j} = \delta_{k,0}.
\eeq

Les fonctions \`{a} un et deux points sont alors obtenues sous la forme:
\beq
W_1(p,1) = -2 {p^2-f(p)f(-p)\over f(p)-f(-p)}\, p dp
= -2p^2dp + {dp\over (2p)^4} + {9!!\,dp\over 3^2\,(2p)^{10}}+{15!! dp \over 3^4\,(2p)^{16}}\dots
\eeq
et
\beq
W_2(p,p',1) = - 4\,{(f(p)-f(p'))\,(f(-p)-f(-p'))\over (p^2-p'^2)^2\,(f(p)-f(-p))\,(f(p')-f(-p'))}\, pdp\, p'dp'.
\eeq
En particulier
\beq
W_2(p,p,1) = {f'(p) f'(-p)\over (f(p)-f(-p))^2}\, dp^2
\eeq
d'o\`u la fonction \`a deux points non corr\'el\'ee
\beq
W_2(p,p,1)+W_1(p,1)^2  = 4p^4\, dp^2 = x\, dx^2.
\eeq

De mani\`ere analogue, on trouve par exemple la fonction \`{a} trois points
\bea
&& W_3(p_1,p_2,p_3,1) \cr
&= &{dx_1 dx_2 dx_3 \over (p_3^2-p_2^2)(p_3^2-p_1^2)(p_2^2-p_1^2)} \times \cr
&& \;\; \times {\sum_{i=1}^3 f(p_i)f(-p_i)(f(p_{i-1})+f(-p_{i-1})-f(p_{i+1})-f(-p_{i+1}))
\over (f(p_1)-f(-p_1))(f(p_2)-f(-p_2))(f(p_3)-f(-p_3))} \cr
&=& {dp_1 \,dp_2 \,dp_3 \over 2\, p_1^2\, p_2^2\, p_3^2} + {dp_1 \,dp_2 \,dp_3 \over 2^6\, p_1^8\, p_2^8\, p_3^8} + \dots  \cr
\eea
ainsi que des expressions similaires pour n'importe quel $W_k$.

\medskip

En fait, on sait par ailleurs que toutes les fonctions de corr\'elations peuvent s'\'ecrire sous forme de d\'eterminant
\cite{Dysondet, eynmetha} en utilisant le noyau de Tracy-Widom \cite{TWlaw}:
\beq
K(x,x')= {Ai(x)Ai'(x')-Ai'(x)Ai(x')\over x-x'},
\eeq
ce qui a un fort lien avec les propri\'et\'es d'int\'egrabilit\'e de ces syst\`{e}mes.
En effet, le fait que la fonction d'Airy satisfasse l'\'equation diff\'erentielle $Ai''=x Ai$ peut \^etre vu comme une cons\'equence
du fait que $A_i$ soit une fonction de Baker--Akhiezer ainsi que des \'equations de Hirota d\'emontr\'ees dans le th\'eor\`eme
\ref{thHirota} de la partie suivante.

\bigskip

\br
Pour cette courbe, l'\'energie libre est nulle:
\beq
\forall g\quad F^{(g)}=0 .
\eeq
\er

\subsection{Exemple: la gravit\'e pure (3,2)}\label{sectpure}

Etudions maintenant en d\'etails le mod\`ele minimal $(3,2)$ souvent appel\'e mod\`{e}le de la gravit\'e pure \cite{ZJDFG}.
Pour cela, consid\'erons la courbe alg\'ebrique:
\beq
\curve_{(3,2)}= \;
\left\{
\begin{array}{l}
x(z) = z^2-2v \cr
y(z) = z^3-3v z \cr
t_1=3 v^2
\end{array}\right. .
\eeq
Remarquons tout d'abord que, puisque ce mod\`ele est unitaire, l'uniformisation est obtenue par les polyn\^omes de
Tchebychev $T_2$ et $T_3$, qui satisfont la relation de Poisson \eq{Poissonpq}.
Par changement d'\'echelle $z=\sqrt{v}\, p$, on peut \'ecrire la courbe comme:
\beq
\curve_{(3,2)} = \;
\left\{
\begin{array}{l}
x(p) = v(p^2-2) \cr
y(z) = v^{3/2} (p^3-3 p) \cr
t_1=3 v^2
\end{array}\right. .
\eeq
Il n'y a alors qu'un seul point de branchement situ\'e en $p=0$ et le point conjugu\'e est alors d\'efini par $\overline{p}=-p$.
Comme pr\'ec\'edemment le noyau de Bergmann est celui de la sph\`ere:
\beq
B(p,q)={dp \, dq\over (p-q)^2}
\virg
dE_q(p) = {q \,dp\over q^2-p^2}
\eeq
\beq
\om(q) = (y(q)-y(\qbar))dx(q) = 4v^{5/2}\, (q^2-3)\, q^2\, dq
\eeq
\beq
\Phi(q) = v^{5/2}({2\, q^5\over 5} - 2 q^3)
\eeq
La variation de $ydx$ correspondant \`a une variation de $t_1$ est donn\'ee par:
\beq
\Omega_1(p) = - \left.{\partial y(p)dx(p)\over \partial t_1}\right|_{x(p)} = v^{1/2}\,dp = -v^{1/2}\,\Res_\infty q B(p,q)
\virg \L_1(q)= -v^{1/2}\,q
\eeq
donc l'effet de $\partial/\partial t_1$ sur les fonctions de corr\'elations est d\'ecrit par:
\beq\label{ddt32}
\left.{\partial \over \partial t_1} W_k^{(g)}\right|_{x} = - v^{1/2}\, \Res_{q\to \infty} q  W_{k+1}^{(g)}.
\eeq

\subsubsection{Quelques fonctions de corr\'elation et \'energies libres.}

On trouve les premi\`eres fonctions de corr\'elations:
\beq\label{W30pq32}
W_{3}^{(0)}({p}_1,{p}_2,{p}_3) = - {v^{-5/2}\over 6}\,\,{d{p}_1 d{p}_2 d{p}_3\over {p}_1^2 {p}_2^2 {p}_3^2}
\eeq

\beq\label{W11pq32}
W_{1}^{(1)}(p) = -{v^{-5/2}\over (12)^2}\,{p^2+3\over p^4}\,dp
\eeq

\beq\label{W21pq32}
W_{2}^{(1)}({p},{q}) = v^{-5}\,\,{15{q}^4 + 15{p}^4 + 6{p}^4{q}^2 + 2{p}^4{q}^4 + 9{p}^2{q}^2 + 6{p}^2{q}^4\over 2^5\,3^3\,\,{p}^6\,{q}^6}\,dp\,dq
\eeq

\beq\label{W12pq32}
W_{1}^{(2)}(p) = - v^{-15/2}\,\,
7 {135 + 87 p^2 + 36 p^4 + 12 p^6 + 4 p^8\over 2^{10}\, 3^5\,p^{10}}\, dp
\eeq

\beq
W_{4}^{(0)}({p}_1,{p}_2,{p}_3,{p}_4) = {v^{-5}\over 9 {p}_1^2 {p}_2^2 {p}_3^2 {p}_4^2}\,\left(1+3\sum_i {1\over {p}_i^2}\right)\,dp_1 dp_2 dp_3 dp_4
\eeq

\beq
{W_{5}^{(0)}({p}_1,{p}_2,{p}_3,{p}_4,{p}_5) \over dp_1 dp_2 dp_3 dp_4 dp_5} = {v^{-15/2}\over 9 {p}_1^2 {p}_2^2 {p}_3^2 {p}_4^2 {p}_5^2}\,\left(1+3\sum_i {1\over {p}_i^2}+6\sum_{i<j} {1\over {p}_i^2{p}_j^2}+5\sum_i {1\over {p}_i^4}\right)
\eeq

Encore une fois, il est ais\'e d'obtenir de cette fa\c{c}on n'importe quelle fonction de corr\'elation.

On peut \'egalement calculer les diff\'erents termes du d\'eveloppement topologique de l'\'energie libre
Par exemple, le th\'eor\`eme \ref{variat} et \eq{ddt32}, permettent de calculer:
\beq
{\partial^3 F^{(0)}\over \partial t_1^3} = -{1\over 6v} = -{1\over 2\sqrt{3 t_1}}
\longrightarrow
{\partial^2 F^{(0)}\over \partial t_1^2} = -{t_1^{1/2}\over \sqrt{3}}
\eeq
en utilisant  eq.\ref{W30pq32}.

De plus eq.\ref{W11pq32} implique:
\beq
{\partial F^{(1)}\over \partial t_1} = -{1\over (12)^2 v^2}=-{1\over 48 t_1}
\longrightarrow
{\partial^2 F^{(1)}\over \partial t_1^2} = {1\over 48 \,t_1^2}
\eeq
et eq.\ref{W12pq32} donne:
\beq
{\partial F^{(2)}\over \partial t_1} =  - v^{-7}\,\,{ 7 \over 2^8\, 3^5} = - { 7\over 2^8\, 3^{3/2}\, t_1^{7/2}}
\longrightarrow
{\partial^2 F^{(2)}\over \partial t_1^2} = {49\over 2^9\,3^{3/2}\,t_1^{9/2}} .
\eeq
On peut alors v\'erifier que la d\'eriv\'ee seconde de l'\'energie libre:
\beq
u = {\partial^2 F\over \partial t_1^2} = \sum_{g=0}^\infty t_1^{(1-5 g)/2} u^{(g)}
\virg
F^{(g)}={4 u^{(g)}\over 5(1-g)(3-5g)}\, .
\eeq
satisfait l'\'equation de Painlev\'e I ordre par ordre:
\beq
u^2+{1\over 6}u'' = {1\over 3}t_1 .
\eeq
Il est bien connu que cette \'equation doit \^etre satisfaite \`{a} tous les ordres \cite{Brezindsl,Douglas,Migdal,ZJDFG}.
Ceci peut une nouvelle fois \^etre vu comme une cons\'equence des \'equations de Hirota pr\'esent\'ees dans le th.\ref{thHirota}.


\section{Int\'egrabilit\'e des invariants alg\'ebriques.}
\label{sectintegrability}

Comme  nous l'avons vu, il exste plusieurs exemples identifiant des fonctions de partitions de mod\`eles de matrices
comme les fonctions $\tau$ d'un syst\`eme int\'egrable. Nous allons montrer dans cette partie comment diff\'erents indices
laissent supposer que notre construction donne acc\`es \`a une fonction $\tau$ (formelle pour le moment)
associ\'ee \`a une courbe alg\'ebrique quelconque.

\subsection{Fonction de Baker-Akhiezer.}

Etant donn\'es deux points $\xi$ et $\eta$ du domaine fondamental, on d\'efinit le noyau suivant comme une s\'erie formelle
en ${1 \over N}$:
\bea
K_N(\xi,\eta) &=& { \ee{- N\int_\eta^\xi y dx}\over E(\xi,\eta)\,\sqrt{dx(\xi)dx(\eta)}} \cr
&& \; \exp{\left( - \sum_{g=0}^\infty \sum_{l=1, 2-2g-l<0}^\infty {1\over l!}\,\,N^{2-2g-l}\,\,\int_\eta^\xi\int_\eta^\xi\dots\int_\eta^\xi W^{(g)}_l(p_1,\dots,p_l) \right)}
\cr
\eea
o\`u le chemin d'int\'egration reste enti\`erement dans le domaine fondamental. Il satisfait les propri\'et\'es suivantes:
\begin{itemize}

\item On peut voir que $(x(\xi)-x(\eta))K_N(\xi,\eta) \to 1$ quand $\eta\to\xi$.

\item On a
\beq
\mathop{{\rm lim}}_{\eta\to\xi}\, \Big( K_N(\xi,\eta) - {1\over (x(\xi)-x(\eta))} \Big) = -Ny(\xi) + {W_1(\xi)\over dx(\xi)}
\eeq
o\`u $W_1={\displaystyle \sum_{g=1}^\infty} N^{1-2g} W_1^{(g)}$.

\item L'\'echange des deux arguments se traduit par:
\beq
K_N(\xi,\eta) = K_{-N}(\eta,\xi).
\eeq

\item A priori, $K_N$ pourrait avoir des p\^oles aux points de branchement puisque son logarithme $\ln K_N$ en a. Cependant,
d'apr\`es le th\'eor\`eme de passage \`a la limite th.\ref{thsinglimit}, on peut voir que le terme dominant de tous les $W_l^{(g)}$ est donn\'e par
les $W_l^{(g)}$ correspondant \`{a} la courbe d'Airy $y^2 = x$ \'etudi\'ee dans la partie \ref{sectairy} du chapitre pr\'ec\'edent.
Ainsi, au voisinage d'un point de branchement $a$, le noyau $K_N$ se comporte comme le noyau de Tracy-Widom \cite{TWlaw}
quand $\xi,\eta \to a$:
\bea
K_N(\xi,\eta) \sim {Ai(\hat\xi)Ai'(\hat\eta)-Ai'(\hat\xi)Ai(\hat\eta)\over \hat\xi-\hat\eta} \cr
\hat\xi=N^{2/3}(x(\xi)-x(a)) \virg \hat\eta=N^{2/3}(x(\eta)-x(a))
\eea
Donc $K_N$ n'est pas singulier aux points de branchements.

\item Les seules singularit\'es de $K_N(\xi,\eta)$ sont: d'une part, des singularit\'es essentielles aux p\^oles de $ydx$ avec une partie singuli\`ere
\'egale \`{a} $\exp{(-N\int^\xi_\eta ydx)}$ et un p\^ole simple en $\xi = \eta$ d'autre part.

\end{itemize}

\bigskip

Consid\'erons maintenant un p\^ole $\alpha$ de $ydx$. Pour $\xi$ dans un voisinage de $\alpha$, on d\'efinit:
\bea
\psi_{\alpha,N}(\xi) &=& {\ee{-N\, V_\alpha(\xi)} \ee{-N\int_\alpha^\xi (ydx-d V_\alpha+t_\alpha {dz_\alpha\over z_\alpha})} \over E(\xi,\alpha)\,\sqrt{dx(\xi) d\zeta_\alpha(\alpha)}}\,\left(z_\alpha(\xi)\right)^{N t_\alpha}  \cr
&& \exp{\left( - \sum_{g=0}^\infty \sum_{l=1, 2-2g-l<0}^\infty {1\over l!}\,\,N^{2-2g-l}\,\,\int_\alpha^\xi\int_\alpha^\xi\dots\int_\alpha^\xi W^{(g)}_l(p_1,\dots,p_l)\right)} \cr
&=&\mathop{{\rm lim}}_{\eta\to\alpha} \left( K(\xi,\eta)\,\sqrt{dx(\eta)\over d\zeta_\alpha(\eta)}\,\ee{-NV_\alpha(\eta)}\,\left(z_\alpha(\eta)\right)^{Nt_\alpha}\,\right)
\eea
o\`{u} $\zeta_\alpha$ est encore un param\`etre local au voisinage de $\alpha$,
$\zeta_\alpha={1\over z_\alpha}$,
et
$
\phi_{\alpha,N}(\xi) = \psi_{\alpha,-N}(\xi)
$.

On appelle ces fonctions, {\bf fonctions de Baker-Akhiezer} car elles poss\`edent les propri\'et\'es suivantes tr\`es similaires \`a
celles des fonctions de Baker-Akhiezer introduites dans le cadre des syst\`emes int\'egrables classiques (cf \cite{BBT}):
\begin{itemize}

\item $\psi_{\alpha,N}$ est d\'efinie seulement dans le voisinage de $\alpha$ mais on peut facilement la prolonger analytiquement
sur la courbe enti\`ere en choisissant un point de base arbitraire $o$ dans le voisinage de $\alpha$ et en \'ecrivant:
\bea
&& \int_{\alpha}^\xi (ydx-dV_\alpha+t_\alpha {dz_\alpha\over z_\alpha}) + V_\alpha(\xi) - t_\alpha\ln{(z_\alpha(\xi))} \cr
&=&
\int_o^\xi ydx + \int_{\alpha}^o (ydx-dV_\alpha+t_\alpha {dz_\alpha\over z_\alpha}) + V_\alpha(o) - t_\alpha\ln{(z_\alpha(o))}.
\eea

\item Comme $K_N$, ces fonctions pourraient \`a priori avoir des p\^oles aux points de banchements. Mais de fa\c{c}on similaire,
le th\'eor\`eme \ref{thsinglimit} montre que lorsque les arguments approchent un point de branchement $a$, $\xi \to a$,
les fonctions de Baker-Akhiezer se comportent suivant:
\bea
\psi_{\alpha,N}(\xi) \sim  C\, Ai(\hat\xi) \virg \hat\xi=N^{2/3}(x(\xi)-x(a))
\eea
o\`u $C={\psi_{\alpha,N}(a)/ Ai(0)}$ est une constante de normalisation.

\item Les seules singularit\'es de $\psi_{\alpha,N}$ sont des singularit\'es essentielles aux p\^oles de $ydx$ o\`u la partie singuli\`ere est donn\'ee par
$\exp{(-N\int^\xi ydx)}$.

\end{itemize}

\br
En fait, ces fonctions $\psi_{\alpha,N}$ ne sont pas \`a proprement parler des fonctions de Bakher-Akhiezer. En effet,
elle ont des int\'egrales non nulles autour de cycles non-triviaux: elles ne reprennent pas la m\^eme valeur apr\`es que
leur argument ai fait un tour autour de l'un de ces cycles. Les fonctions de Baker-Akhizer, quant \`a elles,  ne doivent
pas changer de valeur apr\`es un tour de leur variable autour de n'importe quel cycle. On peut donner cette propri\'et\'e
aux fonctions $\psi$ par normalisation par des fonctions $\theta$ appropri\'ees mais d\'etruisant ainsi le d\'eveloppement
en ${1 \over N^2}$. L'\'etude de cette normalisation est tr\`es int\'eressante mais sort largement du propos de cette th\`ese,
c'est pourquoi nous n'aborderons pas plus le sujet.

Nous nous restreindrons donc aux $\psi_{\alpha,N}$ eux m\^emes qui peuvent \^etre vus comme des "fonctions de Baker-Akhiezer
formelles", c'est-\`{a}-dire qu'elles satisfont des \'equations de Hirota ordre par ordre en ${1 \over N}$ comme nous le montrons
dans le prochain paragraphe.
\er

\subsection{Relation de Sato.}

Etant donn\'es deux points $\xi$ et $\eta$ de la surface de Riemann compacte $\Sigma$ et un nombre complexe $r$, on d\'efinit
la courbe:
\beq
\curve + r[\xi,-\eta] = \left\{ \left(x(p),y(p)+r{dS_{\xi,\eta}(p)\over dx(p)}\right)\quad , \,\, p\in \Sigma  \right\}
\eeq
La diff\'erentielle $ydx + rdS_{\xi,\eta}$ a les m\^emes int\'egrales autour des cycles $\acycle$ que $ydx$. Elle a \'egalement
les m\^emes p\^oles avec la m\^eme partie singuli\`ere plus deux nouveaux p\^oles simples, l'un en $p = \xi$ avec r\'esidu $r$
et l'autre en $p = \eta$ avec r\'esidu $-r$.

On peut alors d\'emontrer la {\bf relation de Sato}:
\bt\label{thSato}
\beq
K_N(\xi,\eta) = {Z_N(\curve+{1\over N}[\xi,-\eta])\over Z_N(\curve)}
\virg
\psi_{\alpha,N}(\xi) = {Z_N(\curve+{1\over N}[\xi,\alpha])\over Z_N(\curve)}.
\eeq
\et
Encore une fois, cette relation est \`{a} comprendre au sens des s\'eries formelles en ${1 \over N}$.

\subsection{Equations de Hirota}
Consid\'erons deux courbes alg\'ebriques $\curve(x,y)$ et $\td\curve(x,y)$ avec la m\^eme structure conforme,
on a alors les relations bilin\'eaires:
\bt\label{thHirota}
\beq
\Res_{\eta\to\zeta} dx(\eta)\,K_N(\xi,\eta)\,\td{K}_{\td{N}}(\eta,\zeta) = K_N(\xi,\zeta)
\eeq
et
\beq\label{hirotapsi}
\Res_{\xi\to \alpha} dx(\xi)\,\psi_{\alpha,N}(\xi)\,\td{\psi}_{\alpha,-\td{N}}(\xi) = 0 \qquad \qquad {\rm if}\,\, \td{N}\td{t}_\alpha >  N t_\alpha+1
\eeq
qui prennent la m\^eme forme que l'\'equation de Hirota des syst\`emes int\'egrables classiques \cite{BBT, kostovhirota}
en les consid\'erant comme des relations entre s\'eries formelles en ${1 \over N}$.
\et

\section{Conclusion du chapitre.}

Nous avons montr\'e dans ce chapitre comment construire, \`a partir de n'importe quelle courbe alg\'ebrique, un ensemble infini
de nombres et de formes diff\'erentielles qui coincident avec les d\'eveloppement topologiques de plusieurs mod\`eles de matrices
lorsque la courbe consid\'er\'ee est une courbe spectrale. Non seulement ceci permet d'unifier ces diff\'erents mod\`eles
dans une m\^eme classe mais de plus, connaissant les propri\'et\'es d'invariance de ces objets sous les d\'eformations
de la courbe alg\'ebrique consid\'er\'ee, nous avons ainsi construit un objet puissant pour comparer diff\'erents mod\`eles:
pour comparer deux mod\`eles il suffit de voir si leurs courbes spectrales appartiennent \`a la m\^eme classe d'\'equivalence
modulo les transformations conservant les $F^{(g)}$.
De plus, nous avons vu que ces propri\'et\'es d'invariance sont un outil tr\`es puissant pour le calcul, comme le montre
l'exemple de l'int\'egrale de Kontsevich o\`u des r\'esultats classiques sont retrouv\'es sans effort.

Nous avons \'egalement fait un premier pas en direction d'un syst\`eme int\'egrable sous-jacent en construisant les briques
\'el\'ementaires d'un syst\`eme int\'egrable classique sur la courbe alg\'ebrique consid\'er\'ee. Cependant, si ce d\'eveloppement
formel semble donner tous les termes d'un d\'eveloppement semi-classique, il n'en donne pas pour autant acc\`es au syst\`eme
quantique sous jacent. En effet, en fixant les fractions de remplissage (et donc en emp\^echant les instantons dans le cas des
mod\`eles de matrices), il semble que l'on a bris\'e une sym\'etrie globale qu'il faut restaurer pour revenir au syst\`eme quantique
correspondant \`a une valeur de $N$ finie. Il reste donc beaucoup \`a explorer dans cette direction.

Une autre direction \`a explorer consiste \`a se demander si cette construction peut s'\'etendre \`a des vari\'et\'es autres
que des surfaces de Riemann obtenues par l'interm\'ediaire d'une courbe alg\'ebrique. En effet, il semblerait que la construction
d\'evelopp\'ee ici ne d\'epende que des points de branchements et pourrait par exemple s'\'etendre \`a certaines
courbes de la forme:
\beq
\CE(x,y) = 0
\eeq
o\`u $\CE$ n'est plus un polyn\^ome mais une fonction analytique en ses deux variables.


\chapter{Mod\`eles de matrices et th\'eorie des cordes.}

Les mod\`{e}les de matrices ont souvent jou\'e un r\^ole, ou du moins tent\'e, dans l'\'etude de diff\'erentes th\'eories des cordes.
Au cours des derni\`eres d\'ec\'ennies, avec l'\'evolution de la compr\'ehension de ces th\'eories, l'interaction entre cordes
et matrices al\'eatoires a pris de nombreuses formes dont nous allons donner ici quelques exemples. Le lecteur int\'eress\'e pourra se
reporter aux revues de Mari\~{n}o sur le sujet \cite{MMbook,MMtopo}.

\section{Th\'eorie des cordes critiques et doubles limites d'\'echelle.}
Les mod\`eles de matrices tels qu'ils sont d\'ecrits dans cette th\`{e}se sont par exemple apparus de mani\`ere d\'etourn\'ee
dans l'\'etude de ce que l'on appelle la th\'eorie des cordes critiques (voir la partie qui leur est consacr\'ee dans le chapitre 1).

Consid\'erons une th\'eorie des cordes correspondant \`a mettre une th\'eorie des champs conforme sur une surface de Riemann
$\Sigma_g$. L'objet principal de cette th\'eorie est alors l'\'energie libre:
\beq
F = \sum_g g_s^{2g-2} F_g \virg F_g = \int {\cal D} h \, {\cal D} \phi e^{-\CS [\phi,h]}
\eeq
o\`u $g_s$ est la constante de couplage des cordes, $h$ la m\'etrique bidimensionnelle sur $\Sigma_g$, $\phi$ les champs de
mati\`ere coupl\'es \`a la gravit\'e par l'action $\CS$ et l'int\'egrale de chemin dans $F_g$ porte sur les configurations de
champs sur $\Sigma_g$. On peut g\'en\'eraliser cette \'energie libre en perturbant l'action de la th\'eorie conforme par
un ensemble d'op\'erateurs \`a l'aide de modules $t_n$:
\beq
\CS [\phi,h,t] : = \CS [\phi,h] + \sum_n t_n \CO_n.
\eeq
Le calcul de l'\'energie libre, tr\`es compliqu\'e en g\'en\'eral, se simplifie beaucoup lorsque la th\'eorie conforme a une charge
centrale \'egale \`a la charge centrale critique $c=26$. En effet, dans ce cas, la m\'etrique se d\'ecouple et il ne reste plus
qu'\`a int\'egrer sur l'espace des modules $\CM_g$ des surfaces de genre $g$:
\beq
F_g = \int_{\CM_g} d\tau \int {\cal D} \phi e^{-\CS [\phi,\tau,t]}
\eeq
o\`{u} $\tau$ est un ensemble de $3g-3$ coordonn\'ees param\'etrisant l'espace des modules. On voit qu'il est alors n\'ecessaire
de d\'ecrire toutes les surfaces $\Sigma_g$, ce qui peut \^etre fait en utilisant les mod\`eles de matrices et leur limite continue.

On peut aller plus loin puisque lorsque la charge centrale satisfait $c<1$, bien que la th\'eorie ne soit plus critique et donc compliqu\'ee
par la pr\'esence du champ de Liouville -- un degr\'e de libert\'e dynamique --, on peut calculer tous les $F_g$
gr\^ace aux mod\`{e}les de matrices.

L'id\'ee de base est assez naturelle. Puisque l'on veut d\'ecrire la feuille d'univers de la corde, on va la discr\'etiser
pour pouvoir la g\'en\'erer par une int\'egrale de matrice formelle comme d\'efinie dans les chapitres 2 et 3 de cette th\`ese.
Les doubles limites d'\'echelle \'etudi\'ees dans la partie \ref{DSL} du chapitre 2 permettent alors d'atteindre des surfaces continues
en imposant une charge centrale $c<1$ \`a la th\'eorie conforme. La partie divergente des diff\'erents termes du d\'eveloppement topologique
du mod\`ele de matrices sont alors identifiables avec les coefficients du d\'eveloppement en $g_s$ de l'\'energie libre
de la th\'eorie des cordes:
\beq
F_g = F^{(g)}(\CE_{DSL})
\eeq
o\`u $\CE_{DSL}$ est la courbe obtenue en prenant la double limite d'\'echelle de la courbe spectrale du mod\`ele de matrices.
De m\^eme, les $W_k^{(g)}$ peuvent \^etre interpr\'et\'es comme des fonctions de corr\'elation de la th\'eorie des cordes: elles
correspondent \`a des amplitudes de cordes ouvertes consistant \`a int\'egrer sur un ensemble de surfaces avec bord.

Notons que, jusqu'\`a maintenant, ce calcul s'effectuait en \'evaluant l'\'energie libre du mod\`ele de matrice lui m\^eme avant
de prendre la double limite d'\'echelle du r\'esultat obtenu. Le r\'esultat de \EOinvariants r\'esum\'e dans la partie
\ref{partsing} du chapitre 4, montre que ce passage \`a la limite n'est pas n\'ecessaire. L'\'energie libre de la th\'eorie des cordes
peut directement \^etre obtenue en calculant les invariants $F^{(g)}$ de la courbe alg\'ebrique $\CE_{(p,q)}$ du mod\`ele
minimal $(p,q)$ voulu sans qu'elle ne vienne d'une quelconque limite!
Ce r\'esultat est un premier indice r\'ev\'elant un lien plus profond entre les invariants alg\'ebriques $F^{(g)}$
et l'\'energie libre de th\'eories des cordes ind\'ependamment de l'existence d'un quelconque mod\`ele de matrice interm\'ediaire.

\section{Th\'eories de cordes topologiques.}

Cette approche des th\'eories des cordes par les mod\`eles de matrices \'etait, jusqu'il y a peu, limit\'ee \`a ce secteur particulier
correspondant aux doubles limites d'\'echelle. Mais l'engouement r\'ecent, en math\'ematiques et en physique, pour les th\'eories des cordes topologiques
a permis de montrer que l'interaction entre les deux mod\`eles est bien plus profonde.

Les th\'eories des cordes topologiques correspondent au cas o\`u la th\'eorie conforme coupl\'ee \`a la gravit\'e
est un mod\`ele Sigma bidimensionnel non lin\'eaire dont l'espace cible $X$ est une vari\'et\'e de Calabi-Yau\footnote{Cette propri\'et\'e
est tr\`es  importante pour que la th\'eorie soit topologique. Nous ne la d\'ecrirons pas ici et le lecteur int\'eress\'e
pourra se r\'eferer aux revues \cite{MMbook,MMtopo} pour plus de d\'etail.} \cite{Laba,Witten1,Witten2}. Cela signifie que les champs de base sont des applications:
\beq
\phi \, : \, \Sigma_g \to X.
\eeq
Il existe en fait deux versions de telles th\'eories: le mod\`ele A li\'e \`a la th\'eorie des invariants de Gromov-Witten et
 le mod\`ele B li\'e aux d\'eformations de la structure complexe de la Calabi-Yau $X$. C'est ce dernier mod\`ele que nous allons
 \'etudier dans la suite.

Explicitons d'abord la structure complexe d'une vari\'et\'e de Calabi-Yau $X$ de dimension complexe $3$.
La structure complexe d'une Calabi-Yau $X$ est donn\'ee par un ensemble de modules formant un espace $\CM$ de dimension
$h^{1,2}(X)$ param\'etrisable comme suit. On commence par choisir une base de $H_3(X,\mathbb{Z})$:
\beq
(A_I , B_I) \virg I = 0,1, \dots , h^{1,2} \; \; \hbox{tels que} \;\; A_I \cap B_J = \delta_{IJ}.
\eeq
On d\'efinit alors les p\'eriodes de la Calabi-Yau par:
\beq
X_I := \int_{A_I} \Omega \; \; \hbox{et} \;\; \CF_I:= \int_{B_I} \Omega
\eeq
o\`{u} $\Omega$ est une section holomorphe non nulle de la fibr\'ee canonique $\Omega^{3,0}(X)$\footnote{L'existence d'une telle
section est assur\'ee par la propri\'et\'e de Calabi-Yau de la vari\'et\'e $X$.}. On peut montrer que les $X_I$ sont des coordonn\'ees projectives
de $\CM$, et donc que les $\CF_I = \CF_I(X)$ en sont des fonctions. On peut m\^eme montrer que leur d\'ependance provient d'une unique
fonction $\CF(X)$ selon:
\beq
\CF_I(X)  = {\partial \CF(X) \over \partial X_I}.
\eeq
Cette fonction est homog\`ene de degr\'e deux en $X_I$, on peut donc introduire des nouvelles coordonn\'ees
\beq
t_I := {X_I \over X_0} \virg I = 1, \dots, h^{1,2}(X)
\eeq
de mani\`ere \`a d\'efinir le {\bf pr\'epotentiel}:
\beq
F_0(t) = {1 \over X_0^2} \CF(X)
\eeq
qui s'av\`ere coincider avec l'\'energie libre de genre 0 des cordes topologiques de type B sur $X$ que nous d\'efinissons
maintenant.

L'action d'une th\'eorie topologique de type B sur $X$ est obtenue par une d\'eformation de l'action standard:
\beq\label{actionS}
S_{\CN = 2} := S(0) + \sum_{I =1}^n t_I \int_{\Sigma_g} \phi_I^{(2)} + \sum_{I =1}^n \overline{t}_I \int_{\Sigma_g} \overline{\phi}_I^{(2)}
\eeq
par les op\'erateurs chiraux $\phi_I$ et anti-chiraux $\overline{\phi}_I$ de la th\'eorie correspondant aux directions marginales.

Notons que la d\'eformation de l'action par les $\overline{t}$ peut s'\'ecrire comme un terme BRST-exact par rapport \`a la sym\'etrie
topologique de la th\'eorie. On s'attendrait alors \`a ce que l'\'energie libre de genre $g$, $F^{(g)}$, ainsi que les fonctions de corr\'elation
\beq
C_{I_1 \dots I_k}^{(g)}:= \left< \int_{\Sigma_g} \phi_{I_1}^{(2)} \dots \int_{\Sigma_g} \phi_{I_k}^{(2)} \right>
\eeq
ne d\'ependent pas de ces param\`etres non-holomorphes $\overline{t}$. Cependant, le couplage \`a la gravit\'e d\'etruit cette
ind\'ependance en introduisant une anomalie connue sous le nom d'{\bf anomalie holomorphe}. Cette anomalie provenant de termes
de bord peut \^etre enti\`erement caract\'eris\'ee par des \'equations de r\'ecurrence: les {\bf \'equations d'anomalie holomorphe} \cite{BCOV}.
Celles-ci peuvent \^etre \'ecrites en utilisant les notations suivantes.
La m\'etrique de Zamolodchikov sur l'espace des modules peut \^etre d\'efinie en termes d'une fonction de K\"ahler $K$ par
\beq
G_{I\overline{J}} = \partial_I \partial_{\overline{J}} K
\eeq
o\`{u} l'on a utilis\'e la notation
\beq
\partial_{I} := {\partial \over \partial t_I} \virg \partial_{\overline{I}} := {\partial \over \partial \overline{t}_I}.
\eeq
L'\'energie libre de genre 0, $F^{(0)}$, donne acc\`es aux fonctions holomorphes \`a deux et trois points:
\beq
\tau_{IJ}  := \partial_I \partial_J F^{(0)} \virg C_{IJK} := \partial_I \partial_J \partial_K F^{(0)}.
\eeq
Les \'equations d'anomalie holomorphes peuvent alors s'\'ecrire:
\beq\label{eqanomalie}
\forall g \geq 2 \, , \;  \partial_{\overline{K}} F^{(g)} = {1 \over 2} C_{\overline{K}}^{IJ}\left( D_I D_J F^{(g-1)}
+ \sum_{h=1}^{g-1} D_I F^{(h)} D_J F^{(g-h)} \right)
\eeq
o\`u la d\'eriv\'ee covariante est d\'efinie par
\beq
D_I = \partial_I - \Gamma_I +(2-2g) \partial_I K
\eeq
avec le symbole de Christoffel
\beq
\Gamma_{IJ}^K:= G^{K\overline{M}} \partial_I G_{J \overline{M}}
\eeq
et le tenseur
\beq
C_{\overline{K}}^{IJ}:=e^{2K} G^{I\overline{M}} G^{J \overline{N}} \overline{C}_{\overline{M}\overline{N}\overline{K}}.
\eeq
Pour le genre 1, l'\'equation s'\'ecrit:
\beq
\partial_I \partial_{\overline{J}} F^{(1)} = {1 \over 2} C_{IKL}C_{\overline{J}}^{KL} - \left( {\chi \over 24} - 1\right) G_{I \overline{J}}.
\eeq

Notons \'egalement que les fonctions de corr\'elation peuvent \^etre obtenues par action de la d\'eriv\'ee covariante sur les
\'energies libres:
\beq
C_{I_1 \dots I_k}^{(g)}  = D_{I_1} \dots D_{I_k} F^{(g)}.
\eeq
On peut donc \'ecrire des \'equations d'anomalie pour elles aussi:
\bea\label{anomalieC}
\partial_{\overline{K}} C_{I_1 \dots I_k}^{(g)} &=&
{1 \over 2} C_{\overline{K}}^{MN} \left( C_{MN I_1 \dots I_k}^{(g-1)} \right. \cr
&& \qquad \left. + \sum_{r=0}^g \sum_{s=0}^k {1 \over s! (n-s)!}
\sum_{\sigma \in S_k} C_{MI_{\sigma(1)} \dots I_{\sigma(s)}}^{(r)} C_{NI_{\sigma(s+1)} \dots I_{\sigma(k)}}^{(g-r)}\right) \cr
&& - (2g-2+n-1)  \sum_{s=1}^k G_{\overline{K}I_s} C_{I_1 \dots I_{s-1} I_{s+1} \dots I_k}^{(g)}.\cr
\eea

Une fois les \'equations d'anomalie holomorphes r\'esolues, il reste \`{a} fixer une constante d'int\'egration, i.e. une fonction
holomorphe des param\`etres. Pour ce faire, on cherche \`{a} d\'eterminer la limite holomorphe de $F^{(g)}(t, \overline{t})$
d\'efinie par:
\beq
F^{(g)}(t):= \lim_{\overline{t}\to \infty} F^{(g)}(t,\overline{t}).
\eeq
Ce passage \`a la limite n'est pas anodin. En effet, si l'on transforme ainsi l'\'energie libre en une fonction holomorphe
des modules, on perd une propri\'et\'e de sym\'etrie: l'invariance modulaire de $F^{(g)}(t,\overline{t})$ est bris\'ee.
Nous avons vu dans le chapitre pr\'ec\'edent que ce comportement se retrouve dans le cadre des invariants alg\'ebriques losque l'on
fait varier $\kappa$.

\br
Les amplitudes d\'efinies ici ont pu \^etre interpr\'et\'ees comme les amplitudes d'une th\'eorie quantique des champs
six dimensionnelle sur la Calabi-Yau $X$: la {\bf th\'eorie de Kodaira-Spencer} \cite{BCOV}. Dans la suite, nous ferons
donc r\'eguli\`erement r\'ef\'erence \`a cette th\'eorie pour parler des amplitudes du mod\`ele de type B.
\er

\section{Conjecture de Dijkgraaf-Vafa et mod\`ele B local.}

\subsection{Vari\'et\'es de Calabi-Yau locales.}

 Parmi les diff\'erentes classes de vari\'et\'es de Calabi-Yau possibles pour l'espace cible, il en existe une qui permet de
rendre le lien entre mod\`ele B et mod\`eles de matrices (ou invariants alg\'ebriques) plus explicite encore. On appelle
les vari\'et\'es non compactes correspondantes des {\bf vari\'et\'es de Calabi-Yau locales} \cite{local}.
Il a \'et\'e monr\'e dans \cite{local} que leurs propri\'et\'es sont alors encod\'ees
dans une surface de Riemann compacte $\overline{\Sigma}_n$ de genre $n$\footnote{Attention, cette surface ne doit pas \^etre confondue
avec la feuille d'univers $\Sigma_g$ sur laquelle portent les int\'egrales du paragraphe pr\'ec\'edent.} par l'interm\'ediaire
des modules
\beq
t_I \propto \int_{\CA_I} ydx \virg I = 1, \dots , n
\eeq
avec $\{(\CA_I,\CB_I)\}$ une base de cycles de $\overline{\Sigma}_n$ et la forme $ydx$ d\'efinie par une param\'etrisation
de $\overline{\Sigma}_n$ comme \'equation alg\'ebrique. Ainsi, au lieu de travailler sur une vari\'et\'e de dimension
3, on travaille sur une vari\'et\'e de dimension 1, une surface de Riemann, tout comme dans le cadre des matrices al\'eatoires.
Dans le paragraphe suivant, nous montrons sur un exemple pr\'ecis comment passer de la Calabi-Yau \`a la surface de Riemann.

Dans ce contexte, la m\'etrique de Zamolodchikov est donn\'ee par
\beq
G_{I \overline{J}} = -i (\tau - \overline{\tau})_{IJ}
\eeq
o\`{u} la fonction \`{a} deux points $\tau$ est simplement la matrice des p\'eriodes de Riemann de $\overline{\Sigma}_n$.
Le symbole de Christoffel et le tenseur $C_{\overline{K}}^{IJ}$ sont quant \`{a} eux donn\'es par:
\beq
\Gamma_{IJ}^L = \left[(\tau - \overline{\tau})^{-1}\right]^{LM} \partial_I \tau_{JM}
\eeq
et
\beq
C_{\overline{K}}^{IJ} = - \left[(\tau - \overline{\tau})^{-1}\right]^{IM} \left[(\tau - \overline{\tau})^{-1}\right]^{JN}
C_{\overline{MNK}}.
\eeq

Les \'equations d'anomalie holomorphes sont alors toujours donn\'ees par \eq{eqanomalie} pour les \'energies libres en consid\'erant
la d\'eriv\'ee covariante:
\beq
D_I:= \partial_I - \Gamma_I.
\eeq
De m\^eme, les fonctions de corr\'elation satisfont toutjours \eq{anomalieC} mais sans le terme de la seconde ligne, i.e:
\beq
\partial_{\overline{K}} C_{I_1 \dots I_k}^{(g)} =
{1 \over 2} C_{\overline{K}}^{MN} \left( C_{MN I_1 \dots I_k}^{(g-1)} + \sum_{r=0}^g \sum_{s=0}^k {1 \over s! (n-s)!}
\sum_{\sigma \in S_k} C_{MI_{\sigma(1)} \dots I_{\sigma(s)}}^{(r)} C_{NI_{\sigma(s+1)} \dots I_{\sigma(k)}}^{(g-r)}\right).
\eeq

\subsection{Conjecture de Dijkgraaf-Vafa.}
Consid\'erons un cas particulier de Calabi-Yau locale donn\'ee par l'\'equation\footnote{C'est bien une vari\'et\'e de dimension
3 complexe: une \'equations polynomiale liant 4 variables complexes $u$, $v$, $x$ et $z$!}
\beq
u^2 + v^2 = H(x,z) \;\;\;\; \hbox{avec} \;\;\;\; H(x,z) = z^2 - (W'(x))^2+f(x)
\eeq
o\`{u} $W(x)$ et $f(x)$ sont deux polyn\^omes de degr\'es respectifs $n+1$ et $n-1$.
La section holomorphe est alors donn\'ee par
\beq
\Omega = {dx dz du \over 2 \pi v}.
\eeq
Pour cette g\'eom\'etrie particuli\`ere, montrons que ses p\'eriodes sur les cycles d'une base symplectique de la Calabi-Yau
peuvent se ramener \`a des int\'egrales sur les cycles de la courbe $\overline{\Sigma}_n$
\beq
y^2(x) = (W'(x))^2+f(x).
\eeq
 Supposons que cette courbe est g\'en\'erique,
c'est-\`a-dire qu'il y a $2 n$ points de branchments distincts:
\beq
(W'(x))^2 - f(x) = \prod_{i = 1}^{2n} (x-x_i).
\eeq
Consid\'erons les cycles $\{\CA_i\}_{i=1 \dots n}$ d\'efinis par
\beq
\forall i = 1, \dots , n \, , \; \; \CA_i:= \left[x_{2i-1},x_{2i} \right].
\eeq
On construit alors les cycles $\tilde{\CA}_i$ de la Calabi-Yau par fibrations sur ces intervalles par l'\'equation:
\beq
\forall j = 1, \dots, n \, , \; \forall x \in \CA_j \, , \; \; u^2 + v^2 + z^2 = \prod_{i=1}^{2n} (x-x_i).
\eeq
Notons que ces cycles sont $n$ 3-sph\`eres de la vari\'et\'e de Calabi-Yau se projetant sur les cycles $A_i$ de la courbe
$H(x,y)=0$ dans le plan de $x$. De m\^{e}me, on d\'efinit les cycles $\tilde{\CB}$ par fibration sur les cycles
$\CB_i$ orthogonaux aux cycles $\CA_i$. Calculons maintenant les p\'eriodes de la section $\Omega$ sur ces cycles.

Soit $\CC$ un cycle $\CA$ ou $\CB$ et $\tilde{\CC}$ le cycle corespondant sur la vari\'et\'e de Calabi-Yau.
Alors
\beq
\int_{\tilde{\CC}} \Omega = \int_{\tilde{\CC}} {dx dz du \over 2 \pi v}.
\eeq
Puisque $\tilde{\CC}$ peut \^etre vu comme une $2$-sph\`ere
\beq
u^2 + v^2 + z^2 = y(x)^2
\eeq
param\'etris\'ee par $x$ se d\'epla\c{c}ant sur $\CC$, on peut d\'ecomposer cette int\'egrale en une int\'egrale
sur cette $2$-sph\`ere de rayon $y(x)$ et une int\'egrale sur $\CC$\footnote{On a suppos\'e ici que $u$, $v$ et $z$ sont
r\'eels lorsque $x \in \CC$. On peut faire exactement la m\^eme chose dans le cas o\`u ils ne le sont pas}:
\bea
\int_{\tilde{\CC}} \Omega &=& \int_{{\CC}}dx \int_{0}^{y(x)} {r dr \over \sqrt{y(x)^2-r^2}}\cr
&=& \int_{\CC} y(x)dx.\cr
\eea
Nous avons ainsi montr\'e que, en int\'egrant la partie "triviale" de la g\'eom\'etrie de la vari\'et\'e de Calabi-Yau,
on peut se ramener \`a \'etudier la projection sur une courbe hyperelliptique.

Jusqu'ici, nous avions simplement vu une tr\`es forte similitude entre les propri\'et\'es de transformation des \'energies libres
sous les d\'eformations de la Calabi-Yau et les propri\'et\'es de variations des invariants alg\'ebriques associ\'es aux
surfaces $\overline{\Sigma}_n$ sans pour autant aller plus loin.
Dans ce cas pr\'ecis, on peut aller plus loin. On  reconnait en effet dans l'\'equation de $\overline{\Sigma}_n$, la forme
de l'\'equation de la courbe spectrale classique d'un mod\`ele \`a une matrice avec un potentiel $W(x)$.
Par ailleurs, Dijgraaf et Vafa ont conjectur\'e dans \cite{DV} que les \'energies libres holomorphes $F^{(g)}(t)$ de cette th\'eorie
sont donn\'ees par les termes du d\'eveloppement topologique du mod\`{e}le \`a une matrice hermitienne. Cette conjecture
a \'et\'e v\'erifi\'ee aux ordres 0 \cite{DV}, 1 \cite{DST,KMT} et 2 \cite{HK}.

\subsection{Miroirs de g\'eom\'etries toriques.}
Il existe une autre classe de Calabi-Yau int\'eressante. Celle-ci est li\'ee \`{a} la {\bf sym\'etrie miroir} \cite{Hori} qui est une sym\'etrie tr\`es importante puisqu'elle
permet de relier les mod\`{e}les de type A et de type B, i.e. la th\'eorie des invariants de Gromov-Witten et la th\'eorie des
d\'eformations de vari\'et\'es de Calabi-Yau. Elle dit qu'\'etant donn\'ee une Calabi-Yau $\widetilde{X}$, il existe
une autre Calabi-Yau $X$ telle que le mod\`ele A sur $\widetilde{X}$ et le mod\`ele B sur $X$ co\"{i}ncident:
\beq
F_A^{(g)}(t;\widetilde{X}) = F_B^{(g)}(t, X).
\eeq
Dans le cas o\`{u} $\widetilde{X}$ a une g\'eom\'etrie torique \cite{Bouchard}, son image par la sym\'etrie miroir est caract\'eris\'ee par
une surface de Riemann $\overline{\Sigma}_n$ d\'efinie par une courbe
alg\'ebrique du type\footnote{En fait, il faut effectuer une transformation exponentielle sur les variables pour obtenir ce
r\'esultat.}:
\beq\label{eqtor}
y^2 = M^2(x) \prod_{i=1}^l (x-x_i)
\eeq
o\`u $M(x)$ est une fonction transcendentale.
De r\'ecents travaux \cite{MMtor}, semblent indiquer que les amplitudes holomorphes des cordes ouvertes et ferm\'ees de telles mod\`eles de type B
sont \'egalement donn\'ees par les \'energies libres et fonctions de corr\'elations associ\'ees \`a la courbe \eq{eqtor}.

\section{Th\'eorie de Kodaira-Spencer et anomalie holomorphe.}

Nous montrons dans cette partie que, pour un choix\footnote{Dans toute cette partie nous allons utiliser une notation vectorielle
pour d\'ecrire les fractions de remplissage $\epsilon$ et les cycles $\CA$ et $\CB$ sous forme de vecteurs de taille $\CG$. Il sera ainsi plus ais\'e de
manipuler leur lien avec les matrices de taille $\CG \times \CG$, $\tau$ et $\kappa$ sans avoir \`a faire appelle \`a de fastidieuses
sommations sur des indices.} particulier de
\beq
\kappa = - (\tau - \overline{\tau})^{-1},
\eeq
les invariants $F^{(g)}$ satisfont les \'equations
d'anomalie holomorphe. Rappelons que pour ce choix, ce sont des invariants modulaires et ils poss\`edent ainsi une premi\`ere
propri\'et\'e n\'ecessaire pour les identifier aux fonction de partitions de la th\'eorie de Kodaira-Spencer.

Pour ce choix de $\kappa$, les fonctions de corr\'elation et \'energies libres ne sont plus des fonctions holomorphes des
modules. En effet, par l'interm\'ediaire de $\kappa$, ils ont maintenant une partie non holomorphe. Nous nous int\'eressons donc
\`a cette nouvelle d\'ependance en \'etudiant particuli\`ement leurs variations par rapport aux $\overline{\epsilon}_i$.
Elles sont obtenues en utilisant les variations par rapport \`{a} $\kappa$ calcul\'ees dans le chapitre 4:
\beq
{\partial W_k^{(g)} \over \partial \overline{\epsilon}} = {\partial \kappa \over \partial \overline{\epsilon}}
{\partial W_k^{(g)} \over \partial \kappa}
\quad , \quad
{\partial \kappa \over \partial \overline{\epsilon}} = -\kappa {\partial \overline\tau\over \partial\overline\epsilon} \kappa
\quad , \quad
{\partial^2 F^{(0)}\over \partial \epsilon^2} = 2\ri \pi \tau
\eeq
d'o\`u
\beq
{\partial W_k^{(g)} \over \partial \overline{\epsilon}} =
-{1\over 2\ri \pi}\, \kappa {\partial^3 {\overline{F}}^{(0)} \over \partial {\overline\epsilon}^3} \kappa
{\partial W_k^{(g)} \over \partial \kappa}.
\eeq
Ainsi, en utilisant le th\'eor\`eme \ref{thdWdkappa} du chapitre 4:
\bea
{\partial W_k^{(g)} \over \partial \overline{\epsilon}} &=&
-{1\over (2\ri \pi)^2}\, \kappa {\partial^3 {\overline{F}}^{(0)} \over \partial {\overline\epsilon}^3} \kappa \,\,
 {1\over 2}\,\oint_{r\in\bcycle_j}\oint_{s\in\bcycle_i} \Big( W^{(g-1)}_{k+2}(p_K,r,s) \cr
& & + \,\sum_h \sum_{L\subset K}  W^{(h)}_{|L|+1}(p_L,r) W^{(g-h)}_{k-|L|+1}(p_{K/L},s) \Big).
\eea

Or, avec le choix $\kappa=1/(\overline\tau-\tau)$, on a la propri\'et\'e:
\beq
\kappa {\partial \tau\over \partial \epsilon}  \kappa  = {\partial \kappa\over \partial \epsilon},
\eeq
ce qui implique que l'op\'erateur diff\'erentiel $D$\footnote{Attention \`a ne pas confondre cet op\'erateur avec la d\'eriv\'ee
covariante introduite dans cette partie.} introduit dans \eq{devcov} du chapitre 4 se r\'eduit \`a la
d\'eriv\'ee usuelle:
\beq
D_{\epsilon_I} = \partial/\partial \epsilon_I.
\eeq
On peut alors calculer son action sur les fonctions de corr\'elations:
\beq
{\partial W_k^{(g)}\over \partial\epsilon} = - \oint_{\bcycle}  W_{k+1}^{(g)} = - \oint_{\underline\bcycle}  W_{k+1}^{(g)} +\tau \oint_{\underline\acycle}  W_{k+1}^{(g)},
\eeq
et, en d\'erivant une seconde fois, on trouve:
\beq
{\partial^2 W_k^{(g)}\over \partial\epsilon^2}
+ {\partial \tau\over \partial\epsilon} \kappa {\partial W_k^{(g)}\over \partial\epsilon} = \oint_{\bcycle} \oint_{\bcycle}  W_{k+2}^{(g)}.
\eeq

On obtient finalement l'\'equation:
\bea
{\partial W_k^{(g)} \over \partial \overline{\epsilon}} &=&
-{1\over (2 \ri\pi)^2}\, \kappa {\partial^3 {\overline{F}}^{(0)} \over \partial {\overline\epsilon}^3} \kappa \,\,
 {1\over 2}\,\Big( {\partial^2 W_k^{(g-1)}\over \partial\epsilon^2} + {\partial \tau\over \partial\epsilon} \kappa {\partial W_k^{(g-1)}\over \partial\epsilon}  \cr
& & + \,\sum_h \sum_{L\subset K}  {\partial W_l^{(h)}\over \partial\epsilon}\,{\partial W_{k-l}^{(g-h)}\over \partial\epsilon} \Big).
\eea
Pour la comparer \`{a} l'\'equation d'anomalie holomorphe, utilisons les notations de th\'eorie des cordes:
\beq
t^I =(2\ri \pi)^{1\over 2} \epsilon^I.
\eeq
et la derni\`ere \'equation s'\'ecrit:
\bea
{\partial W_k^{(g)} \over \partial \overline{t}} &=&
-\, \kappa {\partial^3 {\overline{F}}^{(0)} \over \partial {\overline{t}}^3} \kappa \,\,
 {1\over 2}\,\Big( {\partial^2 W_k^{(g-1)}\over \partial t^2} + {\partial \tau\over \partial t} \kappa {\partial W_k^{(g-1)}\over \partial t}  \cr
& & + \,\sum_h \sum_{L\subset K}  {\partial W_l^{(h)}\over \partial t}\,{\partial W_{k-l}^{(g-h)}\over \partial t} \Big).
\eea
Pour les \'energies libres, on prend le cas o\`u $k=0$ et on obtient:
\bea
{\partial F^{(g)} \over \partial \overline{t}} &=&
-\, \kappa {\partial^3 {\overline{F}}^{(0)} \over \partial {\overline{t}}^3} \kappa \,\,
 {1\over 2}\,\Big( {\partial^2 F^{(g-1)}\over \partial t^2} + {\partial \tau\over \partial t} \kappa {\partial F^{(g-1)}\over \partial t}  + \,\sum_{h=1}^{g-1}  {\partial F^{(h)}\over \partial t}\,{\partial F^{(g-h)}\over \partial t} \Big)
\eea
qui n'est rien d'autre que l'\'equation d'anomalie holomorphe pour les \'energies libres \eq{eqanomalie} puisque
\beq
-\kappa^{IM} {\partial^3 {\overline{F}}^{(0)} \over \partial {\overline{t}}^{\overline M}\partial {\overline{t}}^{\overline N} \partial {\overline{t}}^{\overline K}  } \kappa^{JN}=C^{IJ}_{\overline K}, \qquad {\partial \tau_{JM}\over \partial t^I} \kappa^{KM}=-\Gamma^K_{IJ}.
\eeq
Remarquons que l'on a \'egalement obtenu une \'equation similaire pour les fonctions de corr\'elation elles-m\^emes:
\beq
\label{openstring}
\partial_{\overline K} W_k^{(g)} =
 {1\over 2}C_{\overline K}^{IJ} \Big( D_I D_J W_k^{(g-1)}+ \sum_h \sum_{L\subset K}  D_I W_l^{(h)} D_J W_{k-l}^{(g-h)} \Big)
\eeq
qui correspondent dans le cadre de la th\'eorie des cordes topologiques aux amplitudes de cordes ouvertes par opposition
aux cordes ferm\'ees d\'ecrites par les \'energies libres (cf \cite{MMtor} pour des r\'esutats explicites dans le cas de vari\'et\'es miroires
de Calabi-Yau toriques). On peut alors voir cette \'equation comme une \'equation d'anomalie holomorphe pour
les cordes ouvertes. Ce type d'\'equations a d\'ej\`a \'et\'e discut\'e dans un cadre g\'en\'eral \cite{BCOV}
o\`u elles sont obtenues en \'etudiant les d\'egenerescences possibles des surfaces de Riemann avec bords: c'est-\`{a}-dire que
l'on regarde ce qu'il se passe lorsque la surface est "pinc\'ee" suivant l'un des cycles non triviaux qui est alors ramen\'e
\`a un point. Partant d'une surface $\Sigma_{g,k}$ de genre $g$ \`{a} $k$ bords, les deux termes du membre de droite de
\eq{openstring} correspondent d'une part au cas o\`{u} ce pincement coupe la surface en deux surfaces disconnect\'ees:
\beq
\begin{array}{l} {\includegraphics[width=13cm]{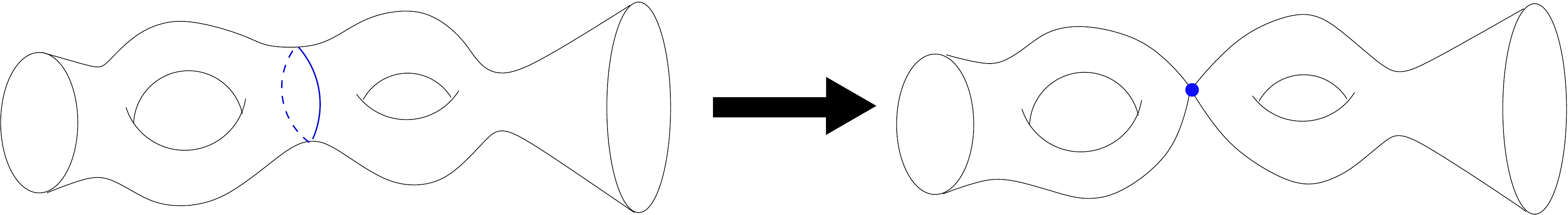}}
\end{array}
\eeq
et au cas o\`{u} l'on a pinc\'e un noeud interne  \`{a} la surface:
\beq
\begin{array}{l} {\includegraphics[width=13cm]{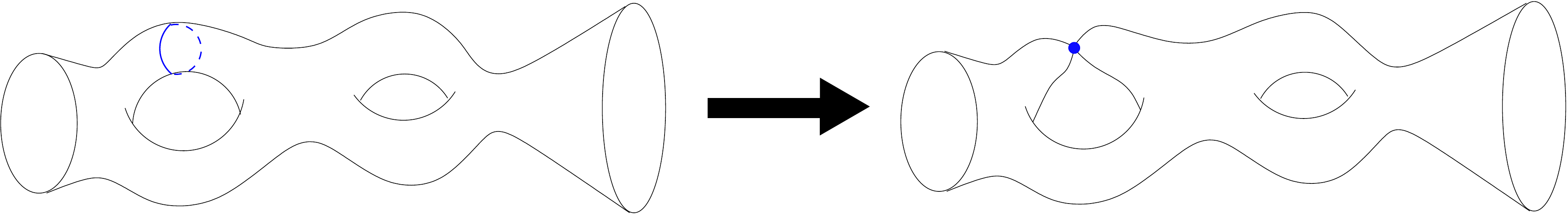}}
\end{array}
\eeq

On peut \'egalement montrer que les amplitudes $C^{(g)}_{I_1 \cdots I_k}$ sont donn\'ees par
\beq
C^{(g)}_{I_1 \cdots I_k}={(-1)^k \over (2\ri \pi)^{k \over 2}} \oint_{\bcycle_{I_1}}  \cdots \oint_{\bcycle_{I_k}} \, W^{(g)}_{k}(p_K)
\eeq
puisque la d\'eriv\'ee covariante $ D_{I} = \partial_{I} - \Gamma_I$ commute avec les int\'egrales sur les cycles $\CB$
lorsque l'on agit sur une diff\'erentielle dont les int\'egrales sur les cycles $\CA$ sont nulles.

Ainsi, l'int\'egration de toutes les variables dans \eq{openstring} sur les cycles $\CB$ redonne l'\'equation d'anomalie holomorphe pour les
fonctions de corr\'elation dans le cas local. Cela montre que l'\'equation d'anomalie holomorphe n'implique que le secteur
des cordes ferm\'ees dans les \'etats interm\'ediaires.

\section{Conclusion du chapitre.}

Nous avons montr\'e dans ce chapitre que les \'energies libres et fonctions de corr\'elation calcul\'ees sur une courbe alg\'ebrique
$\CE$ satisfont toujours les \'equations d'anomalie holomorphe \eq{eqanomalie} pour le choix de $\kappa = - (\tau - \overline{\tau})^{-1}$.
Cela montre que leur partie non holomorphe coincide bien avec la partie non-holomorphe d'un mod\`ele de type B d\'efini
sur une Calabi-Yau locale caract\'eris\'ee par la surface de Riemann $\CE$. Par ailleurs, la limite holomorphe $\overline{t} \to \infty$
corespondant \`a $\kappa = 0$ fixe sans ambiguit\'e la partie holomorphe des invariants alg\'ebriques. Entre autres, si la courbe
alg\'ebrique consid\'er\'ee est la courbe spectrale d'un mod\`ele de matrices, cette limite correspond aux \'energies libres et
fonctions de corr\'elations dudit mod\`ele.

Ainsi, la d\'emonstration de la conjecture de Dijgraaf-Vafa est ramen\'ee \`a montrer que la partie holomorphe de l'\'energie
libre des mod\`eles locaux de type B sont bien les invariants alg\'ebriques d\'efinis dans le chapitre 4 pour $\kappa = 0$.


\chapter{Conclusion.}

\section{R\'esultats principaux: du d\'eveloppement topologique des mod\`eles de matrices \`a la th\'eorie des cordes topologiques.}

\subsection{Calcul d'int\'egrales \`a deux matrices formelles.}

Dans ce m\'emoire, nous avons tout d'abord montr\'e comment d\'efinir proprement un mod\`ele \`a deux matrices formel
comme fonction g\'en\'eratrice des surfaces discr\'etis\'ees portant une structure de spin de type Ising. Nous avons
particuli\`erement insist\'e sur le fait que ce mod\`ele ne coinc\"{i}de pas, en g\'en\'eral, avec le d\'eveloppement d'une quelconque int\'egrale
convergente quand la taille des matrices consid\'er\'ees tend vers l'infini.

Nous avons \'egalement montr\'e commment associer une courbe alg\'ebrique \`a diff\'erents mod\`eles de matrices formels
ne d\'ependant que de l'ordre dominant du d\'eveloppement topologique du mod\`ele consid\'er\'e, c'est-\`a-dire de ce
que l'on appelle habituellement limite classique ou limite planaire. Nous avons alors d\'ecrit une proc\'edure unique
pour cacluler tous les termes du d\'eveloppement topologique des observables du mod\`ele consid\'er\'e.

On peut r\'esumer cette proc\'edure par le graphique suivant:

\begin{center}

\begin{tabular}{|c|}
\hline
  Mod\`ele de matrices formel \\ \hline
\end{tabular}

$\downarrow$

\begin{tabular}{|c|}
\hline
Limite planaire: Courbe spectrale classique $\CE$ \\ \hline
\end{tabular}

$\downarrow$

\begin{tabular}{|c|}
\hline
 Invariants alg\'ebriques sur $\CE$\\
= D\'eveloppement topologique des fonctions de corr\'elation \\\hline
\end{tabular}

\end{center}

Nous avons donc montr\'e que ces diff\'erentes int\'egrales formelles ne sont qu'un seul et m\^eme objet en tant
que fonctionnelle d'une courbe alg\'ebrique et propos\'e une m\'ethode simple et efficace pour les calculer.

Le fait qu'une telle proc\'edure existe peut peut-\^etre s'expliquer par la combinatoire. Un mod\`ele
de matrice formel correspond \`a compter des cartes. Ces cartes peuvent \^etre vues comme recollement de disques entre eux.
 Pour construire de telles carte il faut deux \'el\'ements:
le poids d'un disque (obtenu par $W_1^{(0)}$ et donc $ydx$), et le poids d'un recollement, encod\'e dans l'\'equation
de la courbe spectrale classique. De ce point de vue, il n'est pas surprenant que la courbe spectrale classique poss\`ede toute
l'information n\'ecessaire au calcul de tout le d\'eveloppement topologique des observables du mod\`ele et que
la proc\'edure \`a mettre en oeuvre pour passer de cette courbe aux fonctions de corr\'elation ne d\'epende pas du mod\`ele.

\subsection{Invariants et d\'eformations de courbes alg\'ebriques.}

Nous avons \'egalement montr\'e comment g\'en\'eraliser cette proc\'edure au del\`a des mod\`ele de matrices en \'etendant
cette construction \`a n'importe quelle courbe alg\'ebrique $\CE$. Nous avons alors pu montrer que
les nombres $F^{(g)}$ et les formes diff\'erentielles $W_k^{(g)}$ ainsi obtenues ont un comportement facile \`a d\'ecrire sous les d\'eformations
de $\CE$:
\begin{itemize}

\item {\bf Invariance symplectique:} les \'energies libre $F^{(g)}$ sont invariantes sous de nombreuses transformations conservant
la forme $dx \wedge dy$;

\item {\bf Transformations modulaires:} les transformations modulaires de la courbe peuvent \^etre totalement absorb\'ees
dans la normalisation de la fonction \`a deux points $W_{2,0}^{(0)}$ par l'ajout d'un nouveau param\`etre:
une matrice sym\'etrique $\kappa$;

\item {\bf Variation de $ydx$:} toute variation de la forme diff\'erentielle $ydx$ du type:
\beq
\delta ydx(p):= \int_{\CC} B(\xi,p) \L(\xi)
\eeq
se traduit par une variation des fonctions de corr\'elation:
\beq
\delta W_{k}^{(g)}({\bf p_K}) = \int_{\CC} W_{k+1}^{(g)}(\xi,{\bf p_K}) \L(\xi).
\eeq

\end{itemize}

Ces invariants sont donc fortement li\'es \`a la th\'eorie des d\'eformations des courbes alg\'ebriques. Entre autres choses,
il est tr\`es int\'eressant de noter que ces r\'esultats permettent d'interpr\'eter la fonction de corr\'elation \`a $k$ points
 de genre $g$, $W_{k,0}^{(g)}(p_1, \dots ,p_k)$, calcul\'ee sur une courbe $\CE$ comme l'\'energie libre de genre $g$, $F(\widetilde{\CE})^{(g)}$,
calcul\'ee sur une courbe $\widetilde{\CE}$ obtenue \`a partir de $\CE$ en ajoutant des p\^oles en $k$ points marqu\'es $p_1, \dots, p_k$.
Cette remarque fait \'echo avec l'interpr\'etation combinatoire des fonctions de corr\'elations dans le cadre des mod\`eles
de matrices. En effet, ces propri\'et\'es semblent faire un lien entre la courbe spectrale du mod\`ele de matrices et les
surfaces discr\'etis\'ees g\'en\'er\'ees par les fonctions de corr\'elation: ajouter un bord aux surfaces g\'en\'er\'ees
revient \`a ajouter un p\^ole \`a la courbe spectrale.

\subsection{Th\'eories conformes et int\'egrabilit\'e.}

Nous avons utilis\'e cette proc\'edure g\'en\'erale pour retrouver des r\'esultats d\'ej\`a connus de mani\`ere simple et rigoureuse.
Nous avons ainsi montr\'e comment les doubles limites d'\'echelles de mod\`eles \`a une et deux matrices co\"{i}ncident avec
des mod\`eles de matrices en champ ext\'erieur (obtenu sans prendre de limite). Nous avons donc montr\'e comment avoir acc\`es
directement aux observables de mod\`eles minimaux de type $(p,q)$ sans avoir besoin de faire un quelconque passage \`a la limite.
Il suffit de consid\'erer la courbe alg\'ebrique $\CE_{(p,q)}$ associ\'ee.

Ceci nous a \'egalement permis de montrer \`a nouveau que les int\'egrales de matrices, et donc nos invariant alg\'ebriques, fournissent une r\'ealisation
explicite des fonctions $\tau$ des hierarchies int\'egrables KP et KdV par une simple comparaison des courbes alg\'ebriques associ\'ees.

Ce lien entre les invariants d\'efinis et l'int\'egrabilit\'e a \'et\'e rendu plus explicite, de mani\`ere g\'en\'erale, par la
construction de fonction, de Baker-Akhiezer associ\'ees \`a une courbe alg\'ebrique quelconque. Ces fonctions permettent en effet
de d\'efinir des \'equations de Hirota et des relations de Sato formelles associ\'ees \`a n'importe quelle courbe
alg\'ebrique.

\subsection{Conjecture de Dijkgraaf-Vafa.}

Nous avons montr\'e comment, en fixant le nouveau param\`etre $\kappa = {1 \over \overline{\tau} - \tau}$, on peut rendre
les $F^{(g)}$ et $W_k^{(g)}$ invariants modulaires en leur faisant perdre leur propri\'et\'e d'holomorphicit\'e. Nous avons
ainsi reproduit le comportement des fonctions de partition de la th\'eorie de Kodaira-Spencer d\'ecrit dans \cite{ABK}.
Nous avons alors pu montrer que les $F^{(g)}$ sont solutions des \'equations d'anomalies holomorphes des cordes ferm\'ees de
cette th\'eorie appuyant ainsi un peu plus la conjecture de Dijkgraaf-Vafa liant th\'eorie des cordes topologique et
mod\`eles de matrices.
Gr\^ace \`a ce r\'esultat, il ne reste plus qu'\`a identifier la partie holomorphe de la fonction de partition de la th\'eorie
de Kodaira-Spencer et les invariants alg\'ebriques correspondants pour $\kappa = 0$ (qui est la valeur \`a prendre pour
pouvoir obtenir les observables de mod\`eles de matrices).

Nous avons \'egalement montr\'e que les $W_k^{(g)}$ satisfont des \'equations d'anomalie du m\^eme type que l'on aimerait
interpr\'eter comme \'equations d'anomalies holomorphes pour des cordes ouvertes cette fois-ci.

\section{Perspectives.}

La construction g\'en\'erale d'invariants d\'ecrite dans cette th\`ese pose beaucoup de questions quant \`a son interpr\'etation
d'une part et appelle des g\'en\'eralisations naturelles d'autre part.

\subsection{Etude d'autres mod\`eles de matrices.}

Deux questions viennent naturellement \`a l'esprit suite \`a cette construction. Etant donn\'ee une courbe alg\'ebrique
quelconque, existe-t-il un mod\`ele de matrice dont les observables sont donn\'ees par les invariants associ\'es \`a la courbe?
Au contraire, \'etant donn\'e un mod\`ele de matrice, dans quelle mesure peut-on utiliser la m\^eme proc\'edure pour
calculer le d\'eveloppement topologique de ses observables?

La r\'eponse \`a la premi\`ere question semble \`a priori n\'egative. Il semble, en effet, difficile de construire un mod\`ele de
matrices dont la courbe alg\'ebrique n'aurait qu'un seul point \`a l'infini (i.e. tel que $ydx$ n'ait qu'un seul p\^ole).
Cependant, la question semble plus ouverte si l'on consid\`ere \'egalement les "limites" de mod\`eles de matrices dans les th\'eories acceptables.
 Se pose alors la question de savoir comment reconstruire l'action de ce mod\`ele \`a partir de la courbe
alg\'ebrique. Une partie de la r\'eponse \`a cette question a d\'ej\`a \'et\'e donn\'ee dans cette th\`ese. En effet, nous
avons montr\'e comment associer un "potentiel" $V_{i}$ \`a chaque p\^ole $\alpha_i$ de $ydx$, faisant ainsi un premier pas vers
la construction d'une action. Cependant, il reste beaucoup \`a \'etudier pour pouvoir r\'epondre \`a ce probl\`eme.

D'un autre c\^ot\'e, cette proc\'edure semble tr\`es prometteuse pour l'\'etude d'autres mod\`eles de matrices. Les
\'etudes d\'ej\`a men\'ees sur le mod\`ele de matrices coupl\'ees en cha\^ine ouverte \cite{eynmultimat} semblent, en effet,
indiquer que les fonctions de corr\'elation prennent la m\^eme forme. On peut alors se demander \`a quels autres types de mod\`eles
on peut \'etendre cette construction. En premier lieu, il serait tr\`es int\'eressant d'\'etudier le mod\`ele $O(n)$ \cite{ID}
pour ses applications ainsi que pour le degr\'e de complexit\'e suppl\'ementaire qu'il semble faire appara\^itre. Il est
pour le moment impossible de dire si notre proc\'edure pourra se g\'en\'eraliser \`a ce cadre l\`a sans une r\'esolution explicite du
mod\`ele. On voit qu'il serat tr\`es utile d'avoir un crit\`ere permettant de d\'eterminer si l'on peut calculer une int\'egrale
de matrice formelle donn\'ee par cette m\'ethode.

Enfin, nous n'avons \'etudi\'e ici que des mod\`eles de matrices hermitiennes. Peut-on trouver une proc\'edure \'equivalente
pour des ensembles de matrices pr\'esentant une sym\'etrie diff\'erente? Existe-t-il une telle structure sous-jacente
permettant de calculer des int\'egrales formelles sur l'ensemble des matrices r\'eelles sym\'etriques ou quaternioniques?
Des premiers pas ont \'et\'e faits dans cette direction par Chekhov et Eynard qui ont perturb\'e le mod\`ele \`a une matrice
hermitienne dans \cite{ChEynbeta} pour \'etudier la combinatoire de surfaces discr\'etis\'ees pas forc\'ement orientables.

\subsection{Lien avec l'int\'egrale de matrice convergente.}

Comme nous l'avons montr\'e dans ce m\'emoire, cette proc\'edure donne acc\`es au calcul d'int\'egrales formelles et non
\`a celui d'int\'egrales convergentes sur l'ensemble des matrices hermitiennes. Il serait cependant int\'eressant de faire le
lien entre ces r\'esultats et les int\'egrales convergentes. Dans le cadre du mod\`ele \`a une matrice, Bonnet, David et Eynard \cite{BDE}
avaient en effet pu faire le lien explicite entre int\'egrale formelle et int\'egrale convergente en sommant sur les instantons.
Ils ont en quelque sorte montr\'e comment obtenir un \'etat physique de moindre \'energie comme superposition de
tous les \'etats possibles.

Il serait donc int\'eressant de faire de m\^eme pour le mod\`ele \`a deux matrices et dans le cadre g\'en\'eral
d\'evelopp\'e ici. On aurait ainsi acc\`es, par exemple, aux asymptotiques des polyn\^omes biorthogonaux \'etudi\'es dans \cite{Eynhab}
ainsi qu'aux termes correctifs quand la taille des matrices int\'egr\'ees tend vers l'infini.

\subsection{Lien avec un syst\`eme int\'egrable quantique.}

Ce lien avec l'int\'egrale convergente semble \'egalement \^etre indispensable \`a une meilleure compr\'ehension du syst\`eme
int\'egrable cach\'e derri\`ere cette construction. En effet, si, pour certaines courbes, on retrouve des r\'eductions de
hierarchies int\'egrables et si nous avons \'et\'e capable de construire des s\'eries formelles dont le comportement rappelle
les \'el\'ements de bases des syst\`emes int\'egrables classiques, la structure int\'egrable sous-jacente n'est pas comprise
pour le moment. Hors, dans le cadre des mod\`eles de matrices, on conna\^it tr\`es bien le syst\`eme int\'egrable, et ce quelque soit
la taille $N$ de la matrice. Ce syst\`eme est en g\'en\'eral quantique associ\'e \`a une \'equation des cordes du type
\beq
[P,Q] = {1 \over N}
\eeq
qui devient classique dans la limite des grandes matrices $N\to \infty$. On s'attend donc \`a avoir d\'ecrit le d\'eveloppement semi-classique
d'un tel syst\`eme par notre proc\'edure et il serait int\'eressant de pouvoir reconstruire le syst\`eme quantique
total associ\'e dans ce cadre plus g\'en\'eral. Cependant, il faudra d'abord \^etre capable de reconstruire l'\'equivalent
d'une int\'egrale convergente puisque c'est cet objet qui est vraiment fonction $\tau$ du syst\`eme int\'egrable.
Pour ce faire, la nouvelle variable $\kappa$ semble avoir un r\^ole tr\`es important \`a jouer dans une proc\'edure consistant
\`a resommer sur les fractions de remplissage.

\subsection{Interpr\'etation des invariants alg\'ebriques: calcul du volume de l'espace des modules de surfaces.}

Le plus gros probl\`eme soulev\'e par cette construction consiste \`a comprendre ce que l'on a calcul\'e dans un cadre
g\'en\'eral. Quels sont donc ces objets qui sont \`a la fois fonctions g\'en\'eratrices de surfaces discr\'etis\'ees, fonctions de partitions
de th\'eorie des cordes topologique et fonction $\tau$ d'un syst\`eme int\'egrable? Si il \'etait observ\'e depuis longtemps
que ces objets ont des points communs, il est surprenant qu'ils ne soient en fait qu'un seul et m\^eme objet param\'etr\'e
par une courbe alg\'ebrique quelconque.

Cette unification des diff\'erents mod\`eles de matrices avait d\'ej\`a \'et\'e pr\'essentie dans les travaux de Alexandrov, Mironov et Morozov
\cite{Mth1,Mth2} dans l'optique de d\'efinir une M-th\'eorie matricielle.

Il me semble cependant qu'il existe un ingr\'edient, plus simple, commun aux dif- f\'erentes fonctions retrouv\'ees par cette proc\'edure
faisant en plus le lien entre les inter\^ets des math\'ematiciens et des physiciens. En effet, en regardant de plus pr\`es
les exemples de fonctions $F^{(g)}$ retrouv\'ees par cette m\'ethode, on peut voir qu'elles correspondent toutes
\`a int\'egrer une certaine fonction (ou mesure) sur l'espace des modules de surfaces. En  effet, reprenons quelques exemples:
\begin{itemize}

\item Mod\`eles \`a une et deux matrices: $F^{(g)}$ est la fonction g\'en\'eratrice des surfaces discr\'etis\'ees de
genre $g$. On peut donc voir les \'energies libres comme une int\'egrale sur l'espace des modules des surfaces avec une mesure
donnant un poids non nul \`a un ensemble discret de surfaces seulement. En reprenant les notations introduites dans le chapitre
2:
\beq
F^{(g)} = \int_{\Sigma_{g,0}} \chi_{\CS_{g,0}}(\Sigma_{g,0}) \CW(\Sigma_{g,0})
\eeq
o\`u l'int\'egrale porte sur toutes les surfaces connexes ferm\'ees de genre $g$, $\chi_{\CS_{g,0}}$ est la fonction caract\'eristique
de l'ensemble des surfaces discr\'etis\'ees $\CS_{g,0}$ d\'efini par def.1.3 du chapitre 2 et $\CW$ est le poids associ\'e.

\item Int\'egrale de Kontsevich: l'introduction de ce mod\`ele particulier de matrice en champ ext\'erieur a \'et\'e
originellement motiv\'e par le calcul des nombres d'intersection de surfaces de Riemann et donc du volume
de l'espace des modules des surfaces de Riemann par rapport \`a la mesure de Weyl-Petersson. Les fonctions de corr\'elation
sont donn\'ees dans ce cas par:
\beq
W_{k,0}^{(g)} = \int_{\CM_{g,n}} d\omega_{WP}
\eeq
o\`u $\omega_{WP}$ est la mesure de Weil-Petersson sur l'espace des modules\footnote{Une approche g\'eom\'etrique du probl\`eme
a permis \`a Mirzakhani \cite{Mirz1,Mirz2} de montrer que ces volumes satisfont une relation de r\'ecurrence. Nous avons montr\'e
dans \EOvolume que cette relation de r\'ecurrence n'est rien d'autre que l'equation \ref{defWkgrecursive} du chapitre 4
d\'efinissant les fonctions de corr\'elations pour la courbe $y = sin 2 \pi \sqrt{x}$.}.

\item Mod\`ele minimal $(p,q)$: la fonction de partition est une int\'egrale sur l'espace des modules des surfaces de Riemann
avec l'action d'Einstein-Polyakov:
\beq
F_g = \int_{\CM_g} e^{-i {\CS_p \over \hbar}}.
\eeq

\item Th\'eorie des cordes topologique de type B: la fonction de partition est \'egalement une int\'egrale sur l'espace des
modules des surfaces de genre $g$ de la forme:
\beq
F_g := \int_{\CM_g} d\tau \int {\cal D} \phi e^{-\CS_{\CN = 2}}
\eeq
o\`u $\CS_{\CN = 2}$ est l'action du mod\`ele sigma d\'efinie par \eq{actionS} du chapitre 5.

\end{itemize}

Il semblerait donc que les objets $W_{k,0}^{(g)}[\CE]$ que nous avons d\'efinis peuvent \^etre vus comme des int\'egrales
sur l'espace $\CM_{g,n}$ des modules des surfaces de genre $g$ avec $n$ bords \`a l'aide d'une m\'etrique $d\mu[\CE]$
d\'ependant de la courbe alg\'ebrique $\CE$:
\beq
W_{k,0}^{(g)}[\CE] = \int_{\CM_{g,n}} d\mu[\CE].
\eeq
Une telle interpr\'etation aurait de nombreuses applications que ce soit en physique pour effectuer des calculs de th\'eorie
des cordes et comprendre certaines dualit\'es ou en math\'ematiques o\`u elle permettrait de faire plus explicitement le lien
entre hierarchies int\'egrables et volumes de $\CM_{g,n}$ dans la continuit\'e directe des travaux de Kontsevich. On aurait
\'egalement l\`a un outil pour \'etudier pr\'ecis\'ement les propri\'et\'es topologiques des cartes de grandes tailles.

\subsection{Sym\'etrie miroir, th\'eories topologiques de type A et courbe de Seiberg-Witten.}

Dans cette th\`ese nous n'avons \'etudi\'e qu'un type de th\'eorie des cordes topologique: le type B. Or, on sait
que dans de nombreux cas, une th\'eorie de type B a un dual \'equivalent de type A par la sym\'etrie miroir. Il serait tr\`es
int\'eressant de pouvoir identifier l'\'equivalent fermionique de notre construction alg\'ebrique pour les th\'eories de type A, donnant
ainsi acc\`es au calcul des invariants de Gromov-Witten.

Des travaux r\'ecents de Mari\~{n}o \cite{MMtor} ont montr\'e que si l'on applique notre m\'ethode \`a l'image
par la sym\'etrie miroir de th\'eories de type A avec une Calabi-Yau de g\'eom\'etrie torique, on retrouve bien les r\'esultats
attendus par ailleurs. Notons que dans ce cas, le mod\`ele B correspondant ne fait appara\^itre aucune courbe alg\'ebrique mais
une courbe bien plus compliqu\'ee. Ceci semble indiquer que notre m\'ethode de construction d'invariants peut s'\'etendre encore au
del\`a des courbes alg\'ebriques.

Par ailleurs, il serait tr\`es int\'eressant de comprendre le lien profond entre les \'energies libres calcul\'ees sur
les courbes hypperelliptiques apparaissant dans les th\'eories de Seiberg-Witten \cite{SeibWit} et les couplages gravitationnels introduits par Nekrasov \cite{Nekrasov} puisque ces derniers peuvent \^etre d\'eform\'es en des objets non holomorphes satisfaisant les \'equations d'anomalie holomorphe associ\'ees \`a cette courbe.


\chapter{Appendices.}

\section{Mod\`{e}le Gaussien et th\'eor\`{e}me de Wick.} \label{appwick}

Consid\'erons la mesure gaussienne sur l'ensemble $H_N$ des matrices hermitiennes $M$ de taille $N \times N$:
\beq
d\mu = \left( {g \over 2 \pi} \right)^{N^2 \over 2} e^{-{ g \over 2} \Tr M^2} dM,
\eeq
o\`{u} $dM$ est le produit des mesures de Lebesgues des \'el\'ements r\'eels de $M$:
\beq
dM = \prod_i dM_{ii} \prod_{i<j} dRe(M_{ij}) dIm(M_{ij})
\eeq
et le facteur de normalisation $\CZ_{gauss}:= \left({2 \pi \over g}\right)^{N^2 \over 2}$ est obtenu par le calcul de la fonction
de partition:
\beq
\CZ_{gauss}:=\int_{H_N} e^{-{ g \over 2} \Tr M^2} dM.
\eeq

On peut ais\'ement calculer la corr\'elation entre deux \'el\'ements de matrices:
\beq\label{correl2mat}
\left< M_{ab} M_{cd} \right> :={1 \over \CZ_{gauss}} \int_{H_N} M_{ab} M_{cd} e^{-{ g \over 2} \Tr M^2} dM = {\delta_{ad} \delta_{bc} \over g}.
\eeq
Cette quantit\'e tr\`{e}s simple est en fait la seule dont le calcul est n\'ecessaire pour obtenir la valeur moyenne
du produit d'un nombre quelconque d'\'el\'ements de matrices gr\^{a}ce au th\'eor\`{e}me de Wick:

\bt
Th\'eor\`{e}me de Wick:

La valeur moyenne du produit de $2n$ \'el\'ements de matrices est \'egale \`{a} la somme sur tous les appariements
des \'el\'ements deux \`{a} deux du produit des valeurs moyennes des paires d'\'el\'ements correspondantes:
\beq
\left< \prod_{i=1}^{2n} M_{a_i b_i} \right> = \sum_{\bigcup_{j} (k_j,l_j) = \{1,2, \dots, 2n\}} \prod_{j=1}^n
\left< M_{a_{k_j} b_{k_j}} M_{a_{l_j} b_{l_j}} \right>,
\eeq
o\`{u} la somme dans le membre de droite porte sur tous les appariements $(k_j,l_j)$ deux \`{a} deux des \'el\'ements de
$\{1, \dots ,2n\}$.
\et

Dans les faits, les objets int\'eressants sont les corr\'elations de fonctions
invariantes sous l'action du groupe $U(N)$, de mani\`{e}re g\'en\'erique, on aimerait calculer
\beq
\left< \prod_i (\Tr M^i)^{k_i}\right>,
\eeq
o\`{u} la somme porte sur un nombre fini de termes et les $k_i$ sont des nombres entiers arbitraires.
Puisque l'on peut r\'e\'ecrire les diff\'erentes traces comme un produit d'\'el\'ements de matrice:
\beq
\Tr M^k = \sum_{\alpha_i=1}^N  M_{\alpha_1 \alpha_2} M_{\alpha_2 \alpha_3} M_{\alpha_3 \alpha_4} \dots M_{\alpha_k \alpha_1},
\eeq
on peut appliquer le th\'eor\`{e}me de Wick et d\'ecomposer cette fonction de corr\'elation sur la base des
corr\'elations de deux \'el\'ements de matrices.
Par exemple, on obtient:
\beq
\left<\Tr M^2 \right> = \sum_{i,j = 1}^N \left< M_{ij} M_{ji} \right> = {N^2 \over g},
\eeq
ou encore
\bea
\left<\Tr M^4 \right> &=& \sum_{i,j,k,l} \left< M_{ij} M_{jk}M_{kl}M_{li} \right> \cr
&=& \sum_{i,j,k,l} \left( \left< M_{ij} M_{jk} \right> \left< M_{kl} M_{li} \right> + \left< M_{ij} M_{kl} \right> \left< M_{jk} M_{li} \right>
+\left< M_{ij} M_{li} \right> \left< M_{kl} M_{jk} \right> \right) \cr
&=& {1 \over g^2} \sum_{i,j,k,l} \left( \delta_{ik} + \delta_{il} \delta_{jk} \delta_{ij} \delta_{lk} + \delta_{jl} \right) \cr
&=& {1 \over g^2}( N^3 + N + N^3) = {N \over g^2} (1 + 2 N^2).\cr
\eea
De mani\`{e}re g\'en\'erale, chaque terme non nul dans la d\'ecomposition de $\left< \prod_i (\Tr M^i)^{k_i}\right>$
par le th\'eor\`{e}me de Wick apporte la m\^eme contribution $\left({1 \over g}\right)^{\sum_i i k_i \over 2}$ si
$\sum_i i k_i$ est paire  et $0$ sinon. Ainsi, il faut donc compter le nombre de termes non nuls entrant dans ce d\'eveloppement.
Le r\'esultat prend alors la forme:
\beq
\left< \prod_i (\Tr M^i)^{k_i}\right> = \left({1 \over g}\right)^{\sum_i i k_i \over 2} \sum_{j\geq 0} c_j N^j,
\eeq
o\`{u} les $c_j$ sont des nombres entiers qui appellent donc une interpr\'etation combinatoire.

La c\'el\`ebre repr\'esentation de Feynmann de ces fonctions de corr\'elation sous forme de diagramme nous fournit
en effet une interpr\'etation combinatoire de ces nombres $c_j$.
Pour cela, on repr\'esente chaque objet matriciel par un \'el\'ement graphique. Ainsi on repr\'esente
chaque \'el\'ement de matrice $M_{ab}$ comme un ruban issu d'un point et marqu\'e par les deux indices $a$ et $b$:
\beq
M_{ab}:=\begin{array}{l} {\includegraphics[width=2.5cm]{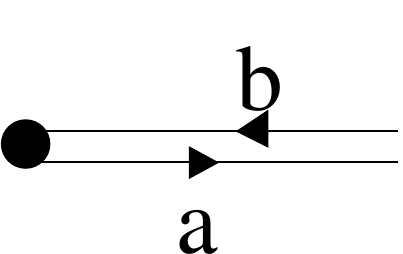}}\end{array}.
\eeq
De m\^eme on repr\'esente $\Tr M^k$ par un
vertex $k$ valent portant des indices de sommation:
\bea
\Tr M^k &=& \sum_{\alpha_i=1}^N  M_{\alpha_1 \alpha_2} M_{\alpha_2 \alpha_3} M_{\alpha_3 \alpha_4} \dots M_{\alpha_k \alpha_1} \cr
&=& \sum_{\alpha_i=1}^N \begin{array}{l} {\includegraphics[width=4cm]{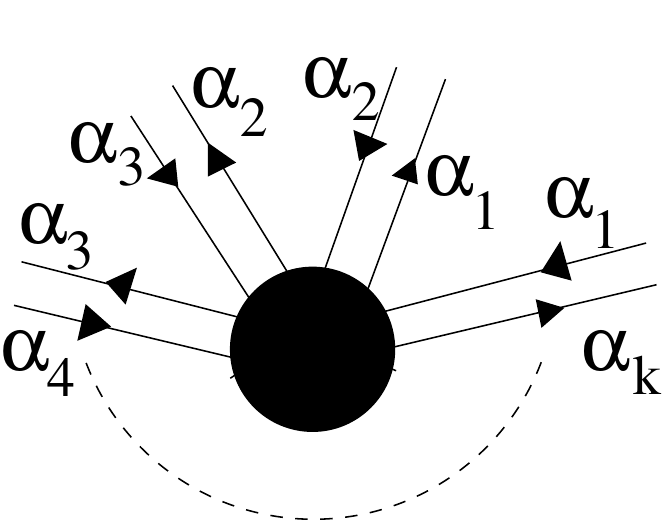}}
\end{array}.\cr
\eea

Sous cette repr\'esentation, l'expression de la corr\'elation de deux \'el\'ements de matrices \eq{correl2mat} s'\'ecrit comme une ar\`{e}te \'epaisse:
\bea
\left< M_{ab} M_{cd} \right> &=& {\delta_{ad} \delta_{bc} \over g} \cr
&=& {1 \over g} \begin{array}{l} {\includegraphics[width=4cm]{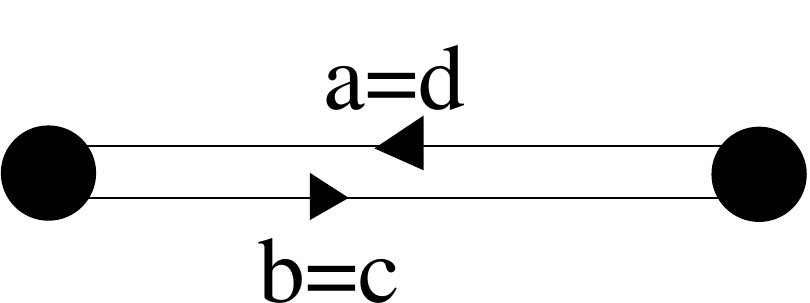}}
\end{array} \cr
\eea
c'est-\`{a}-dire que prendre la valeur moyenne de deux \'el\'ements de matrices correspond \`{a} recoller les rubans les
repr\'esentant et de compter l'ar\^{e}te ainsi form\'ee avec un poids
${1 \over g}$ si les indices associ\'es aux bords sont les m\^{e}mes et un poids nul sinon.

Le th\'eor\`{e}me de Wick nous dit alors que le nombre de termes non nuls dans le d\'eveloppement de $\left< \prod_i (\Tr M^i)^{k_i}\right>$
est \'egal au nombre de diagramme form\'e de ${\sum_i i k_i \over 2}$ ar\^{e}tes \'epaisses li\'ees par $\sum_i k_i$ vertex
dont $k_i$ ont une  valence $i$. A cela, il faut ajouter un facteur pour chaque diagramme venant du fait qu'il doit \^etre
\'etiquet\'e de toutes les mani\`{e}res possibles selon les prescriptions du th\'eor\`{e}me de Wick. Deux facteurs entrent en compte:
\begin{itemize}
\item Deux \'etiquetages peuvent \^{e}tre \'equivalents dans le sens o\`{u} l'un peut \^{e}tre r\'eobtenu \`{a} partir de
l'autre par un simple changement du nom des \'etiquettes muettes\footnote{Par \'etiquettes muettes, j'entends les indices sur lesquels
la sommation de $1$ \`{a} $N$ est efectu\'ee}. Ce facteur est le nombre d'automorphismes du graphe, not\'e \#Aut.

\item L'op\'erateur de trace induit une sommation sur les indices de $1$ \`{a} $N$. Les symboles de Kronecker apparaissant
dans th\'eor\`{e}me de Wick r\'eduisent certaines de ces sommes \`{a} un seul \'el\'ements. Il est facile de se convaincre
que le nombre de sommes restant \`{a} effectuer apr\`{e}s ces prescriptions est \'egal au nombre de boucles ind\'ependantes\footnote{Le
nombre de boucles ind\'ependantes dans un graphe est le nombre de chemin diff\'erents que l'on parcours en suivant le bord
de tous les rubans.} dans le graphe. Ainsi, on doit lui associer un facteur $N$ \`{a} la puissance le nombre de boucles
du graphe not\'e $\chi$\footnote{Cette notation prend son sens par l'observation de 't Hooft \cite{thooft} selon laquelle ce
nombre est la carct\'eristique d'Euler de la surface la plus simple sur laquelle on peut dessiner le graphe de mani\`{e}re
\`{a} le rendre planaire}.

\end{itemize}

On obtient finalement l'expression pour les fonctions de corr\'elation gaussiennes:
\bt
\beq
\left< \prod_i (\Tr M^i)^{k_i}\right> = \left({1 \over g}\right)^{\sum_i i k_i \over 2} \sum_{G \in \CG_{\{k_i\}_{i=1}^N}} \, \hbox{\#Aut}(G) \; N^{\chi(G)}
\eeq
o\`{u} $\CG_{\{k_i\}_{i=1}^N}$ est 'ensemble des graphes \`{a} ${\sum_i i k_i \over 2}$ ar\^{e}tes et $k_i$ vertex de valence $i$.
\et

Pour notre exemple pr\'ec\'edent, $\left<\Tr M^4 \right>$, on doit composer les graphes \`{a} 2 ar\^{e}tes et un seul vertex
tetravalent:
\bea
\left<\Tr M^4 \right> &=& \sum_{i,j,k,l} \left( \left< M_{ij} M_{jk} \right> \left< M_{kl} M_{li} \right> + \left< M_{ij} M_{kl} \right> \left< M_{jk} M_{li} \right>
+\left< M_{ij} M_{li} \right> \left< M_{kl} M_{jk} \right> \right) \cr
& = & \begin{array}{l} {\includegraphics[width=10cm]{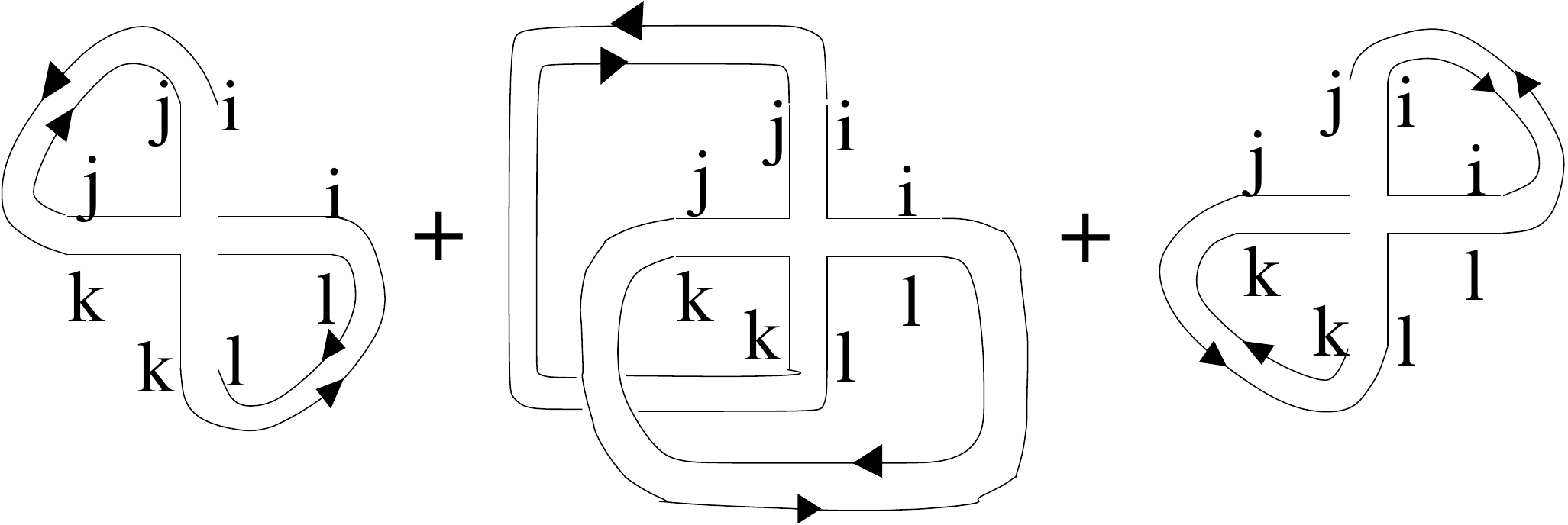}}
\end{array} \cr
&=& \left({1\over g}\right)^2 \left[\begin{array}{l} {\includegraphics[width=5cm]{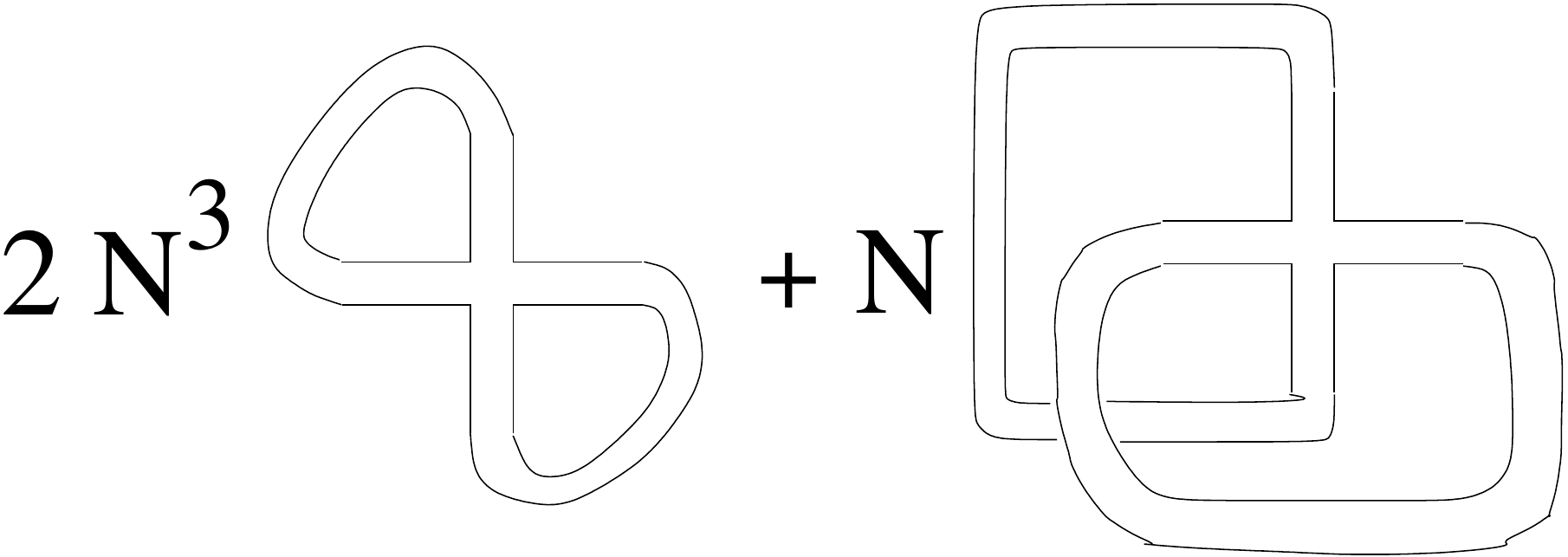}}
\end{array} \right]. \cr
\eea

\br
Cette proc\'edure diagrammatique peut facilement \^{e}tre \'etendue \`{a} un mod\`{e}le gaussien avec plusieurs matrices
en coloriant les vertex et ar\`{e}tes suivant la matrice \`{a} laquelle ils se r\'ef\`{e}rent.
\er

\section{Fonction de partition formelle et surfaces discr\'etis\'ees.}\label{appsurf}

Nous explicitons ici l'expression de la fonction de partition formelle, $\CZ_{form}$, en termes de
surfaces discr\'etis\'ees. $\CG$ d\'esigne ici l'ensemble des cartes d\'efini par la d\'efinition \ref{definterpretsurf} du chapitre 2.

\bd
Soit une carte $G \in CG$. On note:
\begin{itemize}
\item $n_{k,i}(G)$ := nombre $k$-gones de spin + et de couleur $i$ dont le centre est libre\footnote{Un $k$-gone peut \^etre li\'e
\`a un autre par son centre. Son centre est dit libre si il n'est li\'e \`a aucun autre centre de cette fa\c{c}on.};

\item $\tilde{n}_{k,i}(G)$ := nombre $k$-gones de spin - et de couleur $i$ dont le centre est libre;

\item $n_{++,i}(G)$ := nombre de polygones de spins + et de couleur $i$ coll\'es par un bord;

\item $n_{--,i}(G)$ := nombre de polygones de spins - et de couleur $i$ coll\'es par un bord;

\item $n_{+-,i}(G)$ := nombre de polygones de spins diff\'erents et de couleur $i$ coll\'es par un bord;

\item $n_{a,i}(G) = n_{+-,i}(G) + n_{++,i}(G) + n_{--,i}(G)$ := nombre total d'arr\^etes communes \`{a} deux polygones de couleur $i$.

\item $n_{k,i;l,j}(G)$ := nombre de paires de polygones compos\'ees d'un $k$-gone de spin + et couleur $i$
li\'e par son centre \`{a} un $l$-gone de spin + et couleur $j$;

\item $\tilde{n}_{k,i;l,j}(G)$ := nombre de paires de polygones compos\'ees d'un $k$-gone de spin - et couleur $i$
li\'e par son centre \`{a} un $l$-gone de spin - et couleur $j$;

\item $G_i$ := composante de couleur $i$ de la surface obtenue en brisant tous les liens entre centres de polygones;

\item $l_i(G)$ := nombre de sommet de couleur $i$;

\item $\hbox{\#Aut}(G)$ := cardinal du groupes des automorphismes de $G$;

\item $\chi(G)$ := caract\'ristique d'Euler-Poincar\'e de $G$:
\beq
\chi(G) = \sum_i \left[ l_i(G) - n_{a,i}(G) + \sum_k \left(n_{k,i}(G) + \tilde{n}_{k,i}(G)\right) \right].
\eeq

\end{itemize}
\ed

\bd
A chaque surface $G \in \CG$, on associe un poids:
\beq
\begin{array}{rcl}
\CW(G) &:=& {N^{\chi(G)} \over \hbox{\#Aut(G)}} T^{n_T(G)} \prod_{i=1}^{d_1 d_2} \prod_{k=3}^{d_1+1} (t_{k,i}+Th_{k,i})^{n_{k,i}(G)}
\prod_{k=3}^{d_1+1} (\tilde{t}_{k,i}+T\tilde{h}_{k,i})^{\tilde{n}_{k,i}(G)} \cr
&& \prod_{i= 1}^{d_1d_2} h_{1,i}^{n_{1,i}(G)} \tilde{h}_{1,i}^{\tilde{n}_{1,i}(G)} h_{2,i}^{n_{2,i}(G)} \tilde{h}_{2,i}^{\tilde{n}_{2,i}(G)}
\prod_{k>d_1+1} h_{k,i}^{n_{k,i}(G)} \prod_{l>d_2+1} \tilde{h}_{l,i}^{\tilde{n}_{l,i}(G)} \cr
&& \prod_{i=1}^{d_1 d_2} \epsilon_i^{l_i(G)} t_{2,i}^{n_{++,i}(G)} \tilde{t}_{2,i}^{n_{--,i}(G)} (t_{2,i} \tilde{t}_{2,i}-1)^{-n_{v,i}(G)}
\prod_{j>i} \prod_k \prod_l h_{k,i;l,j}^{n_{k,i;l,j}(G)} \tilde{h}_{k,i;l,j}^{\tilde{n}_{k,i;l,j}(G)} \cr
\end{array}
\eeq
o\`{u}
\beq
\begin{array}{rcl}
n_T(G)&:=& \sum_i\left( \sum_k {k \over 2} (n_{k,i}(G)+\tilde{n}_{k,i}(G)) - \sum_{k=3}^{d_1+1} n_{k,i}(G) - \sum_{l=3}^{d_12+1} \tilde{n}_{l,i}(G)\right) \cr
&& + \sum_i \sum_{j>i} \sum_k \sum_l {k+l \over 2} (n_{k,i;l,j}(G) + \tilde{n}_{k,i;l,j}(G)) .\cr
\end{array}
\eeq
\ed

Associer un tel poids \`{a} une carte $G$ correspond \`{a} associer un poids \`{a} chacun des \'el\'ements composant le graphe
comme suit:

\begin{tabular}{|c|c|}
\hline
El\'ement & Poids \\
\hline
\begin{tabular}{c}
$k$-gone de spin + et couleur $i$ \\ dont le centre est libre\\ \end{tabular}&
  $\left\{ \begin{array}{l} {N \over T} t_{k,i} + N h_{k,i} \; \hbox{si} \; k \in [3,d_1 +1]  \cr N h_{k,i} \; \hbox{sinon} \end{array} \right.$\\
\hline\begin{tabular}{c}
$k$-gone de spin - et couleur $i$ \\ dont le centre est libre \\ \end{tabular}&
  $\left\{ \begin{array}{l} {N \over T} \tilde{t}_{k,i} + N h_{k,i} \; \hbox{si} \; k \in [3,d_2 +1]  \cr N h_{k,i} \; \hbox{sinon} \end{array} \right.$\\
\hline
\begin{tabular}{c} ar\`{e}te commune \`{a} deux polygones \\
de spin + et couleur $i$ \end{tabular}& ${1 \over N}{t_{2,i} \over t_{2,i} \tilde{t}_{2,i} -1}$ \\
\hline
\begin{tabular}{c} ar\`{e}te commune \`{a} deux polygones \\
de spin - et couleur $i$ \end{tabular} & ${1 \over N}{\tilde{t}_{2,i} \over t_{2,i} \tilde{t}_{2,i} -1}$ \\
\hline
\begin{tabular}{c}
ar\`{e}te commune \`{a} deux polygones \\
de spin diff\'erents et couleur $i$ \end{tabular}  & ${1 \over N}{1 \over t_{2,i} \tilde{t}_{2,i} -1}$ \\
\hline
\begin{tabular}{c}
centre commun \`{a} un $k$-gone\\ de couleur $i$ et de spin + \\
et un $l$-gone\\ de couleur $j$ et de spin + \\
\end{tabular} & $h_{k,i;l,j}$ \\
\hline
\begin{tabular}{c}
centre commun \`{a} un $k$-gone\\ de couleur $i$ et de spin - \\
et un $l$-gone\\ de couleur $j$ et de spin - \\
\end{tabular} & $h_{k,i;l,j}$ \\
\hline
Sommet de couleur $i$ & $N \epsilon_i$ \\
\hline
\end{tabular}

\vs

La bijection entre les graphes et les surfaces permet d'\'ecrire que:
\bt
La {\bf fonction de partition du mod\`{e}le \`{a} deux matrices hermitiennes formel} est \'egale \`{a}:
\beq\encadremath{
\CZ_{form} := \prod_i e^{-{N^2 \over T} \epsilon_i ( V_1(\xi_i) + V_2(\eta_i) - \xi_i \eta_i)} \prod_{j>i} \left[(x_i-x_j)(y_i-y_j)\right]^{N^2 \epsilon_i \epsilon_j}
\sum_{G\in\CG} \CW(G).}
\eeq
\et

\section{Int\'egrale formelle et \'equations de boucles.} \label{apploop}

Dans cet appendice, nous allons montrer que les fonctions de corr\'elations du mod\`{e}le \`{a} deux matrices satisfont les
\'equations de boucles obtenues avec les r\`{e}gles Split et Merge.

\bd

\ed

\bl
Pour tout
\beq
\begin{array}{l}
G_{k_1,\dots,k_l;m;n}(S_1,S_2, \dots, S_l;x_1 ,\dots, x_m;y_1,\dots,y_n) := \cr
{1 \over x_{1,1}-M}{1 \over y_{1,1}-\tilde{M}} {1 \over x_{1,2}-M}{1 \over y_{1,2}-\td{M}}
\dots {1 \over x_{1,k_1}-M}{1 \over y_{1,k_1}-\td{M}}
 \; \; \prod_{j=1}^m \Tr { 1 \over x_j-M} \prod_{s=1}^n \Tr {1 \over y_s-\td{M}} \cr
\prod_{i=2}^l \Tr \left({1 \over x_{i,1}-M}{1 \over y_{i,1}-\td{M}} {1 \over x_{i,2}-M}{1 \over y_{i,2}-\td{M}}
\dots {1 \over x_{i,k_i}-M}{1 \over y_{i,k_i}-\td{M}}\right), \cr
\end{array}
\eeq
les \'equations
\beq
\begin{array}{rcl}
0&=& \sum_{\alpha \leq \beta}
\int dM_1 \dots dM_d d\td{M}_1 \dots d\td{M}_d \left( {\partial \over \partial Re (M_k)_{\alpha,\beta}} + {\partial \over \partial Im (M_k)_{\alpha,\beta}} \right) \cr
&& \Big[ (\sum_i \Tr \delta V_{1,i}(M_i)+ \delta V_{2,i}(\td{M}_i))^l G_{k_1,\dots,k_l;m;n}(S_1,S_2, \dots, S_l;x_1 ,\dots, x_m;y_1,\dots,y_n) \cr
&& \,\, \prod_{i=1}^{d} \ee{-{N \over T}\left(\Tr {V_1''(\xi_i)\over 2}(M_i-\xi_i\,{\bf 1}_{n_i})^2+{V_2''(\eta_i)\over 2}(\td{M}_i-\eta_i\,{\bf 1}_{n_i})^2 - (M_i-\xi_i\,{\bf 1}_{n_i})(\td{M}_i-\eta_i\,{\bf 1}_{n_i}) \right)} \cr
&& \,\,\,\prod_{i>j} \det(M_i\otimes {\bf 1}_{n_j} - {\bf 1}_{n_i} \otimes M_j)
\,\,\prod_{i>j} \det(\td{M}_i\otimes {\bf 1}_{n_j} - {\bf 1}_{n_i} \otimes \td{M}_j) \Big]\cr
\end{array}
\eeq

\el

\section{Variations de  \texorpdfstring{$F^{(0)}$}{F0} par rapport aux modules.}\label{appF0}

Nous pr\'esentons ici un formulaire des d\'eriv\'ees premi\`{e}res et secondes de $F^{(0)}$ par rapport aux
modules de la courbe alg\'ebrique. Ces r\'esultats avaient d\'ej\`{a} \'et\'e d\'eriv\'es dans le cadre des mod\`{e}les de matrices
de nombreuses fois \cite{Kri, Marco2,MarcoF}. Pour retrouver ces formules, nous prenons $\kappa = 0$.

\subsection{D\'eriv\'ees premi\`{e}res de $F^{(0)}$.}

\beq
{\partial {F}^{(0)}\over \partial t_{k,i}} = \Res_{\alpha_i} z_{\alpha_i}^k \, y dx
\eeq

\beq
{\partial {F}^{(0)} \over \partial t_{\alpha_i,\alpha_j}}:=\left({\partial\over \partial t_{0,i}}-{\partial\over \partial t_{0,j}}\right)  F^{(0)} = \mu_{\alpha_i}-\mu_{\alpha_j}
\eeq

\beq
{\partial F^{(0)}\over \partial \epsilon_i} = - \oint_{\bcycle_i} ydx
\eeq

\bigskip

\subsection{D\'eriv\'ees secondes de $F^{(0)}$.}

\beq
{\partial^2 F^{(0)}\over \partial t_{k,i} \partial t_{l,j} } = (\delta_{i,j}-1)\, \Res_{p\to \alpha_i}\Res_{q\to \alpha_j} z_{\alpha_i}(p)^{k} B(p,q) z_{\alpha_j}(q)^{l}
\eeq
\beq
{\partial^2 F^{(0)}\over \partial t_{k,i} \partial t_{\alpha_j,\alpha_k}} = \Res_{\alpha_i} z_{\alpha_i}^k dS_{\alpha_j,\alpha_k}
\eeq
\beq
{\partial^2 F^{(0)}\over \partial t_{k,j} \partial \epsilon_i } = 2i\pi \Res_{\alpha_j} z_{\alpha_j}^{k} du_i = - \oint_{\bcycle_i} B_{\alpha_j,k}
\eeq
\beq
{\partial^2 F^{(0)}\over \partial \epsilon_i \partial t_{\alpha_j,\alpha_k}} = 2i\pi (u_i(\alpha_k)-u_i(\alpha_j))
\eeq
\beq
{\partial^2 F^{(0)}\over \partial \epsilon_i \partial \epsilon_j} = 0 \;\; (-2i\pi \tau_{ij} \; for \; \underline{F}^{(0)} )
\eeq
\beq
{\partial^2 F^{(0)}\over \partial t_{\alpha_i,\alpha_j}^2} = \ln{(d\zeta_{\alpha_i}(\alpha_i)d\zeta_{\alpha_j}(\alpha_j)\primef(\alpha_i,\alpha_j)^2)}
\eeq
\beq
{\partial^2 F^{(0)}\over \partial t_{\alpha_i,\alpha_j} \partial t_{\alpha_i,\alpha_k}} = \ln{\left(d\zeta_{\alpha_i}(\alpha_i)\primef(\alpha_i,\alpha_j)\primef(\alpha_i,\alpha_k)\over \primef(\alpha_j,\alpha_k)\right)}
\eeq
\beq
{\partial^2 F^{(0)}\over \partial t_{\alpha_i,\alpha_j} \partial t_{\alpha_k,\alpha_l}} = \ln{\left(\primef(\alpha_k,\alpha_j)\primef(\alpha_i,\alpha_l)\over \primef(\alpha_i,\alpha_k)\primef(\alpha_j,\alpha_l)\right)}
\eeq
o\`{u} $\zeta_\alpha = {1 \over z_\alpha}$ est une variable locale au voisinage de $\alpha$.


\end{document}